\documentclass{article}
\newcommand{\longpage}{\enlargethispage{\baselineskip}}

\usepackage{url} 
\usepackage{enumerate, stmaryrd,mathrsfs}
\usepackage{amsmath,amssymb,latexsym}
\usepackage{times} 

\newcounter{zaehler_enumerate_kurz}

\newcommand*{\twodots}{.\,.\,}

\newcommand*{\raus}[1]{ }

\newcommand*{\parno}{\par\noindent}
\newcommand*{\ov}[1]{\overline{#1}}
\newcommand*{\vek}[1]{\ensuremath{\vec{#1}}}

\newcommand*{\bs}[1]{\ensuremath{\boldsymbol{#1}}}

\newcommand*{\und}{\ensuremath{\wedge}}
\newcommand*{\Und}{\ensuremath{\bigwedge}}
\newcommand*{\oder}{\ensuremath{\vee}}
\newcommand*{\Oder}{\ensuremath{\bigvee}}
\newcommand*{\nicht}{\ensuremath{\neg}}

\newcommand*{\gdw}{\ensuremath{\leftrightarrow}}

\renewcommand*{\geq}{\ensuremath{\geqslant}}
\renewcommand*{\leq}{\ensuremath{\leqslant}}

\newcommand*{\set}[1]{\ensuremath{\{ #1 \}}}
\newcommand*{\setc}[2]{\set{#1 \,:\, #2}}

\newcommand*{\bigset}[1]{\ensuremath{ \left\{ #1 \right\} } }
\newcommand*{\bigsetc}[2]{\bigset{#1 \,:\, #2}}

\newcommand*{\interval}{\ensuremath{\textit{int\,}}}
\newcommand*{\cinterval}[1]{\ensuremath{\interval{[} #1 {]}}}
\newcommand*{\ointerval}[1]{\ensuremath{\interval{(} #1 {)}}}
\newcommand*{\cointerval}[1]{\ensuremath{\interval{[} #1 {)}}}
\newcommand*{\ocinterval}[1]{\ensuremath{\interval{(} #1 {]}}}
\newcommand*{\intoc}[1]{\ocinterval{#1}}
\newcommand*{\intco}[1]{\cointerval{#1}}
\newcommand*{\intcc}[1]{\cinterval{#1}}

\newcommand*{\intoo}[1]{\ointerval{#1}}

\newcommand*{\intro}[1]{\intco{#1}}

\newcommand*{\links}{\ensuremath{\scriptscriptstyle {-}}}
\newcommand*{\rechts}{\ensuremath{\scriptscriptstyle {+}}}

\newcommand*{\FktsFont}[1]{\ensuremath{\textup{#1}}}

\newcommand*{\lcm}{\ensuremath{\FktsFont{lcm}}}

\newcommand*{\abgerundet}[1]{\ensuremath{
    \left\lfloor {#1} \right\rfloor }}
\newcommand*{\aufgerundet}[1]{\ensuremath{\left\lceil {#1} \right\rceil }}
\newcommand*{\betrag}[1]{\ensuremath{\left| {#1} \right| }}

\newcommand*{\rep}{\ensuremath{\textit{rep}}}
\newcommand*{\repcan}{\ensuremath{\textit{rep}}}

\newcommand*{\Color}{\ensuremath{\textit{col\,}}}
\newcommand*{\ktype}{\ensuremath{k\textit{-type\,}}}

\newcommand*{\Char}{\ensuremath{\textit{char\,}}}

\newcommand*{\deff}{:=}
\newcommand*{\ffed}{=:}

\newcommand*{\NN}{\ensuremath{\mathbb{N}}}    
\newcommand*{\NNpos}{\ensuremath{\NN_{\mbox{\tiny $\scriptscriptstyle > 0$}}}}

\newcommand*{\ZZ}{\ensuremath{\mathbb{Z}}}

\newcommand*{\RR}{\ensuremath{\mathbb{R}}}
\newcommand*{\QQ}{\ensuremath{\mathbb{Q}}}
\newcommand*{\RRpos}{\ensuremath{\RR_{\mbox{\tiny $\scriptscriptstyle\geq 0$}}}}
\newcommand*{\UU}{\ensuremath{\mathbb{U}}}
\newcommand*{\VV}{\ensuremath{\mathbb{V}}}

\newcommand*{\A}{\ensuremath{\mathcal{A}}}
\newcommand*{\B}{\ensuremath{\mathcal{B}}}

\newcommand*{\Q}{\ensuremath{\mathcal{Q}}}

\newcommand*{\Abig}{\ensuremath{\mathfrak{A}}}
\newcommand*{\Bbig}{\ensuremath{\mathfrak{B}}}

\newcommand*{\abig}{\ensuremath{\mathfrak{a}}}
\newcommand*{\bbig}{\ensuremath{\mathfrak{b}}}
\newcommand*{\cbig}{\ensuremath{\mathfrak{c}}}

\newcommand*{\ablue}{\ensuremath{a_{\scriptscriptstyle\mathit{blue}}}}
\newcommand*{\bblue}{\ensuremath{b_{\scriptscriptstyle\mathit{blue}}}}

\newcommand*{\K}{\ensuremath{\mathscr{K}}}
\newcommand*{\Cl}{\ensuremath{\mathscr{C}}} 
\newcommand*{\Clfinite}{\ensuremath{\Cl_{\class{fin}}}}
\newcommand*{\Clfinrep}{\ensuremath{\Cl_{\class{fin.rep}}}}
\newcommand*{\Clarb}{\ensuremath{\Cl_{\class{arb}}}}
\newcommand*{\ClNrep}{\ensuremath{\Cl_{\class{$\NN$-rep}}}}
\newcommand*{\ClZrep}{\ensuremath{\Cl_{\class{$\ZZ$-rep}}}}
\newcommand*{\ClNemb}{\ensuremath{\Cl_{\class{$\NN$-emb}}}}
\newcommand*{\ClZemb}{\ensuremath{\Cl_{\class{$\ZZ$-emb}}}}

\newcommand*{\LL}{\ensuremath{\mathscr{L}}}

\newfont{\MyScript}{cmfi10 at 10pt}
\newcommand*{\mF}{\ensuremath{\mathtt{m}}}
\newcommand*{\lF}{\ensuremath{\mathtt{l}}}
\newcommand*{\cF}{\ensuremath{\mathtt{c}}}
\newcommand*{\gF}{\ensuremath{\mathtt{g}}}

\newcommand*{\adom}{\ensuremath{\textsl{adom}}}
\newcommand*{\Cube}[1]{\ensuremath{\textit{Cube}_{#1}}}
\newcommand*{\Type}[1]{\ensuremath{\textit{Type}_{{#1}}}}
\newcommand*{\type}[1]{\ensuremath{\textit{type}_{{#1}}}}
\newcommand*{\types}{\ensuremath{\textbf{\itshape types}}}
\newcommand*{\Types}{\ensuremath{\textbf{\itshape Types}}}

\newcommand*{\ar}{\ensuremath{\textit{ar}}}

\newcommand*{\Problem}[1]{\ensuremath{\textsc{#1}}}

\newcommand*{\Spectrum}{\Problem{Spec}}

\newcommand*{\struc}[1]{\ensuremath{\langle #1 \rangle}}
\newcommand*{\bigstruc}[1]{\ensuremath{
    \left\langle #1 \right\rangle}}

\newcommand*{\praedikat}[1]{\ensuremath{\textsl{#1}}}
\newcommand*{\Bit}{\praedikat{Bit}}
\newcommand*{\Squares}{\praedikat{Squares}}

\newcommand*{\Exp}{\praedikat{Exp}}
\newcommand*{\Inv}{\praedikat{Inv}}

\newcommand*{\Succ}{\praedikat{succ}}
\newcommand*{\SuccP}{\Succ^P}
\newcommand*{\Pred}{\praedikat{pred}}
\newcommand*{\PredP}{\Pred^P}

\newcommand*{\Praedikatklasse}[1]{\ensuremath{\mathfrak{#1}}}
\newcommand*{\Mon}{\Praedikatklasse{Mon}}
\newcommand*{\Arb}{\Praedikatklasse{Arb}}
\newcommand*{\Bip}{\Praedikatklasse{Bip}} 
\newcommand*{\Num}{\Bip} 
\newcommand*{\Rel}{\Bip} 

\newcommand*{\Groups}{\Praedikatklasse{Groups}}

\newcommand*{\dwins}{\ensuremath{\approx}} 
\newcommand*{\dwinsBCEFO}{\ensuremath{\sim}} 
\newcommand*{\dwinsBCeFO}{\dwinsBCEFO} 

\newcommand*{\Nx}{\ensuremath{n_{\vek{x}}}}
\newcommand*{\Ny}{\ensuremath{n_{\vek{y}}}}

\newcommand*{\class}[1]{\ensuremath{\textsl{#1}}}

\newcommand*{\FO}{\class{FO}}

\newcommand*{\OrderGen}{\ensuremath{{<}\class{-generic}\;}}
\newcommand*{\OrderGenFO}{\ensuremath{{<}\class{-generic}\;\FO}}
\newcommand*{\FOadom}{\FO_{\class{adom}}}

\newcommand*{\FOKleiner}{\ensuremath{\FOadom(<)}}
\newcommand*{\FONum}{\ensuremath{\FO(<,\Num)}}
\newcommand*{\FONumi}{\ensuremath{\FO(<,\allowbreak\Num')}}
\newcommand*{\FOMon}{\ensuremath{\FO(<,\Mon)}}

\newcommand*{\FOPlus}{\ensuremath{\FO(<,\allowbreak +)}}

\newcommand*{\EFO}{\class{EFO}}
\newcommand*{\BCeFO}{\ensuremath{\class{BC}(\EFO)}}
\newcommand*{\BCeFOadom}{\BCeFO_{\class{adom}}}
\newcommand*{\BCeFOKleiner}{\ensuremath{\BCeFOadom(<)}}

\newcommand*{\BCEFO}[1]{\BCeFO(#1)}

\newcommand*{\Linf}{\ensuremath{{L}_{\infty\omega}}}

\usepackage{amsthm} 

\swapnumbers  

\newcommand*{\fertig}{{}\hfill \ensuremath{\square}}
\renewcommand*{\qed} {\null\hfill\ensuremath{\blacksquare}}

\newtheorem{quasi_theorem}{Theorem}[section] 
\newtheorem*{quasi_theorem_nn}{Theorem} 

\newtheorem{quasi_lemma}[quasi_theorem]{Lemma}
\newtheorem*{quasi_lemma_nn}{Lemma} 

\newtheorem{quasi_corollary}[quasi_theorem]{Corollary}
\newtheorem*{quasi_corollary_nn}{Corollary} 

\newtheorem{quasi_proposition}[quasi_theorem]{Proposition}
\newtheorem*{quasi_proposition_nn}{Proposition} 

\newtheorem{quasi_maintheorem}[quasi_theorem]{Theorem}
\newtheorem*{quasi_maintheorem_nn}{Theorem} 

\newtheorem{quasi_definition}[quasi_theorem]{Definition}
\newtheorem*{quasi_definition_nn}{Definition} 

\newtheorem{quasi_example}[quasi_theorem]{Example}
\newtheorem*{quasi_example_nn}{Example} 

\newtheorem{quasi_remark}[quasi_theorem]{Remark}
\newtheorem*{quasi_remark_nn}{Remark} 

\newtheorem{quasi_remarks}[quasi_theorem]{Remarks}
\newtheorem*{quasi_remarks_nn}{Remarks} 

\newtheorem{quasi_question}[quasi_theorem]{Question}
\newtheorem*{quasi_question_nn}{Question} 

\newtheorem{quasi_fact}[quasi_theorem]{Fact}
\newtheorem*{quasi_fact_nn}{Fact} 

\newtheorem{quasi_facts}[quasi_theorem]{Facts}
\newtheorem*{quasi_facts_nn}{Facts} 

\newenvironment{theorem_ohne}{\begin{quasi_theorem}}{\end{quasi_theorem}\noindent}
\newenvironment{theorem_nn_ohne}{\begin{quasi_theorem_nn}}{\end{quasi_theorem_nn}\noindent}

\newenvironment{lemma_ohne}{\begin{quasi_lemma}}{\end{quasi_lemma}\noindent}
\newenvironment{lemma_nn_ohne}{\begin{quasi_lemma_nn}}{\end{quasi_lemma_nn}\noindent}

\newenvironment{corollary_ohne}{\begin{quasi_corollary}}{\end{quasi_corollary}\noindent}
\newenvironment{corollary_nn_ohne}{\begin{quasi_corollary_nn}}{\end{quasi_corollary_nn}\noindent}

\newenvironment{proposition_ohne}{\begin{quasi_proposition}}{\end{quasi_proposition}\noindent}
\newenvironment{proposition_nn_ohne}{\begin{quasi_proposition_nn}}{\end{quasi_proposition_nn}\noindent}

\newenvironment{maintheorem_ohne}{\begin{quasi_maintheorem}}{\end{quasi_maintheorem}\noindent}
\newenvironment{maintheorem_nn_ohne}{\begin{quasi_maintheorem_nn}}{\end{quasi_maintheorem_nn}\noindent}

\newenvironment{definition_ohne}{\begin{quasi_definition}}{\end{quasi_definition}\noindent}
\newenvironment{definition_nn_ohne}{\begin{quasi_definition_nn}}{\end{quasi_definition_nn}\noindent}

\newenvironment{example_ohne}{\begin{quasi_example}}{\end{quasi_example}\noindent}
\newenvironment{example_nn_ohne}{\begin{quasi_example_nn}}{\end{quasi_example_nn}\noindent}

\newenvironment{remark_ohne}{\begin{quasi_remark}}{\end{quasi_remark}\noindent}
\newenvironment{remark_nn_ohne}{\begin{quasi_remark_nn}}{\end{quasi_remark_nn}\noindent}

\newenvironment{remarks_ohne}{\begin{quasi_remarks}}{\end{quasi_remarks}\noindent}
\newenvironment{remarks_nn_ohne}{\begin{quasi_remarks_nn}}{\end{quasi_remarks_nn}\noindent}

\newenvironment{question_ohne}{\begin{quasi_question}}{\end{quasi_question}\noindent}
\newenvironment{question_nn_ohne}{\begin{quasi_question_nn}}{\end{quasi_question_nn}\noindent}

\newenvironment{fact_ohne}{\begin{quasi_fact}}{\end{quasi_fact}\noindent}
\newenvironment{fact_nn_ohne}{\begin{quasi_fact_nn}}{\end{quasi_fact_nn}\noindent}

\newenvironment{facts_ohne}{\begin{quasi_facts}}{\end{quasi_facts}\noindent}
\newenvironment{facts_nn_ohne}{\begin{quasi_facts_nn}}{\end{quasi_facts_nn}\noindent}

\newenvironment{proof_ohne}{{\rm \bf Proof.}}{}
\newenvironment{proofc_ohne}[1]{{\rm \bf Proof {#1}.}}{}
\newenvironment{proofsketch_ohne}{{\rm \bf Proof (sketch).}}{}

\newenvironment{theorem_mit}{\begin{theorem_ohne}}{\fertig \end{theorem_ohne}}

\newenvironment{lemma_mit}{\begin{lemma_ohne}}{\fertig \end{lemma_ohne}}

\newenvironment{corollary_mit}{\begin{corollary_ohne}}{\fertig \end{corollary_ohne}}

\newenvironment{proposition_mit}{\begin{proposition_ohne}}{\fertig \end{proposition_ohne}}

\newenvironment{definition_mit}{\begin{definition_ohne}}{\fertig \end{definition_ohne}}

\newenvironment{example_mit}{\begin{example_ohne}}{\fertig \end{example_ohne}}

\newenvironment{remark_mit}{\begin{remark_ohne}}{\fertig \end{remark_ohne}}

\newenvironment{fact_mit}{\begin{fact_ohne}}{\fertig \end{fact_ohne}}

\newenvironment{proof_mit}{\begin{proof_ohne}}{\qed \end{proof_ohne}}
\newenvironment{proofc_mit}[1]{\begin{proofc_ohne}{#1}}{\qed \end{proofc_ohne}}
\newenvironment{proofsketch_mit}{\begin{proofsketch_ohne}}{\qed \end{proofsketch_ohne}}

\newenvironment{summary}{}{}%

\usepackage{pstricks,pst-node,pstcol}
\usepackage[dvips]{graphicx}         

\newpsobject{showgrid}{psgrid}{subgriddiv=1,griddots=10,gridlabels=6pt}

\definecolor{rot}{named}{Red}%
\definecolor{hellrot}{named}{Red}
\definecolor{gelb}{named}{Goldenrod}%
\definecolor{hellgrau}{named}{Gray}%

\begin{document}

\title{\bf An Ehrenfeucht-Fra\"\i{}ss\'{e} Game Approach to Collapse Results 
     in Database Theory}
\author{Nicole Schweikardt\thanks{This research was performed while the 
author was employed at the Johannes Gutenberg-Universit\"at Mainz, Germany.} \\
Laboratory for Foundations of Computer Science\\
University of Edinburgh, Scotland, U.K.\\
Email: \url{nisch@informatik.uni-mainz.de} }

\maketitle 
\begin{abstract}
We present a new Ehrenfeucht-Fra\"\i{}ss\'{e} game approach to collapse results 
in database theory. We show that, in principle, \emph{every} natural generic collapse result
may be proved via a translation of winning strategies for the duplicator in an 
Ehrenfeucht-Fra\"\i{}ss\'{e} game. Following this approach we can deal with certain
infinite databases where previous, highly involved methods fail.
We prove the natural generic collapse for $\ZZ$-embeddable databases over any linearly ordered
context structure with arbitrary \emph{monadic} predicates, and for $\NN$-embeddable databases
over the context structure $\struc{\RR,<,+,\Mon_Q,\Groups}$, where $\Groups$ is the collection
of all subgroups of $\struc{\RR,+}$ that contain the set of integers and $\Mon_Q$ is the
collection of all subsets of a particular infinite set $Q$ of natural numbers. 
This, in particular, implies the collapse for arbitrary databases over 
$\struc{\NN,<,+,\Mon_Q}$ and for $\NN$-embeddable databases over $\struc{\RR,<,+,\ZZ,\QQ}$.
I.e., first-order logic with $<$ can express the same order-generic queries as first-order logic
with $<$, $+$, etc.
\\
Restricting the complexity of the formulas that may be used to formulate queries to
Boolean combinations of purely existential first-order formulas, we even obtain the collapse 
for $\NN$-embeddable databases over any linearly ordered context structure with \emph{arbitrary}
predicates.
\\
Finally, we develop the notion of \emph{$\NN$-representable} databases, which is a natural 
generalization of the notion of \emph{finitely representable} databases.
We show that natural generic 
collapse results for $\NN$-embeddable databases can be lifted to the larger class of 
$\NN$-representable databases.  
\par
To obtain, in particular, the collapse result for $\struc{\NN,<,+,\Mon_Q}$, we 
explicitly construct a winning strategy for the duplicator in the presence of the built-in
addition relation $+$. This, as a side product, also leads to an Ehrenfeucht-Fra\"\i{}ss\'{e} game
proof of the theorem of Ginsburg and Spanier, stating that the spectra of $\FO(<,+)$-sentences 
are {semi-linear}.
\\
\end{abstract}
{\small
 \noindent
 {\bf ACM-classification:} 
 F.4.1 [Mathematical Logic and Formal Languages]: Computational Logic; 
 H.2.3 [Database Management]: Query Languages
 \\
 {\bf Keywords:} 
 Logic in Computer Science, Database Theory, 
 Ehrenfeucht-Fra\"\i{}ss\'{e} Games, Natural Generic Collapse Results
}

\setcounter{tocdepth}{2}

{\small
\tableofcontents
}

%
%


\section{Introduction}\label{section:Introduction}
One of the issues in database theory that have attracted much
interest in recent years is the study of relational databases that
are embedded in a fixed, infinite \emph{context structure}. This
occurs, e.g., in
current applications such as spatial or temporal databases, where
data are represented by (natural or real) numbers, and where databases
can be modelled as \emph{constraint databases}.
For a recent comprehensive survey see~\cite{KLP00}. 
\\
In many applications the numerical values only serve as identifiers that are
exchangeable. 
If this is the case, queries commute with
any permutation of the context universe; such queries are 
called {\em generic}.
If the context universe is linearly ordered, a query may refer to the ordering. In this
setting it is more appropriate to consider queries which commute with every 
\emph{order-preserving} (i.e., strictly increasing) mapping.
Such queries are called \emph{order-generic}.
A basic way of expressing order-generic queries is by first-order formulas that make use
of the linear ordering and of the database relations.
\par
It is a reasonable question whether the use of the additional predicates of the
context structure allows first-order logic to express
more order-generic queries than the linear ordering alone.
In some situations this question can be answered ``yes'', e.g., if the context
structure is $\struc{\NN,<,+,\times}$. In other situations
the question must be answered ``no'', e.g., if the context structure is $\struc{\NN,<,+}$ 
--- such results are then
called \emph{collapse results}, because first-order logic with the
additional predicates collapses to first-order logic with linear
ordering alone.
A recent and comprehensive overview of this area of research is given in \cite{Libkin_DIMACS}.
In classical database theory attention usually is restricted to \emph{finite}
databases. In this setting Benedikt et al.\ \cite{BDLW} have
obtained a strong collapse result:
\emph{First-order logic has the natural generic collapse for finite
databases over o-minimal context structures}. This means that if the context structure
has a certain property called \emph{o-minimality}, then for every order-generic first-order
formula $\varphi$ which uses the additional predicates, there is a
formula with linear ordering alone which is equivalent to $\varphi$ on
all \emph{finite} databases. 
In \cite{BB98} this result was generalized to context structures that have
\emph{finite VC-dimension}, a property that, e.g., the structures 
$\struc{\NN,<,+}$, $\struc{\QQ,<,+}$, $\struc{\RR,<,+,\times,\Exp}$
have. 
The proofs for these results are rather involved; in particular the proof of \cite{BB98} uses 
non-standard models and hyperfinite structures.
\par
The present paper proposes a new \emph{Ehrenfeucht-Fra\"\i{}ss\'{e} game} approach to 
collapse results. 
We show that, in principle, \emph{every} natural generic collapse result
can be proved via a translation of winning strategies for the duplicator in an 
Ehrenfeucht-Fra\"\i{}ss\'{e} game. Following this approach we can deal with certain
\emph{infinite} databases where previous, highly involved methods fail.
We prove the natural generic collapse for $\ZZ$-embeddable databases over linearly ordered
context structures with arbitrary \emph{monadic} predicates, and for $\NN$-embeddable databases
over the context structure $\struc{\RR,<,+,\Mon_Q,\Groups}$, where $\Groups$ is the collection
of all subgroups of $\struc{\RR,+}$ that contain the set of integers and $\Mon_Q$ is the
collection of all subsets of a particular infinite set $Q$ of natural numbers. 
This, in particular, implies the collapse for arbitrary databases over 
$\struc{\NN,<,+,\Mon_Q}$ and for $\NN$-embeddable databases over $\struc{\RR,<,+,\ZZ,\QQ}$.
I.e., first-order logic with $<$ can express the same order-generic queries as first-order logic
with $<$, $+$, etc.
\\
Restricting the complexity of the formulas that may be used to formulate queries to
Boolean combinations of purely existential first-order formulas, we also obtain the collapse 
for $\NN$-embeddable databases over linearly ordered context structures with \emph{arbitrary}
predicates.
\\
Finally, we develop the notion of \emph{$\NN$-representable} databases, which is a natural 
generalization of the notion of \emph{finitely representable} databases of \cite{BST99} (also
known as \emph{dense order constraint databases}).
We show that natural generic 
collapse results for $\NN$-embeddable databases can be lifted to the larger class of 
$\NN$-representable databases.  
\par
Apart from the collapse results obtained with the method of the translation of winning strategies, 
the exposition of explicit strategies for the duplicator in the Ehrenfeucht-Fra\"\i{}ss\'{e} game 
is interesting in its own right.
In particular, to obtain the collapse result for $\struc{\NN,<,+,\Mon_Q}$, we 
explicitly construct a winning strategy for the duplicator in the presence of the built-in
addition relation $+$. This, as a side product, also leads to an Ehrenfeucht-Fra\"\i{}ss\'{e} game
proof of the theorem of Ginsburg and Spanier, stating that the spectra of $\FO(<,+)$-sentences 
are {semi-linear}.
\\
\parno
The present paper contains results of the author's dissertation \cite{Schweikardt_Diss}. 
It combines and extends the results of the conference contributions 
\cite{LS_STACS01,Schweikardt_CSL01}.
\\
The paper is structured as follows:
In Section~\ref{section:Preliminaries} we fix the basic notations used throughout the paper.
Section~\ref{section:Databases} gives a brief introduction to collapse considerations in 
database theory, recalls known results of 
\cite{BDLW,BST99,BB98,BL00a,BL00b,Libkin_DIMACS}, 
and summarizes the collapse results obtained in the present paper. 
In Section~\ref{section:EF-game} we present the \emph{translation of strategies for the
Ehrenfeucht-Fra\"\i{}ss\'{e} game} as a new method for obtaining collapse results 
and we show that, at least in principle, all natural generic collapse results
can be proved by this method. In the Sections~\ref{section:Monadic} and \ref{section:Addition}
we show that the translation of strategies is indeed possible for context
structures with monadic built-in predicates and for context structures with the 
addition relation $+$ and some particular monadic predicates, respectively. Restricting
attention to Boolean combinations of purely existential first-order logic, we show in 
Section~\ref{section:BCEFO} that the translation of strategies is possible even for 
\emph{arbitrary} context structures. 
Section~\ref{section:Lift} proves that natural generic collapse results for 
$\NN$-embeddable databases can be lifted 
to the larger class of $\NN$-representable databases. Finally, in
Section~\ref{section:Conclusion} we summarize our results and point out further questions.
\\
\parno
\textbf{Acknowledgements:}
I want to thank Clemens Lautemann for his guidance and for many valuable discussions and
suggestions concerning the research presented in this paper. I thank Thomas Schwentick for
helpful advice especially in the early stages of my research and for drawing my attention to
collapse considerations in database theory.




\section{Basic Notations}\label{section:Preliminaries}
We use $\ZZ$ for the set of integers, $\NN\deff \set{0,1,2,\twodots}$
\index{N-@$\NN$}\index{Z@$\ZZ$}\index{Q1@$\QQ$}\index{R@$\RR$}%
for the set of natural numbers,
$\NNpos$ for the set of positive natural numbers, 
$\QQ$ for the set of rational numbers, $\RR$ for the set of real numbers, and $\RRpos$ for
the set of nonnegative real numbers.
\\
For $a,b\in\ZZ$ we write \,$a\mid b$\, to express that $a$ divides
$b$, \index{$\equiv_n$t@$\mid$ \,($a\mid b$)}%
i.e., that $b = c\cdot a$ for some $c\in\ZZ$.
For $n\in\NNpos$ the symbol $\equiv_n$ \index{$\equiv_n$ ($a\equiv_n b$)}%
denotes the \emph{congruence relation modulo $n$}, i.e., for $a,b\in\ZZ$ we have 
$a\equiv_n b$ \,iff \,$n \mid a{-}b$. 
This relation can be extended to real numbers $r,s\in\RR$
via \,$r\equiv_n s$\, iff \,$r{-}s= z\cdot n$\, for some $z\in\ZZ$. 
For $r\in\RR$ we write $\abgerundet{r}$ to denote the
\index{$\abgerundet{r}$, $\aufgerundet{r}$}%
largest integer $\leq r$, and we write
$\aufgerundet{r}$ for the smallest integer $\geq r$.
$\betrag{r}$ \index{$\betrag{r}$}%
denotes the \emph{absolute value} of $r$, i.e., $\betrag{r} = r$ if $r\geq 0$, and
${-}r$ otherwise.
For $r,s\in\RR$ we write $\intcc{r,s}$ \index{int@$\intco{r,s}$}%
to denote the closed
interval
$\setc{x\in\RR}{r\leq x\leq s}$. Analogously, we write
$\intoo{r,s}$ for the open interval $\intcc{r,s}\setminus\set{r,s}$,
$\intco{r,s}$ for the half open interval $\intcc{r,s}\setminus\set{s}$,
and $\intoc{r,s}$ for the half open interval $\intcc{r,s}\setminus\set{r}$.
We write \,$a_1,\twodots,a_m\mapsto b_1,\twodots,b_m$\, to denote the
mapping $f$ with domain $\set{a_1,\twodots,a_m}$ and range
$\set{b_1,\twodots,b_m}$ which satisfies $f(a_i)=b_i$ for all $i\in\set{1,\twodots,m}$.
Depending on the particular context, we use $\vek{a}$ as abbreviation for a
sequence $a_1,\twodots,a_m$ or a tuple $(a_1,\twodots,a_m)$.
Accordingly, if $f$ is a mapping defined on all elements in $\vek{a}$,
we write $f(\vek{a})$ to denote the sequence
$f(a_1),\twodots,f(a_m)$ or the tuple $(f(a_1),\twodots,f(a_m))$.
If $R$ is an $m$-ary relation on the domain of $f$,  we write 
$f(R)$ to denote the relation $\setc{f(\vek{a})}{\vek{a}\in R}$.
Instead of $\vek{a}\in R$ we often write $R(\vek{a})$.
\par
A \emph{signature} $\sigma$ consists of 
constant symbols and relation symbols. Each relation
symbol $R\in \sigma$ has a fixed arity 
$\ar(R)\in\NNpos$.
Whenever we refer to some ``$R\in\sigma$'' we implicitly assume that $R$ is a
\emph{relation} symbol. Analogously, ``$c\in\sigma$'' 
means that $c$ is a \emph{constant} symbol.
Throughout this paper we adopt the convention that whenever a signature is denoted by the
symbol $\tau$, then it is a \emph{finite} set of relation symbols and constant symbols.
\par
A \emph{$\sigma$-structure} 
$\A = \struc{A,\sigma^{\A}}$ consists of an arbitrary set $A$ which
is called the \emph{universe} of $\A$, and a
set $\sigma^{\A}$ that contains an interpretation $c^{\A} \in A$ for each $c\in\sigma$, and
an interpretation $R^{\A} \subseteq A^{\ar(R)}$ for each $R\in\sigma$.
Sometimes we want to restrict our attention to $\sigma$-structures over a particular
universe $\UU$. In these cases we speak of $\struc{\UU,\sigma}$-structures.
An \emph{isomorphism} $\pi$ between two $\sigma$-structures 
$\A = \struc{A,\sigma^{\A}}$ and
$\B = \struc{B,\sigma^{\B}}$ is a bijective mapping $\pi:A\rightarrow B$ such that
$\pi(c^{\A}) = c^{\B}$ for each $c\in\sigma$, and 
$R^{\A}(\vek{a})$  iff  $R^{\B}\big(\pi(\vek{a})\big)$ for each
$R\in\sigma$.
An \emph{automorphism} of $\A$ \index{automorphism}%
is an isomorphism between $\A$ and $\A$.
A \emph{partial isomorphism} \index{partial isomorphism}%
between $\A$ and $\B$ is a mapping $\pi':A'\rightarrow B'$ such that 
$A'\subseteq A$ and $B'\subseteq B$ contain all constants of $\A$ and $\B$ and
$\pi'$ is an isomorphism between the induced substructures obtained by restricting $\A$ and
$\B$ to the universes $A'$ and $B'$.
\par
We use the usual notation concerning \emph{first-order logic} (cf., e.g., \cite{Immerman,EbbinghausFlum,Libkin_DIMACS}). In particular, we write $\FO(\sigma)$ to denote the set of all formulas of
first-order logic over the signature $\sigma$. Note that our notion of signatures does not
include the use of function symbols. Therefore, when used in the
context of formulas, arithmetic predicates such as $+$, $\times$, $\Exp$, $\Bit$ are always 
interpreted by \emph{relations}, i.e., $+$ (respectively, $\times$, $\Exp$) denotes the ternary 
relation consisting of
all triples $(a,b,c)$ such that $a{+}b{=}c$ (respectively, $a{\cdot}b{=}c$, $a^b{=}c$); 
and $Bit$ is the 
binary relation consisting of all tuples $(a,i)$ such that the $i$-th bit in the binary expansion
of $a$ is 1, i.e., $\abgerundet{\frac{a}{2^i}}$ is odd.



\section{Collapse Results in Database Theory}\label{section:Databases}
Detailed information on the foundations of databases can be found in the 
textbook \cite{AHV}. For a well-written concise survey on database theory we refer to the
paper \cite{VandenBussche_CSL01}. A detailed and very recent overview of collapse results on
finite databases is given in \cite{Libkin_DIMACS}. More information can also be found in the
book \cite{KLP00}.
Not aiming at comprehensiveness, the present section of this paper gives a brief
introduction to concepts, questions, and results in database theory that are related to
collapse considerations. Furthermore, we summarize the collapse results that are obtained in 
this paper (see Section~\ref{subsection:CollapseResultsPaper}).
\subsection{Databases and Queries}\label{subsection:DBsAndQueries}
In relational database theory \index{database theory}%
a database is modelled as a relational structure over a fixed
possibly infinite universe $\UU$.
A \emph{database} \index{database}%
over $\UU$ hence is a $\struc{\UU,\rho}$-structure $\A = \struc{\UU,\rho^{\A}}$,  
for some signature $\rho$ that
consists of a finite number of relation symbols.
In database theory such a relational signature $\rho$ is often called the \emph{database schema}. \index{database schema}
The \emph{active domain} \index{active domain}\index{adom@$\adom(\A)$}%
of $\A$, $\adom(\A)$ for short, is the set of all elements in $\UU$ that belong to (at least) one
of $\A$'s relations. I.e., $\UU$ is the set of all \emph{potential} database elements, whereas
$\adom(\A)$ is the set of all elements that \emph{indeed occur} in the database relations.
\par
A \emph{Boolean query} \index{query}\index{Boolean query}%
is a ``question'' or, more formally, a mapping $\Q$ that assigns to each
$\struc{\UU,\rho}$-structure $\A$ an answer ``yes'' or ``no''. Examples of such queries
are
\begin{itemize}
 \item[1.]\label{query:three}
   Does the unary relation $R_1$ contain at least 3 elements?
 \item[2.]\label{query:even}
   Does the active domain have even cardinality?
 \item[3.]\label{query:orderI}
   Are all elements in $R_1$ smaller than all elements in $R_2$?
\end{itemize}
Often one also considers \emph{$k$-ary queries} which yield as answers $k$-ary relations over $\UU$.
Examples of such queries are
\begin{itemize}
 \item[4.]\label{query:adom}
   What are the elements in the active domain?
 \item[5.]\label{query:tc}
   What is the transitive closure of the binary relation $R_3$?  
 \item[6.]\label{query:orderII}
   Which elements belong to $R_1$ and are smaller than all elements in $R_2$?
\end{itemize} 
For \emph{well-defined} queries one demands the follwing consistency criterion
\begin{enumerate}[(CC):\ ]
 \item
  \em On identical databases, a query must produce identical answers.
\end{enumerate}
Usually, two databases $\A=\struc{\UU,\rho^{\A}}$ and $\B=\struc{\UU,\rho^{\B}}$ are assumed to
be ``identical'' iff they are isomorphic, i.e., iff there is a permutation $\pi$ of $\UU$ such
that $\pi(\rho^{\A})=\rho^{\B}$.
Queries that satisfy (CC) are called \emph{generic} queries. \index{generic}%
If $\Q$ is a Boolean query, this means that $\Q(\A)=\Q(\pi(\A))$; if
$\Q$ is a $k$-ary query it means that $\pi(\Q(\A))= \Q(\pi(\A))$, for
all permutations $\pi$ of $\UU$.
\par
A basic way of expressing queries is by first-order formulas. I.e., a 
\emph{$\FO(\rho)$-sentence} expresses a Boolean query, and a $\FO(\rho)$-formula with $k$ free
variables expresses a $k$-ary query. For example, if $\rho$ consists of two unary relations
$R_1$ and $R_2$ and a binary relation $R_3$, then the queries 1.\ and 4.\ can be expressed as
follows:
\[
\begin{array}{ccl}
  \varphi_1 & \deff & 
  \exists x_1\,\exists x_2\,\exists x_3\;\Big( 
   \begin{array}[t]{l}
      R_1(x_1)\,\und\,  R_1(x_2)\, \und\, R_1(x_3)\ \und
    \\
      x_1\neq x_2\,\und\,x_2\neq x_3\,\und\,x_3\neq x_1\;\Big)
  \end{array}   
\medskip\\
  \varphi_4(x) & \deff & 
  R_1(x)\ \oder\ R_2(x)\ \oder\ \exists y\; R_3(x,y)\ \oder\ \exists y\; R_3(y,x)\;.
\end{array}
\]
To avoid the distinction between Boolean queries and $k$-ary queries, we will not consider 
relational database schemas $\rho$ but, instead, \emph{signatures}
$\tau$ that consist of a finite number of relation symbols {and constant symbols}.
This allows us to restrict our attention to \emph{Boolean} queries in the following way:
A $\FO(\rho)$-formula $\varphi(x_1,\twodots,x_k)$ with $k$ free variables $x_1,\twodots,x_k$
can be viewed as a \emph{$\FO(\tau)$-sentence} for $\tau\deff \rho\cup\set{x_1,\twodots,x_k}$.
In general, any $k$-ary query $\Q$ on $\struc{\UU,\rho}$-structures corresponds to the
Boolean query $\Q'$ on $\struc{\UU,\tau}$-structures that yields the anwer ``yes'' for a
$\struc{\UU,\tau}$-structure \,$\A=\struc{\UU,\rho^{\A},a_1,\twodots,a_k}$\, if and only if
\,$(a_1,\twodots,a_k)$\, belongs to the $k$-ary relation that $\Q$ defines on the 
structure $\struc{\UU,\rho^{\A}}$.
\par
From now on we will, without loss of generality, consider Boolean queries rather than
$k$-ary queries. Signatures $\tau$ will always consist of a finite number of relation symbols and
constant symbols. The name \emph{$\struc{\UU,\tau}$-database} will be used as a synonym for 
\emph{$\struc{\UU,\tau}$-structure}.
\index{database!utdatabase@$\struc{\UU,\tau}$-database}%
\index{utdatabase@$\struc{\UU,\tau}$-database}%
\index{structure!utstructure@$\struc{\UU,\tau}$-structure}%
\index{utstructure@$\struc{\UU,\tau}$-structure}%
The \emph{active domain} of a $\struc{\UU,\tau}$-structure is defined as follows:
\begin{definition_mit}[Active Domain $\adom(\A)$]\label{definition:adom}\index{active domain}\index{adom@$\adom(\A)$}
\mbox{ }\\
Let $\tau$ be a signature and let $\A=\struc{\UU,\tau^{\A}}$ be a
$\tau$-structure.
The \emph{active domain} of $\A$, for short: $\adom(\A)$, is the set
of all elements in $\UU$ that occur in $\tau^{\A}$. I.e., $\adom(\A)$
is the smallest set $A\subseteq\UU$ that contains the constants
$c^{\A}$, for all $c\in\tau$, and that satisfies $R^{\A}\subseteq A^{\ar(R)}$,
for all $R\in\tau$.
\end{definition_mit}%
%
%
%
It is obvious that all $\FO(\tau)$-definable queries are generic. However,
there are generic queries, e.g., the queries 
2.\ and 5.\ above, that are \emph{not} expressible in $\FO(\tau)$ 
(cf., e.g., \cite{EbbinghausFlum} or \cite{Immerman}).
To express \emph{more} queries, one may allow the formulas to use extra information which is
not explicitly part of the database, such as a linear ordering $<$ or arithmetic predicates
$+$ and $\times$. \label{query:Qeven}%
In this framework, for example query 2.\ for \,$\UU\deff \NN$\, can be expressed in
$\FO(<,+,\times,\tau)$ via the formula
\begin{eqnarray*}
 \varphi_2 & \deff &
 \exists y\,\exists z\ \Big( 
   \big( \forall x\;\varphi_4(x)\gdw\varphi_{\Bit}(y,x)\big)\,\und\;
   \varphi_{\praedikat{BitSum}}(y,z)\;\und\; \big(\exists u\; u{+}u{=}z\big)\Big)\,.
\end{eqnarray*}
Here, $y$ encodes the active domain and $z$ is the cardinality of the active domain; 
$\varphi_{\Bit}(y,x)$ and $\varphi_{\praedikat{BitSum}}(y,z)$ are $\FO(+,\times)$-formulas 
expressing that the $x$-th bit of the 
binary representation of $y$ is 1, and that $z$ is the number of ones in the binary representation
of $y$, respectively (cf., e.g., the textbook \cite{Immerman}).
\\
The additional predicates such as $<$, $+$, \ldots\ are viewed as
\emph{built-in predicates} \index{built-in predicate}%
associated with the universe $\UU$ of potential database elements.
In other words, $\struc{\UU,<,\allowbreak +,\ldots}$ is viewed as the 
\emph{context structure} \index{context structure}%
in which the $\struc{\UU,\tau}$-databases live.
In general, we use the following notation:
\begin{definition_mit}[Context Structure $\bs{\struc{\UU,<,\Num}}$]\label{definition:context_structure}\index{context structure}
\mbox{ }\\
A \emph{context structure} consists of an infinite universe $\UU$, a linear ordering $<$
(i.e., $<$ is a binary relation on $\UU$ that is transitive, total, and antisymmetric), and
a (possibly infinite) class $\Num$ of relations on $\UU$. 
\end{definition_mit}%
Given a context structure $\struc{\UU,<,\Num}$ and a set $S\subseteq\UU$, we shortly write
$\struc{S,<,\allowbreak \Num}$ to denote the induced substructure of $\struc{\UU,<,\Num}$ with 
universe $S$. 
%
%
\subsection{Finding an Adequate Notion of Genericity}\label{subsection:Genericity}\index{genericity}
When dealing with $\struc{\UU,\tau}$-databases that live in a context structure $\struc{\UU,<,\Num}$
one has to revisit the concept of \emph{genericity}. 
Paredaens, Van den Bussche, and Van Gucht \cite{PVV} (see also \cite{KV00}) 
\index{Paredaens, Jan}\index{Van den Bussche, Jan}\index{Van Gucht, Dirk}
pointed out that the {adequate} notion of genericity
depends on the particular context (or, the geometry) in which the information stored in 
a database is interpreted.
For the particular context considered in the present paper, this can be explained as follows:
Recall that the consistency criterion (CC) demands that a generic Boolean query produces the same
answer for a database $\A=\struc{\UU,\tau^{\A}}$ as for its isomorphic image
$\pi(\A) = \struc{\UU,\pi(\tau^{\A})}$, for every permutation $\pi$ of $\UU$.
Under this restrictive view, the above queries 3.\ and 6.\ are \emph{not} generic.
Nevertheless, these queries do make sense when having in mind \emph{temporal databases} 
\index{temporal database}
that store,
e.g., the chronological order of events. With this interpretation, query 3.\ asks whether the 
task $R_1$ was finished before the task $R_2$ began.
\par
When the linear ordering of the database elements is relevant, it seems adequate to call 
two $\struc{\UU,\tau}$-databases $\A$ and $\B$ ``identical'' iff they are isomorphic via a
${<}$-preserving mapping. Precisely, several different notions are conceivable: 
\begin{itemize}
 \item[(1.)\ ] 
  $\A$ and $\B$ are called \emph{order-isomorphic} \index{order-isomorphic}
  \index{isomorphic!order-isomorphic}
  iff the linearly ordered structures
  \,$\struc{\UU,<,\allowbreak \tau^{\A}}$\, and \,$\struc{\UU,<,\tau^{\B}}$\, are isomorphic in the 
  usual sense. 
  Queries that produce identical answers for order-isomorphic databases are
  known as \emph{order-generic} queries (cf., e.g., \cite{BST99,BDLW,BL00a}).
  \index{order-generic}\index{generic!order-generic}
\end{itemize}
For \emph{dense} linear orderings such as $\struc{\RR,<}$ or $\struc{\QQ,<}$ this notion of 
genericity seems adequate. However, for \emph{discrete} orderings like
$\struc{\ZZ,<}$ or $\struc{\NN,<}$ the above notion of order-genericity is too liberal and,
equivalently, the notion of order-isomorphy is too restrictive: The identity function is the
only order-isomorphism on $\struc{\NN,<}$, and consequently no two different databases
are assumed to be ``identical with respect to the linear ordering''.
For a good formalization of what it means to be ``identical with respect to the linear ordering''
it seems reasonable to consider the active domain of the databases rather than the whole
context universe $\UU$:
\begin{itemize}
 \item[(2.)\ ]
  $\A$ and $\B$ are called \emph{locally order-isomorphic}\index{locally order-isomorphic} 
  \index{isomorphic!locally order-isomorphic}
  iff the linearly ordered structures
  $\struc{\adom(\A),{<,} \allowbreak \tau^{\A}}$ and $\struc{\adom(\B),<,\tau^{\B}}$ 
  are isomorphic in the 
  usual sense. 
  Queries that produce identical answers for locally order-isomorphic databases are
  known as \emph{locally generic} queries (cf., \cite{BST99,BDLW,BL00a}).
  \index{locally generic}\index{generic!locally generic}
\end{itemize}
When restricting attention to databases whose active domain is \emph{finite}, the above notion
of \emph{local order-isomorphy} perfectly catches the intuitive understanding of being
``identical with respect to the linear ordering''.
Moreover, it is not difficult to see (cf., \cite[Proposition\;1]{BDLW}) that for the context
structures $\struc{\RR,<}$, $\struc{\QQ,<}$, and in general for any doubly transitive linear
ordering $\struc{\UU,<}$, the notions of \emph{order-isomorphy} and 
\emph{local order-isomorphy} coincide.
\par
But what about databases with an \emph{infinite} active domain?
For example, let $\UU\deff \RR$ and let $\tau$ consist of a single unary relation symbol $S$.
Consider the $\struc{\RR,\tau}$-structures $\A$ and $\B$ with 
\,$S^{\A}\deff \set{a_1<a_2<\cdots}$\, where $a_n\deff 1-\frac{1}{n}$,\,  
and 
\,$S^{\B}\deff \set{b_1<b_2<\cdots}$\, where $b_n\deff n$,\, for all $n\in\NNpos$.
Clearly, \,$\struc{\adom(\A),<,\allowbreak \tau^{\A}}$\, is isomorphic to \,$\struc{\adom(\B),<,\tau^{\B}}$,
and thus $\A$ and $\B$ are locally order-isomorphic. But would we intuitively say that
$\A$ and $\B$ are ``identical with respect to the linear ordering''?
--- Not really, since $\adom(\A)$ has an upper bound in the context universe $\RR$ 
whereas $\adom(\B)$ has not.
Here it seems adequate to take into account not $\adom(\A)$ but its \emph{closure}
{\begin{eqnarray*}
 \ov{\adom(\A)} & \deff & 
 \adom(\A)\;\cup\;\setc{x\in\RR}{x\mbox{ is an accumulation point of }\adom(\A)}\,.
\end{eqnarray*}}%
To catch the intuitive meaning of being ``identical with respect to the linear ordering'' we 
therefore propose the following formalization: 
Two $\struc{\RR,\tau}$-structures $\A$ and $\B$ are called 
\emph{${<}$-isomorphic} \index{isomorphic!${<}$-isomorphic} 
iff the linearly ordered structures
\,$\struc{\ov{\adom(\A)},<,\tau^{\A}}$\, and \,$\struc{\ov{\adom(\B)},<,\tau^{\B}}$\, are 
isomorphic in the usual sense. 
For the context universe $\QQ$ rather than $\RR$ it seems appropriate to demand that
\,$\struc{\ov{\adom(\A)},<,\tau^{\A}}$\, and \,$\struc{\ov{\adom(\B)},<,\tau^{\B}}$\, are 
isomorphic via a mapping that maps accumulation points in $\QQ$ on accumulation points in $\QQ$, 
and that maps accumulation points in $\RR\setminus\QQ$ on accumulation points in $\RR\setminus\QQ$.
\par
For an \emph{arbitrary} linearly ordered context universe $\UU$ we propose the following
generalization:
Let $\struc{\ov{\UU},<}$ \index{$\struc{\ov{\UU},<}$}\index{UuDedekind@$\struc{\ov{\UU},<}$}
be a \emph{Dedekind completion} \index{Dedekind completion}%
of $\struc{\UU,<}$. I.e., $\UU\subseteq \ov{\UU}$, and every set $A\subseteq \UU$ that has
an upper bound (respectively, a lower bound) in $\UU$ with respect to $<$, has a unique
\emph{least} upper bound (respectively, \emph{greatest} lower bound) in $\ov{\UU}$.
For example, $\struc{\RR,<}$ is a Dedekind completion of $\struc{\QQ,<}$ and of $\struc{\RR,<}$, and 
$\struc{\NN,<}$ is a Dedekind completion of $\struc{\NN,<}$.
The \emph{closure} \index{closure $\ov{A}$}
of a set $A\subseteq\UU$ is the set $\ov{A}$ that consists of all 
elements of $A$ and all elements $x\in\ov{\UU}$ which are a least upper bound or a 
greatest lower bound of some subset $A'\subseteq A$.
\begin{definition_mit}[$\bs{<}$-isomorphy, $\bs{<}$-genericity]\label{definition:kleiner-generic}
\index{isomorphic!${<}$-isomorphic}\index{generic!<generic@${<}$-generic}%
\mbox{ }\\
Let $\struc{\UU,<}$ be a linearly ordered context structure, and let $\struc{\ov{\UU},<}$ be its
Dedekind completion. Let $\tau$ be a signature.
Two $\struc{\UU,\tau}$-structures $\A$ and $\B$ are called \,\emph{${<}$-isomorphic}\, iff the 
structures
{
\begin{eqnarray*}
  \big\langle\;\ov{\adom(\A)},\;<,\;\tau^{\A},\;\ov{\adom(\A)}\setminus\UU\;\big\rangle
& \mbox{ and }
& \big\langle\;\ov{\adom(\B)},\;<,\;\tau^{\B},\;\ov{\adom(\B)}\setminus\UU\;\big\rangle
\end{eqnarray*}}%
are isomorphic in the usual sense. 
\\
A Boolean query $\Q$ is called
\emph{${<}$-generic} on a $\struc{\UU,\tau}$-structure $\A$ if and only if
\,$\Q(\A) = \Q(\B)$\, for all $\struc{\UU,\tau}$-structures $\B$ that are ${<}$-isomorphic to $\A$.
Accordingly, if $\K$ is a class of $\struc{\UU,\tau}$-structures, then we say that
$\Q$ is ${<}$-generic on $\K$ iff it is ${<}$-generic on every $\A\in\K$.
\end{definition_mit}%
In particular, the notions \emph{${<}$-isomorphy} and \emph{${<}$-genericity} 
coincide with the notions
\emph{order-isomorphy} and \emph{order-genericity} if $\UU$ is $\RR$ or $\QQ$,
and they coincide with the notions
\emph{local order-isomorphy} and \emph{local genericity} if $\UU$ is $\NN$ or $\ZZ$. 
This further indicates that these notions are adequate and uniform formalizations of what
it means for databases to be ``identical with respect to the ordering'' and what it means
for queries to produce consistent answers for ``identical'' databases.
The following notion gives us an alternative characterization of ${<}$-isomorphy and
${<}$-genericity:
\begin{definition_mit}[$\bs{<}$-preserving mapping]\label{definition:KleinerPreserving}
\index{preserving mapping@${<}$-preserving mapping}
\mbox{ }\\
Let $\struc{\UU,<}$ and $\struc{\VV,<}$ be linearly ordered structures, and let
$\struc{\ov{\UU},<}$ and $\struc{\ov{\VV},<}$ be their Dedekind completions.
Let \,$U\subseteq\UU$, \,let \,$\alpha : U\rightarrow \VV$, \,and let \,$V\deff \alpha(U)$.
\\
The mapping $\alpha$ is called \emph{${<}$-preserving} \,iff\, it can be extended to an isomorphism 
between the structures \,$\struc{\,\ov{U},\,<,\,\ov{U}\setminus \UU\,}$\, and 
\,$\struc{\,\ov{V},\,<,\,\ov{V}\setminus \VV\,}$\,.
\end{definition_mit}%
It is straightforward to see the following:
\begin{remarks_ohne}[$\bs{<}$-isomorphy, $\bs{<}$-genericity]\label{remarks:kleiner-generic}\hspace{3cm}\index{isomorphic!${<}$-isomorphic}\index{generic!<generic@${<}$-generic}
\begin{enumerate}[(a)\ ]
\item
Two $\struc{\UU,\tau}$-structures $\A$ and $\B$ are ${<}$-isomorphic if and only if
there is a ${<}$-preserving mapping 
\,$\alpha:\adom(\A)\rightarrow \UU$\, such that 
\,$\alpha(\tau^{\A})=\tau^{\B}$.\\
Consequently, a Boolean query $\Q$ is ${<}$-generic on \,$\A = \struc{\UU,\tau^{\A}}$\, 
if and only if
\ $\Q\big(\struc{\UU,\tau^{\A}}\big) = \allowbreak\Q\big(\struc{\UU,\alpha(\tau^{\A})}\big)$ \  
for all ${<}$-preserving mappings \,{$\alpha:\adom(\A)\rightarrow \UU$}.  
\item
If, in particular, $\UU$ and $\VV$ are $\NN$ or $\ZZ$, then a mapping
\,$\alpha:U\rightarrow \VV$\, is \emph{${<}$-preserving} if and only if it is
\emph{strictly increasing}, i.e., \,$u<u'$\, iff \,$\alpha(u)<\alpha(u')$ 
\,(for all $u,u'\in U$). 
\mbox{ }\fertig
\end{enumerate}
\end{remarks_ohne}%
%
%
%
\subsection[Collapse Results for ${<}$-Generic Queries]{Collapse Results for $\bs{<}$-Generic Queries}\label{subsection:CollapseResults}
\begin{figure}[!htbp]
\begin{center}
\fbox{
\scalebox{0.6}{
\begin{pspicture}(-1.5,0)(6,4)
\rput(2,3.5){\large$\FO(<,\Num)$}%
\rput(0.25,2){\large$\FO(<)$}\rput(4,2){\large$\FOadom(<,\Num)$}%
\rput(2,0.5){\large$\FOadom(<)$}%
\rput{135}(1,1.25){\Large $\subseteq$}%
\rput{45}(3,1.25){\Large $\subseteq$}%
\rput{45}(1,2.75){\Large $\subseteq$}%
\rput{135}(3,2.75){\Large $\subseteq$}%
\end{pspicture} 
}
}
\caption{\small Expressive power of the above query languages.}\label{figure:Inclusion_Diagram_QueryLanguages}
\end{center}
\end{figure}
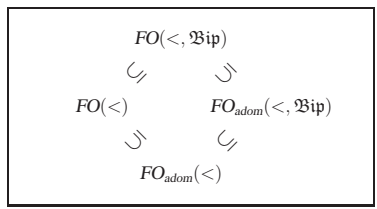

Given a context structure \,$\struc{\UU,<,\Num}$\, and a signature $\tau$ we will 
consider the following query languages: \index{query language}
Let $\Q$ be a Boolean query on $\struc{\UU,\tau}$-structures and let $\Cl$ be a class of 
$\struc{\UU,\tau}$-structures. We say that, on structures in $\Cl$, $\Q$ is expressible in
\begin{enumerate}[$\bullet$]
\item
  $\FO(<,\Num)$ \ iff \ there is a $\FO(<,\Num,\tau)$-sentence $\varphi$ such that
  \begin{eqnarray*}
    \struc{\,\UU,\,<,\,\Num,\,\tau^{\A}\,}\ \models\ \varphi
  & \mbox{ \ iff \ }
  & \Q(\A) = \mbox{``yes''} 
  \end{eqnarray*}
is true for all $\A = \struc{\UU,\tau^{\A}}$ in $\Cl$.
One speaks of \emph{natural semantics}, \index{natural semantics}
since quantification 
ranges, in the natural way, over the whole universe $\UU$.
If $\Num$ is empty we simply write $\FO(<)$.
\item
  active domain $\FO(<,\Num)$, for short: $\FOadom(<,\Num)$, \index{FOadom<Bip@$\FOadom(<,\Num)$}
  \ iff \ there is a 
  $\FO(<,\allowbreak\Num,\tau)$-sentence $\varphi$ such that
  \begin{eqnarray*}
    \struc{\,\adom(\A),\,<,\,\Num,\,\tau^{\A}\,}\,\models\,\varphi
  & \mbox{ \ iff \ }
  & \Q(\A) = \mbox{``yes''} 
  \end{eqnarray*}
is true for all $\A = \struc{\UU,\tau^{\A}}$ in $\Cl$.
One speaks of \emph{active domain semantics}, \index{active domain semantics}
since quantification 
is restricted to the active domain.
If $\Num$ is empty we simply write $\FOadom(<)$.
\end{enumerate}
It should be clear that all queries expressible in $\FOadom(<)$ are ${<}$-generic.
Figure~\ref{figure:Inclusion_Diagram_QueryLanguages} illustrates the obvious inclusions 
concerning the expressive power of the above query languages.
%
\\
It is an interesting question whether, for a particular context structure
\,$\struc{\UU,<,\allowbreak\Num}$\, and a particular class $\Cl$ of $\struc{\UU,\tau}$-structures
\begin{enumerate}[$\bullet$]
\item
  the predicates in $\Num$ allow to express {more} ${<}$-generic queries
  than the linear ordering alone
\item
  the quantification over all elements in the context universe $\UU$ allows to express
  {more} ${<}$-generic queries than the active domain quantification alone.
\end{enumerate}
One speaks of a \emph{collapse result} if the apparently stronger language is no more 
expressive than the apparently weaker one.\index{collapse result}
For the particular class $\Clfinite$ \index{C1fin@$\Clfinite$} 
of all structures whose active domain
is \emph{finite}, strong collapse results are known. A comprehensive overview of
such results can be found in \cite{BL00a,Libkin_DIMACS}.
Here we in particular want to mention the following:
\\
  In \cite{BDLW} it was shown that for all linearly ordered context universes $\UU$ and
  for the class $\Arb$ of \emph{arbitrary}, i.e., all, predicates on $\UU$, we have
  \begin{eqnarray*}
   \OrderGen\FOadom(<,\Arb) & = & \FOadom(<)
   \quad\mbox{on}\quad\Clfinite\mbox{ \ over \ }\UU\,.
  \end{eqnarray*}
  This means that every query $\Q$ that is ${<}$-generic on
  $\Clfinite$ and that is \allowbreak{}
  $\FOadom(<,\allowbreak \Arb)$-expressible on $\Clfinite$, is also $\FOadom(<)$-expressible on
  $\Clfinite$. I.e., when quantification is restricted to the active domain, arbitrary
  built-in predicates do not help first-order logic to express ${<}$-generic queries over 
  finite databases. 
  This result is known as the \emph{active generic collapse} (over finite databases).
  \index{active generic collapse}%
\par
  Also the so-called \emph{natural generic collapse} \index{natural generic collapse}
  has been investigated. Various
  different conditions on the context structure \,$\struc{\UU,<,\Num}$\, are known which
  guarantee that
  \begin{eqnarray*}
   \OrderGen\FO(<,\Rel) & = & \OrderGen\FO(<)
   \quad\mbox{on}\quad\Clfinite\mbox{ \ over \ }\UU\,.\footnotemark
  \end{eqnarray*}%
  \footnotetext{Considering $\Clfinite$, one even obtains the collapse to $\FOadom(<)$.}%
  In particular, Benedikt et al., \index{Benedikt, Michael}
  Belegradek et al., \index{Belegradek, Oleg V.}
  and Baldwin and Benedikt  \index{Baldwin, John T.}
  have shown that the natural generic collapse over finite databases 
  holds if\footnote{but not necessarily ``only if''}  the context structure is
  \emph{o-minimal} \cite{BDLW}, \index{o-minimal}
  has the \emph{Isolation Property} \cite{BST99}, \index{Isolation Property}
  or has
  \emph{finite Vapnik-Chervonenkis (VC) dimension} \cite{BB98} 
  (see also \cite{BaldwinBenedikt}).
  \index{Vapnik-Chervonenkis dimension|see{VC-dimension}}\index{VC-dimension}
  Context structures which satisfy (at least) one of these conditions are, for example,
  \,$\struc{\NN,<,+}$, \,$\struc{\QQ,<,+}$,
  \,$\struc{\RR,<,\allowbreak +,\times,\Exp}$, and
  \,$\struc{\UU,<,\Mon}$\, for any linearly ordered $\UU$ and the
  class $\Mon$ of all \emph{monadic} predicates on $\UU$ (cf., e.g., the
  survey \cite{BL00a}).
  Indeed, in \cite{BL00a} it is mentioned that the notion of 
  \emph{finite VC-dimension} coincides with the notion \emph{NIP} \index{NIP@\textit{NIP}}
  of structures that \emph{lack the 
  independence property},\footnote{%
      In fact, the correspondence between \emph{NIP} and 
      \emph{finite VC-dimension} easily follows from the definition of \emph{NIP} in
      \cite[Definition\;2.2]{BB98} and the definition of \emph{finite VC-dimension} as
      presented in Definition~\ref{definition:VC-dim}.%
  } and that these two notions include all context structures for which the 
  natural generic collapse over $\Clfinite$ is known by now (and, in particular, they
  include all o-minimal structures and all structures that have the Isolation Property).
  The following definition of \emph{finite VC-dimension} is basically taken from \cite{BL00a}:
\begin{definition_ohne}[Finite VC-Dimension]\label{definition:VC-dim}\index{VC-dimension}
Let \,$\struc{\UU,<,\Num}$\, be a context structure.
\begin{enumerate}[(a)\ ]
 \item
   Let \,$\varphi(\vek{x},\vek{y})$\, be a $\FO(<,\Num)$-formula, and let
   $\Nx$ and $\Ny$ be the lengths of the tuples $\vek{x}$ and $\vek{y}$, respectively.
   \begin{enumerate}[$\bullet$\ ] 
   \item
     For every \,$\vek{a}\in\UU^{\Ny}$\, the formula \,$\varphi(\vek{x},\vek{y})$\, defines
     the relation \\
       $
       R_{\varphi(\vek{x},\vek{a})}  \deff 
       \setc{\,\vek{b}\in\UU^{\Nx}\,}{\,\struc{\UU,<,\Num} \models \varphi(\vek{b},\vek{a})\,}.
       $ 
   \item
     The formula \,$\varphi(\vek{x},\vek{y})$\, defines
     the following family of relations on $\UU$: \\
$
       F_{\varphi(\vek{x},\vek{y})}  \deff 
       \setc{\,R_{\varphi(\vek{x},\vek{a})}\,}{\,\vek{a}\in\UU^{\Ny}\,}.
$ 
   \item 
     A set \,$B\subseteq \UU^{\Nx}$\, is \emph{shattered} by \,$F_{\varphi(\vek{x},\vek{y})}$ 
     \ iff \ \ 
     $\setc{B\cap R}{R\in F_{\varphi(\vek{x},\vek{y})}} = $
     $ \setc{X}{X\subseteq B}$.
     I.e., for every $X\subseteq B$ there is a $\vek{a}_X\in\UU^{\Ny}$ such that
     for all $\vek{b}\in B$ we have \\ 
$
       \vek{b}\in X  \mbox{ \ iff \ } 
       \struc{\UU,<,\Num} \models \varphi(\vek{b},\vek{a}_X)
$.
   \item
     The family \,$F_{\varphi(\vek{x},\vek{y})}$\, has \emph{finite VC-dimension} \ iff \
     there exists a number \,$m_{\varphi(\vek{x},\vek{y})}\in\NN$\, such that the following is true 
     for all \,$B\subseteq \UU^{\Nx}$: 
\\
     $
       \mbox{If \,$B$\, is shattered by \,$F_{\varphi(\vek{x},\vek{y})}$,\,
       then \,$|B|\leq m_{\varphi(\vek{x},\vek{y})}$.}
     $
   \end{enumerate}   
 \item
   $\struc{\UU,<,\Num}$\, has \emph{finite VC-dimension} if and only if  
   $F_{\varphi(\vek{x},\vek{y})}$\, has finite VC-dimension, for every $\FO(<,\Num)$-formula 
   $\varphi(\vek{x},\vek{y})$.
  \mbox{ } \fertig
\end{enumerate}
\end{definition_ohne}%
According to \cite{Libkin_DIMACS}, the following result of \cite{BB98,BaldwinBenedikt} is
the most general \emph{natural generic collapse} theorem that is known by now 
for the class $\Clfinite$ of all \emph{finite} databases.
\begin{theorem_mit}[Baldwin, Benedikt]\label{theorem:collapse_VCdim}
\index{Baldwin, John T.}\index{Benedikt, Michael} 
\mbox{ }\\
If \,$\struc{\UU,<,\Num}$\, is a context structure that has \emph{finite VC-dimension}\, then 
\\
 \(
   \OrderGen\FO(<,\Num) \ = \ \FOadom(<)
   \mbox{ \ on \ }\Clfinite\mbox{ \ over \ }\UU\,.
 \)%
\end{theorem_mit}%
On the other hand, it is straightforward to see (cf., e.g., \cite{BL00a}) 
that the natural generic collapse does {not} hold for \emph{all} context structures: 
\begin{facts_ohne}[No Collapse for $\bs{\struc{\NN,<,+,\times}}$ and $\bs{\struc{\NN,<,+,\Squares}}$]\label{facts:NoCollapse} \hspace{3cm}\rm
\begin{enumerate}[(a)\ ]
\item
 For the context structure \,$\struc{\NN,<,+,\times}$\, and any class 
 \,$\Cl\supseteq\Clfinite$\, we have \\
 \(
   \OrderGen\FO(<,+,\times) \ \neq \ \OrderGen\FO(<)
   \mbox{ on }\Cl\mbox{ over }\NN\,.
 \)%
 \\
 To see this consider the query 
   $\Q_{\class{even}}$ :  
   \emph{``Does the active domain have even cardinality?''}.
 Obviously, this query is ${<}$-generic on all structures.
 Furthermore, in Section~\ref{subsection:DBsAndQueries} we already saw that $\Q_{\class{even}}$
 is expressible in $\FO(<,+,\times)$ but not in $\FO(<)$ over $\NN$.
\item\rm
 For the context structure \,$\struc{\NN,<,+,\Squares}$, where
 \,$\Squares\deff \setc{n^2}{n\in\NN}$,\, 
 we have
 \(
   \OrderGen\FO(<,+,\Squares)\ \neq\ \OrderGen\FO(<)
   \mbox{ on } \Cl\mbox{\ over\ }\NN\,.
 \)%
 \\
 To see this use (a) and recall that $\times$ is first-order definable in $\struc{\NN,<,+,\Squares}$ 
 (cf., e.g., the survey \cite{Bes}).
 \mbox{} \fertig
\end{enumerate}
\end{facts_ohne}%
%
%
The collapse results mentioned so far all deal with the class $\Clfinite$ of databases
whose active domain is \emph{finite}.
Belegradek et al.\ \cite{BST99} \index{Belegradek, Oleg V.}
investigated \emph{finitely representable} \index{finitely representable}%
databases, i.e., databases whose relations, essentially, consist of a {finite} number
of multidimensional rectangles in the context universe $\UU$. They showed, for
every context structure \,$\struc{\UU,<,\Rel}$,\, that a natural generic collapse
on $\Clfinite$ over $\UU$ can be lifted to a natural generic collapse on the larger
class $\Clfinrep$ \index{C1finrep@$\Clfinrep$}
of all finitely representable databases over $\UU$.
We will further concentrate on this result in Section~\ref{section:Lift}.
\par
But what happens for the class $\Clarb$ \index{C1arb@$\Clarb$} 
of arbitrary, i.e., \emph{all}, structures?
Can collapse results be lifted from $\Clfinite$ to $\Clarb$? --- Not in general!
Recall from the (already mentioned) result of \cite{BDLW} for
o-minimal structures that
\begin{eqnarray*}
  \OrderGen\FO(<,+) & = & \OrderGen\FO(<)\quad
  \mbox{on}\quad \Clfinite\mbox{ \ over \ }\QQ\,.
\end{eqnarray*}
However, in \cite[Theorems~3.3~and~3.4]{BST99} it was shown that 
\begin{eqnarray*}
  \OrderGen\FO(<,+) & \neq & \OrderGen\FO(<)\quad
  \mbox{on}\quad \Clarb\mbox{ \ over \ }\QQ\,.
\end{eqnarray*}
%
%
I.e., the natural generic collapse is valid for \emph{finite} but not for 
\emph{arbitrary} databases over the context structure $\struc{\QQ,<,+}$.
\\
On the other hand, in \cite{LS_STACS01} it was shown that
the collapse does hold for \emph{arbitrary} databases over the context structure
\,$\struc{\NN,<,\allowbreak +}$. 
To the author's knowledge this is the only collapse result known so far for the class
$\Clarb$ of \emph{arbitrary} databases, and there are no publications
other than \cite{LS_STACS01,Schweikardt_CSL01} that show the 
{natural generic collapse} for classes
of databases larger than $\Clfinite$ and $\Clfinrep$.
In the subsequent sections of this paper we will obtain these and other collapse results for
such larger classes of databases.
Precisely, our collapse results are of the following kind:
\begin{definition_mit}[Collapse Result]\label{definition:Natural-Generic-Collapse}
\index{collapse result}
\mbox{ }\\
Let \,$\struc{\UU,<,\Num}$\, be a context structure, and 
let $\Cl$ be a class of structures over the universe $\UU$.
We write \index{generic!<generic@${<}$-generic}\index{FOadom<@$\FOadom(<)$} 
\begin{quote} 
$\OrderGenFO(<,\Num) \ \; = \; \ \FOadom(<)$\quad on\quad $\Cl$ \ over \ $\UU$
\end{quote}
if and only if the following is true: 
\begin{quote} 
For every signature\footnote{Recall that signatures $\tau$ always consist of a 
\emph{finite} number of relation symbols and constant symbols.}
$\tau$ and every $\FO(<,\Num,\tau)$-sentence $\varphi$ there
is a $\FO(<,\tau)$-sentence $\varphi'$ such that
\begin{eqnarray*}
   \bigstruc{\,\UU,\,<,\,\Num,\,\tau^{\A}\,}\,\models\,\varphi
 & \mbox{ iff }
 & \bigstruc{\,\adom(\A),\,<,\,\tau^{\A}\,}\,\models\,\varphi'
\end{eqnarray*}
is true for all $\struc{\UU,\tau}$-structures $\A\in\Cl$ on which the query defined by
$\varphi$ is ${<}$-generic.
For convenience we will henceforth say \emph{$\varphi$ is ${<}$-generic on $\A$} to 
express that \emph{the query defined by $\varphi$ is ${<}$-generic on $\A$}.
\end{quote}
The collapse result for any logic $F$ other than $\FO$ is defined in the analogous way, 
replacing $\FO$ with $F$ in the above definition.
\end{definition_mit}%
Let us mention a technical detail:
The ``traditional'' definition of collapse for the class $\Clfinite$ states the following:
If a sentence $\varphi\in\FO(<,\Rel,\tau)$ is ${<}$-generic on $\Clfinite$, then it
can be replaced by a $\varphi'\in\FO(<,\tau)$ that is equivalent to $\varphi$ on
$\Clfinite$. Just replacing $\Clfinite$ with $\Clarb$ in this definition would 
reduce the set of formulas $\varphi$ to which the collapse applies, because there
certainly are formulas $\varphi$ that are ${<}$-generic on $\Clfinite$ but not on
$\Clarb$.
The above Definition~\ref{definition:Natural-Generic-Collapse} circumvents this
problem by stating that {any} $\varphi$ can be replaced by a $\varphi'$ that is
equivalent to $\varphi$ on all databases {on which} $\varphi$ is ${<}$-generic.
\\
\parno
We will in particular deal with the following classes of databases:
\begin{definition_mit}[finite, $\NN$-embeddable, $\ZZ$-embeddable]\label{definition:NN-embeddable}
\index{finite (database/structure)}\index{N-embeddable@$\NN$-embeddable}
\index{Z-embeddable@$\ZZ$-embeddable} 
\mbox{ }\\
Let $\struc{\UU,<}$ be a linearly ordered structure and let $\struc{\ov{\UU},<}$ be its
Dedekind completion.
Let $\tau$ be a signature.
A $\struc{\UU,\tau}$-structure $\A$ is called
\begin{enumerate}[$\bullet$\ ]
      \item
        \emph{finite} \,iff\, $\adom(\A)$ is finite.
      \item
        \emph{$\NN$-embeddable} \index{N-embeddable@$\NN$-embeddable}%
        \,iff\, 
        there is a ${<}$-preserving mapping
        \,$\alpha:\adom(\A)\rightarrow\NN$.
        I.e., $\adom(\A)$ is finite or $\adom(\A)$ is of the form 
        \,$\set{a_1<a_2<\cdots}$\, and has
        no accumulation points in $\ov{\UU}$. 
        In particular, all $\struc{\NN,\tau}$-structures are $\NN$-embeddable.
      \item
        \emph{$\ZZ$-embeddable} \index{Z-embeddable@$\ZZ$-embeddable}%
        \,iff\, there is a ${<}$-preserving mapping
        \,$\alpha:\adom(\A)\rightarrow\ZZ$.
        I.e., $\adom(\A)$ has no accumulation points and is $\NN$-embeddable or of one of 
        the forms \,$\set{\,\cdots < a_{-2}<a_{-1}<a_1<a_2<\cdots\,}$\, or 
        \,$\set{\,a_{-1}>a_{-2}>\cdots\,}$\,. 
        In particular, all $\struc{\ZZ,\tau}$-structures are $\ZZ$-embeddable.
\end{enumerate}%
We use $\Clfinite$, $\ClNemb$, $\ClZemb$, and $\Clarb$, respectively, 
\index{C1fin@$\Clfinite$}\index{C1Nemb@$\ClNemb$}\index{C1Zemb@$\ClZemb$}\index{C1arb@$\Clarb$}
to denote the
classes of all finite, $\NN$-embeddable, $\ZZ$-embeddable, and arbitrary (i.e., all)
structures, respectively.
\end{definition_mit}%
%
%
%
\subsection{Collapse Results Obtained in this Paper}\label{subsection:CollapseResultsPaper}
In Section~\ref{section:Mon} we will consider 
context structures which have as built-in predicates the class $\Mon$ of all
\emph{monadic}, i.e., unary, relations over the context universe. Our result is
\begin{enumerate}[$\bullet$]
\item
  $\OrderGen\FO(<,\Mon) \ = \ \FOadom(<)$\, on \,$\ClZemb$\, over \,$\UU$, \\
  for any linearly ordered infinite context universe $\UU$.
\end{enumerate}
In particular, for $\UU=\NN$ and $\UU=\ZZ$ this implies the collapse over $\Clarb$.
\par
In Section~\ref{section:FOPlus_game} we will investigate context structures with built-in 
\emph{addition} $+$,
and we will prove the result of \cite{LS_STACS01} and several extensions of that result.
Precisely, we will expose an infinite set $Q\subseteq \NN$ 
(which is not $\FO(<,+)$-definable) and show, for the class $\Mon_Q$ of all subsets of $Q$, that
\begin{enumerate}[$\bullet$]
\item \index{FO<+QMonQ@$\FO(<,+,Q,\Mon_Q)$} 
  $\OrderGen\FO(<,+,Q,\Mon_Q) \ = \ \FOadom(<)$\, on \,$\Clarb$\, over \,$\NN$, \ and
\item \index{FO<+QMonQGroups@$\FO(<,+,Q,\Mon_Q,\Groups)$}  
  $\OrderGen\FO(<,+,Q,\Mon_Q,\Groups) \ = \ \FOadom(<)$\, on \,$\ClNemb$\, over \,$\RR$,\,  
  where $\Groups$ is the class of all subsets of $\RR$ that contain the number $1$ and that
  are groups with respect to $+$.
\end{enumerate}
In particular, this implies the natural generic collapse on $\ClNemb$ for the 
context structures \,$\struc{\NN,<,\allowbreak +,Q}$, \,$\struc{\QQ,<,+}$, \,$\struc{\QQ,<,+,\ZZ}$,\, and
\,$\struc{\RR,<,+,\ZZ,\QQ}$.
The collapse for the context structure \,$\struc{\NN,<,+,Q}$\, is remarkable since we
know from Fact~\ref{facts:NoCollapse}\;(b) that the collapse does \emph{not} hold when 
replacing the set $Q$ with the set $\Squares$ of all square numbers.
\par
In Section~\ref{section:BCEFO} we will look at the restriction of first-order logic
to the class $\BCeFO$, i.e., to Boolean combinations of purely \emph{existential}
$\FO$-formulas. As built-in predicates we will consider the class $\Arb$ of
\emph{arbitrary}, i.e., all, relations. We will show that
\begin{enumerate}[$\bullet$]
\item
  $\OrderGen\BCeFO(<,\Arb) \ = \ \BCeFOadom(<)$\, on \,$\ClNemb$\, over \,$\UU$,\\
  for any linearly ordered infinite context universe $\UU$.
\end{enumerate}
In particular, for $\UU=\NN$ this implies the collapse on $\Clarb$.
\par
In Section~\ref{section:Lift} we will present the result from 
\cite{Schweikardt_CSL01} which, in the spirit of
\cite{BST99}'s lifting \index{lifting theorem}
from $\Clfinite$ to $\Clfinrep$, \index{C1finrep@$\Clfinrep$} 
allows to lift collapse
results from $\ClNemb$ to a class $\ClNrep$ \index{C1Nrep@$\ClNrep$}
that is a proper extension of the class $\Clfinrep$.
\par
The proof method used in this paper for obtaining the collapse results over 
$\ClNemb$ and $\ClZemb$ is considerably different from the
methods used so far for proving collapse results in database theory:
\begin{enumerate}[$\bullet$]
\item
 The proofs in \cite{BDLW,BST99,BB98} are via model theory and use
 non-standard, hyperfinite structures. So far, no \emph{elementary} proof of 
 Theorem~\ref{theorem:collapse_VCdim}, stating that the collapse is valid over all 
 context structures that have \emph{finite VC-dimension}, is known.
\item
 An elementary and constructive proof of the results of \cite{BDLW} for
 \emph{o-minimal} context structures was given by Benedikt \index{Benedikt, Michael}
 and Libkin \index{Libkin, Leonid} in \cite{BL00b} (see\ also \cite{BL00a,Libkin_DIMACS}).
 There, the \emph{natural generic collapse} over \emph{o-minimal} context structures is 
 proved by a combination of the \emph{natural active collapse} and the 
 \emph{active generic collapse}.
 In \cite[Proposition\,6.10]{Libkin_DIMACS} also an elementary proof for the particular 
 (non o-minimal) context structure $\struc{\NN,<,+}$ is sketched.
\end{enumerate}
In Section~\ref{section:EF-game} we will present a specific 
notion of the \emph{tranlation of strategies for the Ehrenfeucht-Fra\"\i{}ss\'{e} game} 
 which allows us to prove collapse results. 
Apart from the collapse results obtained with this method, 
the exposition of explicit strategies for the Ehrenfeucht-Fra\"\i{}ss\'{e} game is interesting 
in its own right.
Let us emphasize that the present paper investigates collapse results from the
point of view of \emph{mathematical logic}. That is, we want to gain a deeper understanding
of the expressive power, or the expressive weakness, of first-order logic with
certain built-in predicates, and we want to construct explicit winning strategies for the
Ehrenfeucht-Fra\"\i{}ss\'{e} game in the presence of built-in predicates.
\\
Those readers who are mainly interested in database theory or computer science
as such, may have the objection that $\NN$-embeddable structures in general cannot be 
represented in the \emph{finite} and thus cannot be used as input for an algorithm.
In this context we want to mention a line of research that considers 
\emph{recursive structures} \index{recursive structures}
\cite{Harel98}, i.e., structures where every relation
is computable by an algorithm that decides whether or not an input tuple belongs to the
respective relation. 
Of course, our collapse results for the classes $\Clarb$ or $\ClNemb$ are still 
applicable when restricting attention to {recursive structures} in $\Clarb$ or $\ClNemb$.
\label{subsection:CollapseResultsPaper:Ende}%
%
%
%



%
\section{An 
Ehrenfeucht-Fra\"\i{}ss\'{e} Game Approach}\label{section:EF-game}
\index{Ehrenfeucht-Fra\"\i{}ss\'{e} game}\index{EF-game|see{Ehrenfeucht-Fra\"\i{}ss\'{e} game}}
\index{game|see{Ehrenfeucht-Fra\"\i{}ss\'{e} game}}
%
%
In this section we present  
the \emph{translation of strategies for the Ehrenfeucht-Fra\"\i{}ss\'{e} game} 
as a method for proving 
collapse results in database theory.
We show that, in principle, all collapse results of the
kind fixed in Definition~\ref{definition:Natural-Generic-Collapse} can be proved via 
Ehrenfeucht-Fra\"\i{}ss\'{e} games.
%
%
%
 
%
\subsection{The Ehrenfeucht-Fra\"\i{}ss\'{e} Game for $\FO$}\label{subsection:EF-game}
Ehrenfeucht-Fra\"\i{}ss\'{e} games, for short: \emph{EF-games}, 
\index{Ehrenfeucht-Fra\"\i{}ss\'{e} game}\index{EF-game}
were invented by Ehrenfeucht \index{Ehrenfeucht, A.}
and Fra\"\i{}ss\'{e} \index{Fra\"\i{}ss\'{e}, R.}
in \cite{Ehrenfeucht,Fraisse}. These combinatorial games
are particularly useful for investigating what can, and what cannot,
be expressed in various logics. A well-written survey on EF-games is, e.g., given by
Fagin \index{Fagin, Ronald} in \cite{Fagin_EasierWays}. 
More details can be found in the textbooks \cite{Immerman,EbbinghausFlum}. 
In the present section we will concentrate on the classical, first-order $r$-round
EF-game, which is defined as follows.
%
%
\index{rround EF-game@$r$-round EF-game ($\dwins_r$)}
\index{Ehrenfeucht-Fra\"\i{}ss\'{e} game!rround EF-game@$r$-round EF-game ($\dwins_r$)}
\par
Let $\tau$ be a signature and let $r$ be a natural number.
The \emph{$r$-round EF-game} 
is played by two players, \emph{the spoiler} \index{spoiler}
and \emph{the duplicator}, \index{duplicator}
on two $\tau$-structures $\A$ and $\B$.
The spoiler's intention is to show a difference between the two
structures, while the duplicator tries to make them look alike.\\
There is a fixed number $r$ of rounds. 
Each round $i\in\set{1,\twodots,r}$ is played as follows:
First, the spoiler chooses either an element $a_i$ in the universe of $\A$
or an element $b_i$ in the universe of $\B$. Afterwards, the
duplicator chooses an element in the other structure. I.e., she
chooses either an element $b_i$ in the universe of $\B$, if the
spoiler's move was in $\A$, or an element $a_i$ in the universe of
$\A$, if the spoiler's move was in $\B$.
After $r$ rounds the game finishes with elements $a_1,\twodots,a_r$
chosen in $\A$ and $b_1,\twodots,b_r$ chosen in $\B$. 
\par
\emph{The duplicator has won the game} if, \index{winning condition}
restricted to the chosen elements and the interpretations of the
constant symbols, the structures $\A$ and $\B$ are indistinguishable
with respect to $\set{=}\cup\tau$. Precisely, this means that
the mapping $\pi$ defined via
\[
  \pi \ : \ 
    \left\{
       \begin{array}{rcll}
          c^{\A} & \mapsto & c^{\B} & \mbox{for all constant symbols $c\in\tau$} \\
           a_i   & \mapsto & b_i    & \mbox{for all $i\in\set{1,\twodots,r}$}
       \end{array}
    \right\}
\]
is a \emph{partial isomorphism}
between $\A$ and $\B$. 
Otherwise, \emph{the spoiler has won the game}.
\par
Since the game is finite, one of the two players must have a
\emph{winning strategy}, \index{winning strategy}
i.e., he or she can always win the game, no
matter how the other player plays.
We say that \emph{the duplicator wins the $r$-round EF-game on $\A$ and $\B$} 
and we write \,$\A\dwins_r\B$ \ 
\index{$\dwins_r$ ($\A\dwins_r\B$)}%
iff
the duplicator has a winning strategy in the
$r$-round EF-game on $\A$ and $\B$.
It is straightforward to see that, for every signature $\tau$, the relation $\dwins_r$ is an
{equivalence relation} on the set of all $\tau$-structures.
%
%
\par
The fundamental use of the game comes from the fact that it
characterizes first-order logic as follows (cf., e.g.,
\cite{Fagin_EasierWays,Immerman,EbbinghausFlum}): 
\begin{theorem_ohne}[Ehrenfeucht, Fra\"\i{}ss\'{e}]\label{theorem:E-F}
Let $\tau$ be a signature.
\begin{enumerate}[(a)\ ]
 \item\label{theorem:E-F:structure}  
   Let $r\in\NN$ and let $\A$ and $\B$ be $\tau$-structures.
   $\A \dwins_r \B$ if and only if $\A$ and $\B$ satisfy the same
   $\FO(\tau)$-sentences of quantifier depth at most $r$.
 \item\label{theorem:E-F:class}
   Let $\K$ be a class of $\tau$-structures and let $\LL\subseteq \K$. 
   The following are equivalent:
   \begin{enumerate}[(i)]
    \item
      $\LL$ is \emph{not} $\FO(\tau)$-definable in $\K$, i.e., there is
      no $\FO(\tau)$-sentence $\varphi$ such that ``$\A\models\varphi$
      \ iff \ $\A\in \LL$'' is true for all $\A\in\K$.
    \item 
      For each $r\in\NN$ there are $\A,\B\in\K$ such that
      $\A\in \LL$, $\B\not\in \LL$, and $\A \dwins_r \B$.
      \mbox{ }\fertig
   \end{enumerate}  
\end{enumerate}
\end{theorem_ohne}%
\begin{remark_mit}\label{remark:E-F-classes}\rm
It is well-known (cf., e.g., \cite[Exercise~6.11]{Immerman}) that for
a fixed (finite) signature $\tau$ there are only \emph{finitely} many 
inequivalent $\FO(\tau)$-sentences of quantifier depth at most $r$.
Consequently, due to Theorem~\ref{theorem:E-F}\,(\ref{theorem:E-F:structure}),
the relation $\dwins_r$ has only \emph{finitely} many equivalence classes 
on the \index{equivalence classes}
set of all $\tau$-structures --- and each equivalence class can be defined by
a $\FO(\tau)$-sentence of quantifier depth at most $r$.
More precisely: Let $c=c(r,\tau)\in\NN$ be the number of equivalence classes.
There are $\FO(\tau)$-sentences $\varphi_1,\twodots,\varphi_c$ of quantifier depth
at most $r$, such that 
\begin{enumerate}[$\bullet$]
 \item
   each $\tau$-structure $\A$ satisfies \emph{exactly one} of the sentences
   $\varphi_1,\twodots,\varphi_c$,\quad and
 \item
   two $\tau$-structures $\A$ and $\B$ satisfy the same sentence from 
   $\varphi_1,\twodots,\varphi_c$ if and only if $\A\dwins_r\B$.
\end{enumerate}
The formulas defining the equivalence classes are also known as 
\emph{Hintikka formulas}. \index{Hintikka formula}
\end{remark_mit}%
\subsection{Using EF-Games for Collapse Results}\label{section:DB_EF-game}
\index{collapse result!EF-game}
\begin{definition_ohne}[Translation of Strategies]\label{definition:TranslatingStrategies}
\index{translation of strategies!on $\bs{\Cl}$ over $\bs{\UU}$}
\mbox{}\\
Let \,$\struc{\UU,<,\Num}$\, be a context structure
and let $\Cl$ be a class of structures over the universe $\UU$.
We say that
\begin{quote}
the duplicator can translate strategies for the $\FOKleiner$-game into 
strategies for the $\FONum$-game on $\Cl$ over $\UU$
\index{Ehrenfeucht-Fra\"\i{}ss\'{e} game!FOadom<@$\FOadom(<)$-game}
\index{Ehrenfeucht-Fra\"\i{}ss\'{e} game!FO<Bip@$\FO(<,\Bip)$-game}
\end{quote}
if and only if the following is true:
\begin{quote}
   For every finite set \,$\Num'\subseteq\Num$,\, for every
   signature\footnote{Recall that signatures $\tau$ always consist of a {finite} number
   of relation symbols and constant symbols.} 
   $\tau$,\, and 
   for every number $k\in\NN$ 
   there is a number $r(k)\in\NN$ 
   such that the following is true for all
   $\struc{\UU,\tau}$-structures $\A,\B\in\Cl$:
   If the duplicator wins the $r(k)$-round
   $\FOKleiner$-game on $\A$ and $\B$, i.e., if 
   \,$\struc{\adom(\A), <, \tau^{\A}} 
      \dwins_{r(k)} 
      \struc{\adom(\B), <, \tau^{\B}}$,\, 
   then there are $<$-preserving mappings \,$\alpha : \adom(\A)
   \rightarrow \UU$\, and \,$\beta : \adom(\B) \rightarrow \UU$\, such that
   the duplicator wins the $k$-round
   $\FONumi$-game on $\alpha(\A)$ and $\beta(\B)$, i.e., 
    $\struc{\UU, <, \Num', \alpha\big(\tau^{\A}\big)} 
      \dwins_{k} 
     \struc{\UU, <, \allowbreak \Num', \beta\big(\tau^{\B}\big)}$.
\end{quote}
\mbox{ }\vspace{-5ex}\\ \mbox{}\fertig
\end{definition_ohne}%
\noindent{}Due to the specific notion of \emph{collapse result} fixed in 
Definition~\ref{definition:Natural-Generic-Collapse}, we obtain that {all}
collapse results can be proved via the
translation of strategies:
\begin{theorem_ohne}[Translation of Strategies $\bs{\Leftrightarrow}$ Collapse Result]\label{theorem:EF-Collapse}
\mbox{ }\\%
Let \,$\struc{\UU,<,\Num}$\, be a context structure, 
and let $\Cl$ be a class of structures over the universe $\UU$.
The following are equivalent:
\begin{enumerate}[(a)\ ]
 \item 
   The duplicator can translate strategies for the $\FOKleiner$-game into 
   strategies for the 
   \linebreak[4] 
   $\FONum$-game on $\Cl$ over $\UU$.
 \item  
    $\OrderGenFO(<,\Num) \, = \, \FOadom(<)$\, on \,$\Cl$\, over \,$\UU$. 
    \mbox{ }\fertig
\end{enumerate} 
\end{theorem_ohne}%
%
%
\begin{proof_mit}
\emph{(a)$\Rightarrow$(b): }
Let $\tau$ be a signature, let $\varphi$ be a $\FO(<,\Num,\tau)$-sentence, and
let $\K$ be the set of all $\struc{\UU,\tau}$-structures in $\Cl$ on which $\varphi$ is
${<}$-generic. 
\\
We need to show that there is a $\FO(<,\tau)$-sentence $\varphi'$ such that
\begin{eqnarray*}
   \struc{\UU,\, <,\,\Num,\,\tau^{\A}}\,\models\,\varphi
 & \mbox{ iff }
 & \struc{\adom(\A),\,<,\,\tau^{\A}}\,\models\,\varphi'
\end{eqnarray*}
is true for all structures $\A=\struc{\UU,\tau^{\A}}$ in $\K$.
\emph{For the sake of contradiction, we assume that such a $\FO(<,\tau)$-sentence
$\varphi'$ does not exist.}
This means that the class
\begin{eqnarray*}
 \LL' & \deff &
 \bigsetc{\,\bigstruc{\,\adom(\A),\,<,\,\tau^{\A}\,}}{\A\in\K \mbox{ and } 
    \bigstruc{\,\UU,\,<,\,\Num,\,\tau^{\A}\,}\models\varphi\,}
\end{eqnarray*}
is \emph{not} $\FO(<,\tau)$-definable in 
$\K'\deff\bigsetc{\bigstruc{\,\adom(\A),<,\tau^{\A}\,}}{\A\in\K}$.
Hence, for every $r\in\NN$, Theorem~\ref{theorem:E-F}\,(\ref{theorem:E-F:class}) gives us
structures $\A'_r,\B'_r\in\K'$ such that $\A'_r\in \LL'$, $\B'_r\not\in \LL'$, and 
$\A'_r \dwins_r \B'_r$.
I.e., for every $r\in\NN$, there are structures $\A_r,\B_r\in\K$ such that
$\struc{\UU,<,\Num,\tau^{\A_r}}\models \varphi$,
$\struc{\UU,<,\allowbreak \Num,\tau^{\B_r}}\not\models \varphi$, and
$\struc{\adom(\A_r),<,\allowbreak \tau^{\A_r}}\dwins_r \struc{\adom(\B_r),<,\tau^{\B_r}}$. 
\par
Let us now make use of the presumption  
that the duplicator can translate strategies for the $\FOKleiner$-game into 
strategies for the $\FONum$-game on structures in $\Cl$. 
Let $\Num'$ be the finite set of relations from $\Num$ that occur in $\varphi$, 
let $k$ be the quantifier depth of $\varphi$, and 
let $r\deff r(k)$ be chosen according to 
Definition~\ref{definition:TranslatingStrategies}.
Thus, there are ${<}$-preserving mappings $\alpha:\adom(\A_r)\rightarrow \UU$ and
$\beta:\adom(\B_r)\rightarrow \UU$ such that\quad
$\bigstruc{\UU,<,\Num',\alpha\big(\tau^{\A_r}\big)} 
 \dwins_{k} 
 \bigstruc{\UU,<,\Num',\beta\big(\tau^{\B_r}\big)}
$.
\\
However, since $\varphi$ is ${<}$-generic on $\A$ and on $\B$, we have that
\[
  \bigstruc{\,\UU,\,<,\,\Num',\,\alpha\big(\tau^{\A_r}\big)\,}\ \models\ \varphi
  \quad\mbox{ and }\quad
  \bigstruc{\,\UU,\,<,\,\Num',\,\beta\big(\tau^{\B_r}\big)\,}\ \not\models\ \varphi\,. 
\]
This is a 
contradiction to Theorem~\ref{theorem:E-F}\,(\ref{theorem:E-F:structure}) which states
that structures that are equivalent with respect to $\dwins_k$ do satisfy the same 
first-order sentences of quantifier depth $k$.
\\
Altogether, the proof of \,``\emph{(a)}$\Rightarrow$\emph{(b)}''\, is complete.
\\
\par
\emph{(b)$\Rightarrow$(a): } 
Let $\Num'$ be a finite subset of $\Num$, let $\tau$ be a signature, and let
$k\in\NN$. 
From Remark~\ref{remark:E-F-classes} we know that the relation $\dwins_k$ has only
a finite number $c\in\NN$ of equivalence classes on the set of all
$(<,\Num',\tau)$-structures; and 
these equivalence classes can be described by $\FO(<,\allowbreak \Num',\tau)$-sentences
$\varphi_1,\twodots,\varphi_c$ of quantifier depth at most $k$. I.e., 
   each structure $\A$ satisfies \emph{exactly one} of the sentences
   $\varphi_1,\twodots,\varphi_c$, and
   two structures $\A$ and $\B$ satisfy the same sentence from 
   $\varphi_1,\twodots,\varphi_c$ \,iff\, $\A\dwins_k\B$.
\\
We will consider all possible disjunctions of the formulas $\varphi_i$. 
I.e., for each \,$I\subseteq\set{1,\twodots,c}$\, we define 
\,$\varphi_I\deff\Oder_{i\in I}\varphi_i$.
\\
From the presumption we know that 
$\OrderGenFO(<,\Num)\, = \,\FOadom(<)$\, on \,$\Cl$\, over \,$\UU$.  
I.e., for each sentence $\varphi_I$ there is a
$\FO(<,\tau)$-sentence $\varphi'_I$ such that
\begin{eqnarray*}
   (*)\qquad \bigstruc{\,\UU,\,<,\,\Num,\,\tau^{\A}\,}\,\models\,\varphi_I
 & \mbox{ iff }
 & \bigstruc{\,\adom(\A),\,<,\,\tau^{\A}\,}\,\models\,\varphi'_I
\end{eqnarray*}
is true for all $\struc{\UU,\tau}$-structures $\A\in\Cl$ on which 
$\varphi_I$ is ${<}$-generic.
\\
Choose $r(k)\in\NN$ to be the maximum quantifier depth of the sentences 
$\varphi_I'$.
Let $\A = \struc{\UU,\tau^{\A}}$ and $\B= \struc{\UU,\tau^{\B}}$ be structures in $\Cl$ 
with\quad
$\bigstruc{\adom(\A),<,\tau^{\A}} 
 \dwins_{r(k)} 
 \bigstruc{\adom(\B),<,\tau^{\B}}$.
Our aim is now to
find $<$-preserving mappings $\alpha : \adom(\A)
   \rightarrow \UU$ and $\beta : \adom(\B) \rightarrow \UU$ such that\quad
$\bigstruc{\UU,<,\Num',\alpha\big(\tau^{\A}\big)} 
 \dwins_{k} 
 \bigstruc{\UU,<,\Num',\beta\big(\tau^{\B}\big)}$.
\\
To this end, let $I$ be the set of all those \,$i\in\set{1,\twodots,c}$\, for which there exists
a ${<}$-preserving mapping $\alpha_i:\adom(\A)\rightarrow\UU$ such that\quad
\(
  \bigstruc{\UU,<,\Num',\alpha_i\big(\tau^{\A}\big)} \models \varphi_i
\).
\\
Furthermore, let $J$ be the according set for $\B$ instead of $\A$.
\par
If $I\cap J\neq \emptyset$, then there exists an $i\in\set{1,\twodots,c}$ and 
${<}$-preserving mappings $\alpha:\adom(\A)\rightarrow \UU$ and $\beta:\adom(\B)
\rightarrow\UU$ such that
\,$\bigstruc{\UU,<,\Num',\alpha\big(\tau^{\A}\big)}\ \models \varphi_i$\, and 
\,$\bigstruc{\UU,<,\Num',\beta\big(\tau^{\B}\big)}\ \models \varphi_i$.\, 
From the choice of $\varphi_1,\twodots,\varphi_c$ we know that
\,$\bigstruc{\UU,<,\Num',\alpha\big(\tau^{\A}\big)}$\, and
\,$\langle\UU,<,\Num',\beta\big(\tau^{\B}\big)\rangle$\, must belong to the same
equivalence class of $\dwins_k$. I.e.,
$\bigstruc{\UU,<,\Num',\alpha\big(\tau^{\A}\big)} 
 \dwins_{k} 
 \bigstruc{\UU,<,\allowbreak \Num',\beta\big(\tau^{\B}\big)}$.
\\
All that remains to show is that indeed $I\cap J\neq\emptyset$.
\par
\emph{For the sake of contradiction, let us assume that $I\cap J=\emptyset$.}
Note that the set $I$ is defined in such a way that the formula $\varphi_I$ is
${<}$-generic on $\A$. Furthermore, if $I\cap J = \emptyset$, then $\varphi_I$ is
${<}$-generic on $\B$, too, and we have\quad
$\bigstruc{\UU,<,\Num',\tau^{\A}}\models \varphi_I$ \ and \ 
$\bigstruc{\UU,<,\Num',\tau^{\B}}\not\models \varphi_I$.
Thus, from $(*)$ we obtain a $\FO(<,\tau)$-formula $\varphi'_I$ of quantifier depth
at most $r(k)$, such that\quad
$\bigstruc{\adom(\A),<,\tau^{\A}}\models \varphi'_I$ 
\ and \ 
$\bigstruc{\adom(\B),<,\tau^{\B}}\not\models \varphi'_I$.
However, $\A$ and $\B$ were chosen in such a way that \ 
$\bigstruc{\adom(\A),<,\tau^{\A}} 
 \dwins_{r(k)} 
 \bigstruc{\adom(\B),<,\tau^{\B}}$, \ 
which is a contradiction to Theorem~\ref{theorem:E-F}\,(\ref{theorem:E-F:structure}).
Altogether, this completes the proof of Theorem~\ref{theorem:EF-Collapse}.
\end{proof_mit}%
\\
\parno
In the following two sections we will show how the duplicator can translate
strategies for the $\FOKleiner$-game into strategies for the
$\FO(<,\allowbreak +)$-game and the
$\FOMon$-game, where $\Mon$ is the class of all
\emph{monadic} relations.
Via Theorem~\ref{theorem:EF-Collapse} 
these translations of strategies will directly give
us the according collapse results.
Apart from the results themselves, the exposition of explicit strategies for 
the EF-game will be interesting in its own right.
%
%
%
\subsection{A Lemma Useful for the Sections~\ref{section:Monadic} 
  and \ref{section:Addition}}\label{section:Useful_Lemma}
Before concentrating on the translation proofs for $\FO(<,\Mon)$ and $\FO(<,+)$,
we first show the following easy lemma that will help us avoid some annoying
case distinctions within our proofs.
\begin{lemma_mit}\label{lemma:nuetzlich}
Let \,$P \deff \set{p_1<p_2<p_3<\cdots}$\, be a countable, infinitely
increasing sequence.
Let $\tau$ be a signature, and let $\A$ and $\B$ be two $\NN$-embeddable 
$\tau$-structures over linearly ordered universes.
Furthermore, let \,$\alpha:\adom(\A)\rightarrow P$\, and
\,$\beta:\adom(\B)\rightarrow P$\,
map, for every $j$, the $j$-th smallest
element in $\adom(\A)$ and $\adom(\B)$, respectively, onto the position $p_j$.
Let $r\in\NN$ and $r\geq 2$.
\\
If \ 
$\struc{\adom(\A),<,\tau^{\A}}  \dwins_r 
  \struc{\adom(\B),<,\tau^{\B}}$, \ 
then also \
$\Abig \deff \struc{P,<,\alpha(\tau^{\A})}  \dwins_r 
  \struc{P,<,\beta(\tau^{\B})} \ffed \Bbig$. 
\end{lemma_mit}%
\begin{proof_mit}
Since $r\geq 2$, one can easily seee that $\adom(\A)$ and $\adom(\B)$ are 
{either} both finite {or} both infinite. 
\raus{
This can be seen as follows: For the sake of contradiction, assume that $\adom(\A)$ is finite and
$\adom(\B)$ is infinite. Then the spoiler has the following winning strategy in the
$2$-round EF-game: In the first round he chooses $a_1$ to be the maximum element in $\adom(\A)$.
The duplicator will answer some $b_1$ in $\adom(\B)$. Since $\adom(\B)$ is infinite, the spoiler 
can choose in the second round some $b_2$ that is strictly larger than $b_1$, whereas
the duplicator cannot find an according element in $\adom(\A)$ that is strictly larger than
$a_1$. Hence, the spoiler wins the game. This is a contradiction to our presumption that
\,$\struc{\adom(\A),<,\tau^{\A}} \dwins_r \struc{\adom(\B),<,\tau^{\B}}$.
}
\par
First consider the case where $\adom(\A)$ and $\adom(\B)$ are both infinite.
Then, $\alpha$ is an {isomorphism} between 
\,$\struc{\adom(\A),<,\tau^{\A}}$\, and \,$\Abig$, and $\beta$
is an {isomorphism} between 
\,$\struc{\adom(\B),<,\tau^{\B}}$\, and \,$\Bbig$.
This obviously implies that 
\,$\Abig\dwins_r\Bbig$.
\par
There remains the case where $\adom(\A)$ and $\adom(\B)$ are both finite.
Let $m$ and $n$ denote the cardinalities of $\adom(\A)$ and $\adom(\B)$, respectively.
From our presumption we know that the duplicator has a winning strategy in the 
$r$-round EF-game on $\struc{\adom(\A),<,\tau^{\A}}$ and 
$\struc{\adom(\B),<,\tau^{\B}}$. Henceforth, this game will be called \emph{the small game}.
\\
We now describe a winning strategy for the duplicator in \emph{the big game}, i.e., 
in the $r$-round EF-game on $\Abig$ and $\Bbig$. An illustration of this strategy is
given in Figure~\ref{figure:LemmaNuetzlich}.
%
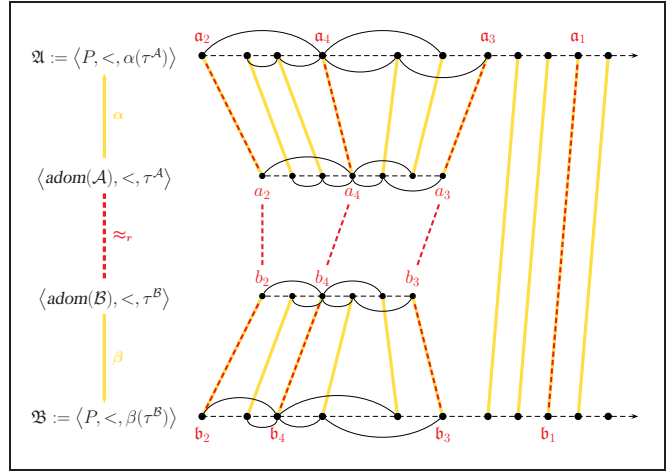
\begin{figure}[!htbp]
\bigskip
\begin{center}
\fbox{
\scalebox{0.4}{
\begin{pspicture}(-5.9,-7.5)(15,+7.5)
\psset{linewidth=0.8pt,linecolor=black,linestyle=solid,dotsize=6pt,arrowsize=5pt}%
%
%
\rput(-3.25,2){\rnode{SA}{\LARGE$\bigstruc{\adom(\A),<,\tau^{\A}}$\vphantom{\Huge$\frac{A}{A}\mid$}}}%
\rput(-3.25,6){\rnode{SAA}{\LARGE$\Abig\deff\bigstruc{P,<,\alpha(\tau^{\A})}$\vphantom{\Huge$\frac{A}{A}$}}}%
\ncline[linecolor=gelb,linewidth=3pt]{->}{SA}{SAA}\nbput{\Large\color{gelb}$\boldsymbol{\alpha}$}%
%
\rput(-3.25,-2){\rnode{SB}{\LARGE$\bigstruc{\adom(\B),<,\tau^{\B}}$\vphantom{\Huge$\frac{A}{A}$}}}%
\rput(-3.25,-6){\rnode{SBB}{\LARGE$\Bbig\deff\bigstruc{P,<,\beta(\tau^{\B})}$\vphantom{\Huge$\mid$}}}%
\ncline[linecolor=gelb,linewidth=3pt]{->}{SB}{SBB}\naput{\Large\color{gelb}$\boldsymbol{\beta}$}%
%
\ncline[linecolor=hellrot,linewidth=3pt,linestyle=dashed]{SA}{SB}\naput{\Large\color{hellrot}$\boldsymbol{\dwins_r}$}%
%
%
%
\dotnode(2,2){UA1}\dotnode(3,2){UA2}\dotnode(4,2){UA3}\dotnode(5,2){UA4}%
\dotnode(6,2){UA5}\dotnode(7,2){UA6}\dotnode(8,2){UA7}%
\ncline[linestyle=dashed,linewidth=0.2pt]{UA1}{UA7}%
%
\dotnode(2,-2){UB1}\dotnode(3,-2){UB2}\dotnode(4,-2){UB3}%
\dotnode(5,-2){UB4}\dotnode(6,-2){UB5}\dotnode(7,-2){UB6}%
\ncline[linestyle=dashed,linewidth=0.2pt]{UB1}{UB6}%
%
\pnode(0,6){UAA0}\pnode(14.5,6){UAA8}%
\ncline[linestyle=dashed,linewidth=0.2pt]{->}{UAA0}{UAA8}%
\psset{dotsize=7pt}%
\dotnode(0,6){UAA1}\dotnode(1.5,6){UAA2}\dotnode(2.5,6){UAA3}\dotnode(4,6){UAA4}%
\dotnode(6.5,6){UAA5}\dotnode(8,6){UAA6}\dotnode(9.5,6){UAA7}\dotnode(10.5,6){UAA8}%
\dotnode(11.5,6){UAA9}\dotnode(12.5,6){UAA10}\dotnode(13.5,6){UAA11}%
\psset{linecolor=black}%
%
\pnode(0,-6){UBB0}\pnode(14.5,-6){UBB7}%
\ncline[linestyle=dashed,linewidth=0.2pt]{->}{UBB0}{UBB7}%
\dotnode(0,-6){UBB1}\dotnode(1.5,-6){UBB2}\dotnode(2.5,-6){UBB3}\dotnode(4,-6){UBB4}%
\dotnode(6.5,-6){UBB5}\dotnode(8,-6){UBB6}\dotnode(9.5,-6){UBB7}\dotnode(10.5,-6){UBB8}%
\dotnode(11.5,-6){UBB9}\dotnode(12.5,-6){UBB10}\dotnode(13.5,-6){UBB11}%
\psset{linecolor=black,dotsize=6pt}%
%
\psset{linewidth=3pt,linecolor=gelb}%
\ncline{UA1}{UAA1}\ncline{UA2}{UAA2}\ncline{UA3}{UAA3}\ncline{UA4}{UAA4}%
\ncline{UA5}{UAA5}\ncline{UA6}{UAA6}\ncline{UA7}{UAA7}%
%
\psset{linewidth=3pt,linecolor=gelb}%
\ncline{UB1}{UBB1}\ncline{UB2}{UBB2}\ncline{UB3}{UBB3}\ncline{UB4}{UBB4}%
\ncline{UB5}{UBB5}\ncline{UB6}{UBB6}%
%
\psset{linewidth=3pt,linecolor=gelb}%
\ncline{UAA8}{UBB7}\ncline{UAA9}{UBB8}\ncline{UAA10}{UBB9}\ncline{UAA11}{UBB10}%
%
\psset{linecolor=black,linewidth=0.8pt}%
\psset{linecolor=black}%
\psset{arcangle=50}\ncarc{UA1}{UA4}%
\psset{arcangle=60}\ncarc{UA4}{UA6}%
\psset{arcangle=-85}\ncarc{UA2}{UA3}\ncarc{UA3}{UA4}\ncarc{UA4}{UA5}%
\psset{arcangle=-60}\ncarc{UA5}{UA7}%
%
\psset{linecolor=black}%
\psset{arcangle=60}\ncarc{UB1}{UB3}%
\psset{arcangle=60}\ncarc{UB3}{UB5}%
\psset{arcangle=-85}\ncarc{UB2}{UB3}\ncarc{UB3}{UB4}%
\psset{arcangle=-60}\ncarc{UB4}{UB6}%
%
\psset{linecolor=black}%
\psset{arcangle=50}\ncarc{UAA1}{UAA4}%
\psset{arcangle=50}\ncarc{UAA4}{UAA6}%
\psset{arcangle=-85}\ncarc{UAA2}{UAA3}
\psset{arcangle=-50}\ncarc{UAA3}{UAA4}\ncarc{UAA4}{UAA5}%
\psset{arcangle=-60}\ncarc{UAA5}{UAA7}%
%
\psset{linecolor=black}%
\psset{arcangle=60}\ncarc{UBB1}{UBB3}%
\psset{arcangle=40}\ncarc{UBB3}{UBB5}%
\psset{arcangle=-85}\ncarc{UBB2}{UBB3}
\psset{arcangle=-50}\ncarc{UBB3}{UBB4}%
\psset{arcangle=-40}\ncarc{UBB4}{UBB6}%
%
%
\psset{nodesep=5pt}%
\rput(12.5,6.6){\rnode{LAA10}{\color{hellrot}\LARGE$\abig_1$}}%
\rput(0,6.6){\rnode{LAA1}{\color{hellrot}\LARGE$\abig_2$}}%
\rput(9.5,6.6){\rnode{LAA7}{\color{hellrot}\LARGE$\abig_3$}}%
\rput(4,6.6){\rnode{LAA4}{\color{rot}\LARGE${\abig_4}$}}%
%
\rput(2,1.4){\rnode{LA1}{\color{hellrot}\LARGE$a_2$}}%
\rput(8,1.4){\rnode{LA7}{\color{hellrot}\LARGE$a_3$}}%
\rput(5,1.4){\rnode{LA4}{\color{rot}\LARGE${a_4}$}}%
%
\rput(2,-1.4){\rnode{LB1}{\color{hellrot}\LARGE$b_2$}}%
\rput(7,-1.4){\rnode{LB6}{\color{hellrot}\LARGE$b_3$}}%
\rput(4,-1.4){\rnode{LB3}{\color{rot}\LARGE${b_4}$}}%
%
\rput(11.5,-6.6){\rnode{LBB9}{\color{hellrot}\LARGE$\bbig_1$}}%
\rput(0,-6.6){\rnode{LBB1}{\color{hellrot}\LARGE$\bbig_2$}}%
\rput(8,-6.6){\rnode{LBB6}{\color{hellrot}\LARGE$\bbig_3$}}%
\rput(2.5,-6.6){\rnode{LBB3}{\color{rot}\LARGE${\bbig_4}$}}%
%
\ncline[linestyle=dashed,linecolor=hellrot,linewidth=2pt]{UAA1}{UA1}%
\ncline[linestyle=dashed,linecolor=hellrot,linewidth=2pt]{UAA7}{UA7}%
\ncline[linestyle=dashed,linecolor=rot,linewidth=2pt]{UAA4}{UA4}%
\ncline[linestyle=dashed,linecolor=hellrot,linewidth=2pt]{LA1}{LB1}%
\ncline[linestyle=dashed,linecolor=hellrot,linewidth=2pt]{LA7}{LB6}%
\ncline[linestyle=dashed,linecolor=rot,linewidth=2pt]{LA4}{LB3}%
\ncline[linestyle=dashed,linecolor=hellrot,linewidth=2pt]{UB1}{UBB1}%
\ncline[linestyle=dashed,linecolor=hellrot,linewidth=2pt]{UB6}{UBB6}%
\ncline[linestyle=dashed,linecolor=rot,linewidth=2pt]{UB3}{UBB3}%
%
\ncline[linestyle=dashed,linecolor=rot,linewidth=2pt]{UAA10}{UBB9}%
\end{pspicture}
}
}
\caption{\small Visualization of the duplicator's strategy in the big game in
 Lemma~\ref{lemma:nuetzlich}.
 Here, $\tau$ consists of one binary relation $E$.}\label{figure:LemmaNuetzlich}
\end{center}
\end{figure}
%
%
\\
In each round $i\in\set{1,\twodots,r}$ of the big game we proceed as follows:
If the spoiler chooses an element $\abig_i$ in the universe of $\Abig$, we distinguish 
between two cases.
(If he chooses an element $\bbig_i$ in the universe of $\Bbig$, we proceed in the according
way, interchanging the roles of $\Abig$ and $\Bbig$.)
\par
\textbf{Case 1:} $\abig_i\not\in\alpha(\adom(\A))$, i.e., $\abig_i = p_{m+d_i}$ for
some $d_i\in\NNpos$.
In this case the duplicator chooses $\bbig_i\deff p_{n+d_i}$.
\par
\textbf{Case 2:} $\abig_i\in\alpha(\adom(\A))$, i.e., $\abig_i = \alpha(a_i)$ for some
$a_i\in\adom(\A)$.
In this case we define $a_i$ to be a move for a ``virtual spoiler'' in the $i$-th round of the
small game on $\struc{\adom(\A),<,\tau^{\A}}$. 
A ``virtual duplicator'' who plays according to her winning strategy in the
small game will find some answer $b_i$ in $\struc{\adom(\B),<,\tau^{\B}}$.  
We can translate this answer into a move $\bbig_i$ for the duplicator in the big game via
$\bbig_i\deff \beta(b_i)$.
\par
After $r$ rounds, the ``virtual duplicator'' has won the small game; and it is 
straightforward to check that the duplicator has also won the big game.
\raus{
We need to show that the mapping $\pi$ defined via
\[
  \pi\ :\ \left\{
   \begin{array}{rcll}
      \alpha(c^{\A}) & \mapsto & \beta(c^{\B}) & \mbox{for all constant symbols $c\in\tau$} \\
             \abig_i & \mapsto & \bbig_i       & \mbox{for all $i\in\set{1,\twodots,r}$} 
   \end{array}
   \right\}
\]
is a partial isomorphism between $\Abig$ and $\Bbig$.
\\
We already know that the ``virtual duplicator'' has won the small game. Furthermore, from
the duplicator's choice in the big game we know, for all $x$ in the domain of $\pi$, that
\,$x\in\alpha(\adom(\A))$\, iff \,$y\in\beta(\adom(\B))$.
This, together with case~2, directly gives us that $\pi$ is a partial isomorphism between
\,$\struc{P,\alpha(\tau^{\A})}$\, and \,$\struc{P,\beta(\tau^{\B})}$. 
\\
It remains to be shown that \ ``$x<y$ \ iff \ $\pi(x)<\pi(y)$'' \ is true for all $x,y$ in the
domain of $\pi$.
If $x,y\in\alpha(\adom(\A))$, then this follows from the fact that the ``virtual 
duplicator'' has won the small game.
\\
If $x,y \not\in \alpha(\adom(\A))$, then, according to case~1, we must have that
$x=p_{m+d_x}$, $\pi(x)= p_{n+d_x}$, $y= p_{m+d_y}$, and $\pi(y)= p_{n+d_y}$, for suitable
$d_x,d_y\in\NNpos$. In particular, this implies that \ $x<y$ \ iff \ $\pi(x)<\pi(y)$.
\\
If $x\in\alpha(\adom(\A))$ and $y\not\in\alpha(\adom(\A))$, then, according to
case~1 and case~2, we must have that $x=p_j$ for some $j\leq m$, $\pi(x) = p_k$ for 
some $k\leq n$, and $y=p_{m+d_y}$ and $\pi(y) = p_{n+d_y}$ for some $d_y\in\NNpos$.
In particular, this gives us that \ $x<y$ \ iff \ $\pi(x)<\pi(y)$.
}
\\
Altogether, this completes the proof of Lemma~\ref{lemma:nuetzlich}.
\end{proof_mit}%
%
%



\section[How to Win the Game for ${\FO(<,\Mon)}$]{How to Win the Game 
for $\bs{\FO(<,\Mon)}$}\label{section:Mon}\label{section:Monadic} 
\index{Ehrenfeucht-Fra\"\i{}ss\'{e} game!FO<Mon@$\FO(<,\Mon)$-game} 
In this section we concentrate on the class $\Mon$ of \emph{monadic},
i.e., unary, built-in predicates.
We consider the context structure
$\struc{\UU,<,\Mon}$, for any linearly ordered infinite universe $\UU$; and we
explicitly describe how the duplicator can
translate strategies for the $\FOKleiner$-game
into strategies for the $\FOMon$-game 
on {$\ZZ$-embeddable} structures over $\UU$. 
The overall proof idea is an adaption and extension of a proof developed 
by several researchers in the context of the \emph{Crane Beach conjecture} \cite{BILST} 
for the specific context of finite strings instead of arbitrary structures. 
\index{Crane Beach Conjecture!FO<Mon@$\FO(<,\Mon)$}%
\begin{theorem_mit}[${\FOMon}$-game for $\ZZ$-embeddable structures]\label{theorem:collapse_Monadic_UU}\mbox{}\\ 
Let $\struc{\UU,<}$ be a linearly ordered {infinite} structure, and
let $\Mon$ be the class of all {monadic} predicates on $\UU$.
\\
The duplicator can translate strategies for the $\FOKleiner$-game into 
strategies for the $\FO(<,\allowbreak \Mon)$-game on $\ZZ$-embeddable structures over $\UU$.
\end{theorem_mit}%
\begin{proof_mit}
Let $\Mon'$ be a finite subset of $\Mon$, and let $\tau$ be a signature. For every number
$k\in\NNpos$ of rounds for the $\FO(<,\Mon')$-game we choose $r(k)\deff k{+}1$ to be the 
according number of rounds for the $\FOKleiner$-game.
\\
Now let $\A = \struc{\UU,\tau^{\A}}$ and $\B = \struc{\UU,\tau^{\B}}$ be two 
\emph{$\ZZ$-embeddable} structures
on which the duplicator wins the $(k{+}1)$-round $\FOKleiner$-game, i.e., 
\begin{eqnarray*}
  (*):\qquad\bigstruc{\,\adom(\A),\,<,\,\tau^{\A}\,}
& \dwins_{k+1}
& \bigstruc{\,\adom(\B),\,<,\,\tau^{\B}\,}.
\end{eqnarray*}
Our aim is to find ${<}$-preserving mappings \,{$\alpha : \adom(\A)\rightarrow \UU$}\, and
\,{$\beta : \adom(\B)\rightarrow \UU$}\, such that the duplicator wins the
$k$-round $\FO(<,\Mon')$-game on $\alpha(\A)$ and $\beta(\B)$, i.e.,
$\bigstruc{\UU,<,\Mon',\alpha\big(\tau^{\A}\big)}
   \allowbreak\dwins_k
   \bigstruc{\UU,<,\Mon',\beta\big(\tau^{\B}\big)}$.
\\
Note that the condition $(*)$ gives us that, in particular, $\adom(\A)$ has a lower
bound (respectively, an upper bound) if and only if $\adom(\B)$ has.
Since $\A$ and $\B$ are $\ZZ$-embeddable, we know that they have no accumulation points and
that exactly one of the following four cases is valid:
{
 \begin{enumerate}[\bf\mbox{Case} I:]
 \item $\adom(\A) = \set{u_1<u_2<\cdots\,}$  and $\adom(\B)$ are infinitely increasing.
 \item $\adom(\A) = \set{u_1>u_2>\cdots\,}$  and $\adom(\B)$ are infinitely decreasing.
 \item $\adom(\A) = \set{\,\cdots<u_{-2}<u_{-1}<u_{1}<u_{2}<\cdots\,}$ and $\adom(\B)$ are
    infinite in both directions.
 \item $\adom(\A)$ and $\adom(\B)$ are finite.
\end{enumerate}
}%
\noindent%
Let us first concentrate on {Case I}, i.e., on the case where
$\adom(\A)$ and $\adom(\B)$ are infinitely increasing.
Let $u_1<u_2<\cdots$ such that $\adom(\A)=\set{u_1,u_2,\twodots}$.
\medskip\\
\textbf{\underline{Step 1:}} \ 
We first choose a 
suitable subsequence \,$p_1<p_2<\cdots$ of \,$u_1<u_2<\cdots$ onto which the active domain
elements of $\A$ and $\B$ will be moved via ${<}$-preserving mappings $\alpha$ and $\beta$.
To find this sequence, we use the following theorem from Ramsey Theory.\index{Ramsey Theory}
A well-presented introduction to Ramsey Theory as well as a proof of the Ramsey Theorem can be
found in Diestel's textbook \cite[Section~9]{Diestel}.\index{Diestel, Reinhard}
\begin{theorem_mit}[Ramsey]\label{theorem:Ramsey}\index{Ramsey Theorem}
Let $G=\struc{V,E}$ be the graph with vertex set $V=\set{u_1,u_2,\twodots}$ and edge set
$E= \setc{(u_i,u_j)\in V^2}{i<j}$.
Let $C$ be a finite set, and let each edge $(u_i,u_j)$ of $G$ be colored with an element
$\Color(u_i,u_j) \in C$.\\
There exists an infinite \emph{monochromatic} path, \index{monochromatic path}
i.e., there is an
infinite sequence $p_1<p_2<\cdots$ in $V$, such that 
$\Color(p_1,p_2) = \Color(p_2,p_3) = \cdots = \Color(p_i,p_j)$, for all $i<j$.
\mbox{ }
\end{theorem_mit}%
We choose the following coloring: 
The edge $(u_i,u_j)$ is colored with the 
\emph{$k$-type} of the substructure of $\struc{\UU,<,\Mon'}$ with universe 
$[u_i,u_j) \deff \setc{u\in\UU}{u_i\leq u < u_j}$.
I.e., we choose \index{ktype@$\ktype[i,j)$} 
\,$\Color(u_i,u_j)  \deff  \ktype[u_i,u_j),$\, 
where $\ktype[u_i,u_j)$ is 
the equivalence class of the structure $\bigstruc{[u_i,u_j), u_i, <, \Mon' }$
with respect to the relation $\dwins_k$. 
According to Remark~\ref{remark:E-F-classes}, the
number of $k$-types is \emph{finite}, and hence the Ramsey 
Theorem~\ref{theorem:Ramsey} gives us
an infinite monochromatic path, i.e., an infinite sequence 
\,$p_1<p_2<\cdots$\, in $V$ such that $\Color(p_1,p_2) = \Color(p_2,p_3) = \cdots =  
\Color(p_i,p_j)$, for all $i<j$.
Note that, by definition, 
\,$\ktype[p_j,p_{j+1}) = \ktype[p_{j'},p_{j'+1})$\, means that 
\begin{eqnarray*}
  (**):\qquad\bigstruc{\,[p_j,p_{j+1}),\,p_j,\,<,\,\Mon'\,}
& \dwins_k
& \bigstruc{\,[p_{j'},p_{j'+1}),\,p_{j'},\,<,\,\Mon'\,}\,.
\end{eqnarray*}
The positions $p_1,p_2,\ldots$ will be called ``special positions'', and
the set of all special positions will be denoted $P$.
We define $\alpha$ and $\beta$ to be the ${<}$-preserving mappings that move the 
active domain elements of $\A$ and $\B$ onto the ``special positions''. 
Precisely, \,$\alpha:\adom(\A)\rightarrow P$\, and
\,$\beta:\adom(\B)\rightarrow P$\,
map, for every $j$, the $j$-th smallest
element of $\adom(\A)$ and $\adom(\B)$, respectively, onto the position $p_j$.
\par
From the presumption $(*)$ and from Lemma~\ref{lemma:nuetzlich} we obtain 
that a ``virtual duplicator'' has a winning strategy for the $k$-round EF-game on
\,$\Abig'\deff \struc{\,P,\,<,\allowbreak\,\alpha(\tau^{\A})\,}$\, and 
\,$\Bbig'\deff \struc{\,P,\,<,\,\beta(\tau^{\B})\,}$. I.e., we know that
\,$\Abig' \deff \bigstruc{P,<,\alpha\big(\tau^{\A}\big)}
   \dwins_k
   \bigstruc{P,<,\beta\big(\tau^{\B}\big)} \ffed \Bbig'$.
Henceforth, this game will be called \emph{the ${<}$-game} (on $\Abig'$ and $\Bbig'$). 
\medskip\\
\textbf{\underline{Step 2:}} \ 
We now describe a winning strategy for the duplicator in the
$k$-round $\FO(<,\allowbreak\Mon')$-game on $\alpha(\A)$ and $\beta(\B)$. 
An illustration of this strategy is given in Figure~\ref{figure:Mon}.
Precisely, we show that
\,$\Abig\deff\bigstruc{\UU,<,\Mon',\alpha\big(\tau^{\A}\big)}
   \dwins_k
   \bigstruc{\UU,<,\Mon',\beta\big(\tau^{\B}\big)}\ffed \Bbig$.
Henceforth, this game will be called \emph{the $\Mon'$-game} (on $\Abig$ and $\Bbig$).
%
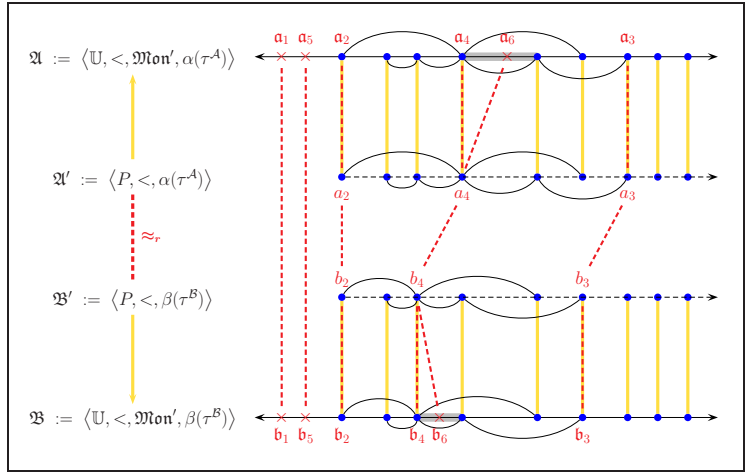
\begin{figure}[!htbp]
\bigskip
\begin{center}
\fbox{
\scalebox{0.4}{
\begin{pspicture}(-10.6,-7.5)(13,+7.5)
\psset{linewidth=0.8pt,linecolor=black,linestyle=solid,dotsize=6pt,arrowsize=7pt}%
%
%
\rput(-6.95,2){\rnode{SA}{\LARGE$\Abig'\;\deff\;\bigstruc{P,<,\alpha(\tau^{\A})}$\vphantom{\Huge$\frac{A}{A}\mid$}}}%
\rput(-6.95,6){\rnode{SAA}{\LARGE$\Abig\;\deff\;\bigstruc{\UU,<,\Mon',\alpha(\tau^{\A})}$\vphantom{\Huge$\frac{A}{A}$}}}%
\ncline[linecolor=gelb,linewidth=3pt]{->}{SA}{SAA}%
%
\rput(-6.95,-2){\rnode{SB}{\LARGE$\Bbig'\;\deff\;\bigstruc{P,<,\beta(\tau^{\B})}$\vphantom{\Huge$\frac{A}{A}$}}}%
\rput(-6.95,-6){\rnode{SBB}{\LARGE$\Bbig\;\deff\;\bigstruc{\UU,<,\Mon',\beta(\tau^{\B})}$\vphantom{\Huge$\mid$}}}%
\ncline[linecolor=gelb,linewidth=3pt]{->}{SB}{SBB}%
%
\ncline[linecolor=hellrot,linewidth=3pt,linestyle=dashed]{SA}{SB}\naput{\Large\color{hellrot}$\boldsymbol{\dwins_r}$}%
%
%
\pnode(4,6){AAIntLinks}\pnode(6.5,6){AAIntRechts}
\ncline[linewidth=8pt,linecolor=lightgray]{AAIntLinks}{AAIntRechts}%
\pnode(2.5,-6){BBIntLinks}\pnode(4,-6){BBIntRechts}
\ncline[linewidth=8pt,linecolor=lightgray]{BBIntLinks}{BBIntRechts}%
%
%
\psset{linecolor=blue}%
\pnode(12.5,2){UA11}\psset{dotsize=7pt}%
\dotnode(0,2){UA1}\dotnode(1.5,2){UA2}\dotnode(2.5,2){UA3}\dotnode(4,2){UA4}%
\dotnode(6.5,2){UA5}\dotnode(8,2){UA6}\dotnode(9.5,2){UA7}\dotnode(10.5,2){UA8}%
\dotnode(11.5,2){UA9}%
\psset{linecolor=black}%
\ncline[linestyle=dashed,linewidth=0.2pt]{->}{UA1}{UA11}%
%
\psset{linecolor=blue}%
\pnode(12.5,-2){UB11}\psset{dotsize=7pt}%
\dotnode(0,-2){UB1}\dotnode(1.5,-2){UB2}\dotnode(2.5,-2){UB3}\dotnode(4,-2){UB4}%
\dotnode(6.5,-2){UB5}\dotnode(8,-2){UB6}\dotnode(9.5,-2){UB7}\dotnode(10.5,-2){UB8}%
\dotnode(11.5,-2){UB9}%
\psset{linecolor=black}%
\ncline[linestyle=dashed,linewidth=0.2pt]{->}{UB1}{UB11}%
%
\pnode(-2.9,6){UAA0}\pnode(12.5,6){UAA11}%
\ncline[linestyle=solid,linewidth=0.5pt]{<->}{UAA0}{UAA11}%
\psset{linecolor=blue}%
\psset{dotsize=7pt}%
\dotnode(0,6){UAA1}\dotnode(1.5,6){UAA2}\dotnode(2.5,6){UAA3}\dotnode(4,6){UAA4}%
\dotnode(6.5,6){UAA5}\dotnode(8,6){UAA6}\dotnode(9.5,6){UAA7}\dotnode(10.5,6){UAA8}%
\dotnode(11.5,6){UAA9}%
\psset{linecolor=black}%
%
\pnode(-2.9,-6){UBB0}\pnode(12.5,-6){UBB11}%
\ncline[linestyle=solid,linewidth=0.5pt]{<->}{UBB0}{UBB11}%
\psset{linecolor=blue}
\dotnode(0,-6){UBB1}\dotnode(1.5,-6){UBB2}\dotnode(2.5,-6){UBB3}\dotnode(4,-6){UBB4}%
\dotnode(6.5,-6){UBB5}\dotnode(8,-6){UBB6}\dotnode(9.5,-6){UBB7}\dotnode(10.5,-6){UBB8}%
\dotnode(11.5,-6){UBB9}%
\psset{linecolor=black,dotsize=6pt}%
%
\psset{linewidth=3pt,linecolor=gelb}%
\ncline{UA1}{UAA1}\ncline{UA2}{UAA2}\ncline{UA3}{UAA3}\ncline{UA4}{UAA4}%
\ncline{UA5}{UAA5}\ncline{UA6}{UAA6}\ncline{UA7}{UAA7}%
\ncline{UA8}{UAA8}\ncline{UA9}{UAA9}\ncline{UA10}{UAA10}%
%
\psset{linewidth=3pt,linecolor=gelb}%
\ncline{UB1}{UBB1}\ncline{UB2}{UBB2}\ncline{UB3}{UBB3}\ncline{UB4}{UBB4}%
\ncline{UB5}{UBB5}\ncline{UB6}{UBB6}\ncline{UB7}{UBB7}%
\ncline{UB8}{UBB8}\ncline{UB9}{UBB9}\ncline{UB10}{UBB10}%
%
\psset{linewidth=3pt,linecolor=gelb}%
%
%
\psset{linecolor=black,linewidth=0.8pt}%
\psset{linecolor=black}%
\psset{arcangle=50}\ncarc{UA1}{UA4}%
\psset{arcangle=50}\ncarc{UA4}{UA6}%
\psset{arcangle=-85}\ncarc{UA2}{UA3}
\psset{arcangle=-50}\ncarc{UA3}{UA4}\ncarc{UA4}{UA5}%
\psset{arcangle=-60}\ncarc{UA5}{UA7}%
%
\psset{linecolor=black}%
\psset{arcangle=60}\ncarc{UB1}{UB3}%
\psset{arcangle=40}\ncarc{UB3}{UB5}%
\psset{arcangle=-85}\ncarc{UB2}{UB3}
\psset{arcangle=-50}\ncarc{UB3}{UB4}%
\psset{arcangle=-40}\ncarc{UB4}{UB6}%
%
\psset{linecolor=black}%
\psset{arcangle=50}\ncarc{UAA1}{UAA4}%
\psset{arcangle=50}\ncarc{UAA4}{UAA6}%
\psset{arcangle=-85}\ncarc{UAA2}{UAA3}
\psset{arcangle=-50}\ncarc{UAA3}{UAA4}\ncarc{UAA4}{UAA5}%
\psset{arcangle=-60}\ncarc{UAA5}{UAA7}%
%
\psset{linecolor=black}%
\psset{arcangle=60}\ncarc{UBB1}{UBB3}%
\psset{arcangle=40}\ncarc{UBB3}{UBB5}%
\psset{arcangle=-85}\ncarc{UBB2}{UBB3}
\psset{arcangle=-50}\ncarc{UBB3}{UBB4}%
\psset{arcangle=-40}\ncarc{UBB4}{UBB6}%
%
%
\psset{nodesep=5pt}%
\rput(0,6.6){\rnode{LAA1}{\color{hellrot}\LARGE$\abig_2$}}%
\rput(9.5,6.6){\rnode{LAA7}{\color{hellrot}\LARGE$\abig_3$}}%
\rput(4,6.6){\rnode{LAA4}{\color{rot}\LARGE${\abig_4}$}}%
\rput(-2,6.6){\rnode{LAA1}{\color{rot}\LARGE${\abig_1}$}}%
\rput(-2,6){\rnode{PAA1}{\color{rot}\LARGE${\times}$}}%
\rput(-1.2,6.6){\rnode{LAA5}{\color{rot}\LARGE${\abig_5}$}}%
\rput(-1.2,6){\rnode{PAA5}{\color{rot}\LARGE${\times}$}}%
\rput(5.5,6.6){\rnode{LAA6}{\color{rot}\LARGE${\abig_6}$}}%
\rput(5.5,6){\rnode{PAA6}{\color{rot}\LARGE${\times}$}}%
%
\rput(0,1.4){\rnode{LA1}{\color{hellrot}\LARGE$a_2$}}%
\rput(9.5,1.4){\rnode{LA7}{\color{hellrot}\LARGE$a_3$}}%
\rput(4,1.4){\rnode{LA4}{\color{rot}\LARGE${a_4}$}}%
%
\rput(0,-1.4){\rnode{LB1}{\color{hellrot}\LARGE$b_2$}}%
\rput(8,-1.4){\rnode{LB6}{\color{hellrot}\LARGE$b_3$}}%
\rput(2.5,-1.4){\rnode{LB3}{\color{rot}\LARGE${b_4}$}}%
%
\rput(0,-6.6){\rnode{LBB1}{\color{hellrot}\LARGE$\bbig_2$}}%
\rput(8,-6.6){\rnode{LBB6}{\color{hellrot}\LARGE$\bbig_3$}}%
\rput(2.5,-6.6){\rnode{LBB3}{\color{rot}\LARGE${\bbig_4}$}}%
\rput(-2,-6.6){\rnode{LBB1}{\color{rot}\LARGE${\bbig_1}$}}%
\rput(-2,-6){\rnode{PBB1}{\color{rot}\LARGE${\times}$}}%
\rput(-1.2,-6.6){\rnode{LBB5}{\color{rot}\LARGE${\bbig_5}$}}%
\rput(-1.2,-6){\rnode{PBB5}{\color{rot}\LARGE${\times}$}}%
\rput(3.25,-6.6){\rnode{LBB6}{\color{rot}\LARGE${\bbig_6}$}}%
\rput(3.25,-6){\rnode{PBB6}{\color{rot}\LARGE${\times}$}}%
%
\ncline[linestyle=dashed,linecolor=hellrot,linewidth=2pt]{UAA1}{UA1}%
\ncline[linestyle=dashed,linecolor=hellrot,linewidth=2pt]{UAA7}{UA7}%
\ncline[linestyle=dashed,linecolor=rot,linewidth=2pt]{UAA4}{UA4}%
\ncline[linestyle=dashed,linecolor=rot,linewidth=2pt]{PAA6}{UA4}%

\ncline[linestyle=dashed,linecolor=hellrot,linewidth=2pt]{LA1}{LB1}%
\ncline[linestyle=dashed,linecolor=hellrot,linewidth=2pt]{LA7}{LB6}%
\ncline[linestyle=dashed,linecolor=rot,linewidth=2pt]{LA4}{LB3}%
\ncline[linestyle=dashed,linecolor=hellrot,linewidth=2pt]{UB1}{UBB1}%
\ncline[linestyle=dashed,linecolor=hellrot,linewidth=2pt]{UB6}{UBB6}%
\ncline[linestyle=dashed,linecolor=rot,linewidth=2pt]{UB3}{UBB3}%
\ncline[linestyle=dashed,linecolor=rot,linewidth=2pt]{UB3}{PBB6}%
%
\ncline[linestyle=dashed,linecolor=rot,linewidth=2pt]{PAA1}{PBB1}%
\ncline[linestyle=dashed,linecolor=rot,linewidth=2pt]{PAA5}{PBB5}%
\end{pspicture}
}
}
\caption{\small Visualization of the duplicator's strategy in the $\FO(<,\Mon')$-game
 on $\Abig$ and $\Bbig$.
 Here, $\tau$ consists of one binary relation $E$. Blue points represent elements in $P$.}\label{figure:Mon}
\end{center}
\end{figure}
%
%
\\
In each round $i\in\set{1,\twodots,k}$ of the $\Mon'$-game we proceed as follows:
If the spoiler chooses an element $\abig_i$ in the universe of $\Abig$, we distinguish
between two cases. (If he chooses an element $\bbig_i$ in the
universe of $\Bbig$, we proceed in the according way, interchanging the roles of $\Abig$ and 
$\Bbig$.)
\par
\textbf{Case 1:} $\abig_i$ is smaller than the smallest ``special position'', i.e., 
$\abig_i < p_1$. In this case, the duplicator chooses the identical element in the universe 
of $\Bbig$, i.e., she chooses $\bbig_i\deff \abig_i < p_1$.
\par
\textbf{Case 2:} $\abig_i \geq p_1$. In this case there exists a $j\in\NNpos$ such
that $\abig_i\in [p_j,p_{j+1})$ (note that we essentially use here that $P$ has no
accumulation points in $\overline{\UU}$). 
The position $p_{j}$ represents 
the interval $[p_{j},p_{j+1})$ to which the spoiler's choice $\abig_i$ belongs.
We define $\abig'_i \deff p_{j}$ to be a move for a ``virtual spoiler'' in the $i$-th
round of the ${<}$-game on $\Abig'$. 
A ``virtual duplicator'' who plays according to her winning strategy in the ${<}$-game will
find some answer $\bbig'_i$ in $\Bbig'$. Let $j'\in\NNpos$ such that
$\bbig'_i = p_{j'}$.
\\
The duplicator in the ${\Mon'}$-game will choose some $\bbig_i$ in $\Bbig$ that lies in the 
interval $[p_{j'},p_{j'+1})$.
But which element in this interval shall she choose? --- Here we make use of the fact that
another ``virtual duplicator'' wins the game $(**)$ on the intervals $[p_{j},p_{j+1})$ 
and $[p_{j'},p_{j'+1})$: 
Let $\abig_{s_1},\twodots,\abig_{s_t}$ be those elements among
$\abig_{1},\twodots,\abig_{i-1}$
that lie in the interval $[p_{j},p_{j+1})$.
By induction we know that
$
   \set{s_1,\twodots,s_t} =
   \{s\in\set{1,\twodots,i{-}1}\,:$ 
$  \abig_s \in [p_{j}, p_{j+1})\} =
   \bigsetc{s\in\set{1,\twodots,i{-}1}}{\bbig_s \in [p_{j'}, p_{j'+1})}.
$
We write $\vek{\abig}_{i-1}$ as abbreviation for $\abig_{s_1},\twodots,\abig_{s_{t}}$ and 
$\vek{\bbig}_{i-1}$ as abbreviation for $\bbig_{s_1},\twodots,\bbig_{s_{t}}$.
By induction with $(**)$ we know that
{
\[
\struc{[p_{j},p_{j+1}), p_{j}, <, \Mon', 
   \vek{\abig}_{i-1}}
\dwins_{k-i+1}
\struc{[p_{j'},p_{j'+1}), p_{j'}, <, \Mon', 
   \vek{\bbig}_{i-1}}.
\]
}%
For $i{=}1$ this is true because of $(**)$; for $i{>}1$ this follows
from the duplicator's choices in the previous rounds.
Since $\abig_i \in [\,p_{j}, p_{j+1})$, a ``virtual duplicator'' 
in the game $(**)$ can choose a suitable $\bbig_i \in
[p_{j'}, p_{j'+1})$, such that
\[
\struc{[p_{j},p_{j+1}), p_{j}, <, \Mon', 
   \vek{\abig}_{i-1},\abig_{i}}
\dwins_{k-i}
\struc{ [p_{j'},p_{j'+1}), p_{j'}, <, \Mon', 
   \vek{\bbig}_{i-1},\bbig_{i}}.
\]
We choose exactly this $\bbig_i$ to be the answer of the duplicator in
the $i$-th round of the $\Mon'$-game on $\Bbig$.
\par
After $k$ rounds we know that the ``virtual duplicator'' has won the $<$-game (on
$\Abig'$ and $\Bbig'$) as well as all the interval games $(**)$. It is straightforward
(although tedious) to check that the duplicator has also won the $\Mon'$-game (on 
$\Abig$ and $\Bbig$).
This completes the proof of Theorem~\ref{theorem:collapse_Monadic_UU} for Case~I, i.e., for 
the case that $\adom(\A)$ and $\adom(\B)$ are infinitely increasing.
\\
\parno
Case II, i.e., the case where $\adom(\A)$ and $\adom(\B)$ are infinitely decreasing,
is symmetric to Case~I.
\\
\parno
Let us now concentrate on Case III, i.e., the case where $\adom(\A)$ and $\adom(\B)$ 
are infinite in both directions. 
Let $\adom(\A) = \set{\,\cdots<u_{-2}<u_{-1}<u_{1}<u_{2}<\cdots\,}$.
The problem here is that the Ramsey Theorem~\ref{theorem:Ramsey} gives us 
one infinite monochromatic increasing path 
$p_{\scriptscriptstyle 1}<p_{\scriptscriptstyle 2}<\cdots$,
and another infinite monochromatic decreasing path 
$p_{\scriptscriptstyle -1}>p_{\scriptscriptstyle -2}>\cdots$.
However, these two paths do not necessarily have the same color. Imagine, e.g., that
all edges on the increasing path are colored ``blue'' and all edges 
on the decreasing path are colored ``red''.
We therefore have to carefully decide which part of the original structure is mapped onto
the ``blue'' path and which part is mapped onto the ``red'' path.
To this end, let a ``virtual spoiler'' choose
an element $\ablue$ in $\adom(\A)$ in the first round of the game $(*)$. 
A ``virtual duplicator'' who wins the game $(*)$ can
answer with an element $\bblue$ in $\adom(\B)$ such that
\begin{eqnarray*}
  (*)'\qquad\bigstruc{\,\adom(\A),\,<,\,\tau^{\A},\,\ablue\,}
& \dwins_{k}
& \bigstruc{\,\adom(\B),\,<,\,\tau^{\B},\,\bblue\,}.
\end{eqnarray*}
The idea is now to map the active domain elements of $\A$ which are $\geq \ablue$ 
(and the active domain elements of $\B$ which are $\geq \bblue$) onto an increasing 
``blue'' path
$p_{\scriptscriptstyle 1}<p_{\scriptscriptstyle 2}<\cdots$. 
Similarly, the active domain elements of $\A$ which are $< \ablue$ 
(and the active domain elements of $\B$ which are $< \bblue$) will be mapped onto 
a decreasing ``red'' path $p_{\scriptscriptstyle -1}>p_{\scriptscriptstyle -2}>\cdots$. 
More precisely:
\medskip\\
\textbf{\underline{Step 1:}} \ 
In the same way as in the proof for Case~I, the 
Ramsey Theorem~\ref{theorem:Ramsey} gives us an
infinite increasing sequence 
\,$p_{\scriptscriptstyle 1}<p_{\scriptscriptstyle 2}<p_{\scriptscriptstyle 3}<\cdots$ 
in $\setc{u\in\adom(\A)}{u\geq \ablue}$ 
such that
{\small\begin{eqnarray*}
  (**)_{\scriptscriptstyle\mathit{blue}}:\quad\
  \bigstruc{\,\big[p_{\scriptscriptstyle j},p_{\scriptscriptstyle j+1}\big),\;p_{\scriptscriptstyle j},\;<,\;\Mon'\,}
& \dwins_k
& \bigstruc{\,\big[p_{\scriptscriptstyle j'},p_{\scriptscriptstyle j'+1}\big),\;p_{\scriptscriptstyle j'},\;<,\;\Mon'\,}\, 
\end{eqnarray*}}%
is true for all $j,j'\in\NNpos$.
Another application of the Ramsey Theorem~\ref{theorem:Ramsey} gives us an
infinite decreasing sequence 
\,$p_{\scriptscriptstyle -1}>p_{\scriptscriptstyle -2}>p_{\scriptscriptstyle -3}>\cdots$ 
in $\setc{u\in\adom(\A)}{u< \ablue}$
such that
{\small \begin{eqnarray*}
  (**)_{\scriptscriptstyle\mathit{red}}:\quad\
  \bigstruc{\,\big(p_{\scriptscriptstyle -(j+1)},p_{\scriptscriptstyle -j}\big],\;p_{\scriptscriptstyle -j},\;<,\;\Mon'\,}
& \dwins_k
& \bigstruc{\,\big(p_{\scriptscriptstyle -(j'+1)},p_{\scriptscriptstyle -j'}\big],\;p_{\scriptscriptstyle -j'},\;<,\;\Mon'\,}\, 
\end{eqnarray*}}%
is true for all $j,j'\in\NNpos$.
Now let \,$\alpha:\adom(\A)\rightarrow\UU$\, and \,$\beta:\adom(\B)\rightarrow\UU$\, be 
${<}$-preserving mappings that move 
\begin{enumerate}[$\bullet$]
 \item
   the active domain elements $\geq \ablue$ of $\A$ and the
   active domain elements $\geq \bblue$ of $\B$, respectively, 
   onto the ``special blue positions'' $p_{\scriptscriptstyle 1},p_{\scriptscriptstyle 2},\twodots$, \,and 
 \item
   the active domain elements $< \ablue$ of $\A$ and the
   active domain elements $< \bblue$ of $\B$, respectively, 
   onto the ``special red positions'' $p_{\scriptscriptstyle -1},p_{\scriptscriptstyle -2},\twodots$.
\end{enumerate}
\textbf{\underline{Step 2:}} \ 
We now describe a winning strategy for the duplicator in the
$k$-round $\FO(<,\allowbreak \Mon')$-game on $\alpha(\A)$ and $\beta(\B)$. 
I.e., we show that
\,$\Abig\deff\bigstruc{\UU,<,\Mon',\alpha\big(\tau^{\A}\big)}
   \dwins_k
   \bigstruc{\UU,<,\Mon',\beta\big(\tau^{\B}\big)} \ffed \Bbig$. 
%
\\
When the spoiler chooses an element
$\abig_i$ in the universe of $\Abig$, we translate this move into a move
$a_i$ for a ``virtual spoiler'' in the game $(*)'$. (The case when the spoiler chooses an
element $\bbig_i$ in the universe of $\Bbig$ is symmetric.)
We distinguish between three cases:
\par
\textbf{Case 1:} 
If $\abig_i \geq p_{\scriptscriptstyle 1}$, then let $j\in\NNpos$ such that $\abig_i\in \big[p_{\scriptscriptstyle j},p_{\scriptscriptstyle j+1}\big)$.
Choose $a_i$ such that $\alpha(a_i)=p_{\scriptscriptstyle j}$.
In particular, $a_i\geq\ablue$.
Now, a ``virtual duplicator'' who wins
the game $(*)'$ will answer with an element $b_i$ in the active domain
of $\B$. Certainly we have $b_i\geq\bblue$, and thus $\beta(b_i)=p_{\scriptscriptstyle j'}$ 
for some $j'\in\NNpos$.
The duplicator in the game on $\Abig$ and $\Bbig$ will choose
an element $\bbig_i \in \big[p_{\scriptscriptstyle j'},p_{\scriptscriptstyle j'+1}\big)$. For her exact choice she
makes use of the fact that another ``virtual duplicator'' can win
the game $(**)_{\scriptscriptstyle\mathit{blue}}$ played on the 
intervals \,$\big[p_{\scriptscriptstyle j},p_{\scriptscriptstyle j+1}\big)$\, and \,$\big[p_{\scriptscriptstyle j'},p_{\scriptscriptstyle j'+1}\big)$.
\par
\textbf{Case 2:} 
If $\abig_i \leq p_{\scriptscriptstyle -1}$, then let $j\in\NNpos$ such that $\abig_i\in \big(p_{\scriptscriptstyle -(j+1)},p_{\scriptscriptstyle -j}\big]$.
Choose $a_i$ such that $\alpha(a_i)=p_{\scriptscriptstyle -j}$.
In particular, $a_i<\ablue$.
Now, a ``virtual duplicator'' who wins
the game $(*)'$ will answer with an element $b_i$ in the active domain
of $\B$. Certainly we have $b_i<\bblue$, and thus $\beta(b_i)=p_{\scriptscriptstyle -j'}$ 
for some $j'\in\NNpos$.
The duplicator in the game on $\Abig$ and $\Bbig$ will choose
an element $\bbig_i$ in the interval 
\,$\big(p_{\scriptscriptstyle -(j'+1)},p_{\scriptscriptstyle -j'}\big]$. 
For her exact choice she
makes use of the fact that another ``virtual duplicator'' can win
the game $(**)_{\scriptscriptstyle\mathit{red}}$ played on the 
intervals \,$\big(p_{\scriptscriptstyle -(j+1)},p_{\scriptscriptstyle -j}\big]$\, and \,$\big(p_{\scriptscriptstyle -(j'+1)},p_{\scriptscriptstyle -j'}\big]$.
\par
\textbf{Case 3:}
If $p_{\scriptscriptstyle -1} < \abig_i < p_{\scriptscriptstyle 1}$, then the duplicator
in the game on $\Abig$ and $\Bbig$ can choose the identical element in the universe of
$\Bbig$, i.e., she chooses $\bbig_i\deff\abig_i$.
\par
Similar to Case~I it is straightforward to check that
after $k$ rounds the duplicator has won
the $\FO(<,\Mon')$-game on $\Abig$ and $\Bbig$. 
This completes the proof of Theorem~\ref{theorem:collapse_Monadic_UU} for Case III, i.e., 
for the case that $\adom(\A)$ and $\adom(\B)$ are infinite in both directions.
\\
\parno
Case~IV, i.e., the case where $\adom(\A)$ and $\adom(\B)$ are finite, 
can be treated in a similar way as Case~I.
However, unlike in the previous cases, we cannot take the ``special positions''
$p_1<p_2<\cdots$ from the active domain of $\A$, since $\adom(\A)$ is \emph{finite}.
However, 
since $\UU$ is \emph{infinite}, there must exist an infinite increasing sequence 
$u_1<u_2<\cdots$ or an infinite decreasing sequence $u_{-1}>u_{-2}>\cdots$ (see  
Fact~\ref{fact:Analysis} below).
If we have an infinite increasing sequence $u_1<u_2<\cdots$, we can proceed in the same way as
in Case~I to obtain an infinite subsequence $p_1<p_2<\cdots$ such that
$\ktype[p_1,p_2) = \ktype[p_j,p_{j+1})$, for all $j\in\NNpos$.
Define $\alpha$ and $\beta$ to be the ${<}$-preserving mappings which move the
active domain elements of $\A$ and $\B$ onto the ``special positions'' $p_1<p_2<\cdots$.
The rest of the proof is identical to the proof for Case~I.
The case where we have an infinite \emph{de}creasing sequence in $\UU$ is symmetric to
the case where we have an infinite \emph{in}creasing sequence in $\UU$. 
\\
Altogether, this completes the proof of Theorem~\ref{theorem:collapse_Monadic_UU}.
\end{proof_mit}%
\\
\parno
In the above proof we used the following well-known fact from \emph{Analysis}:
\begin{fact_mit}\label{fact:Analysis}
Let $\struc{\UU,<}$ be a linearly ordered infinite structure.
There exists an infinitely increasing sequence \index{infinitely increasing sequence}
$u_1<u_2<\cdots$ or \index{infinitely decreasing sequence}
an infinitely decreasing sequence $u_1>u_2>\cdots$ of elements in $\UU$.
\null\mbox{\quad }
\end{fact_mit}%
To conclude the investigation of the class $\Mon$ of \emph{monadic}
predicates, let us mention that 
several generalizations of the notion of \emph{$\ZZ$-embeddable}
structures are conceivable, to which the proof of
Theorem~\ref{theorem:collapse_Monadic_UU} can be
generalized --- e.g.: structures whose active domain is of the form
\,\(
  u_1 <  u_2  <  u_3  <  \cdots  <  v_3  <  v_2  <  v_1
\)\, 
where \,$u_1<u_2<u_3<\cdots$\, is 
infinitely increasing, \,$v_1>v_2>v_3>\cdots$\, is 
infinitely decreasing, and \,$u_i<v_j$\, for all $i,j\in\NNpos$.
\\
It remains open whether Theorem~\ref{theorem:collapse_Monadic_UU}
is still valid when replacing ``\emph{$\ZZ$-embeddable} structures''
with ``\emph{arbitrary} structures''.
%
%
%


\section[How to Win the Game for ${\FO(<,+,Q)}$]{How to Win the Game for $\bs{\FO(<,+,Q)}$}\label{section:FOPlus_game}\label{section:Addition}
\index{Ehrenfeucht-Fra\"\i{}ss\'{e} game!FO<+Q@$\FO(<,+,Q)$-game} 
\begin{summary}
In this section we concentrate on context structures with built-in
\emph{addition} relation $+$. We show that the duplicator can translate
strategies for the $\FOKleiner$-game into strategies for the
$\FOPlus$-game on arbitrary structures over $\NN$.
We even obtain the following extension of this result:
We enrich the context structures \,$\struc{\NN,<,+}$\, and
\,$\struc{\ZZ,<,+}$\, with a set $Q\subseteq\NN$ which is not definable in
$\FO(<,+)$. 
We expose certain conditions $W(\omega)$ and show that the duplicator can translate
strategies for the $\FOKleiner$-game into strategies for the
$\FO(<,+,Q)$-game on arbitrary structures over $\NN$ and on
$\NN$-embeddable structures over $\ZZ$, whenever $Q$ satisfies the
conditions $W(\omega)$.
This possibility of translating strategies for the augmented context structure 
\,$\struc{\NN,<,+,Q}$\, is notable especially in the light of
Fact~\ref{facts:NoCollapse}\;(b) which (together with
Theorem~\ref{theorem:EF-Collapse}) tells us  that the translation is \emph{not} possible when
replacing $Q$ with the set $\Squares$ of all square numbers.
\par
In Section~\ref{subsection:PlusQQRR} we transfer the
translation result to $\NN$-embeddable structures over the context structure 
\,$\struc{\RR,<,+,Q,\Groups}$, where $\Groups$ is the class of
all subsets of $\RR$ that contain the number $1$ and that are groups
with respect to $+$. In particular, this implies the translation
result for the context structures \,$\struc{\QQ,<,+}$,
\,$\struc{\QQ,<,+,\ZZ}$, \,and \,$\struc{\RR,<,+,\ZZ,\QQ}$.
In Section~\ref{subsection:Variations} we present some 
variations and consequences of the translation proofs, including the result that even
all subsets of $Q$ may be added as built-in predicates.
\par
Since the duplicator's strategy in the $\FO(<,+,Q)$-game is rather involved, we first 
concentrate on a basic case which, as a side product, will give us an
EF-game proof of the theorem of Ginsburg and Spanier, stating that the spectra of 
$\FO(<,+)$-sentences are \emph{semi-linear}.
\end{summary}
%
%
%
\subsection[A Basic Case of the $\FOPlus$-Game over $\ZZ$]{A Basic Case of the $\bs{\FOPlus}$-Game over $\bs{\ZZ}$}\label{subsection:FOPlus_game:easy}
\begin{figure}[!htbp]
\bigskip
\begin{center}
\fbox{
\scalebox{0.5}{
\begin{pspicture}(-1.5,-2)(14,+3)
\psset{linewidth=0.8pt,linecolor=black,linestyle=solid,dotsize=6pt,arrowsize=5pt}%
%
\rput(-0.5,2){\LARGE$\Abig$:}%
\rput(-0.5,-1){\LARGE$\Bbig$:}%
%
\psset{dotsize=5pt}%
%
\pnode(0.5,2){NullAA}\pnode(13.5,2){MaxAA}%
\ncline{NullAA}{MaxAA}%
\multirput(1,2)(1,0){13}{$\mid$}%
\dotnode(2,2){s3}\dotnode(5.5,2){s1}\dotnode(7,2){s2}\dotnode(8,2){s4}\dotnode(12.5,2){s5}%
\rput(2,2.5){\Large$\ov{s_3}$}\rput(5.5,2.5){\Large$\ov{s_1}$}\rput(7,2.5){\Large$\ov{s_2}$}%
\rput(8,2.5){\Large$\ov{s_4}$}\rput(12.5,2.5){\Large$\ov{s_5}$}%
%
\pnode(0.5,-1){NullBB}\pnode(13.5,-1){MaxBB}%
\ncline{NullBB}{MaxBB}%
\multirput(1,-1)(1,0){13}{$\mid$}%
\dotnode(3,-1){t3}\dotnode(4.5,-1){t1}\dotnode(8,-1){t2}\dotnode(9,-1){t4}\dotnode(11.5,-1){t5}%
\rput(3,-1.5){\Large$\ov{t_3}$}\rput(4.5,-1.5){\Large$\ov{t_1}$}\rput(8,-1.5){\Large$\ov{t_2}$}%
\rput(9,-1.5){\Large$\ov{t_4}$}\rput(11.5,-1.5){\Large$\ov{t_5}$}%
%
\psset{linestyle=dashed}%
\ncline{s1}{t1}\ncline{s2}{t2}\ncline{s3}{t3}\ncline{s4}{t4}\ncline{s5}{t5}%
\end{pspicture}
}
}
\caption{\small The evaluations of all linear combinations
  $s$ in $S$ and all linear combinations $t$ in $T$. Integers are represented
  by strokes.}\label{figure:Plus_einfach}
\end{center}
\end{figure}
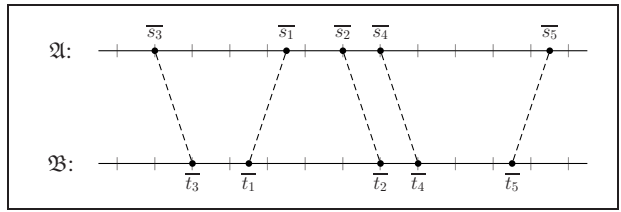

\index{Ehrenfeucht-Fra\"\i{}ss\'{e} game!FO<+@$\FO(<,+)$-game} 
Assume that we are given a number $n\in\NN$ and two structures
\,$\Abig\deff \struc{\ZZ,<,+,\abig_1,\twodots,\abig_n}$\, and
\,$\Bbig \deff \struc{\ZZ,<,+,\bbig_1,\twodots,\bbig_n}$.
The aim of this section is to find, for each $k\in\NNpos$, a list $W(k)$ of conditions
such that the duplicator wins the $k$-round EF-game on $\Abig$ and $\Bbig$ whenever
$\abig_1,\twodots,\abig_n$ and $\bbig_1,\twodots,\bbig_n$ 
satisfy the conditions $W(k)$.
This question has been considered before: 
\begin{enumerate}[$\bullet$]
\item
In the textbook \cite[Section~3.3]{Hodges} such
conditions were formulated, aiming at a proof for the decidability of Presburger arithmetic.
\item
Ruhl \index{Ruhl, Matthias}%
\cite{Ruhl} obtained according conditions for the (more difficult) 
$k$-round EF-game for first-order logic with unary counting quantifiers and addition. 
\item
Lynch \index{Lynch, James F.}%
\cite{Lynch_Addition} developed a winning strategy for the duplicator in the $k$-round 
$\FO(<,\allowbreak +,P_k)$-game, for a suitable set $P_k$ of natural numbers. 
\item
Lautemann and the author of the present paper \cite{LS_STACS01} extended Lynch's method
in order to show that the duplicator can translate strategies for the $\FOKleiner$-game into
strategies for the $\FOPlus$-game; we will prove (an extension of) this in the following
Section~\ref{subsection:FOPlus_game:general}.
\end{enumerate}%
All the above references are written in a \emph{top-down} manner, i.e., they first 
formulate the (very involved) conditions, and afterwards they prove that the conditions
indeed lead to a winning strategy for the duplicator. However, it remains unclear 
{how} one can find such conditions and {why} 
they need to be chosen in the way they are.
In the present section we try to answer this question by developing the conditions in
a \emph{bottom-up} manner.
\\
We start with $k=1$.
In the unique round of the EF-game elements $\abig_{n+1}$ and $\bbig_{n+1}$ are chosen
in $\Abig$ and $\Bbig$ --- and afterwards the duplicator shall have won the game. I.e., 
for all $\mu,\nu,\eta\in\set{1,\twodots,n{+}1}$ we shall have 
\[
\abig_{\mu} < \abig_{\nu}
\mbox{ \ iff \ }
\bbig_{\mu} < \bbig_{\nu} 
\qquad\mbox{ and }\qquad
\abig_{\mu} + \abig_{\nu} = \abig_{\eta} 
\mbox{ \ iff \ }
\bbig_{\mu} + \bbig_{\nu} = \bbig_{\eta}.
\]
%
What conditions do these atoms impose on $\abig_{n+1}$ and $\bbig_{n+1}$?\\
Let us have a look at all atoms that involve $\abig_{n+1}$, and let us solve these atoms
for $\abig_{n+1}$:
\[
\begin{array}{rclcrcl}
  \multicolumn{3}{r}{\mbox{atoms involving }\abig_{n+1}} & \quad &
  \multicolumn{3}{l}{\mbox{solved for }\abig_{n+1}}
\\ \hline
  \abig_{n+1} & = & \abig_{\mu} & \quad & 
  \abig_{n+1} & = & \abig_{\mu} 
\\
  \abig_{n+1} & < & \abig_{\mu} & \quad & 
  \abig_{n+1} & < & \abig_{\mu} 
\\
  \abig_{n+1} & > & \abig_{\mu} & \quad & 
  \abig_{n+1} & > & \abig_{\mu} 
\\
  \abig_{\mu} + \abig_{\nu} & = & \abig_{n+1} &  &
  \abig_{n+1} & = & \abig_{\mu} + \abig_{\nu}
\\
  \abig_{n+1} + \abig_{\nu} & = & \abig_{\mu} &  &
  \abig_{n+1} & = & \abig_{\mu} - \abig_{\nu}
\\
  \abig_{\mu} + \abig_{\mu} & = & \abig_{n+1} &  &
  \abig_{n+1} & = & 2 \abig_{\mu} 
\\
  \abig_{n+1} + \abig_{n+1} & = & \abig_{\mu} &  &
  \abig_{n+1} & = & \frac{1}{2} \abig_{\mu} 
\\
  \abig_{n+1} + \abig_{n+1} & = & \abig_{n+1} &  &
  \abig_{n+1} & = & 0 
\\
  \abig_{n+1} + \abig_{\mu} & = & \abig_{\mu} &  &
  \abig_{n+1} & = & 0 
\\
  \abig_{n+1} + \abig_{\mu} & = & \abig_{n+1} &  &
  \multicolumn{3}{l}{\mbox{no condition on }\abig_{n+1}}\vspace{3ex}
\end{array}
\]
On the righthand side of the equations
\,``$\abig_{n+1}= \cdots$''\, we have terms, or \emph{linear combinations}, 
of the form \,$d_1\abig_{\mu} + d_2 \abig_{\nu}$, where
$\mu,\nu\in\set{1,\twodots,n}$, $\mu\neq \nu$, $d_1,d_2\in
\QQ[2]\deff \setc{\frac{u}{u'}}{u,u'\in\ZZ,\ u'\neq 0,\ |u|,|u'|\leq 2}$.
Each such linear combination $s$ evaluates to a real number $\ov{s}$. 
Let $S$ be the set of all these linear combinations, and let
$T$ be the according set of linear combinations obtained from
replacing $\abig_1,\twodots,\abig_n$ with $\bbig_1,\twodots,\bbig_n$. 
I.e., if $s\in S$ is of the
form $d_1\abig_{\mu} + d_2\abig_{\nu}$, then the linear combination
$t \deff d_1\bbig_{\mu} + d_2\bbig_{\nu}$ is the according element of
$T$ that corresponds to $s$.
The evaluations of these linear combinations are distributed over the
real numbers. An illustration is given in
Figure~\ref{figure:Plus_einfach}.
%
%
%
\\
Certainly, if the spoiler chooses $\abig_{n+1}=\ov{s}$, for some $s\in
S$, then the
duplicator should answer $\bbig_{n+1}\deff \ov{t}$, for the
corresponding $t\in T$. 
Similarly, if the spoiler's choice $\abig_{n+1}$ lies strictly between $\ov{s_{1}}$ and
$\ov{s_{2}}$, for $s_{1}, s_{2}\in S$, then the
duplicator should answer a $\bbig_{n+1}$ that
lies strictly between $\ov{t_{1}}$ and $\ov{t_{2}}$, for
the corresponding $t_{1},t_{2}\in T$. Obviously, the
duplicator wins if the following conditions are satisfied:
\begin{enumerate}[$\bullet$\ ]
  \item 
    The numbers $\ov{s}$, for all $s\in S$, are ordered in the same way as the
    corresponding numbers $\ov{t}$, for all $t\in T$,  
  \item 
    $\ov{s}$ is an integer if and only if the corresponding $\ov{t}$ is, and 
  \item 
    there is an integer between $\ov{s_1}$ and $\ov{s_{2}}$ if
    and only if there is an integer between $\ov{t_1}$ and
    $\ov{t_{2}}$
    (for all $s_1,s_2\in S$ and the corresponding $t_1,t_2\in T$).
\end{enumerate} 
Precisely, the following procedure leads to a winning strategy for the duplicator:
\\
If the spoiler chooses $\abig_{n+1}\in\ZZ$ in $\Abig$ such that $\abig_{n+1}=\ov{s}+f$, 
for some
$s\in S$ and
$f\in\intcc{{-}\frac{1}{2},{+}\frac{1}{2}} \subseteq\RR$, 
then the duplicator answers $\bbig_{n+1}\deff \ov{t}+f$, where $t$ is
the according linear combination that corresponds to $s$. 
The \emph{gap parameter} $f$ is added here to ensure that there is an integer
between $\ov{s_1}$ and $\ov{s_{2}}$ if and only if there is an integer between 
$\ov{t_1}$ and $\ov{t_{2}}$. Indeed, it would suffice to restrict attention to
\emph{rational} $f\in\QQ[2]$. However, later in this section, in the proof of
Theorem~\ref{theo:additionQQRR} we will essentially need that the duplicator's strategy works
for all \emph{real} numbers \,$f\in\intcc{{-}\frac{1}{2},{+}\frac{1}{2}} \subseteq\RR$.
\\
Certainly, the duplicator will win if the following two conditions are satisfied:
\begin{enumerate}[(1.)\ ]
 \item
   $\ov{s_{1}}+f\ < \ \ov{s_{2}}+h$ \quad iff \quad
   $\ov{t_{1}}+f\ < \ \ov{t_{2}}+h$ \\
   for all \,$s_1,s_2\in S$ and the corresponding $t_1,t_2\in T$, and
   all \,$f,h\in\intcc{{-}\frac{1}{2},{+}\frac{1}{2}}$.
 \item
   $\ov{s}+f \in \ZZ$ \quad iff \quad 
   $\ov{t}+f \in \ZZ$ \\
   for all \,$s\in S$ and the corresponding $t\in T$, and
   all \,$f\in\intcc{{-}\frac{1}{2},{+}\frac{1}{2}}$.
\end{enumerate}
Since the denominator of a coefficient $d$ in a linear combination $s$ is either
$1$ or $2$ or $-1$ or $-2$, condition (2.) is equivalent to the condition\footnote{Recall from
Section~\ref{section:Preliminaries} that $\equiv_2$ denotes the congruence relation modulo 2.}
\begin{enumerate}[(2.)'\ ]
  \item
   $\abig_{\nu} \equiv_2 \bbig_{\nu}$ \quad for all \,$\nu\in\set{1,\twodots,n}$.
\end{enumerate}
If the spoiler chooses $\abig_{n+1}\in\ZZ$ in $\Abig$ such that
$\abig_{n+1}\neq \ov{s}+f$ for all $s\in S$ and all 
$f \in \intcc{{-}\frac{1}{2},{+}\frac{1}{2}}$, then determine the interval w.r.t.\
$S$ to which $\abig_{n+1}$ belongs. I.e., choose
$s_{\links}, s_{\rechts} \in S$ such that $\ov{s_{\links}}<\abig_{n+1}<\ov{s_{\rechts}}$ and,
for all $s\in S$,\,
$\ov{s}\leq \ov{s_{\links}}$ \ or \ $\ov{s}\geq\ov{s_{\rechts}}$.
\\
Now, the duplicator takes her answer $\bbig_{n+1}$ from the corresponding interval in 
$\Bbig$. 
I.e., she chooses the linear combinations $t_{\links}, t_{\rechts}\in T$ that correspond to
$s_{\links}, s_{\rechts}$, and 
she answers with an arbitrary $\bbig_{n+1}\in\ZZ$ such that 
$\ov{t_{\links}}<\bbig_{n+1}<\ov{t_{\rechts}}$.
Such an \emph{integer} does really exist, because we
know that $\ov{s_{\links}}+\frac{1}{2}<\abig_{n+1}<\ov{s_{\rechts}}-\frac{1}{2}$ and, due to 
condition (1.), \,$\ov{t_{\links}}+\frac{1}{2}<\ov{t_{\rechts}}-\frac{1}{2}$,\,
i.e., \,$\ov{t_{\rechts}}-\ov{t_{\links}}>1$.
\\
What we have seen is the following:
\longpage
\begin{lemma_ohne}[$\bs{W(1) \Rightarrow\ \dwins_1}$]\label{lemma:Plus_ErsteRunde}\mbox{ }\\
Let $n\in\NN$, let $\abig_1,\twodots,\abig_n,\bbig_1,\twodots,\bbig_n\in\ZZ$, and let
$\Abig\deff\struc{\ZZ,<,+,\abig_1,\twodots,\abig_n}$ and
$\Bbig\deff\struc{\ZZ,<,\allowbreak +,\bbig_1,\twodots,\bbig_n}$. 
The duplicator has a winning strategy in the 1-round EF-game on 
$\Abig$ and $\Bbig$
if the following conditions $W(1)$ are satisfied: \index{conditions $W(1)$} 
\begin{enumerate}[$(**)$]
 \item[$(*)$]
   $\abig_{\nu} \equiv_2 \bbig_{\nu}$\quad for all \,$\nu\in\set{1,\twodots,n}$, \quad and
 \item[$(**)$]
   for all \,$f,h\in\intcc{{-}\frac{1}{2},{+}\frac{1}{2}}\subseteq\RR$,\,
   for all \,$\nu_1,\nu_2,\mu_1,\mu_2\in\set{1,\twodots,n}$\, with \,$\nu_1\neq \nu_2$\, and 
   \,$\mu_1\neq \mu_2$,\, and \index{Q2@$\QQ[2]$} 
   all $d_1,d_2,e_1,e_2\in\QQ[2] \deff 
   \setc{\frac{u}{u'}}{u,u'\in\ZZ,\ u'\neq 0,\ |u|,|u'|\leq 2}$,\, we have
   \[
     \begin{array}{c}
         d_1\abig_{\nu_1}\, + \,d_2\abig_{\nu_2}\, + \,f \ \ < \ \ 
         e_1\abig_{\mu_1}\, + \,e_2\abig_{\mu_2}\, + \,h
       \\[1ex]
         \mbox{ if and only if } 
       \\[1ex]
        d_1\bbig_{\nu_1}\, + \,d_2\bbig_{\nu_2}\, + \,f \ \ < \ \  
        e_1\bbig_{\mu_1}\, + \,e_2\bbig_{\mu_2}\, + \,h\,.
     \end{array}
   \]
   \vspace{-5ex}\\ \mbox{ }\fertig
\end{enumerate}
\end{lemma_ohne}%
Let us now concentrate on the $2$-round EF-game on 
$\Abig\deff\struc{\ZZ,<,+,\abig_1,\twodots,
\abig_n}$ and $\Bbig \deff \struc{\ZZ,{<,} \allowbreak +,\bbig_1,\twodots,\bbig_n}$.
Our aim is to find a list $W(2)$ of conditions that enable the duplicator to play 
the first round in such a way that afterwards the conditions $W(1)$
are satisfied. From Lemma~\ref{lemma:Plus_ErsteRunde} we
then obtain that the duplicator can play the remaining round in such a
way that she wins the game.\\
In general, by induction on $k$, we will find a list $W(k{+}1)$
of conditions that enable the duplicator to play the first round in such a way that
afterwards the conditions $W(k)$ are satisfied.
To this end we consider the following generalization of the conditions
$W(1)$.
\begin{definition_ohne}[$\bs{(l,c,g)}$-Combinations; Conditions
  $\bs{C(m,l,c,g)}$]\label{definition:conditions_Plus_einfach}
\index{lcgcombination@$(l,c,g)$-combination}
\index{conditions $C(m,l,c,g)$}\mbox{ }\\
Let $m,l,c\in\NNpos$ and $g\in\RRpos$. 
Here, 
{
\begin{enumerate}[$\bullet$\ ]
 \item 
   $m$ is the \emph{modulus} with respect to which
   $\abig_{\nu}$ and $\bbig_{\nu}$ shall be congruent, 
 \item 
   $l$ is the maximum \emph{length} of the linear combinations
   under consideration,
 \item  
   $c$ is the maximum size of the numerator and the
   denominator of the \emph{coefficients} occurring in linear
   combinations, and 
 \item 
   $g$ is the maximum size of the \emph{gap parameters} that \index{gap parameter}
   are respected by the linear combinations.
\end{enumerate}}\noindent%
Let $n\in\NNpos$ and let
$\abig_1,\twodots,\abig_n,\bbig_1,\twodots,\bbig_n\in\ZZ$.
\medskip\\
An \emph{$(l,c,g)$-combination over ${\abig_1,\twodots,\abig_n}$}\, is a 
formal sum, or a \emph{linear combination}, of the form \index{linear combination}
\,$\sum_{i=1}^{{l'}}d_{i}\abig_{\nu_{i}} + f$,\, where \,${l'}\leq l$,\,
$\nu_1,\twodots,\nu_{{l'}}$\, are pairwise distinct elements in $\set{1,\twodots,n}$, \index{Q3@$\QQ[c]$}
\,$d_1,\twodots,d_{{l'}} \in \QQ[c] \deff
\setc{\frac{u}{u'}}{u,u'\in\ZZ,\ u'\neq 0,\ |u|,|u'|\leq c}$,\, 
and $f\in\intcc{{-}g,{+}g}\,\subseteq\,\RR$. 
\\
Every $(l,c,g)$-combination $s$ evaluates to a real number $\ov{s}$.
\medskip\\
Given an $(l,c,g)$-combination $s$ over $\abig_1,\twodots,\abig_n$,
the according $(l,c,g)$-combination $t$ over
$\bbig_1,\twodots,\bbig_n$ that \emph{corresponds} to $s$ is obtained
by replacing every $\abig_{\nu}$ in $s$ with $\bbig_{\nu}$. 
I.e., if \,$s = \sum_{i=1}^{{l'}}d_{i}\abig_{\nu_{i}} +
f$,\, then \,$t = \sum_{i=1}^{{l'}}d_{i}\bbig_{\nu_{i}} + f$.
\medskip\\
We say that $\abig_1,\twodots,\abig_n$ and
$\bbig_1,\twodots,\bbig_n$ \emph{satisfy the conditions $C(m,l,c,g)$} if
and only if
\begin{enumerate}[$(**)$]
 \item[$(*)$] 
   $\abig_{\nu} \equiv_m \bbig_{\nu}$\quad for all \,$\nu\in\set{1,\twodots,n}$, \quad and
 \item[$(**)$]
   for all $(l,c,g)$-combinations $s_1$ and $s_2$ over 
   ${\abig_1,\twodots,\abig_n}$\, 
   and the corresponding $(l,c,g)$-combinations $t_1$ and $t_2$ over 
   $\bbig_1,\twodots,\bbig_n$\,
   we have
   \begin{eqnarray*}
      \ov{s_1} < \ov{s_2}
    & \quad\mbox{if and only if}\quad
    & \ov{t_1} < \ov{t_2}\,. 
   \end{eqnarray*}
\end{enumerate}
\mbox{ } \fertig
\end{definition_ohne}%
In particular, the conditions $W(1)$ are exactly the
conditions $C(2,2,2,\frac{1}{2})$.
\\
\parno
Our aim is now to find, for given parameters $m,l,c,g$, new parameters
$\tilde{m},\tilde{l},\tilde{c},\tilde{g}$ such that the following is true: If the
conditions $C(\tilde{m},\tilde{l},\tilde{c},\tilde{g})$ are satisfied at the beginning, then
the duplicator can play one round of the EF-game in
such a way that afterwards the conditions $C(m,l,c,g)$ are satisfied.
\par
To this end, let $n\in\NN$, let $\abig_1,\twodots,\abig_n,
\bbig_1,\twodots,\bbig_n\in\ZZ$, and let
$\Abig\deff\struc{\ZZ,<,\allowbreak +,\abig_1,\twodots,\abig_n}$ and
$\Bbig\deff\struc{\ZZ,<,+,\bbig_1,\twodots,\bbig_n}$.
In one round of the EF-game elements $\abig_{n+1}$ and $\bbig_{n+1}$
are chosen in $\Abig$ and $\Bbig$, and afterwards the conditions
$C(m,l,c,g)$ shall be satisfied by $\abig_1,\twodots,\abig_{n+1}$ and
$\bbig_1,\twodots,\bbig_{n+1}$. I.e.,
\begin{enumerate}[$(**)$]
 \item[$(*)$]
   $\abig_{\nu} \equiv_m \bbig_{\nu}$\quad for all \,$\nu\in\set{1,\twodots,n{+}1}$, \quad and
 \item[$(**)$]
   for all $(l,c,g)$-combinations over ${\abig_1,\twodots,\abig_{n+1}}$ of the form 
   \,$\sum_{i=1}^{{l'}}d_{i}\abig_{\nu_{i}}+f$\, and
   \,$\sum_{i=1}^{{l''}}e_{i}\abig_{\mu_{i}}+h$\,
   we have
   \[
     \begin{array}{c}
         \sum_{i=1}^{{l'}}d_{i}\abig_{\nu_{i}}\, + \,f \ \ < \ \ 
         \sum_{i=1}^{{l''}}e_{i}\abig_{\mu_{i}}\, + \,h
       \\[1ex]
         \mbox{ if and only if } 
       \\[1ex]
        \sum_{i=1}^{{l'}}d_{i}\bbig_{\nu_{i}}\, + \,f \ \ < \ \  
        \sum_{i=1}^{{l''}}e_{i}\bbig_{\mu_{i}}\, + \,h\,.
     \end{array}
   \]
\end{enumerate}
What conditions do the inequalities of $(**)$ impose on $\abig_{n+1}$
and $\bbig_{n+1}$?
To answer this question, we have a look at all inequalities that
involve $\abig_{n+1}$ and we solve them for $\abig_{n+1}$.
Let, for example, $\nu_1=\mu_1=n{+}1$, let $d_1>e_1$, let 
$\nu_2=\mu_2\neq n{+}1$, and let the indices
$\nu_3,\twodots,\nu_{{l'}},\mu_3,\twodots,\mu_{{l'}}$ be pairwise
distinct (and different from $n{+}1$ and $\nu_2$).
In this case, we have
{
\[
 \begin{array}{c}
     {\displaystyle\sum_{i=1}^{{l'}}}d_{i}\abig_{\nu_{i}}\, + \,f \ \ < \ \ 
     {\displaystyle\sum_{i=1}^{{l''}}}e_{i}\abig_{\mu_{i}}\, + \,h
   \\[4ex]\displaystyle
     \mbox{if and only if}
   \\[2ex]
     (d_1{-}e_1)\, \abig_{n+1} \quad < \quad 
     (e_2{-}d_2)\, \abig_{\nu_2} \, + \,
     {\displaystyle\sum_{i=3}^{{l''}}} e_{i} \abig_{\mu_{i}} \ -\,  
     {\displaystyle\sum_{i=3}^{{l'}}} d_{i} \abig_{\nu_{i}} \ + \,
     (h {-} f)
   \\[4ex]\displaystyle
     \mbox{if and only if}
   \\[2ex]
     (*{**}):\qquad\quad \abig_{n+1} \quad < \quad 
     \frac{e_2{-}d_2}{d_1{-}e_1}\, \abig_{\nu_2} \, + \,
     {\displaystyle\sum_{i=3}^{{l''}}} \frac{e_{i}}{d_1{-}e_1}\,
     \abig_{\mu_{i}} \ +\,  
     {\displaystyle\sum_{i=3}^{{l'}}}\frac{{-}d_{i}}{d_1{-}e_1} \,
     \abig_{\nu_{i}} \ + \,
     \frac{h {-} f}{d_1{-}e_1}\,.
 \end{array}
\]}\noindent%
Let us have a close look at the coefficients on the
righthand side of the last inequality $(*{**})$: 
We know that $d_{i},e_{i}\in\QQ[c]$, i.e., that 
$d_{i} = \frac{u_{i}}{u'_{i}}$ and $e_{i} = \frac{v_{i}}{v'_{i}}$ for 
suitable integers
$u_{i},u'_{i},v_{i},v'_{i}$ with $\betrag{u_{i}},\betrag{u'_{i}},
\betrag{v_{i}},\betrag{v'_{i}}\leq c$.
Hence, $d_1{-}e_1 = \frac{u_1}{u'_1}-\frac{v_1}{v'_1} = 
\frac{u_1v'_1 - v_1u'_1}{u'_1v'_1}$. In particular, 
$\frac{e_2-d_2}{d_1-e_1} = 
\frac{v_2u'_2 - u_2v'_2}{u'_2v'_2}\cdot \frac{u'_1v'_1}{u_1v'_1-v_1u'_1} \,\in\,
\QQ[2c^4]$. Obviously, also the other coefficients $\frac{e_{i}}{d_1-e_1}$ and
$\frac{{-}d_{i}}{d_1-e_1}$ belong to $\QQ[2c^4]$.
Similarly, the gap parameter $\frac{h-f}{d_1-e_1}$ belongs to
$\intcc{{-}2gc^2,{+}2gc^2}\subseteq\RR$, because
$\betrag{\frac{h-f}{d_1-e_1}} =
 \betrag{h-f}\cdot\betrag{\frac{u'_1v'_1}{u_1v'_1-v_1u'_1}} \leq
 (\betrag{h}+\betrag{f})\cdot\betrag{u'_1}\cdot\betrag{v'_1} \leq 2gc^2$.
Altogether, the righthand side of the inequality 
$(*{**})$ 
is a
$(2l{-}1,\,2c^4,\,2gc^2)$-combination over ${\abig_1,\twodots,\abig_n}$.
\\
Indeed, one can easily see that \emph{every} inequality of $(**)$ that involves
$\abig_{n+1}$ 
\begin{enumerate}[$\bullet$]
 \item
   is {either}, for $\ltimes\in\set{{<},{>}}$, equivalent to an inequality of the form
   ``$\abig_{n+1}\ltimes\cdots$'', the righthand side of which is a
   $(2l{-}1,2c^4,2gc^2)$-combination over ${\abig_1,\twodots,\abig_n}$,\quad {or}
 \item 
   does not impose any condition on $\abig_{n+1}$ at all.
\end{enumerate}
Let $S$ be the set of all $(2l{-}1,2c^4,2gc^2)$-combinations 
over ${\abig_1,\twodots,\abig_n}$, and let
$T$ be the according set of $(2l{-}1,2c^4,2gc^2)$-combinations for 
$\bbig_1,\twodots,\bbig_n$ instead of $\abig_1,\twodots,\abig_n$. 
If $s\in S$ is of the form \,$\sum_{i=1}^{l'}d_{i}\abig_{\nu_{i}} +
f$,  then \,$t \deff \sum_{i=1}^{l'}d_{i}\bbig_{\nu_{i}} + f$\, is the
according element in $T$ that corresponds to $s$.
The evaluations $\ov{s}$ (for all $s\in S$) and
$\ov{t}$ (for all $t\in T$) of these linear combinations are distributed over the
real numbers.
\\
Certainly, if the spoiler chooses $\abig_{n+1}=\ov{s}$, for some $s\in
S$, then the
duplicator should answer $\bbig_{n+1}\deff \ov{t}$, for the
corresponding $t\in T$. Similarly, if the
spoiler's choice $\abig_{n+1}$ lies strictly between $\ov{s_{1}}$ and
$\ov{s_{2}}$, for $s_{1}, s_{2} \in S$, then the
duplicator should answer a $\bbig_{n+1}$ that
lies strictly between $\ov{t_{1}}$ and $\ov{t_{2}}$ and that belongs to the same
residue class modulo $m$ as $\abig_{n+1}$ (here, $t_{1}$ and $t_{2}$ are the
according linear combinations that correspond to $s_{1}$ and $s_{2}$). Afterwards, 
$\abig_1,\twodots,\abig_{n+1}$ and $\bbig_1,\twodots,\bbig_{n+1}$ satisfy the
conditions $C(m,l,c,g)$, if the following is true:
\begin{enumerate}[$\bullet$\ ]
  \item 
    The numbers $\ov{s}$, for all $s\in S$, are ordered in the same way
    as the corresponding
    numbers $\ov{t}$, for all $t\in T$, 
  \item 
    $\ov{s} \equiv_m \ov{t}$, \ for every $s\in S$ and the
    corresponding $t\in T$, \quad and 
  \item 
    for every $r\in\set{0,\twodots,m{-}1}$, there is 
    an  integer $a$ between $\ov{s_1}$ and $\ov{s_2}$ with $a \equiv_m r$ \,if
    and only if\, there is an integer $b$ between $\ov{t_1}$
    and $\ov{t_2}$ with
    $b \equiv_m r$ \ (for all $s_1,s_2\in S$ and the corresponding
    $t_1,t_2\in T$).
\end{enumerate} 
Precisely, the following procedure leads to a successful strategy for the duplicator:
\\
If the spoiler chooses $\abig_{n+1}\in\ZZ$ in $\Abig$ such that 
$\abig_{n+1}=\ov{s}+f'$, 
for some
$s\in S$ and
$f'\in\intcc{{-}\frac{m}{2},{+}\frac{m}{2}} \subseteq\RR$, 
then the duplicator answers $\bbig_{n+1}\deff \ov{t}+f'$, where
$t\in T$ is the according linear combination that corresponds to $s$. 
The \emph{gap parameter} $f'$ is added here to ensure, for every 
$r\in\set{0,\twodots,m{-}1}$, that there is an integer $a$ 
between $\ov{s_1}$ and $\ov{s_2}$ with $a\equiv_m r$ if and only if there is an 
integer $b$ between $\ov{t_1}$ and $\ov{t_2}$ with $b\equiv_m r$.
\\
Certainly, the conditions $C(m,l,c,g)$ are satisfied if the following 
is true:
\begin{enumerate}[(1.)\ ]
 \item
   $\ov{s_1}+f'\ < \ \ov{s_2}+h'$ \quad iff \quad
   $\ov{t_1}+f'\ < \ \ov{t_2}+h'$ \\
   for all \,$s_1,s_2\in S$ and the corresponding $t_1,t_2\in T$, and
   all \,$f',h'\in\intcc{{-}\frac{m}{2},{+}\frac{m}{2}}$.
 \item
   $\ov{s}+f'\,\equiv_m\,\ov{t}+f'$ \\
   for all \,$s\in S$ and the corresponding $t\in T$, and
   all \,$f'\in\intcc{{-}\frac{m}{2},{+}\frac{m}{2}}$.
\end{enumerate}
As explained below, condition (2.) can be replaced by the condition\footnote{Recall 
that $\lcm\set{n_1,\twodots,n_k}$ denotes the
\emph{least common multiple} of $n_1,\twodots,n_k$.}
\begin{enumerate}[(2.)'\ ]
  \item
   $\abig_{\nu} \equiv_{m\cdot\lcm\set{1,\twodots,2c^4}} \bbig_{\nu}$ \quad 
   for all \,$\nu\in\set{1,\twodots,n}$.
\end{enumerate}
This can be seen as follows:
Let $s$ be of the form $\sum_{i =1}^{l'}d_{i}\abig_{\nu_{i}}+f$. 
We know that all the coefficients $d_{i}$ belong to $\QQ[2c^4]$. I.e.,  
$d_{i}= \frac{u_{i}}{u'_{i}}$ with $u_{i},u'_{i}\in \ZZ$, 
$u'_{i}\neq 0$, and $\betrag{u_{i}},\betrag{u'_{i}} \leq 2c^4$.
By definition of $\equiv_m$ we have $\ov{s}+f' \,\equiv_m\, \ov{t} +f'$\,
if and only if \,there is an integer $z\in\ZZ$ such that $\ov{s}-\ov{t} = m\cdot z$.
\\
Of course, $\ov{s}-\ov{t} = \sum_{i = 1}^{l'}
d_{i}\,(\abig_{\nu_{i}}{-}\bbig_{\nu_{i}}) =
\sum_{i = 1}^{l'} u_{i}\cdot\frac{\abig_{\nu_{i}}-\bbig_{\nu_{i}}}{u'_{i}}$.\\
Now, if $\abig_{\nu_{i}} \equiv_{m\cdot\lcm\set{1,\twodots,2c^4}} \bbig_{\nu_{i}}$, then
$\abig_{\nu_{i}}{-}\bbig_{\nu_{i}} = z_{i} \cdot m\cdot\lcm\set{1,\twodots,2c^4}$ for
a suitable $z_{i}\in\ZZ$.
Thus, $\ov{s}-\ov{t} = \sum_{i = 1}^{l'} u_{i}\cdot
\frac{z_{i} \cdot m\cdot\lcm\set{1,\twodots,2c^4}}{u'_{i}} =
m\cdot
\sum_{i = 1}^{l'} u_{i}\cdot z_{i}\cdot\frac{\lcm\set{1,\twodots,2c^4}}{u'_{i}}$. 
Since $|u'_{i}|\in\set{1,\twodots,2c^4}$, we thus have found the desired integer
$z\deff 
\sum_{i = 1}^{l'} u_{i}{\cdot} z_{i}{\cdot}\frac{\lcm\set{1,\twodots,2c^4}}{u'_{i}}$
with $\ov{s}-\ov{t}= m\cdot z$. Altogether, this gives us that 
$\ov{s}+f' \,\equiv_m\,\ov{t}+f'$. I.e., condition (2.) follows from
condition (2.)'. (Indeed, one can easily see that both conditions are \emph{equivalent}).
\\
\parno
If the spoiler chooses $\abig_{n+1}\in\ZZ$ in $\Abig$ such that
$\abig_{n+1}\neq \ov{s}+f'$ for all $s\in S$ and all 
$f' \in \intcc{{-}\frac{m}{2},{+}\frac{m}{2}}$, then determine the interval w.r.t.\
$S$ to which $\abig_{n+1}$ belongs. I.e., choose
$s_{\links}, s_{\rechts} \in S$ such that $\ov{s_{\links}}<\abig_{n+1}<\ov{s_{\rechts}}$ and,
for all $s\in S$,\,
$\ov{s}\leq \ov{s_{\links}}$ \ or \ $\ov{s}\geq\ov{s_{\rechts}}$.
\\
Now, the duplicator takes her answer $\bbig_{n+1}$ from the corresponding interval in 
$\Bbig$. I.e., let $t_{\links}, t_{\rechts} \in T$ be the
according linear combinations that correspond to $s_{\links},
s_{\rechts}$.  The element $\bbig_{n+1}\in\ZZ$ is chosen such that 
$\ov{t_{\links}}<\bbig_{n+1}<\ov{t_{\rechts}}$ and $\bbig_{n+1}\equiv_m\abig_{n+1}$. 
Such an {integer} does really exist, because we
know that $\ov{s_{\links}}+\frac{m}{2}<\abig_{n+1}<\ov{s_{\rechts}}-\frac{m}{2}$ and, due to 
condition (1.), \,$\ov{t_{\links}}+\frac{m}{2}<\ov{t_{\rechts}}-\frac{m}{2}$,\,
i.e., \,$\ov{t_{\rechts}}-\ov{t_{\links}}>m$.
\\
\parno
What we have seen is that the conditions (1.) and (2.)' enable the duplicator to play
one round of the EF-game in such a way that afterwards the conditions $C(m,l,c,g)$ are
satisfied. Note that the conditions (1.) and (2.)' are exactly the conditions
$C(\tilde{m},\tilde{l},\tilde{c},\tilde{g})$, with parameters 
$\tilde{m},\tilde{l},\tilde{c},\tilde{g}$ as defined in the following lemma that 
sums up what we have obtained so far:
\begin{lemma_mit}[$\bs{C(\tilde{m},\tilde{l},\tilde{c},\tilde{g})\Rightarrow C(m,l,c,g)}$]\label{lemma:Plus_Conditions}
Let $m,l,c\in\NNpos$ and let $g\in\RRpos$. 
Define  
{
\[
 \begin{array}{rcl}
   \tilde{m} & \deff & m\cdot\lcm\set{1,\twodots,2c^4}\,,
 \\[1ex]
   \tilde{l} & \deff & 2\,l - 1\,,
 \\[1ex]
   \tilde{c} & \deff & 2\,c^4\,,
 \\[1ex]
   \tilde{g} & \deff & 2\,g\,c^2 + \frac{m}{2}\,.
 \end{array}
\]
}\noindent%
Let $n\in\NN$ and let
 $\abig_1,\twodots,\abig_n,\bbig_1,\twodots,\bbig_n\in\ZZ$.\\
Let \,$\Abig \deff \struc{\ZZ,<,+,\abig_1,\twodots,\abig_n}$, \,and let \,$\Bbig \deff
\struc{\ZZ,<,+,\bbig_1,\twodots,\bbig_n}$.
\\
If $\abig_1,\twodots,\abig_n$ and $\bbig_1,\twodots,\bbig_n$ satisfy the conditions 
$C(\tilde{m},\tilde{l},\tilde{c},\tilde{g})$, then the duplicator can play 
one round in the EF-game in which integers $\abig_{n+1}$ and $\bbig_{n+1}$ are chosen in 
$\Abig$ and $\Bbig$ in such a way that afterwards the conditions
$C(m,l,c,g)$ are satisfied by $\abig_1,\twodots,\abig_{n+1}$ and 
$\bbig_1,\twodots,\bbig_{n+1}$.
\end{lemma_mit}%
Using the Lemmas~\ref{lemma:Plus_ErsteRunde} and \ref{lemma:Plus_Conditions}, we can easily 
formulate, for every $k\in\NNpos$, conditions $W(k)$ which enable the duplicator to
win the $k$-round EF-game:
\begin{theorem_mit}[$\bs{W(k) \Rightarrow\ \dwins_k}$]\label{theorem:Plus_Wk} 
\index{conditions $W(k)$} 
By induction on $k$ we define the functions
\index{mk@$\mF(k)$}\index{lk@$\lF(k)$}\index{ck@$\cF(k)$}\index{gk@$\gF(k)$}   
{
\[
 \begin{array}{rcllrcl}
    \mF(1)     & \deff & 2\,, & \quad &
    \mF(k{+}1) & \deff & \mF(k)\cdot \lcm\set{1,\twodots,2\cF(k)^4}\,,
 \\[1ex]
    \lF(1)     & \deff & 2\,, & &
    \lF(k{+}1) & \deff & 2\,\lF(k)-1\,,
 \\[1ex]
    \cF(1)     & \deff & 2\,, & &
    \cF(k{+}1) & \deff & 2\,\cF(k)^4\,,
 \\[1ex] 
    \gF(1)     & \deff & \frac{1}{2}\,, & &
    \gF(k{+}1) & \deff & 2\, \gF(k)\, \cF(k)^2 + \frac{\mF(k)}{2}\,. 
 \end{array}
\]
}\noindent%
We define $W(k)$ to be exactly the conditions
$C\big(\mF(k),\,\lF(k),\,\cF(k),\,\gF(k)\big)$.
\\
Let $n\in\NN$, let $\abig_1,\twodots,\abig_n,\bbig_1,\twodots,\bbig_n\in\ZZ$, and
let $\Abig\deff\struc{\ZZ,<,+,\abig_1,\twodots,\abig_n}$ and
$\Bbig\deff\struc{\ZZ,{<,} \allowbreak +,\bbig_1,\twodots,\bbig_n}$.
\\
If $\abig_1,\twodots,\abig_n$ and $\bbig_1,\twodots,\bbig_n$ satisfy the conditions
$W(k)$, then the duplicator has a winning strategy in the $k$-round EF-game
on $\Abig$ and $\Bbig$.
\\
{\rm
The duplicator's strategy is summarized in Figure~\ref{figure:Plus_Strategie}, where
$\mF(0)\deff 1$.}
\end{theorem_mit}%
\begin{proof_mit}
By induction on $k$. The induction start is established in Lemma~\ref{lemma:Plus_ErsteRunde}.
The induction step from $k$ to $k{+}1$ follows from Lemma~\ref{lemma:Plus_Conditions}.
\end{proof_mit}%
\begin{figure}[!htb]
\bigskip
\begin{center}
\fbox{ \ %
\parbox{12cm}{\small 
\rule{0ex}{3ex}%
{\bf How to play the $\bs{i}$-th round 
(for $\bs{i\in\set{1,\twodots,k}}$)}\medskip\\ 
\small
In the $i$-th round, elements $\abig_{n+i}$ and $\bbig_{n+i}$ will be chosen in
$\Abig$ and $\Bbig$.
\medskip\\
Let $S$ be the set of all
\,{
$\big(\, \lF(k{-}i{+}1),\, \cF(k{-}i{+}1),\, 
\gF(k{-}i{+}1){-}\frac{\mF(k{-}i)}{2}\, \big)$}-combinations\,
over $\abig_1,\twodots,\abig_{n+i-1}$, and let $T$ be the according
set of linear combinations over $\bbig_1,\twodots,\bbig_{n+i-1}$.
\medskip\\
We consider the case where the spoiler chooses an element $\abig_{n+i}$ in $\Abig$.\\
(The case where he chooses an element $\bbig_{n+i}$ in $\Bbig$ is symmetric.)
\\
To find her answer $\bbig_{n+i}$, the duplicator distinguishes between two cases:
\begin{enumerate}[$\bullet$]
 \item
   If \,$\abig_{n+i} = \ov{s}+f'$\, for some $s\in S$ and
   $f'\in\intcc{{-}\frac{\mF(k{-}i)}{2}, {+}\frac{\mF(k{-}i)}{2}}$,
   then the duplicator answers $\bbig_{n+i}\deff \ov{t}+f'$, where
   $t$ is the according element in $T$ that corresponds to $s$.
 \item
   If $\abig_{n+i}\neq \ov{s}+f'$ for all $s\in S$ and all
   $f'\in\intcc{{-}\frac{\mF(k{-}i)}{2}, {+}\frac{\mF(k{-}i)}{2}}$,
   then the duplicator determines $s_{\links}, s_{\rechts} \in S$ such that
   $\ov{s_{\links}}<\abig_{n+i}<\ov{s_{\rechts}}$ and, for all $s\in S$, 
   $\ov{s} \leq \ov{s_{\links}}$ or $\ov{s}\geq \ov{s_{\rechts}}$. She
   chooses $t_{\links}, t_{\rechts}$ to be the according elements in $T$ that correspond to
   $s_{\links}, s_{\rechts}$, and she 
   answers an arbitrary $\bbig_{n+i}$ with $\ov{t_{\links}}<\bbig_{n+i}<\ov{t_{\rechts}}$ and
   $\abig_{n+i}\equiv_{\mF(k-i)} \bbig_{n+i}$. 
\end{enumerate}
At the end of the $i$-th round, the duplicator knows that $\abig_1,\twodots,\abig_{n+i}$ and
$\bbig_1,\twodots,\bbig_{n+i}$ satisfy the conditions $W(k{-}i)$.
\rule[-2ex]{0ex}{2ex}}
\ \ }
\caption{\small The duplicator's winning strategy in the $k$-round EF-game on
\,$\Abig = \struc{\ZZ,<,\allowbreak +,\allowbreak\abig_1,\twodots,\abig_n}$\, and 
\,$\Bbig = \struc{\ZZ,<,+,\bbig_1,\twodots,\bbig_n}$, where
$\abig_1,\twodots,\abig_n$ and $\bbig_1,\twodots,\bbig_n$ satisfy the conditions
$W(k) \deff C\big(\, \mF(k),\, \lF(k),\, \cF(k),\, \gF(k)\,\big)$.}\label{figure:Plus_Strategie}
\end{center}
\end{figure}%
\\
\parno
Let us mention that Theorem~\ref{theorem:Plus_Wk} gives us an EF-game proof of the
theorem of Ginsburg and Spanier,  
stating that the spectra of $\FO(<,+)$-sentences are \emph{semi-linear}:
\begin{corollary_mit}[$\bs{\FO(<,+)}$-sentences have semi-linear
  spectra]\label{corollary:semi-linear-spectra}
\index{spectrum}\index{semi-linear}\index{Spec@$\Spectrum$} 
\mbox{ }\\
Let $k\in\NNpos$ and let $\varphi$ be a $\FO(<,+)$-sentence of quantifier depth at most
$k$. \\
The spectrum 
\,$\Spectrum(\varphi) \deff 
\setc{N\in\NNpos}{\struc{\set{0,\twodots,N},<,+}\models\varphi}$\,
is \emph{semi-linear} with parameters $N_0 \deff 2\,\gF(k)\,\cF(k)^2$ and 
$p\deff \mF(k)$. I.e., 
\,``$N\in\Problem{Spec}(\varphi)$\, iff \ $N{+}p\in\Problem{Spec}(\varphi)$''\,
is true for all $N > N_0$.
\end{corollary_mit}%
\begin{proof_mit}
Let $N> N_0 \deff 2\,\gF(k)\,\cF(k)^2$ and let $p\deff \mF(k)$.\\
We use Theorem~\ref{theorem:Plus_Wk} for $n=2$, $\abig_1 = 0$, $\abig_2 = N$,
$\bbig_1 = 0$, and $\bbig_2 = N{+}p = N{+}\mF(k)$.
\\
It is straightforward to verify that $\abig_1,\abig_2$ and $\bbig_1,\bbig_2$ satisfy the
conditions $W(k)$.
Therefore,
Theorem~\ref{theorem:Plus_Wk} gives us a winning strategy for the duplicator in the
$k$-round EF-game on $\struc{\ZZ,<,+,\abig_1,\abig_2}$ and $\struc{\ZZ,<,+,\bbig_1,\bbig_2}$.
I.e., we have
\,$\struc{\ZZ,<,+,0,N}
   \dwins_k
   \struc{\ZZ,<,+,0,N{+}p}$.
This implies that
\,$\struc{\set{0,\twodots,N},<,+,0,N}
   \dwins_k
   \struc{\set{0,\twodots,N{+}p},<,\allowbreak +,0,N{+}p}$. 
Since $\varphi$ is of quantifier depth at most $k$,
Theorem~\ref{theorem:E-F}\,(\ref{theorem:E-F:structure}) gives us that
\,$\struc{\set{0,\twodots,N},<,+} \models \varphi$\, iff
\,$\struc{\set{0,\twodots,N{+}p},<,+}\models \varphi$,
i.e., $N\in \Problem{Spec}(\varphi)$ iff $N{+}p \in\Problem{Spec}(\varphi)$.
\end{proof_mit}%
%
%
%
%
\subsection[The $\FO(<,+,Q)$-Game over $\NN$ and $\ZZ$]{The $\bs{\FO(<,+,Q)}$-Game over $\bs{\NN}$ and $\bs{\ZZ}$}\label{subsection:FOPlus_game:general}\label{subsection:PlusNNZZ} 
The aim of this section is to show that
the duplicator can translate strategies for the $\FOKleiner$-game into 
strategies for the $\FO(<,+,Q)$-game on arbitrary structures over $\NN$.
Here, $Q$ is an infinite subset of $\NN$ that satisfies certain conditions 
$W(\omega)$.
\\ 
Our proof is based on Lynch's proof of his following theorem from \cite{Lynch_Addition}.
%
%
%
\subsubsection{Lynch's Theorem and his Proof Idea}
\begin{theorem_mit}[{\cite[Theorem~3.7]{Lynch_Addition}}]\label{theorem:Lynch}\index{Lynch, James F.}
For every $k\in\NNpos$ there exists a number $d(k)\in\NN$ and an 
infinite set $P_k \subseteq \NN$ such that,
for all sets $A,B\subseteq P_k$, the following holds:
If $|A|=|B|$ or $d(k) < |A|,|B| < \infty$, then
the duplicator wins the $k$-round EF-game on
\,$\struc{\NN,<,+,A}$\, and \,$\struc{\NN,<,+,B}$.
\end{theorem_mit}%
Unfortunately, neither the statement nor the proof of Lynch's theorem 
gives us directly what we need for translating strategies for the $\FOKleiner$-game into
strategies for the $\FO(<,+,Q)$-game.
Going through Lynch's proof in detail, we will modify and extend his
notation and his reasoning in a way appropriate for obtaining our  
translation results.
\par
To illustrate the overall proof idea, let us first
try to explain intuitively Lynch's proof method.
For simplicity, we concentrate on subsets $A,B
\subseteq P_k$ of the same size and discuss what the duplicator has to do in
order to win the $k$-round EF-game on $\Abig\deff\struc{\NN,<,+,A}$ and
$\Bbig\deff\struc{\NN,<,\allowbreak +,B}$.
Assume that after $i{-}1$ rounds, the elements
$\abig_{1},\twodots,\abig_{i-1}$ have been chosen in $\Abig$, and the elements
$\bbig_{1},\twodots,\bbig_{i-1}$ have been chosen in $\Bbig$.  
In the $i$-th round let the spoiler choose some element $\abig_{i}$ in $\Abig$. 
\\
In the previous Section~\ref{subsection:FOPlus_game:easy} we have seen that, in order
to win, the duplicator should play in such a way that after the
$i$-th round the conditions $W(k{-}i)$ are satisfied by $\abig_1,\twodots,\abig_i$ and
$\bbig_1,\twodots,\bbig_i$. I.e., she should follow the strategy described in 
Figure~\ref{figure:Plus_Strategie}. 
In particular, this means that if $\abig_i = \ov{s}+f'$ for a suitable linear combination
$s$ over $\abig_1,\twodots,\abig_{i-1}$, then she should answer
$\bbig_i \deff \ov{t}+f'$, for the corresponding linear combination $t$ over
$\bbig_1,\twodots,\bbig_{i-1}$.
However, in the present situation we also have the sets $A$ and $B$ which must
be respected, i.e., we need that $\abig_i \in A$ if and only if $\bbig_i\in B$.
This means that, for any linear combination $s$, we need to have
\,$\ov{s}+f' \in A$ if and only if \,$\ov{t}+f' \in B$.
\\
To solve this problem we demand that $A,B\subseteq P_k$, where $P_k$ satisfies the 
conditions $W(k)$ in the following uniform way: 
For \emph{all} sequences $p_1<\cdots<p_{\lF(k)}$ and $q_1<\cdots < q_{\lF(k)}$ in $P_k$, 
the conditions $W(k)$ are satisfied by $p_1,\twodots,p_{\lF(k)}$ and
$q_1,\twodots,q_{\lF(k)}$.
Instead of considering linear combinations $s$ only over $\abig_1,\twodots,\abig_{i-1}$, we
now consider linear combinations over $A,\abig_1,\twodots,\abig_{i-1}$.
If $s$ is such a linear combination, in which elements $p_1,\twodots,p_{l'}$ from $A$ occur, 
then the according linear combination $t$ over $B,\bbig_1,\twodots,\bbig_{i-1}$ is obtained by
replacing $\abig_1,\twodots,\abig_{i-1}$ with $\bbig_1,\twodots,\bbig_{i-1}$, and
replacing $p_1,\twodots,p_{l'}$ with $q_1,\twodots,q_{l'}$, where $q_{\nu}$ is the 
$j$-th smallest element in $B$ whenever $p_{\nu}$ is the $j$-th smallest element in $A$.
(Recall that here we assume that $|A|=|B|$; in the more difficult case where $|A|,|B|>d(k)$ we
will make use of an EF-game on $\struc{A,<}$ and $\struc{B,<}$ to find suitable
$q_1,\twodots,q_{l'}$ in $B$ that fit for the elements $p_1,\twodots,p_{l'}$ in $A$.)
\\
Now, the duplicator's strategy in the $i$-th round can be described as 
follows:\label{explanation:Lynch}%
\begin{enumerate}[$\bullet$\ ]
 \item
   If \,$\abig_i = \ov{s} + f'$\, for some   
   \,{
   $\big(\,\lF(k{-}i{+}1),\, \cF(k{-}i{+}1),\, 
   \gF(k{-}i{+}1)-\frac{\mF(k-i)}{2} 
      \,\big)$}-combination
   $s$ over $A,\abig_1,\twodots,\abig_{i-1}$ 
   and some
   \,$f'\in\intcc{{-}\frac{\mF(k-i)}{2},{+}\frac{\mF(k-i)}{2}}$,\, 
   then the duplicator chooses
   $\bbig_i\deff \ov{t}+f'$, where $t$ is the corresponding combination over 
   $B,\bbig_1,\twodots,\bbig_{i-1}$. \\
   In particular, we get that $\abig_i\in A$ iff $\bbig_i\in B$.
 \item
   If no such $s$ and $f'$ exist, then let $s_{\links}$ and $s_{\rechts}$ be the 
   \,{
   $\big(\,\lF(k{-}i{+}1),\,\allowbreak \cF(k{-}i{+}1),\,\allowbreak 
   \gF(k{-}i{+}1)-\frac{\mF(k-i)}{2} 
      \,\big)$}-combinations
   that approximate $\abig_i$ from below and from above as closely as possible; and let
   $t_{\links}$ and $t_{\rechts}$ be the corresponding combinations over  
   $B,\bbig_1,\twodots,\bbig_{i-1}$.
   The duplicator chooses an arbitrary $\bbig_i$ that lies strictly between 
   $\ov{t_{\links}}$ and $\ov{t_{\rechts}}$ with $\bbig_{i} \equiv_{\mF(k-i)}\abig_i$.
   In particular, we know that $\abig_i\not\in A$ and $\bbig_i\not\in B$.
\end{enumerate}%
As we will see below, this leads to a successful strategy for the duplicator.
%
%
%
\subsubsection{The Translation of Strategies}
To formally state our precise translation result, we need the
following generalized version of Definition~\ref{definition:conditions_Plus_einfach}.
%
%
\begin{definition_ohne}[{$\bs{(l,c,g)}$-{Combination},
    {Correspondence}, $\bs{C(m,l,c,g)}$, $\bs{W(k)}$}]\label{definition:conditions_Plus_allgemein}
\mbox{}\\ 
Let $m,l,c\in\NNpos$ and $g\in\RRpos$.\\
Let $P\subseteq\NN$ be infinite, let $n\in\NN$, and let 
$\abig_1,\twodots,\abig_n,\allowbreak\bbig_1,\twodots,\bbig_n,\allowbreak\cbig_1,\twodots,\cbig_n\in\ZZ$.
\begin{enumerate}[(a)\ ]
\item {\bf $\bs{(l,c,g)}$-Combination $\bs{s}$:}
\index{lcgcombination@$(l,c,g)$-combination}
\\
An \emph{$(l,c,g)$-combination $s$ over $P,{\abig_1,\twodots,\abig_n}$}\, is a 
formal sum (or: a \emph{linear combination}) \index{linear combination}
of the form \,$\sum_{i=1}^{l'}d_{i}x_{i} + f$,\, where \,$l'\leq l$,\, 
$x_1,\twodots,x_{l'}$\, are pairwise distinct elements in 
$P\cup\set{\abig_1,\twodots,\abig_n}$,\,
\,$d_1,\twodots,d_{{l'}} \in \QQ[c]$,\, 
and $f\in\intcc{{-}g,{+}g}\,\subseteq\,\RR$.    
\\
Every such linear combination $s$ evaluates to a real number 
$\ov{s}$. \\
The elements $x_1,\twodots,x_{l'}$ are called \emph{the terms} of $s$.
\item {\bf Correspondence $\bs{\pi}$:}
\index{correspondence $\pi$} 
\\
A \emph{correspondence} between $P,\abig_1,\twodots,\abig_n$ and
$P,\bbig_1,\twodots,\bbig_n$ is a partial mapping $\pi$ from
$P\cup\set{\abig_1,\twodots,\abig_n}$ to $P\cup\set{\bbig_1,\twodots,\bbig_n}$ which
satisfies the following conditions:
\begin{itemize}
 \item 
   $\pi$ is ${<}$-preserving on $P$,
 \item 
   $\pi(\abig_{\nu}) = \bbig_{\nu}$,\, for all
   $\nu\in\set{1,\twodots,n}$, \,and
 \item 
   $x\in P$ iff $\pi(x)\in P$,\, for all elements $x$ on which $\pi$ is defined.
\end{itemize}
If $\pi$ is such a correspondence, and if 
$s = \sum_{i=1}^{l'}d_ix_i + f$ is a $(l,c,g)$-combination 
over $P,\abig_1,\twodots,\abig_n$ whose terms are in the domain of $\pi$, then we
write $\pi(s)$ to denote the $(l,c,g)$-combination 
over $P,\bbig_1,\twodots,\bbig_n$ obtained by replacing every term in $s$ with 
its image under $\pi$, i.e., 
\,$\pi(s) \deff \sum_{i=1}^{l'}d_i\,\pi(x_i) + f$.
\item {\bf Conditions $\bs{C(m,l,c,g)}$:}
\index{conditions $C(m,l,c,g)$}
\\
We say that $P,\cbig_1,\twodots,\cbig_n$ 
\emph{satisfy the conditions $C(m,l,c,g)$} if
and only if
\begin{enumerate}[$(**)$]
 \item[$(*)$] 
   $p \equiv_m q$ \ for all $p,q\in P$,\quad and
 \item[$(**)$]
   for all $(l,c,g)$-combinations $s_1$ and $s_2$ over 
   $P,\cbig_1,\twodots,\cbig_n$\, and 
   \,for every correspondence $\pi$ between $P,\cbig_1,\twodots,\cbig_n$ and
   $P,\cbig_1,\twodots,\cbig_n$ which is defined on all the terms of $s_1$ and $s_2$, 
   we have \ $\ov{s_1} < \ov{s_2}$ \ iff \ $\ov{\pi(s_1)} < \ov{\pi(s_2)}$.
\end{enumerate}
\item {\bf Conditions $\bs{W(k)}$ \ (for $\bs{k\in\NNpos}$):}
\index{conditions $W(k)$}\index{Wk@$W(k)$}
\\
We say that
$P,\cbig_1,\twodots,\cbig_n$ 
\emph{satisfy the conditions $W(k)$} if
and only if they satisfy the conditions 
\,$C\big(\, \mF(k),\, \lF(k),\, \cF(k),\, \gF(k)\,\big)$.  Here, the functions
$\mF,\lF,\cF,\gF$ are chosen as defined in Theorem~\ref{theorem:Plus_Wk}.
\mbox{}\fertig
\end{enumerate}
\end{definition_ohne}%
%
%
%
As the following lemma shows, there do exist infinite sets $P\subseteq\NN$ that
satisfy the conditions $C(m,l,c,g)$.
%
%
\begin{lemma_ohne}\label{lemma:Q_satisfying_Conditions}
Let $m,l,c\in\NNpos$ and $g\in\RRpos$.
\begin{enumerate}[(a)\ ]
\item
If \,$p_0<p_1<p_2<\cdots$\, is a sequence of natural numbers that satisfies
\[
\begin{array}{clll}
    p_0 & \geq & 0\,, & 
 \\
    p_{i} & \geq & (2l-1)\cdot 2c^3 \cdot p_{i-1} \, + \, 2gc^2 \,,
    & \mbox{for all $i\in\NNpos$,\quad and}
 \\
    p_i & \equiv_m & p_{i+1}\,, & \mbox{for all $i\in\NNpos$,}
\end{array}
\]
then the set $P\deff \set{p_1,p_2,\ldots}$ satisfies the conditions $C(m,l,c,g)$.
Moreover, the conditions $C(m,l,c,g)$ are satisfied even by $P,\cbig_1,\twodots,\cbig_n$, for
arbitrary $n\in\NN$ and $\cbig_1,\twodots,\cbig_n\in\set{0,\twodots,p_0}$.
\item
There is an infinite set \,$Q = \set{q_0 < q_1 < q_2 < \cdots}$\, 
such that,
for every $k\in\NNpos$, the conditions $W(k)$ are satisfied by $Q,q_0,\twodots,q_{k-1}$.
One example of such a set is given via
{
\[
\begin{array}{clll}
    q_0 & \deff & 0\,,
 \\
    q_{i} & \deff& \mF(i)\cdot \Big( (2\lF(i){-}1)\cdot 2\cF(i)^3\cdot q_{i-1} \ + \
     2\,\gF(i)\,\cF(i)^2\Big)\,,
    & \mbox{for all $i\in\NNpos$\,.}
\end{array}
\]}%
\rm Obviously, this set $Q$ is not semi-linear and hence, due to the theorem of Ginsburg and 
Spanier (cf., Corollary~\ref{corollary:semi-linear-spectra}), not definable in $\FO(<,+)$.
\qquad\mbox{ }\fertig
\end{enumerate}
\end{lemma_ohne}%
\begin{proof_mit}
\emph{(a):} \ It is obvious that the congruence condition $(*)$ 
is satisfied. We thus concentrate on condition $(**)$.
Let $\pi$ be a correspondence between $P,\cbig_1,\twodots,\cbig_n$ and 
$P,\cbig_1,\twodots,\cbig_n$, and let $s_1$ and $s_2$ be $(l,c,g)$-combinations 
over $P,\cbig_1,\twodots,\cbig_n$ whose
terms are in the domain of $\pi$.
We need to show that \ $\ov{s_1}<\ov{s_2}$ \ iff \ $\ov{\pi(s_1)}<\ov{\pi(s_2)}$.
\par
Let \,$s_1 = \sum_{i=1}^{l} d_i x_i + f$\, and 
\,$s_2  = \sum_{j=1}^{l} d'_j x'_j + f'$.
By definition we know that $x_1,\twodots,x_l$ (resp., $x'_1,\twodots,x'_l$) are 
pairwise distinct elements in $P\cup\set{\cbig_1,\twodots,\cbig_n}$.
Hence, $\set{x_1,\twodots,x_l,x'_1,\twodots,x'_l}$ consists of $l'$ pairwise distinct
elements $z_1,\twodots,z_{l'}$, for some $l'\leq 2l$. 
Obviously, 
{
\[
\begin{array}{ccccc}
  \ov{s_2} - \ov{s_1} & \;=\; &
  \sum_{j=1}^{l} d'_j x'_j \ -\ \sum_{i=1}^{l}d_i x_i \ +\  (f'{-}f)
  & \; = \; &
  \sum_{r=1}^{l'}e_r z_r + h\,,
\end{array}
\]}\noindent%
where $h\deff f'{-}f$, and 
if $z_r= x'_j = x_i$ then $e_r \deff d'_j {-} d_i$,
if $z_r = x'_j \neq x_i$ for all $i$, then $e_r \deff d'_j$, and
if $z_r = x_i \neq x'_j$ for all $j$, then $e_r \deff {-}d_i$.
\\
Since $d_i,d'_j\in\QQ[c]$ and $\betrag{f},\betrag{f'}\leq g$, one can easily see that
\[
\textstyle (\bullet):\qquad
l'\leq 2l, \quad \betrag{h} \leq 2g,\quad
\betrag{e_r} \leq 2c, \mbox{ and} \quad 
e_r = 0 \mbox{ \ or \ } \betrag{e_r} > \frac{1}{c^2}.
\]
In case that $e_r=0$ for all $r$, we have
\,$\ov{s_2} - \ov{s_1} = h = \ov{\pi(s_2)}-\ov{\pi(s_1)}$,\,
and thus \ $\ov{s_1}<\ov{s_2}$ \ iff \ $\ov{\pi(s_1)}<\ov{\pi(s_2)}$.
We can hence concentrate on the case where at least one of the coefficients $e_r$ is
different from $0$.
Without loss of generality we may assume that there is an $l''$ with 
$1\leq l'' \leq l'$, such that
$e_r\neq 0$ for all $r\leq l''$, and $e_r = 0$ for all $r>l''$. 
Furthermore, we may assume that
\,$z_1>\cdots >z_{l''}$.
\par
If $z_1$ is not an element in $P$, then all $z_r$ must be elements in
$\set{\cbig_1,\twodots,\cbig_n}$, and hence $\pi(z_r) = z_r$ for all $r$. 
In particular, this means that 
\,$\ov{s_2} - \ov{s_1} = \ov{\pi(s_2)}-\ov{\pi(s_1)}$,\,
and thus \ $\ov{s_1}<\ov{s_2}$ \ iff \ $\ov{\pi(s_1)}<\ov{\pi(s_2)}$.
\par
It remains to consider the case where $z_1$ is an element in $P$.
In this case we have $z_1 = p_i$ for some $i\in\NNpos$, and 
$z_2,\twodots,z_{l''}\leq p_{i-1}$. 
Furthermore, $\pi(z_1) = p_j$ for some $j\in\NN$, and 
$\pi(z_2),\twodots,\pi(z_{l''})\leq p_{j-1}$. 
Of course, we have
\[ \textstyle
\ov{s_2} - \ov{s_1}\quad = \quad
\sum_{r=1}^{l''}e_r z_r + h \quad \leq \quad
e_1 p_i \,+\, \betrag{\sum_{r=2}^{l''}e_r z_r + h}\,.
\]
Moreover, due to $(\bullet)$ we have 
{
\[ \textstyle
  \betrag{\sum_{r=2}^{l''}e_r z_r + h} \ \ \leq \ \
  \sum_{r=2}^{l''}|e_r|\, p_{i-1} \,+\, |h|  \ \ \leq \ \
  (2l-1) \cdot 2c \cdot p_{i-1} \,+\, 2g       \ \ \leq \ \
  \frac{p_i}{c^2}\,.
\]}%
This gives us that \,$\ov{s_2}-\ov{s_1} \leq e_1p_i + \frac{p_i}{c^2}$\,
and \,$\ov{s_2}-\ov{s_1} \geq e_1p_i - \frac{p_i}{c^2}$.
Since $\pi$ is a \emph{correspondence}, the same reasoning leads to the 
anologous result for \,$\pi(s_2)-\pi(s_1)$. I.e., we have
\[
 \begin{array}{rcccl}
    \big(e_1 - \frac{1}{c^2}\big) \cdot p_i 
  & \leq
  & \ov{s_2} - \ov{s_1}
  & \leq
  & \big(e_1 + \frac{1}{c^2}\big) \cdot p_i \qquad\mbox{and}
  \\[1ex]
    \big(e_1 - \frac{1}{c^2}\big) \cdot p_j 
  & \leq
  & \ov{\pi(s_2)} - \ov{\pi(s_1)}
  & \leq
  & \big(e_1 + \frac{1}{c^2}\big) \cdot p_j\,. 
 \end{array}
\]
Due to $(\bullet)$ we know that $\betrag{e_1}>\frac{1}{c^2}$. 
Hence we have \emph{either}
\,$\big( e_1 - \frac{1}{c^2}\big)>0$,\, implying that 
\,$\ov{s_2} - \ov{s_1}>0$ and \,$\ov{\pi(s_2)} - \ov{\pi(s_1)}>0$,\,
\emph{or} 
\,$\big( e_1 + \frac{1}{c^2}\big)<0$,\, implying that 
\,$\ov{s_2} - \ov{s_1}<0$ and \,$\ov{\pi(s_2)} - \ov{\pi(s_1)}<0$.\\
This gives us that \ $\ov{s_1}<\ov{s_2}$ \ iff \ $\ov{\pi(s_1)}<\ov{\pi(s_2)}$,\, 
and the proof of part \emph{(a)} is complete.
\\
\par
\emph{(b):} \ 
Let $k\in\NNpos$.
Define $m\deff \mF(k)$, $l\deff \lF(k)$, $c\deff \cF(k)$, and $g\deff \gF(k)$.
Furthermore, define $n\deff k$ and 
\,$(\cbig_1,\twodots,\cbig_n) \deff (q_0,\twodots,q_{k-1})$.
We consider the sequence $p_0<p_1<p_2<\cdots$ given, for all $i\in\NN$, via
$p_i \deff q_{k-1+i}$.
Let $P\deff \set{p_1,p_2,\ldots} = Q\setminus\set{q_0,\twodots,q_{k-1}}$.
\\
From Definition~\ref{definition:conditions_Plus_allgemein} one can directly see that
$Q,q_0,\twodots,q_{k-1}$ satisfies the conditions $W(k)$ if and only if
$P,\cbig_1,\twodots,\cbig_n$ satisfies the conditions $C(m,l,c,g)$.
We can thus make use of part \emph{(a)}.
Of course, $\cbig_1,\twodots,\cbig_n\leq q_{k-1} = p_0$.
Furthermore, it is straightforward to check that the sequence
$p_0<p_1<p_2<\cdots$ satisfies the conditions formulated in part \emph{(a)}:
The congruence condition is satisfied since, for $i\in\NNpos$, 
\,$p_i = q_{k-1+i}$ is a multiple of $\mF(k{-}1{+}i)$ which itself is a multiple
of $\mF(k) = m$. Hence, $p_i\equiv_m 0$ for all $i\in\NNpos$.
The growth condition formulated in part \emph{(a)} is satisfied since  
the functions $\mF$, $\lF$, $\cF$, and $\gF$ are increasing.
From part \emph{(a)} we therefore obtain that 
$P,\cbig_1,\twodots,\cbig_n$ satisfies the conditions $C(m,l,c,g)$.
Altogether, this completes the proof of Lemma~\ref{lemma:Q_satisfying_Conditions}
\end{proof_mit}%
%
\begin{definition_mit}[Conditions
  $\bs{W(\omega)}$]\label{definition:conditions_W_omega}
\index{Ww@$W(\omega)$}\index{conditions $W(k)$t@conditions $W(\omega)$}\index{Q5@$Q$}%
\hspace{3cm}\\
Let $Q=\set{q_0<q_1<q_2<\cdots}\subseteq\NN$ be an infinite set of
natural numbers.\\
We say that $Q$ satisfies the conditions $W(\omega)$ if and only if
the following is true: 
For every $k\in\NNpos$ there exists an $n_k\in\NNpos$ such that
the conditions $W(k)$ (cf.,
Definition~\ref{definition:conditions_Plus_allgemein}) are satisfied by
\,$Q,q_0,\twodots,q_{n_k-1}$.\\
\rm An example of such a set $Q$ is given in 
Lemma~\ref{lemma:Q_satisfying_Conditions}\,(b).
\end{definition_mit}%
We are now ready to state the main result of this section:
\begin{theorem_mit}[$\bs{\FO(<,+,Q)}$-game over $\NN$ and $\ZZ$]\label{theorem:collapse_AdditionQ}\label{theo:additionQ}
\index{Ehrenfeucht-Fra\"\i{}ss\'{e} game!FO<+Q@$\FO(<,+,Q)$-game}
\mbox{}\\ 
Let $Q=\set{q_0<q_1<q_2<\cdots}\subseteq\NN$ satisfy the conditions 
$W(\omega)$.
\\
The duplicator can translate strategies for
the $\FOKleiner$-game into strategies for the $\FO({<, }\allowbreak +,Q)$-game on 
$\Clarb$ over $\NN$ and on $\ClNemb$ over $\ZZ$.
\end{theorem_mit}%
The above theorem is a direct consequence of the
following technical result: 
\begin{proposition_mit}\label{proposition:PlusGame}
Let $\tau$ be a signature, 
let $k\in\NNpos$ be a number of rounds for the
``${+}$-game''. 
The according number $r(k)$ of rounds for the ``${<}$-game'' is inductively defined
via \ 
$r(0) \deff 1$ \ and, for all $j\in\NN$, \ 
$ r(j{+}1) \deff r(j) +  2 \cdot \lF(j{+}1)$.
\\
Let $n\in\NN$, let $\vek{\cbig} \deff \cbig_1,\twodots,\cbig_n\in\NN$, and let
\,$P\deff \set{p_1<p_2<p_3<\cdots}\subseteq \NN$\, be an infinite set 
such that \,$P,\vek{\cbig}$\, satisfies the conditions 
$W(k)$ and such that, for all $\nu\in\set{1,\twodots,n}$, \,$\cbig_{\nu}$ is smaller 
than the smallest element in $P$.\footnote{From 
Lemma~\ref{lemma:Q_satisfying_Conditions}\,(a) 
we know how to construct such \,$P,\vek{\cbig}$.
}%
\\
Let $\A$ and $\B$ be two $\NN$-embeddable $\tau$-structures, and let
$\alpha:\adom(\A)\rightarrow P$ map, for every $j$, the $j$-th smallest element 
in $\adom(\A)$ onto
the position $p_j$. Accordingly, let $\beta:\adom(\B)\rightarrow P$ map, for every $j$, 
the $j$-th smallest element in $\adom(\B)$
onto the position $p_j$.
\\
If
\ $\struc{\adom(\A),<,\tau^{\A}}
   \dwins_{r(k)}
   \struc{\adom(\B),<,\tau^{\B}}
$,\, then
\ $\struc{\ZZ,<,+,P,\vek{\cbig},\alpha\big(\tau^{\A}\big)}
   \dwins_{k} \linebreak[4]
   \struc{\ZZ,<,+,P,\vek{\cbig}, \beta\big(\tau^{\B}\big)}
$.
\end{proposition_mit}%
Before proving Proposition~\ref{proposition:PlusGame} let us first show that it 
enables us to prove Theorem~\ref{theo:additionQ}.
\\
\parno
\begin{proofc_mit}{of Theorem~\ref{theo:additionQ}}\mbox{ }\\
Let $\tau$ be a signature and let $k$ be the number of rounds for the
$\FO(<,\allowbreak +,Q)$-game.
Choose the number $r(k)$ of rounds for the $\FOKleiner$-game as given in 
Proposition~\ref{proposition:PlusGame}. 
Choose $P\deff Q\setminus\set{q_0,\twodots,q_{n_k-1}}$, choose
$n\deff n_k{+}1$, and $\vek{\cbig}\deff 0,q_0,\twodots,q_{n_k-1}$.
From the presumption we know that $Q$ satisfies the conditions
$W(\omega)$, and thus \,$Q,q_0,\twodots,q_{n_k-1}$\, satisfies the conditions
$W(k)$. From Definition~\ref{definition:conditions_Plus_allgemein} one
can directly see that this implies that also
\,$P,\vek{\cbig}$\, satisfies the conditions $W(k)$.
\\
If $\A$ and $\B$ are two $\NN$-embeddable $\struc{\ZZ,\tau}$-structures with
\ $\struc{\adom(\A),<,\tau^{\A}}
   \dwins_{r(k)}
   \struc{\adom(\B),<, \allowbreak \tau^{\B}}$, \ 
then Proposition~\ref{proposition:PlusGame} gives us ${<}$-preserving mappings $\alpha$,
$\beta$ such that 
\,$\struc{\ZZ,<,+,P,\vek{\cbig},\alpha\big(\tau^{\A}\big)} \allowbreak
   \dwins_{k}
   \struc{\ZZ,<,+,P,\vek{\cbig}, \beta\big(\tau^{\B}\big)}
$.
Since 
\,$\vek{\cbig} = 0,q_0,\twodots,q_{n_k-1}$\, and
\,$Q = P\cup \set{q_0,\twodots,q_{n_k-1}}$,\, this in particular implies
that
\,$\struc{\ZZ,<,+,0,Q,\alpha\big(\tau^{\A}\big)}
   \dwins_{k}
   \struc{\ZZ,<,+,0,Q, \beta\big(\tau^{\B}\big)}
$,\, 
and hence
\,$\struc{\NN,<,\allowbreak +,Q,\alpha\big(\tau^{\A}\big)}
   \dwins_{k}
   \struc{\NN,<,+,Q, \beta\big(\tau^{\B}\big)}
$. 
Altogether, this completes the proof of Theorem~\ref{theo:additionQ} both, for 
$\NN$-embeddable structures over $\ZZ$, and for arbitrary structures over $\NN$. 
\end{proofc_mit}%
\\
\parno
We will now concentrate on the proof of Proposition~\ref{proposition:PlusGame}.
\\
\parno
\begin{proofc_mit}{of Proposition~\ref{proposition:PlusGame}} \mbox{ }\\
Let $\tau$ be a signature, 
let $k\in\NNpos$, and let $r(k)$ be the according number defined in 
Propostion~\ref{proposition:PlusGame}.
Let $n\in\NN$, let $\vek{\cbig} \deff \cbig_1,\twodots,\cbig_n\in\NN$, and let
\,$P\deff \set{p_1<p_2<\cdots}\subseteq \NN$\, be an infinite set 
such that \,$P,\vek{\cbig}$\, satisfies the conditions 
$W(k)$ and such that, for all $\nu\in\set{1,\twodots,n}$, \,$\cbig_{\nu}$ is smaller 
than the smallest element in $P$.
Let $\A$ and $\B$ be two $\NN$-embeddable $\tau$-structures, and let
$\alpha:\adom(\A)\rightarrow P$ map, for every $j$, the $j$-th smallest element 
in $\adom(\A)$ onto
the position $p_j$. Accordingly, let $\beta:\adom(\B)\rightarrow P$ map, for every $j$, 
the $j$-th smallest element in $\adom(\B)$
onto the position $p_j$.
\\
We assume that  
\,$\struc{\adom(\A),<,\tau^{\A}}
   \dwins_{r(k)}
   \struc{\adom(\B),<,\tau^{\B}}
$, i.e., the duplicator wins the $r(k)$-round $\FOKleiner$-game on
$\A$ and $\B$.
%
From Lemma~\ref{lemma:nuetzlich} we obtain that
\ $\Abig' \deff \struc{P,<,\alpha\big(\tau^{\A}\big)}
   \dwins_{r(k)}
   \struc{P,<,\beta\big(\tau^{\B}\big)} \ffed \Bbig'
$.
\\
Our aim is to show that
\ $\Abig \deff \struc{\ZZ,<,+,P,\vek{\cbig},\alpha\big(\tau^{\A}\big)}
   \dwins_{k}
   \struc{\ZZ,<,+,P,\vek{\cbig},\beta\big(\tau^{\B}\big)} \ffed \Bbig
$.
%
%
\\
Henceforth, the game on $\Abig'$ and $\Bbig'$ will be called \emph{the ${<}$-game}, and
the game on $\Abig$ and $\Bbig$ will be called \emph{the ${+}$-game}.
\par
For each round $i\in\set{1,\twodots,k}$ of the ${+}$-game we use
$\abig_i$ and $\bbig_i$, respectively, to denote the element chosen in that round in
$\Abig$ and $\Bbig$. 
We will translate each move of the spoiler in the ${+}$-game, say $\abig_i$ (if he
chooses in $\Abig$), into a number of moves $\abig'_{i,1},\twodots,\abig'_{i,n_i}$ for
a ``virtual spoiler'' in the ${<}$-game in $\Abig'$. 
Then we can find the answers $\bbig'_{i,1},\twodots,\bbig'_{i,n_i}$ of a 
``virtual duplicator'' who plays according to her winning strategy in the ${<}$-game.
Afterwards, we translate these answers into a move $\bbig_i$ for the duplicator in
the ${+}$-game. (The case where the spoiler chooses $\bbig_i$ in $\Bbig$ is symmetric.)
\par
As abbreviation we use $\vek{\abig}'_i$ to denote the sequence 
$\abig'_{i,1},\twodots,\abig'_{i,n_i}$, and 
we use $\vek{\bbig}'_i$ to denote the sequence 
$\bbig'_{i,1},\twodots,\bbig'_{i,n_i}$.
A partial mapping from $\ZZ$ to $\ZZ$ is called \emph{$\tau$-respecting} iff it is 
a partial isomorphism between the structures \index{trespecting@$\tau$-respecting correspondence}
\,$\struc{\ZZ,\,\alpha(\tau^{\A})}$\, and \,$\struc{\ZZ,\,\beta(\tau^{\B})}$.
\parno
We show that the duplicator can play the ${+}$-game in such a way that the following
conditions  hold at the end of each round $i$, for $i\in\set{0,\twodots,k}$:
\begin{enumerate}[(1)\ ]
 \item
   $\struc{\,\Abig',\,\vek{\abig}'_1,\,\twodots,\,\vek{\abig}'_i\,}
   \ \dwins_{r(k-i)}\ 
   \struc{\,\Bbig',\,\vek{\bbig}'_1,\,\twodots,\,\vek{\bbig}'_i\,}$\,.
 \item
   $\abig_{i} \equiv_{\mF(k-i)} \bbig_{i}$ \quad (if $i\neq 0$).
 \item
   The following mapping 
   \[
    \pi_i\ :\ 
    \left\{\ 
     \begin{array}{rcll}
       \alpha(c^{\A}) & \mapsto & \beta(c^{\B}) & \mbox{for all constant symbols $c\in\tau$} \\
       \vek{\cbig} & \mapsto & \vek{\cbig} & \\
       \vek{\abig}'_{\nu} & \mapsto & \vek{\bbig}'_{\nu} & \mbox{for all $\nu\in\set{1,\twodots,i}$} \\
       \abig_{\nu} & \mapsto & \bbig_{\nu} & \mbox{for all $\nu\in\set{1,\twodots,i}$} 
     \end{array}
    \right\}
   \]
   is a \emph{$\tau$-respecting correspondence} between
   \,$P,\vek{\cbig},\abig_1,\twodots,\abig_i$\, and
   \,$P,\vek{\cbig},\bbig_1,\twodots,\bbig_i$.
 \item
   If $i\neq 0$, then for every 
   \,{
    $(\lF(k{-}i{+}1),\,\cF(k{-}i{+}1) ,\,\gF(k{-}i{+}1)-\frac{\mF(k{-}i)}{2})$}-combination
   $t$ over $P,\vek{\cbig},\abig_1,\twodots,\abig_{i-1}$, and 
   for every extension $\pi$ of $\pi_{i}$ which is
   ${<}$-preserving on $P$ and which is defined on all the terms of $t$, we have
   \begin{eqnarray*}
       \abig_i \ <\ \ov{t} & \mbox{ if and only if } & \bbig_i\ <\ \ov{\pi(t)}\,. 
   \end{eqnarray*}
 \item
   If $i\neq k$, then
   for all 
   \,{
   $(\lF(k{-}i),\,\cF(k{-}i) ,\,\gF(k{-}i))$}-combinations $s_1$ and $s_2$ over 
   $P,\vek{\cbig},\abig_1,\twodots,\abig_i$\, and 
   \,for every extension $\pi$ of $\pi_i$ which is ${<}$-preserving on $P$ and
   which is defined on all the terms of 
        $s_1$ and $s_2$, 
        we have
        \begin{eqnarray*}
          \ov{s_1} < \ov{s_2}
        & \quad\mbox{if and only if}\quad
        & \ov{\pi(s_1)} < \ov{\pi(s_2)}\,. 
        \end{eqnarray*}
\end{enumerate}
The following can be seen easily:
\medskip\\
\textbf{Claim~1.}
{\em If the conditions (3) and (4) are satisfied for $i{=}k$ and condition (5) is satisfied for 
$i=k{-}1$, then the mapping $\pi_k$ is a partial isomorphism between
$\Abig$ and $\Bbig$ and hence the duplicator has won the $k$-round ${+}$-game on $\Abig$ and
$\Bbig$.} \mbox{ }\fertig
\medskip\\
\begin{proof_mit}
Recall that 
\,{
$\Abig  = \struc{\,\ZZ,\,<,\,+,\,P,\,\vek{\cbig},\,\alpha(\tau^{\A})\,}$}\, and 
\,{
$\Bbig  = \struc{\,\ZZ,\,<,\,+,\,P,\,\vek{\cbig},\,\beta(\tau^{\B})\,}$}.
From condition (3) (for $i\deff k$) we know that the mapping $\pi\deff\pi_k$ is a
$\tau$-respecting correspondence between $P,\vek{\cbig},\abig_1,\twodots,\abig_k$ and
$P,\vek{\cbig},\bbig_1,\twodots,\bbig_k$.
In particular, this means that
$\pi$ is a partial isomorphism between \,$\struc{\ZZ,\,\alpha(\tau^{\A})}$\, and
\,$\struc{\ZZ,\,\beta(\tau^{\B})}$, \ and that
``$x\in P$ \ iff \ $\pi(x)\in P$'' \ is true for all $x\in\ZZ$ on which $\pi$ is defined.
All that remains to be done is to show that for all $x,y,z$ in the domain of $\pi$ we have \ 
``$x<y$ \ iff \ $\pi(x)<\pi(y)$''\, and \ 
``$x+y=z$ \ iff \ $\pi(x) + \pi(y) = \pi(z)$''.
\par
In order to prove that ``$x<y$ \ iff \ $\pi(x)<\pi(y)$'' we distinguish between three cases:
If $x=y=\abig_k$ then, certainly, $x=y$ and $\pi(x) = \pi(y)$.
If $x$ and $y$ are both different from $\abig_k$, then \,$s_1\deff x$\, and \,$s_2\deff y$\, can be
viewed as 
\,$(1,1,0)$-combinations over 
   $P,\vek{\cbig},\abig_1,\twodots,\abig_{k-1}$.
Hence, condition (5) (for $i\deff k{-}1$) gives us that \ $x<y$ \ iff \ $\pi(x)<\pi(y)$. 
If either $x$ or $y$ is equal to $\abig_k$, then condition (4) (for $i\deff k$) gives us
that \ $x<y$ \ iff \ $\pi(x)<\pi(y)$.
\par
In order to prove that ``$x+y=z$ \ iff \ $\pi(x) + \pi(y) = \pi(z)$'' we distinguish between 
three cases:
If $z=\abig_k$ and either $x$ or $y$ is equal to $\abig_k$, then, certainly, 
$x+y=z$ \ iff \ $\pi(x)+\pi(y)=\pi(z)$.
If $x$, $y$, and $z$ are different from $\abig_k$, then it is straightforward to define
$(2,2,0)$-combinations $s_1$ and $s_2$ over 
$P,\vek{\cbig},\abig_1,\twodots,\abig_{k-1}$ 
such that \ $x+y=z$\, iff \,$\ov{s_1}=\ov{s_2}$ \ and \
$\pi(x)+\pi(y)=\pi(z)$\, iff \,$\ov{\pi(s_1)}=\ov{\pi(s_2)}$.
Hence, condition (5) (for $i\deff k{-}1$) gives us that
$x+y=z$ \ iff \ $\pi(x)+\pi(y)=\pi(z)$.
In all remaining cases it is straightforward
to define a $(2,2,0)$-combination $t$ over $P,\vek{\cbig},\abig_1,\twodots,\abig_{k-1}$ such that
\ $x+y=z$\, iff \,$\abig_k=\ov{t}$ \ and \
$\pi(x)+\pi(y)=\pi(z)$\, iff \,$\bbig_k=\ov{\pi(t)}$.
Condition (4) (for $i\deff k$) then gives us that
$x+y=z$ \ iff \ $\pi(x)+\pi(y)=\pi(z)$.
\\
Altogether, the proof of Claim~1 is complete.
\end{proof_mit}%
\\
\parno
From our presumptions we know that the conditions (1)--(5) are satisfied for $i=0$.
\\
For the induction step from $i{-}1$ to $i\in\set{1,\twodots,k}$ we assume that (1)--(5) 
hold for $i{-}1$.
We show that in the $i$-th round the duplicator can play in such a way that 
(1)--(5) hold for $i$. 
Let us assume that the spoiler chooses $\abig_i$ in $\Abig$. 
(The case where he chooses $\bbig_i$ in $\Bbig$ is symmetric.)
\par
The duplicator's strategy in the $i$-th round is similar to the strategy described in 
Figure~\ref{figure:Plus_Strategie}.
First, she determines two linear combinations $s_{\links}$ and $s_{\rechts}$ over
$P,\vek{\cbig},\abig_1,\twodots,\abig_{i-1}$ which
approximate $\abig_i$ from below and from above as closely as possible. For the precise choice
of $s_{\links}$ and $s_{\rechts}$ she distinguishes between three cases:
\begin{enumerate}[(I)\ ]
 \item 
   If \,$\abig_i\in P\cup\set{\vek{\cbig},\abig_1,\twodots,\abig_{i-1}}$\, then
   \,$s_{\links} \deff s_{\rechts} \deff \abig_i$. 
 \item
   Otherwise, 
   if \,$\abig_i = \ov{s} + f'$\, for some   
   \,{
   $\big(\,\lF(k{-}i{+}1),\, \cF(k{-}i{+}1),\, 
   \gF(k{-}i{+}1)-\frac{\mF(k-i)}{2} 
      \,\big)$}-combination
   $s$ over $P,\vek{\cbig},\abig_1,\twodots,\abig_{i-1}$ 
   and some
   \,{
   $f'\in\intcc{{-}\frac{\mF(k-i)}{2},{+}\frac{\mF(k-i)}{2}}$},\, 
   then \,$s_{\links} \deff s_{\rechts} \deff s+f'$.
 \item
   Otherwise, let $s_{\links}$ and $s_{\rechts}$ be the 
   \,{
   $\big(\,\lF(k{-}i{+}1),\, \cF(k{-}i{+}1),\, 
   \gF(k{-}i{+}1)-\frac{\mF(k-i)}{2} 
      \,\big)$}-com\-bi\-na\-tions  over $P,\vek{\cbig},\abig_1,\twodots,\abig_{i-1}$ 
   that approximate $\abig_i$ from below and from above as closely as possible.\\
   I.e., \,$\ov{s_{\links}} < \abig_i < \ov{s_{\rechts}}$,\, and
   for all \,{
   $\big(\,\lF(k{-}i{+}1),\, \cF(k{-}i{+}1),\, 
   \gF(k{-}i{+}1)-\frac{\mF(k-i)}{2}  
      \,\big)$}-com\-bi\-na\-tions $s$ we have
   \,$\ov{s} \leq \ov{s_{\links}}$\, or \,$\ov{s} \geq
   \ov{s_{\rechts}}$. 
   In particular, since case (II) does not apply, we know that
   \,$\ov{s_{\links}}+\frac{\mF(k-i)}{2} < \abig_i < \ov{s_{\rechts}}-\frac{\mF(k-i)}{2}$,\,
   and hence \,$\ov{s_{\rechts}} - \ov{s_{\links}} > \mF(k{-}i)$.
\end{enumerate}
In all three cases, $s_{\links}$ and $s_{\rechts}$ are 
\,{
$\big(\,\lF(k{-}i{+}1),\, \cF(k{-}i{+}1),\, 
   \gF(k{-}i{+}1) \,\big)$}-combinations over \linebreak[4] 
$P,\vek{\cbig},\abig_1,\twodots,\abig_{i-1}$. 
\\
\parno
Let \,$\vek{\abig}'_i \deff \abig'_{i,1},\twodots,\abig'_{i,n_i}$\, be those pairwise distinct
terms of $s_{\links}$ and $s_{\rechts}$ that belong to $P$.
In particular, we know that \,$n_i \leq 2\cdot \lF(k{-}i{+}1)$.
The elements $\vek{\abig}'_i$ are the moves for a ``virtual spoiler'' in the ${<}$-game.
From condition~(1) (for $i{-}1$) we know that
\ $\struc{\Abig',\vek{\abig}'_1,\twodots,\vek{\abig}'_{i-1}}
   \dwins_{r(k-i+1)} 
   \struc{\Bbig',\vek{\bbig}'_1,\twodots,\vek{\bbig}'_{i-1}}
$.
Thus, a ``virtual duplicator'' can find answers 
\,$\vek{\bbig}'_i \deff \bbig'_{i,1},\twodots,\bbig'_{i,n_i}$\, such that
\ $\struc{\Abig',\vek{\abig}'_1,\twodots,\vek{\abig}'_{i-1},\vek{\abig}'_{i}}
   \dwins_{r(k-i+1)-n_i}
   \struc{\Bbig',\vek{\bbig}'_1,\twodots,\vek{\bbig}'_{i-1},\vek{\bbig}'_{i}}
$.
Since \,$n_i \leq 2\cdot \lF(k{-}i{+}1)$, and since the function $r$ was defined in such
a way that \ $r(k{-}i{+}1) \,=\, r(k{-}i) + 2{\cdot} \lF(k{-}i{+}1)$,\, we know that
\,$r(k{-}i{+}1)-n_i \, \geq \, r(k{-}i)$,\, and hence 
{\bf condition~(1) is satisfied for $\bs{i}$.}
\\
\parno
Let $\hat{\pi}_i$ be the extension of the mapping $\pi_{i-1}$ via
\,$\vek{\abig}'_i \mapsto \vek{\bbig}'_i$. 
It should be clear that, due to condition (3) (for $i{-}1$), 
$\hat{\pi}_i$ is a \emph{$\tau$-respecting correspondence} between
$P,\vek{\cbig},\abig_1,\twodots,\abig_{i-1}$ and
$P,\vek{\cbig},\bbig_1,\twodots,\bbig_{i-1}$.
\medskip\\
For her choice of $\bbig_i$ in $\Bbig$, the duplicator makes use of the following:
\medskip\\
\textbf{Claim 2.}\mbox{ } \\ {\em
(a) \ \,$\ov{s_{\links}} \ \equiv_{\mF(k-i)}\ \ov{\hat{\pi}_i(s_{\links})}$\,. \\
(b) \ \,If\quad $\ov{s_{\rechts}} - \ov{s_{\links}} \, >\, \mF(k{-}i)$\quad then
        \quad $\ov{\hat{\pi}_i(s_{\rechts})} - \ov{\hat{\pi}_i(s_{\links})} \, >\, \mF(k{-}i)$\,. 
\mbox{ } }\fertig
\medskip\\
\begin{proof_mit}
\emph{(a):} \ 
We know that $s_{\links}$ is a 
\,{
$\big(\,\lF(k{-}i{+}1),\, \cF(k{-}i{+}1),\, 
   \gF(k{-}i{+}1) \,\big)$}-combination over $P,\vek{\cbig},\abig_1,\twodots,\abig_{i-1}$.
In particular, 
\,$s_{\links} = \sum_{\nu = 1}^{l'} d_{\nu} x_{\nu} + f$,\,  where 
$d_{\nu} \in \QQ[\cF(k{-}i{+}1)]$, i.e., \,$d_{\nu} = \frac{u_{\nu}}{u'_{\nu}}$\, for 
$u'_{\nu}\neq 0$ and $|u_{\nu}|, |u'_{\nu}| \in\set{0,\twodots,\cF(k{-}i{+}1)}$.
In order to show that \,$\ov{s_{\links}} \ \equiv_{\mF(k-i)}\ \ov{\hat{\pi}_i(s_{\links})}$,\, we need 
to find some $z\in\ZZ$ such that 
\,$\ov{s_{\links}} - \ov{\hat{\pi}_i(s_{\links})} \, = \, z\cdot \mF(k{-}i)$.
\\
Of course, 
\,$\ov{s_{\links}} - \ov{\hat{\pi}_i(s_{\links})} \, = \, 
   \sum_{\nu=1}^{l'} u_{\nu}\cdot \frac{x_{\nu}- \hat{\pi}_i(x_{\nu})}{u'_{\nu}}$.
From the presumption that $P,\vek{\cbig}$ satisfies the conditions $W(k)$ and
from condition (2) (for $i{-}1$) we know 
for all the $x_{\nu}$ that \,$x_{\nu} \, \equiv_{\mF(k-i+1)} \, \hat{\pi}_i(x_{\nu})$.
\\
I.e., there exists $z_{\nu}\in\ZZ$ such that
\,$x_{\nu}-\hat{\pi}_i(x_{\nu}) \,=\, z_{\nu}\cdot \mF(k{-}i{+}1)$.
By the definition of $\mF$ we know that 
\,$\mF(k{-}i{+}1) = \mF(k{-}i)\cdot\lcm\set{1,\twodots,\cF(k{-}i{+}1)}$.\,
Hence, 
\,$\ov{s_{\links}} - \ov{\hat{\pi}_i(s_{\links})} \, = \, 
   \sum_{\nu=1}^{l'} u_{\nu}\cdot z_{\nu} \cdot \mF(k{-}i) \cdot 
      \frac{\lcm\set{1,\twodots,\cF(k{-}i{+}1)}}{u'_{\nu}}$.\,
This gives us the desired integer
\,$z\deff \sum_{\nu=1}^{l'} u_{\nu}\cdot z_{\nu} \cdot 
      \frac{\lcm\set{1,\twodots,\cF(k{-}i{+}1)}}{u'_{\nu}}$\,
such that 
\,$\ov{s_{\links}} - \ov{\hat{\pi}_i(s_{\links})} \, = \, z\cdot \mF(k{-}i)$.
\medskip\\\indent
\emph{(b):} \ 
Since \,$\ov{s_{\rechts}}-\ov{s_{\links}} > \mF(k{-}i)$,\, we know that
$s_{\links}$ and $s_{\rechts}$ must have been chosen according to case (III) and must hence be 
\,{
$\big(\,\lF(k{-}i{+}1),\, \cF(k{-}i{+}1),\, 
   \gF(k{-}i{+}1)-\frac{\mF(k-i)}{2} \,\big)$}-combinations.
Let $h\deff \ov{\hat{\pi}_i(s_{\rechts})}-\ov{\hat{\pi}_i(s_{\links})}$.
We need to show that $h>\mF(k{-}i)$.
\\
Suppose that, on the contrary, $h \leq \mF(k{-}i)$. 
Then, $s_1\deff s_{\links}+\frac{h}{2}$ and $s_2\deff s_{\rechts}-\frac{h}{2}$ are
\,{
$\big(\,\lF(k{-}i{+}1),\, \cF(k{-}i{+}1),\, 
   \gF(k{-}i{+}1) \,\big)$}-combinations with
\,$\ov{\hat{\pi}_i(s_1)} = \ov{\hat{\pi}_i(s_2)}$.
From condition (5) (for $i{-}1$) we obtain that \,$\ov{s_1} = \ov{s_2}$\, and hence
\,$\ov{s_{\rechts}} - \ov{s_{\links}} = h \leq \mF(k{-}i)$.
This is a contradiction to our presumption that 
\,$\ov{s_{\rechts}} - \ov{s_{\links}} > \mF(k{-}i)$.
Altogether, the proof of Claim~2 is complete.
\end{proof_mit}%
\\
\parno
The duplicator chooses $\bbig_i$ in $\Bbig$ as follows:
\begin{enumerate}[$\bullet$\ ]
 \item
   If \,$\abig_i = \ov{s_{\links}}$\, then \,$\bbig_i\deff \ov{\hat{\pi}_i(s_{\links})}$.\\
   According to Claim~2\,(a) we have 
   \,$\abig_{i} \equiv_{\mF(k-i)} \bbig_i$.\,
   In particular, since $\abig_i\in\ZZ$, this implies that $\bbig_i\in\ZZ$.
 \item
   If \,$\abig_i \neq \ov{s_{\links}}$\, then $s_{\links}$ and $s_{\rechts}$ must have
   been chosen according to case (III). In particular, we know that
   \ $\ov{s_{\rechts}} - \ov{s_{\links}} \, >\, \mF(k{-}i)$. \\
   According to Claim~2\,(b) we have 
   \ $\ov{\hat{\pi}_i(s_{\rechts})} - \ov{\hat{\pi}_i(s_{\links})} \, >\, \mF(k{-}i)$.
   Thus there exists a $\bbig_i\in\ZZ$ with 
   \,$\abig_{i} \equiv_{\mF(k-i)} \bbig_i$. 
\end{enumerate}%
In both cases, {\bf condition (2) is satisfied for $\bs{i}$.}
\\
\parno
In order to show that condition (3) is satisfied for $i$, we distinguish between case (I) on the
one hand and the cases (II) and (III) on the other hand, and we make use of the fact that we
already know that $\hat{\pi}_i$ is a $\tau$-respecting correspondence between
$P,\vek{\cbig},\abig_1,\twodots,\abig_{i-1}$ and $P,\vek{\cbig},\bbig_1,\twodots,\bbig_{i-1}$.
\par
In case (I) we know that \,$\abig_i\in P\cup\set{\vek{\cbig},\abig_1,\twodots,\abig_{i-1}}$\, and
that $s_{\links} = \abig_i$.
In particular, $\abig_i$ lies in the domain of $\hat{\pi}_i$.
As described above, the duplicator chooses $\bbig_i\deff \ov{\hat{\pi}_i(s_{\links})}
= \hat{\pi}_i(\abig_i)$.
Hence, $\hat{\pi}_i$ is exactly the mapping $\pi_i$ considered in condition (3); and certainly, 
$\pi_i$ is a $\tau$-respecting correspondence  between
$P,\vek{\cbig},\abig_1,\twodots,\abig_{i}$ and $P,\vek{\cbig},\bbig_1,\twodots,\bbig_{i}$.
\par
In the cases (II) and (III) we know that 
\,$\abig_i\not\in P\cup\set{\vek{\cbig},\abig_1,\twodots,\abig_{i-1}}$.
In particular, $\abig_i$ is \emph{not} in the domain of $\hat{\pi}_i$. Thus we can extend 
$\hat{\pi}_i$ to $\pi_i$ via $\abig_i\mapsto\bbig_i$.
If we can show that $\bbig_i\not\in P$, then $\pi_i$ inherits from $\hat{\pi}_i$ that it 
is $\tau$-respecting, that it is ${<}$-preserving on $P$, and that it satisfies, 
for all elements $x$ on which it is defined, that \,$x\in P$ iff $\pi_i(x)\in P$. 
I.e., we obtain that $\pi_i$ is a $\tau$-respecting correspondence  between
$P,\vek{\cbig},\abig_1,\twodots,\abig_{i}$ and $P,\vek{\cbig},\bbig_1,\twodots,\bbig_{i}$.
\\
It remains to show that $\bbig_i\not\in P$. 
\emph{For the sake of contradiction, assume that $\bbig_i\in P$.}
From condition (1) (for $i$) we know that 
\,$\struc{\Abig',\vek{\abig}'_1,\twodots,\vek{\abig}'_i}
   \dwins_{r(k-i)}
   \struc{\Bbig',\vek{\bbig}'_1,\twodots,\vek{\bbig}'_i}$.\,
Furthermore, $r(k{-}i)\geq r(0) = 1$, and hence the ``virtual duplicator'' can win 
(at least) one more round of the game. 
In this round let the ``virtual spoiler'' choose $\bbig_i$ in $\Bbig'$
(this is possible since we assume that $\bbig_i\in P$). The ``virtual duplicator'' can 
find some $p$ in $\Abig'$ (i.e., $p\in P$) such that
\,$\struc{\Abig',\vek{\abig}'_1,\twodots,\vek{\abig}'_i,p}
   \dwins_{0}
   \struc{\Bbig',\vek{\bbig}'_1,\twodots,\vek{\bbig}'_i,\bbig_i}$.\,
Hence, the extension $\pi$ of $\hat{\pi}_i$ via \,$p\mapsto \bbig_i$\, must be
${<}$-preserving on $P$. In particular, condition (5) (for $i{-}1$) can be applied to
the mapping $\pi$. 
Furthermore, we have $\pi(s_{\links}) = \hat{\pi}_i(s_{\links})$ and 
$\pi(s_{\rechts}) = \hat{\pi}_i(s_{\rechts})$; and 
$p$ can be viewed as a 
   \,{
 $\big(\,\lF(k{-}i{+}1),\, \cF(k{-}i{+}1),\, 
   \gF(k{-}i{+}1)-\frac{\mF(k-i)}{2} 
      \,\big)$}-combination  over $P,\vek{\cbig},\abig_1,\twodots,\abig_{i-1}$.
\\
In case (II) we know that $\abig_i=\ov{s_{\links}}$ and
$\bbig_i=\ov{\pi(s_{\links})}$. I.e., we have 
$\ov{\pi(p)} = \bbig_i = \ov{\pi(s_{\links})}$. From condition (5) (for $i{-}1$) we 
obtain that $p = \ov{s_{\links}} = \abig_i$, which is a 
contradiction to $\abig_i\not\in P$.
\\
In case (III) we know that 
$\ov{s_{\links}} < \abig_i < \ov{s_{\rechts}}$ and 
$\ov{\pi(s_{\links})} < \bbig_i = \ov{\pi(p)} < \ov{\pi(s_{\rechts})}$.
From condition (5) (for $i{-}1$) we obtain that 
$\ov{s_{\links}} < p < \ov{s_{\rechts}}$. 
This is a contradiction to the choice of $s_{\links}$ and $s_{\rechts}$ according to 
case (III).
In the cases (II) and (III) we thus must have $\bbig_i\not\in P$.
\\
Altogether, we have seen that 
{\bf condition (3) is satisfied for $\bs{i}$.}
\\
\parno
In order to show that condition (4) is satisfied for $i$, let 
$t$ be a  
\,{
 $(\lF(k{-}i{+}1),\,\cF(k{-}i{+}1) ,\allowbreak\,\gF(k{-}i{+}1)-\frac{\mF(k{-}i)}{2})$}-combination
over $P,\vek{\cbig},\abig_1,\twodots,\abig_{i-1}$, and 
let $\pi$ be an extension of $\pi_{i}$ which is
${<}$-preserving on $P$ and which is defined on all the terms of $t$. 
We need to show that
\,$\abig_i  < \ov{t}$\, if and only if \,$\bbig_i < \ov{\pi(t)}$. 
\par
For the ``if'' direction we assume that $\abig_i\geq \ov{t}$, and we show that 
$\bbig_i\geq \ov{\pi(t)}$.\\
From the choice of $s_{\links}$ we know that $\ov{s_{\links}}\geq \ov{t}$.
Condition (5) (for $i{-}1$) gives us that $\ov{\pi(s_{\links})}\geq \ov{\pi(t)}$.
Furthermore, from the choice of $\bbig_i$ we know that 
$\bbig_i\geq \ov{\hat{\pi}_i(s_{\links})} = \ov{\pi(s_{\links})}$. Hence,
$\bbig_i\geq \ov{\pi(t)}$.\par
For the ``only if'' direction we assume that $\abig_i < \ov{t}$, and we show that 
$\bbig_i < \ov{\pi(t)}$.\\
In case that $\abig_i = \ov{s_{\links}}$ we know that 
$\bbig_{i} = \ov{\hat{\pi}_i(s_{\links})} = \ov{\pi(s_{\links})}$ and that
$\ov{s_{\links}}<\ov{t}$. Condition (5) (for $i{-}1$) gives
us that $\ov{\pi(s_{\links})}< \ov{\pi(t)}$, and hence 
$\bbig_i < \ov{\pi(t)}$.
\\
In case that $\abig_i \neq \ov{s_{\links}}$ we know that
$s_{\links}$ and $s_{\rechts}$ must have been chosen according to case (III).
This, in particular, implies that $\abig_i < \ov{s_{\rechts}} \leq \ov{t}$. 
Condition (5) (for $i{-}1$) gives us that $\ov{\pi(s_{\rechts})}\leq \ov{\pi(t)}$.
Furthermore, from the choice of $\bbig_i$ we know that 
$\bbig_i < \ov{\hat{\pi}_i(s_{\rechts})} = \ov{\pi(s_{\rechts})}$. 
Hence, $\bbig_i < \ov{\pi(t)}$.
\\
Altogether, we obtain that {\bf condition (4) is satisfied for $\bs{i}$.}
\\
\parno
To show that condition (5) is satisfied for $i$ (if $i\neq k$), 
let $s_1$ and $s_2$ be
\,{
 $\big(\lF(k{-}i),\,\allowbreak\cF(k{-}i) ,\,\allowbreak\gF(k{-}i)\big)$}-combinations over 
$P,\vek{\cbig},\abig_1,\twodots,\abig_i$,\, and let
$\pi$ be an extension of $\pi_i$ which is ${<}$-preserving on $P$ and
which is defined on all the terms of $s_1$ and $s_2$.
We have to show that 
\,$\ov{s_1} < \ov{s_2}$\, if and only if \,$\ov{\pi(s_1)} < \ov{\pi(s_2)}$. 
\par 
Let \,$s_1 = \sum_{i=1}^{l} d_i x_i + f$\, and 
\,$s_2  = \sum_{j=1}^{l} d'_j x'_j + f'$.
By definition we know that $x_1,\twodots,x_l$ (resp., $x'_1,\twodots,x'_l$) are 
pairwise distinct elements in 
$P\cup\set{\vek{\cbig},\abig_1,\twodots,\abig_i}$.
Hence, $\set{x_1,\twodots,x_l,x'_1,\twodots,x'_l}$ consists of $l'$ pairwise distinct
elements $z_1,\twodots,z_{l'}$, for some $l'$ with $l'\leq 2l\leq 2\lF(k{-}i)$.
Obviously, 
{
\[
\begin{array}{ccccc}
  \ov{s_1} - \ov{s_2} & \; =\; &
  \sum_{i=1}^{l} d_i x_i \ -\ \sum_{j=1}^{l}d'_j x'_j \ +\  (f{-}f')
  & \; =\; &
  \sum_{r=1}^{l'}e_r z_r \,+\, h\,,
\end{array}
\]}\noindent%
where $h\deff f{-}f'$, and 
if $z_r= x_i = x'_j$ then $e_r \deff d_i {-} d'_j$,
if $z_r = x_i \neq x'_j$ for all $j$, then $e_r \deff d_i$, and
if $z_r = x'_j \neq x_i$ for all $i$, then $e_r \deff {-}d'_j$.
\\
Since $d_i,d'_j\in\QQ[\cF(k{-}i)]$ and $\betrag{f},\betrag{f'}\leq \gF(k{-}i)$, one can easily 
see that
{
\[ 
 (\bullet):\quad\
   \begin{array}{l} 
       l'\leq 2\lF(k{-}i), \quad \betrag{h} \leq 2\gF(k{-}i),\quad\mbox{and}
     \\[1ex]
       e_r = \frac{u_r}{u'_r}\mbox{ \ for }u_r,u'_r\in\ZZ\mbox{ with }
       \betrag{u_r} \leq 2\cF(k{-}i)^2\mbox{ \ and }
       \betrag{u'_r} \leq \cF(k{-}i)^2.
   \end{array}
\]}%
In case that $e_r=0$ for all $r$, we have
\,$\ov{s_1} - \ov{s_2} = h = \ov{\pi(s_1)}-\ov{\pi(s_2)}$,\,
and thus \,$\ov{s_1}<\ov{s_2}$ \ iff \ $\ov{\pi(s_1)}<\ov{\pi(s_2)}$.
\par
We can hence concentrate on the case where at least one of the coefficients $e_r$ is
different from $0$.
Without loss of generality we may assume that there is an $l''$ with 
$1\leq l'' \leq l'$, such that
$e_r\neq 0$ for all $r\leq l''$, and $e_r = 0$ for all $r>l''$. 
Furthermore, we may assume that if $\abig_i\in\set{z_1,\twodots,z_{l''}}$, then 
$\abig_i = z_1$. 
\\
Define 
\,$t_1\deff z_1$\, and 
\,$t_2\deff \sum_{r=2}^{l''}\frac{-e_r}{e_1}\cdot z_r +\frac{-h}{e_1}$.
It is straightforward to see that $t_2$ is a 
\,{
 $\big(\lF(k{-}i{+}1),\,\cF(k{-}i{+}1) ,\allowbreak\,\gF(k{-}i{+}1)-\frac{\mF(k{-}i)}{2}\big)$}-combination over
$P,\vek{\cbig},\abig_1,\twodots,\abig_{i-1}$:
From $(\bullet)$ we obtain 
\,$l''{-}1\leq 2\lF(k{-}i)-1 =\allowbreak \lF(k{-}i{+}1)$,\, and
\,$\frac{-e_r}{e_1}\in\QQ[2\cF(k{-}i)^4]$, where $2\cF(k{-}i)^4 = \cF(k{-}i{+}1)$,\, and
\,$\betrag{\frac{-h}{e_1}}\leq 2\gF(k{-}i)\, \cF(k{-}i)^2$, where
\,$2\gF(k{-}i)\, \cF(k{-}i)^2 = \gF(k{-}i{+}1)-\frac{\mF(k-i)}{2}$.
\\
In case that $t_1 = \abig_i$ we can apply condition (4) (for $i$); and otherwise we can
apply condition (5) (for $i{-}1$) to obtain that
\ $\ov{t_1} < \ov{t_2}$ \ iff \ $\ov{\pi(t_1)} < \ov{\pi(t_2)}$.
Of course, this in particular gives us
\begin{quote}
  \mbox{(a):}\qquad $e_1\cdot\ov{t_1}\ <\ e_1\cdot\ov{t_2}$  
  \quad iff \quad
  $e_1\cdot\ov{\pi(t_1)}\ <\ e_1\cdot\ov{\pi(t_2)}$\,.
\end{quote} 
Furthermore, we know that \ $\ov{s_1}<\ov{s_2}$ \ iff \ $\ov{s_1}-\ov{s_2} < 0$ \
iff \ $\sum_{r=1}^{l''}e_r\,z_r + h \,<\,0$ \ iff 
\ $e_1 z_1 < \sum_{r=2}^{l''}({-}e_r)\, z_r + ({-}h)$\,. In other words, we have
\begin{quote}
   \mbox{(b):}\qquad $\ov{s_1}\ <\ \ov{s_2}$
   \quad iff \quad
   $e_1\cdot \ov{t_1}\ <\ e_1\cdot \ov{t_2}$\,.
\end{quote}
Analogously, \ $\ov{\pi(s_1)}<\ov{\pi(s_2)}$ \ iff \ $\ov{\pi(s_1)}-\ov{\pi(s_2)} < 0$ \
iff \ $\sum_{r=1}^{l''}e_r\, \pi(z_r) + h \,<\,0$ \ iff 
\ $e_1\, \pi(z_1) < \sum_{r=2}^{l''}({-}e_r)\, \pi(z_r) + ({-}h)$\,. I.e., we have
\begin{quote}
   \mbox{(c):}\qquad $\ov{\pi(s_1)}\ <\ \ov{\pi(s_2)}$
   \quad iff \quad
   $e_1\cdot \ov{\pi(t_1)}\ <\ e_1\cdot \ov{\pi(t_2)}$\,.
\end{quote}
Altogether, (a), (b), and (c) give us that \ $\ov{s_1}<\ov{s_2}$ \ iff \ 
$\ov{\pi(s_1)}<\ov{\pi(s_2)}$.
\\
We hence obtain that {\bf condition (5) if satisfied for $\bs{i}$.}
\\
\parno 
Summing up, we have shown that the conditions (1)--(5) hold for $i{=}0$. Furthermore, we have
shown for each $i\in\set{1,\twodots,k}$, that if they hold for $i{-}1$, then the duplicator can
play in such a way that they hold for $i$.
In particular, we conclude that the duplicator can play in such a way that the conditions (1)--(5)
hold for all $i\in\set{0,\twodots,k}$. According to Claim~1 she thus has a winning strategy 
in the $k$-round ${+}$-game on $\Abig$ and $\Bbig$.
\\ 
This completes our proof of Proposition~\ref{proposition:PlusGame}.
\end{proofc_mit}%
\\
\parno
In fact, the proof of Proposition~\ref{proposition:PlusGame} shows the following result 
which is stronger but also more technical than Theorem~\ref{theo:additionQ}.
We will use this result in the following Section~\ref{subsection:PlusQQRR} in order to
transfer the translation result to context structures whose universe is the set $\RR$
of real numbers.
\begin{proposition_ohne}\label{proposition:PlusGame_stronger}
Let \,$Q = \set{q_0 < q_1 < q_2 <\cdots}\subseteq \NN$\, satisfy the
conditions $W(\omega)$ (cf., Definition~\ref{definition:conditions_W_omega}).
Let $m,l,c,g\in\NNpos$ and $g\in \RRpos$.
\\
For every number $k\in\NNpos$ of rounds for the
$\FO(<,+,Q)$-game there is a number $r_{(m,l,c,g)}(k)\in\NN$ of rounds for the
$\FOKleiner$-game such that the following is true for 
every signature $\tau$ and for all $\NN$-embeddable $\tau$-structures
$\A$ and $\B$:
If 
\begin{eqnarray*}
  \bigstruc{\,\adom(\A),\,<,\,\tau^{\A}\,}
& \dwins_{r_{(m,l,c,g)}(k)}
& \bigstruc{\,\adom(\B),\,<,\,\tau^{\B}\,}\,,
\end{eqnarray*} 
then
there are ${<}$-preserving mappings \,$\alpha:\adom(\A)\rightarrow Q$\, and
\,$\beta:\adom(\B)\rightarrow Q$\, such that the duplicator wins the $k$-round EF-game
on
{
\begin{eqnarray*}
  \Abig\ \deff\ \bigstruc{\,\ZZ,\;<,\;+,\;0,\;Q,\;\alpha\big(\tau^{\A}\big)\,}
& \mbox{ and }
& \Bbig\ \deff\ \bigstruc{\,\ZZ,\;<,\;+,\;0,\;Q,\; \beta\big(\tau^{\B}\big)\,}
\end{eqnarray*}}%
in such a way that after the $k$-th round the following holds true:
\\
Let, for every $i\in\set{1,\twodots,k}$, $\abig_i$ and $\bbig_i$ be the elements chosen in
the $i$-th round in $\Abig$ and $\Bbig$. Furthermore, let $\pi$ be the mapping defined via
\[
  \pi\ :\ \left\{
  \begin{array}{rcll}
     \alpha(c^{\A})& \mapsto & \beta(c^{\B}) & \mbox{ for all constant symbols $c\in\tau$} \\
     \abig_i       & \mapsto & \bbig_i       & \mbox{ for all $i\in\set{1,\twodots,k}$} 
  \end{array} \right\}\,.
\]
Then we have 
\begin{enumerate}[$\bullet$\ ]
\item
  $x\ \equiv_{m}\pi(x)$\,, \quad for every $x$ in the domain of $\pi$,\quad and
\item
  $\ov{s_1}<\ov{s_2}$ \ iff \ $\ov{\pi(s_1)}<\ov{\pi(s_2)}$\,, \quad
  for all $(l,c,g)$-combinations $s_1$ and $s_2$ over the domain of $\pi$.
  \\ \null\mbox{ }\fertig
\end{enumerate}
\end{proposition_ohne}%
\begin{proof_mit}
Since the functions $\lF$, $\cF$, $\gF$ are {increasing}, we can find some
$k_0\in\NNpos$ such that $\lF(k_0)\geq l$, $\cF(k_0)\geq c$, $\gF(k_0)\geq g$, and
$\cF(k_0)\geq m$. In particular, this also gives us that \,$m\mid \mF(k_0)$\,, because
$\mF(k_0) = \mF(k_0{-}1)\cdot\lcm\set{1,\twodots,\cF(k_0)}$. 
I.e., we have
{
\[
  (*):\qquad\qquad
  \lF(k_0)\geq l,\quad
  \cF(k_0)\geq c,\quad
  \gF(k_0)\geq g,\quad\mbox{and}\quad
  m \mid \mF(k_0)\,.
\]}%
Let $r$ be the function defined in Proposition~\ref{proposition:PlusGame}.
Define the function $r_{(m,l,c,g)}$ via
\,$r_{(m,l,c,g)}(k)\deff r(k{+}k_0)$,\, for every $k\in\NN$.
Let $\vek{\cbig}\deff 0,q_0,\twodots,q_{n_{k+k_0}-1}$ and 
$P\deff Q\setminus \set{\vek{\cbig}}$.
From the presumption we know that \,$P,\vek{\cbig}$\, satisfies the conditions $W(k{+}k_0)$ and
that all elements in $\vek{\cbig}$ are smaller than the smallest element in $P$.
\par
Let $\tau$ be a signature and let $\A$ and $\B$ be two $\NN$-embeddable $\tau$-structures such that
\,$\struc{\adom(\A),{<, }\allowbreak \tau^{\A}}
   \dwins_{r(k+k_0)}
   \struc{\adom(\B),<,\tau^{\B}}
$.
Let \,$\alpha:\adom(\A)\rightarrow P$\, and 
\,$\beta:\adom(\B)\rightarrow P$\, map, for every
$j$, the $j$-th smallest element of $\A$ respectively $\B$ onto the $j$-th smallest element in $P$.
In the proof of Proposition~\ref{proposition:PlusGame} we have seen that the duplicator can 
win the $(k{+}k_0)$-round EF-game on 
\ $\Abig \deff \struc{\ZZ,<,+,P,\vek{\cbig},\alpha\big(\tau^{\A}\big)}$ \ 
and 
\ $\Bbig \deff \struc{\ZZ,<,+,P,\vek{\cbig},\beta\big(\tau^{\B}\big)}$ \ 
in such a way that after the $k$-th
round condition (5) is satisfied for $i=k$ and condition (2) is satisfied for all
$i\in\set{1,\twodots,k}$. In particular, for the mapping $\pi$ defined in the 
formulation of Proposition~\ref{proposition:PlusGame_stronger} this means that
\begin{enumerate}[$\bullet$\ ]
\item
  $x\ \equiv_{\mF(k_0)}\pi(x)$\,, \quad for every $x$ in the domain of $\pi$,\quad and
\item
  $\ov{s_1}<\ov{s_2}$ \ iff \ $\ov{\pi(s_1)}<\ov{\pi(s_2)}$\,, \ 
  for all $\big(\lF(k_0),\,\cF(k_0),\,\gF(k_0)\big)$-combinations $s_1$ and $s_2$ over 
  the domain of $\pi$.
\end{enumerate}
Due to $(*)$, this completes the proof of 
Proposition~\ref{proposition:PlusGame_stronger}.
\end{proof_mit}%
%
%
%
\subsection[The $\FO(<,+,Q,\Groups)$-Game over $\RR$]{The
  $\bs{\FO(<,+,Q,\Groups)}$-Game over $\bs{\RR}$}\label{subsection:PlusQQRR}
\index{Ehrenfeucht-Fra\"\i{}ss\'{e} game!FO<+QGroupsgame@$\FO(<,+,Q,\Groups)$-game}%
In the previous Section~\ref{subsection:PlusNNZZ} we investigated the
context universes $\NN$ and $\ZZ$,
and showed that the duplicator can translate strategies for the
$\FOKleiner$-game into strategies for the $\FO(<,+,Q)$-game on
arbitrary structures over $\NN$ and on $\NN$-embed\-dable structures over
$\ZZ$ (cf., Theorem~\ref{theo:additionQ}).
In the present section we transfer these results to the context
universes $\QQ$ and $\RR$. 
As a consequence of Proposition~\ref{proposition:PlusGame_stronger} we obtain the 
following:
\begin{theorem_mit}[$\bs{\FO(<,+,Q,\Groups)}$-game over $\bs{\RR}$]\label{theo:additionQQRR}\mbox{ }\\
Let \,$Q\subseteq \NN$\, satisfy the
conditions $W(\omega)$. 
Let $\Groups$ consist of all sets $G\subseteq\RR$ where $1\in G$ and $\struc{G,+}$
is a subgroup of $\struc{\RR,+}$.\\
The duplicator can translate strategies for the $\FOKleiner$-game into strategies for the
$\FO(<,\allowbreak +,Q,\Groups)$-game on $\ClNemb$ over $\RR$. 
\medskip\\
\rm
In particular, this implies that
the duplicator can translate strategies for the $\FOKleiner$-game into strategies for the
$\FO(<,+,Q,\ZZ,\QQ)$-game on $\NN$-embeddable structures over $\QQ$
and over $\RR$.
\mbox{ }
\end{theorem_mit}%
\begin{proof_mit}
Let $k\in\NNpos$ be a number of rounds for the $\FO(<,+,Q,\Groups)$-game.
We define the according number $r(k)$ of rounds for the $\FOKleiner$-game via
\,$r(k)\deff r_{(1,2,2,2)}(k)$,\, where $r_{(1,2,2,2)}$ is the function from
Proposition~\ref{proposition:PlusGame_stronger} for \,$m = 1$\, and 
\,$l = c = g = 2$.
\par
Let $\tau$ be a signature and let $\A$ and $\B$ be two $\NN$-embeddable 
$\struc{\RR,\tau}$-structures such that
\, $\struc{\adom(\A),<,\tau^{\A}}
    \dwins_{r(k)}
    \struc{\adom(\B),<,\tau^{\B}}$.
From Proposition~\ref{proposition:PlusGame_stronger} we obtain ${<}$-preserving
mappings \,$\alpha:\adom(\A)\rightarrow Q$\, and \,$\beta:\adom(\B)\rightarrow Q$\,
such that the duplicator can win the $k$-round EF-game on
\ $\Abig_{\ZZ} \deff \struc{\ZZ,<,+,Q,\alpha\big(\tau^{\A}\big)}$ \ 
and
\ $\Bbig_{\ZZ} \deff \struc{\ZZ,<,+,Q,\beta\big(\tau^{\B}\big)}$ \ 
in such a way that after the $k$-th round the conditions formulated in 
Proposition~\ref{proposition:PlusGame_stronger} are satisfied.
Henceforth, this game on $\Abig_{\ZZ}$ and $\Bbig_{\ZZ}$ will be called \emph{the $\ZZ$-game}.
\\
Our aim is to show that the duplicator wins the $k$-round EF-game on
\ $\Abig_{\RR} \deff \struc{\RR,<,\allowbreak +,Q,\Groups, \allowbreak \alpha\big(\tau^{\A}\big)}$ \ 
and
\ $\Bbig_{\RR} \deff \struc{\RR,<,+,Q,\Groups,\beta\big(\tau^{\B}\big)}$.
Henceforth, the game on $\Abig_{\RR}$ and $\Bbig_{\RR}$ will be called \emph{the $\RR$-game}.
\par
In order to win the $\RR$-game, the duplicator plays according to the strategy illustrated
in Figure~\ref{figure:PlusQQRR}.
%
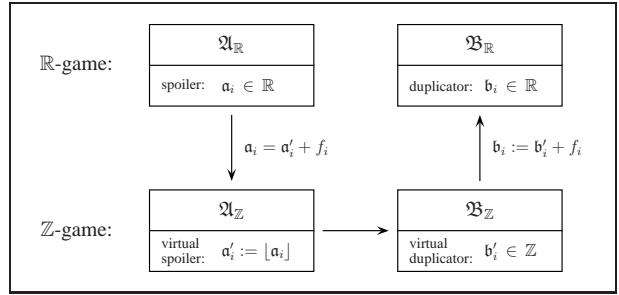
\begin{figure}[!htbp]
\bigskip
\begin{center}
\fbox{
\scalebox{0.55}{
\begin{pspicture}(-1,-3.3)(13,+3.3)
\psset{arrowsize=6pt}%
\rput(0.25,2){\Large $\RR$-game:}\rput(0.25,-2){\Large $\ZZ$-game:}%
\psframe(2,1)(6,3)\psline(2,2)(6,2)%
\psframe(2,-1)(6,-3)\psline(2,-2)(6,-2)%
\psframe(8,1)(12,3)\psline(8,2)(12,2)%
\psframe(8,-1)(12,-3)\psline(8,-2)(12,-2)%
\psline{->}(4,0.8)(4,-0.8)\rput[l](4.3,0){\large $\abig_i = \abig'_i + f_i$}
\psline{->}(10,-0.8)(10,0.8)\rput[l](10.3,0){\large $\bbig_i\deff \bbig'_i+f_i$}%
\psline{->}(6.2,-2)(7.8,-2)%
\rput(4,2.5){\Large $\Abig_{\RR}$}\rput(10,2.5){\Large $\Bbig_{\RR}$}%
\rput(4,-1.5){\Large $\Abig_{\ZZ}$}\rput(10,-1.5){\Large $\Bbig_{\ZZ}$}%
\rput[l](2.3,1.5){spoiler:}\rput[l](3.75,1.5){\large $\abig_i\,\in\,\RR$}%
\rput[l](8.3,1.47){duplicator:}\rput[l](10.1,1.5){\large $\bbig_i\,\in\,\RR$}%
\rput[l](2.3,-2.3){virtual}\rput[l](2.3,-2.7){spoiler:}%
\rput[l](3.75,-2.5){\large $\abig'_i\deff\abgerundet{\abig_i}$}%
\rput[l](8.3,-2.3){virtual}\rput[l](8.3,-2.7){duplicator:}%
\rput[l](10.1,-2.5){\large $\bbig'_i\,\in\,\ZZ$}%
\end{pspicture}
}
}
\caption{\small The duplicator's strategy in the $i$-th round of the $\RR$-game. 
Here, the spoiler chooses an element $\abig_i$ in $\Abig_{\RR}$ and the duplicator answers
with a $\bbig_i$ in $\Bbig_{\RR}$. The case where the spoiler chooses an 
element $\bbig_i$ in $\Bbig_{\RR}$ can be treated analogously.}\label{figure:PlusQQRR}
\end{center}
\end{figure}%
%
%
\\
For the $i$-th round (for every $i\in\set{1,\twodots,k}$) this precisely means the following:
\\
Assume that the spoiler chooses an element $\abig_i$ in $\Abig_{\RR}$ (the case where he
chooses $\bbig_i$ in $\Bbig_{\RR}$ is symmetric).
We translate the spoiler's move $\abig_i$ into a move $\abig'_i$ in $\Abig_{\ZZ}$ for a 
``virtual spoiler'' in the 
$\ZZ$-game via \,$\abig'_i \deff \abgerundet{\abig_i}$. In particular, we know that
$\abig_i = \abig'_i + f_i$ for some $f_i \in\intco{0,1}\subseteq\RR$.\\
Now, let $\bbig'_i$ in $\Bbig_{\ZZ}$ be the answer of a ``virtual duplicator'' who plays according 
to her winning strategy in the $\ZZ$-game. 
We can translate this answer into a move $\bbig_i$ for the duplicator
in the $\RR$-game via 
\,$\bbig_i\deff \bbig'_i + f_i$.
%
\par
It is straightforward to see that after $k$ rounds the duplicator has won the $\RR$-game:
We need to show that the mapping $\pi$
defined via  
\[
  \pi\ :\ \left\{
  \begin{array}{rcll}
     \alpha(c^{\A})& \mapsto & \beta(c^{\B}) & \mbox{ for all constant symbols $c\in\tau$} \\
     \abig_i       & \mapsto & \bbig_i       & \mbox{ for all $i\in\set{1,\twodots,k}$} 
  \end{array} \right\}\,
\]
is a partial isomorphism between $\Abig_{\RR}$ and $\Bbig_{\RR}$.
We already know that the ``virtual duplicator'' has won the $\ZZ$-game and that even
the conditions formulated in Proposition~\ref{proposition:PlusGame_stronger} are satisfied.
I.e., for the mapping $\pi'$ defined via
\[
  \pi'\ :\ \left\{
  \begin{array}{rcll}
     \alpha(c^{\A})& \mapsto & \beta(c^{\B}) & \mbox{ for all constant symbols $c\in\tau$} \\
     \abig'_i       & \mapsto & \bbig'_i       & \mbox{ for all $i\in\set{1,\twodots,k}$} 
  \end{array} \right\}
\]
we know that 
\begin{enumerate}[$(**)$\ ]
\item[$(*)$\ ]
  $\pi'$ is a partial isomorphism between $\Abig_{\ZZ}$ and $\Bbig_{\ZZ}$,\quad and
\item[$(**)$\ ]
  $\ov{s_1}<\ov{s_2}$ \ iff \ $\ov{\pi'(s_1)}<\ov{\pi'(s_2)}$ \quad is true 
  for all $(2,2,2)$-combinations $s_1$ and $s_2$ over the domain of $\pi'$.
\end{enumerate}
Furthermore, we know that \,$\abig_i = \abig'_i + f_i$\, and 
\,$\bbig_i = \bbig'_i + f_i$\, for $\abig'_i,\bbig'_i\in\ZZ$ and $f_i\in\intco{0,1}\subseteq\RR$.
This, in particular, gives us that 
\ $\abig_i\in\ZZ$ \ iff \ $\bbig_i\in\ZZ$\, and, in general, for every
$G\in\Groups$, that 
\ $\abig_i\in G$ \ iff \ $\bbig_i\in G$.
Together with $(*)$ we furthermore obtain that 
\ $\abig_i\in Q$ \ iff \ $\bbig_i\in Q$\,,  and \ that
$\pi$ is a partial isomorphism between 
\,$\struc{\RR,\,Q,\,\Groups,\,\alpha(\tau^{\A})}$\, and
\,$\struc{\RR,\,Q,\,\Groups,\,\beta(\tau^{\B})}$.
\par
All that remains to be done is to show that \,``$x<y$ \ iff \ $\pi(x)<\pi(y)$''\,
and ``$x+y=z$ \ iff \ $\pi(x)+\pi(y)=\pi(z)$''\, are true for all $x,y,z$ in the domain of 
$\pi$.
In order to show this, consider the integers $x'\deff \abgerundet{x}$, $y'\deff \abgerundet{y}$, and
$z'\deff \abgerundet{z}$, and choose $f,g,h\in\intco{0,1}\subseteq\RR$ such that
$x=x'+f$, $y=y'+g$, and $z=z'+h$.
Obviously, $x',y',z'$ must belong to the domain of $\pi'$, and we must have
$\pi(x) = \pi'(x') + f$, \ $\pi(y) = \pi'(y') + g$, \ and \ $\pi(z) = \pi'(z') + h$.
Due to $(**)$ we know that
\ $x'+f \,<\,y'+g$ \ iff \ $\pi'(x')+f \,<\,\pi'(y')+g$.\,
This, in particular, gives us that \ $x<y$ \ iff \ $\pi(x) < \pi(y)$.
Furthermore, $(**)$ gives us that
\ $x'+ y' + (f{+}g) = z' + h$ \ iff \ $\pi'(x')+ \pi'(y') + (f{+}g) = \pi'(z') + h$.
In other words, we obtain that
\ $x+y=z$ \ iff \ $\pi(x)+\pi(y) = \pi(z)$.
Altogether, the proof of Theorem~\ref{theo:additionQQRR} is complete. 
\end{proof_mit}
%
%
%
\subsection{Variations}\label{subsection:Variations}
%
%
%
%
\subsubsection[More Built-In Predicates: $\Mon_Q$]{More Built-In Predicates: $\bs{\Mon_Q}$}\label{subsection:MonQ}\index{MonAQ@$\Mon_Q$} 
With Theorem~\ref{theo:additionQ} we obtained the translation result for every
context structure $\struc{\ZZ,<,\allowbreak +,Q}$ where $Q$ satisfies the 
conditions $W(\omega)$.
Making use of the 
method for \emph{monadic} predicates described in
Section~\ref{section:Monadic}, we may add all subsets of $Q$ as built-in predicates:
\begin{theorem_ohne}\label{theorem:Variation_Plus_MonQ}
\mbox{ }\\
Let \,$Q\subseteq \NN$\, satisfy the
conditions $W(\omega)$. 
Let $\Mon_Q$ be the class of all subsets of $Q$.
\begin{enumerate}[(a)]
 \item 
   The duplicator can translate strategies for the $\FOKleiner$-game
   into 
   strategies for the $\FO(<,\allowbreak +,Q,\Mon_Q)$-game on arbitrary structures over
   $\NN$ and on $\NN$-embeddable structures over $\ZZ$.
   \index{Ehrenfeucht-Fra\"\i{}ss\'{e} game!FO<+QMonQgame@$\FO(<,+,Q,\Mon_Q)$-game}
 \item 
   The duplicator can translate strategies for the $\FOKleiner$-game
   into 
   strategies for the $\FO(<,\allowbreak +,Q,\Mon_Q,\Groups)$-game on
   $\NN$-embeddable 
   structures over $\RR$.
   \mbox{ }\fertig 
\end{enumerate} 
\end{theorem_ohne}%
\begin{proofsketch_mit}
\emph{(b)} can be obtained from \emph{(a)} (respectively, from the
according variant of Proposition~\ref{proposition:PlusGame_stronger})
in the same way as Theorem~\ref{theo:additionQQRR} was obtained from
Theorem~\ref{theo:additionQ}.
Part \emph{(a)} is a direct consequence of the following variant of 
Proposition~\ref{proposition:PlusGame}:
\begin{proposition_mit}\label{proposition:VariantPlusGame}
Let $k,n\in\NN$, let $\vek{\cbig} \deff \cbig_1,\twodots,\cbig_n\in\NN$, and let
\,$P\deff \set{p_1<p_2<p_3<\cdots}\subseteq \NN$\, be an infinite set 
such that \,$P,\vek{\cbig}$\, satisfies the conditions 
$W(k)$ and such that, for all $\nu\in\set{1,\twodots,n}$, \,$\cbig_{\nu}$ is smaller 
than the smallest element in $P$.
Let $\Mon_P$ be the class of all subsets of $P$.
There is a number \,$r(k)\in\NN$\, such that the following is true for all
finite subsets $\Mon'_P$ of $\Mon_P$, and for all signatures $\tau$:
\\
If $\A$ and $\B$ are $\NN$-embeddable $\tau$-structures with
$\struc{\adom(\A),<,\tau^{\A}}
   \dwins_{r(k)}
   \struc{\adom(\B),<,\allowbreak{}\tau^{\B}}$,\,
then there are ${<}$-preserving mappings \,$\alpha:\adom(\A)\rightarrow P$\, and
\,$\beta:\adom(\B)\rightarrow P$\, such that
\,$\struc{\ZZ,<,+,P,\vek{\cbig},\Mon'_P,\alpha(\tau^{\A})}
   \dwins_{r(k)}
   \struc{\ZZ,<,+,P,\vek{\cbig},\Mon'_P,\beta(\tau^{\B})}$. 
\end{proposition_mit}%
For the proof of Proposition~\ref{proposition:VariantPlusGame} choose
\,$r(k)\deff r_1(r_2(k))$, where $r_2$ is the function $r$ obtained from 
Proposition~\ref{proposition:PlusGame}, and $r_1$ is the function $r$ obtained from 
Theorem~\ref{theorem:collapse_Monadic_UU}.\\
Assume we are given two $\NN$-embeddable $\tau$-structures $\A$ and $\B$ with
\,$\struc{\adom(\A),<,\allowbreak \tau^{\A}}
   \dwins_{r_1(r_2(k))}
   \struc{\adom(\B),<,\tau^{\B}}$. 
Theorem~\ref{theorem:collapse_Monadic_UU} (for $\UU\deff P$) gives us 
${<}$-preserving mappings \,$\alpha:\adom(\A)\rightarrow P$\, and 
\,$\beta:\adom(\B)\rightarrow P$\, such that
\,$\struc{P,<,\Mon'_P,\alpha(\tau^{\A})}
   \allowbreak \dwins_{r_2(k)}
   \struc{P,<,\Mon'_P,\beta(\tau^{\B})}$.
\\
In the proof of Proposition~\ref{proposition:PlusGame} we considered the structures
\ $\Abig'\deff\struc{P,<,\alpha(\tau^{\A})}
   \allowbreak \dwins_{r_2(k)} \linebreak[4]
   \struc{P,<,\beta(\tau^{\B})}\ffed \Bbig'$.
Instead of $\Abig'$ and $\Bbig'$ we now use the structures
\,$\Abig'' \deff \struc{P,<,\Mon'_P,\alpha(\tau^{\A})}
   \allowbreak \dwins_{r_2(k)}
   \struc{P,<,\Mon'_P,\beta(\tau^{\B})}\ffed \Bbig''$.
In the proof of Proposition~\ref{proposition:PlusGame} we replace $\Abig'$ and $\Bbig'$  
with $\Abig''$ and $\Bbig''$. This gives us the desired result that
\,$\struc{\ZZ,<,+,P,\vek{\cbig},\Mon'_P,\alpha(\tau^{\A})}
   \dwins_{k}
   \struc{\ZZ,<, \allowbreak +,P,\vek{\cbig},\Mon'_P,\beta(\tau^{\B})}$.
%
Altogether, this proves Proposition~\ref{proposition:VariantPlusGame} and completes the 
proof sketch of Theorem~\ref{theorem:Variation_Plus_MonQ}.
\end{proofsketch_mit}%
%
%
%
\subsubsection{A Question of Belegradek et al.}\label{subsection:conjectureBST}
Considering the natural generic collapse over \emph{finite} databases, Belegradek et al.\ asked
in the conclusion of \cite{BST99}:\index{Belegradek, Oleg V.}
\emph{``How much higher than $+$ in $\struc{\ZZ,<}$ can we go?''}
Our Theorem~\ref{theorem:Variation_Plus_MonQ} gives an answer: 
We still obtain the natural generic collapse\footnote{even on all 
\emph{$\NN$-embeddable} databases over $\ZZ$} when adding a set $Q$ that satisfies the 
conditions $W(\omega)$ and when adding all subsets of $Q$ as built-in predicates.
\par
Furthermore, Belegradek et al.\ conjectured the following: 
\emph{``If, for some class $\Rel$ of built-in predicates, 
\(
  \OrderGen\FO(<,+,\Rel) \neq \FOadom(<) \mbox{ on } \Clfinite \mbox{ over } \ZZ,
\)
then the first-order theory of $\struc{\ZZ,<,+,\Rel}$ is undecidable.''}
Our result shows that the \emph{converse} of this conjecture is not
true: Let $Q$ be the set obtained in Lemma~\ref{lemma:Q_satisfying_Conditions}\;(b), 
and let $\tilde{Q}$ be an 
undecidable subset of $Q$. E.g., $\tilde{Q}$ can be chosen to contain, for every $n\in\NNpos$, 
the $n$-th largest 
element in $Q$ if and only if the $n$-th Turing machine \index{Turing machine}
halts with empty input.
Clearly, the first-order theory of \,$\struc{\ZZ,<,+,\tilde{Q}}$\, is undecidable. 
On the other hand,
$\tilde{Q}$ satisfies the conditions $W(\omega)$, and hence
Theorem~\ref{theo:additionQ} gives us that
\(
  \OrderGen\FO(<,+,\tilde{Q}) = \FOadom(<) \mbox{ on } \Clfinite \mbox{ over } \ZZ.
\)
%
%
%
\subsubsection[More Structures: $\ZZ$-embeddable Structures?]{More Structures: $\bs{\ZZ}$-embeddable Structures?}\label{subsection:ZZ-emb}\index{Z-embeddable@$\ZZ$-embeddable}
Theorem~\ref{theo:additionQ} states the translation result for \emph{$\NN$-embeddable}
structures over the context structure \,$\struc{\ZZ,<,+,Q}$.
It remains open whether the translation is possible also for \emph{$\ZZ$-embeddable}
structures.
The main reason why our proof does not work for all $\ZZ$-embeddable structures 
is that there does not exist a set $P$ which satisfies the
conditions $W(k)$ and which is infinite in \emph{both} directions (this easily follows
from the definition of the conditions $W(k)$).
\\
However, with some modification, our proof of Proposition~\ref{proposition:PlusGame} 
shows the following:
\begin{theorem_ohne}\label{theorem:addition_ZZ-emb}
Let \,$Q\subseteq \NN$\, satisfy the conditions $W(\omega)$. 
Let $\Inv$ be the binary relation which connects each number with its additive inverse, 
i.e., \,$\Inv(x,y)$ \,iff\, $x\geq 0$ \,and\, $y={-}x$. \index{Inv@$\Inv$} 
\begin{enumerate}[(a)\ ]
\item
  The duplicator can translate strategies for the \,$\FO(<,\Inv)$-game into 
  strategies for the \,$\FO({<, }\allowbreak +,Q)$-game on arbitrary structures over
  $\ZZ$. 
\item
  The duplicator can translate strategies for the \,$\FO(<,\Inv)$-game into 
  strategies for the \,$\FO({<, }\allowbreak +,Q,\Groups)$-game on $\ZZ$-embeddable structures over
  $\RR$. \mbox{ }\fertig
\end{enumerate}
\end{theorem_ohne}%
\begin{proofsketch_mit}
It should be clear that \emph{(b)} can be obtained from \emph{(a)} (respectively, from the
according variant of Proposition~\ref{proposition:PlusGame_stronger}) in the same way as
Theorem~\ref{theo:additionQQRR} was obtained from Theorem~\ref{theo:additionQ}.
\\
Part \emph{(a)} can be proved as follows:
Let $k$ be a number of rounds for the $\FO(<,+,Q)$-game, and let $r(k)$ and \,$P,\vek{\cbig}$\, 
be chosen as in the proof of Theorem~\ref{theo:additionQ}. 
Assume we are given two $\struc{\ZZ,\tau}$-structures $\A$ and $\B$ with
\,$\struc{\ZZ,<,\Inv,\tau^{\A}}
   \dwins_{r(k)}
   \struc{\ZZ,<,\Inv,\tau^{\B}}$.
In the proof of Proposition~\ref{proposition:PlusGame} we considered structures
\ $\Abig'\deff\struc{P,<,\alpha(\tau^{\A})}
   \dwins_{r(k)}
   \struc{P,<,\beta(\tau^{\B})}\ffed \Bbig'$. 
Instead, we now consider the following ${<}$-preserving mappings $\alpha$ and $\beta$: 
The mapping $\alpha$ is defined via 
\,$\alpha(0)\deff 0$, \,$\alpha(n) \deff p_n$, and \,$\alpha({-}n)\deff {-}p_n$, 
for all $n\in\NNpos$. Here, we assume that \,$P=\set{p_1<p_2<\cdots}\subseteq\NNpos$.
We define $\beta$ to be identical to $\alpha$, and we use $\hat{P}$ for the range of $\alpha$ 
and $\beta$, i.e., $\hat{P}$ is the set 
\,$\set{\,\cdots<{-}p_{2}<{-}p_1 < 0 < p_1 < p_2 <\cdots\,}$\,.
Instead of $\Abig'$ and $\Bbig'$ we now use the structures
\ $\Abig'' \deff \struc{\hat{P},<,\Inv,\alpha(\tau^{\A})}
   \dwins_{r(k)}
   \struc{\hat{P},<,\Inv,\beta(\tau^{\B})}\ffed \Bbig''$.
\\
In the proof of Proposition~\ref{proposition:PlusGame} we replace \,$\Abig'$, $\Bbig'$, $P$\, with
\,$\Abig''$, $\Bbig''$, $\hat{P}$. This will give us the desired result that
\,$\struc{\ZZ,<,+,\hat{P},\vek{\cbig},\alpha(\tau^{\A})}
   \dwins_{k}
   \struc{\ZZ,<,+,\hat{P},\vek{\cbig},\beta(\tau^{\B})}$.
It is tedious, but straightforward to check that all the details of the proof 
remain correct.
\end{proofsketch_mit}%
%
%
%
%
%
%



\section[How to Win the Game for ${\BCEFO{<,\Arb}}$]{How to Win the Game
  for $\bs{\BCEFO{<,\Arb}}$}\label{section:BCEFO}
\index{Ehrenfeucht-Fra\"\i{}ss\'{e} game!BCEFO<Arb@$\BCeFO(<,\Arb)$-game} 
\index{BCEFO@$\BCeFO$} 
In the previous sections we concentrated on the EF-game for $\FO$. 
In the present section we restrict our attention to the sublogic $\BCeFO$, consisting
of the Boolean combinations of purely \emph{existential} first-order formulas.
We introduce the 
\emph{single-round $r$-move game} as a variant of the ``classical'' EF-game that is 
suitable for characterizing the logic $\BCeFO$. 
The main result of this section is that the duplicator can translate strategies
for the $\BCeFOadom(<)$-game into strategies for the 
$\BCeFO(<,\Bip)$-game on $\NN$-embeddable structures over \emph{every} context
structure $\struc{\UU,<,\allowbreak \Bip}$.  
%
%
\subsection[The EF-Game for $\BCeFO$]{The EF-Game for $\bs{\BCeFO}$}\label{section:EF-game_BCEFO} 
In the same way as the ``classical'' EF-game characterizes the logic $\FO$,  
the following variant of the EF-game characterizes the logic $\BCeFO$.
\par
Let $\tau$ be a signature and let $r$ be a natural number.
The \emph{single-round $r$-move game} on two $\tau$-structures $\A$ and $\B$ is played
as follows:
First, the \emph{spoiler} chooses either $r$ elements $a_1,\twodots,a_r$ in the universe of $\A$,
or $r$ elements $b_1,\twodots,b_r$ in the universe of $\B$.
Afterwards, the \emph{duplicator} chooses $r$ elements in the other structure. I.e., she
chooses either $r$ elements $b_1,\twodots,b_r$ in the universe of $\B$, if the spoiler's move
was in $\A$, or she chooses $r$ elements $a_1,\twodots,a_r$ in the universe of $\A$, if
the spoiler's move was in $\B$.
\par
The \emph{winning condition} is identical to the winning condition in the \index{winning condition}
``classical'' $r$-round EF-game for $\FO$. 
%
We say that \emph{the duplicator wins the single-round $r$-move game on $\A$
  and $\B$}, and we write \,$\A \dwinsBCEFO_r \B$,
\index{$\dwinsBCEFO_r$ ($\A\dwinsBCEFO_r\B$)}%
if and only if the duplicator has a winning strategy in the
single-round $r$-move game on $\A$ and $\B$.
It is straightforward to see that, for every signature $\tau$, the relation $\dwinsBCEFO_r$ is an
\emph{equivalence relation} on the set of all $\tau$-structures.
By the standard argumentation 
(see, e.g., the textbooks \cite{Immerman,EbbinghausFlum}) one obtains the according
variants of Theorem~\ref{theorem:E-F} and Remark~\ref{remark:E-F-classes}. I.e.:
\begin{enumerate}[(1.)\ ]
\item
  $\A\dwinsBCEFO_r\B$\,
  if and only if \,$\A$ and $\B$ cannot be distinguished by $\BCeFO(\tau)$-sentences of
  quantifier depth $\leq r$.
\item
  A class $\LL$ of $\tau$-structures is \emph{not} $\BCeFO(\tau)$-definable in $\K$\, 
  if and only if\, for every $r\in\NN$ there are structures \,$\A_r,\B_r\in\K$\, with
  \,$\A_r\in\LL$\, and \,$\B_r\not\in\LL$\, and \,$\A_r\dwinsBCEFO_r\B_r$.
\item
  The relation $\dwinsBCEFO_r$ has only \emph{finitely} many equivalence classes 
  \index{equivalence classes}
  on the set
  of all $\tau$-structures; and each such equivalence class is definable by a
  $\BCeFO(\tau)$-sentence of quantifier depth $\leq r$.
\end{enumerate}%
It is straightforward to modify Definition~\ref{definition:TranslatingStrategies}
in such a way that it serves for proving a collapse 
of the form
\,$\OrderGen\BCeFO(<,\Num) = \BCeFOadom(<) \mbox{ on $\Cl$ over $\UU$}$.
%
%
\begin{definition_ohne}[{Translation of Strategies for $\bs{\BCeFO}$}]\label{definition:BCEFO_TranslatingStrategies}
\index{translation of strategies!for $\BCeFO$} 
\mbox{}\\
Let \,$\struc{\UU,<,\Num}$\, be a context structure, 
and let $\Cl$ be a class of structures over the universe $\UU$.
We say that
\begin{quote}
the duplicator can translate strategies for the {$\BCeFOKleiner$}-game into 
strategies for the {$\BCeFO(<,\Num)$}-game on $\Cl$ over $\UU$
\index{Ehrenfeucht-Fra\"\i{}ss\'{e} game!BCEFOadom<@$\BCeFOadom(<)$-game}
\index{Ehrenfeucht-Fra\"\i{}ss\'{e} game!BCEFO<Bip@$\BCeFO(<,\Bip)$-game}
\end{quote}
if and only if the following is true:
\begin{quote}
   For every finite set \,$\Num'\subseteq\Num$,\, for every
   signature
   $\tau$,\, and for every number $k\in\NN$ 
   there is a number $r(k)\in\NN$ 
   such that the following is true for all
   $\struc{\UU,\tau}$-structures $\A,\B\in\Cl$:
   If the duplicator wins the single-round $r(k)$-move {$\BCeFOKleiner$}-game
   on $\A$ and $\B$, i.e., if 
   \,$\struc{\adom(\A), <, \tau^{\A}} 
      \dwinsBCEFO_{r(k)} 
     \struc{\adom(\B), <, \tau^{\B}}$,
   then there are $<$-preserving mappings \,$\alpha : \adom(\A)
   \rightarrow \UU$\, and \,$\beta : \adom(\B) \rightarrow \UU$\, such that
   the duplicator wins the single-round $k$-move $\BCeFO(<, \allowbreak \Num')$-game on 
   $\alpha(\A)$ and $\beta(\B)$, i.e., 
   $\struc{\UU, <, \allowbreak \Num',\alpha\big(\tau^{\A}\big)} 
      \dwinsBCEFO_{k} 
      \struc{\UU, <,\allowbreak\Num', \beta\big(\tau^{\B}\big)}$. 
\end{quote}
\mbox{ }\vspace{-5ex}\\ \mbox{}\fertig
\end{definition_ohne}%
Replacing $\FO$ with $\BCeFO$ and replacing $\dwins_r$ with $\dwinsBCeFO_r$ in the proof of Theorem~\ref{theorem:EF-Collapse}, we 
directly obtain the following:
\begin{theorem_ohne}[Translation of Strategies $\bs{\Leftrightarrow}$ Collapse Result]\label{theorem:BCEFO_EF-Collapse}
\index{collapse result!EF-game}
\mbox{}\\%
Let \,$\struc{\UU,<,\Num}$\, be a context structure, 
and let $\Cl$ be a class of structures over the universe $\UU$.
The following are equivalent:
\begin{enumerate}[(a)\ ]
 \item 
   The duplicator can translate strategies for the $\BCeFOKleiner$-game into 
   strategies for the $\BCeFO(<,\Num)$-game on $\Cl$ over $\UU$.
 \item  
    $\OrderGen\BCeFO(<,\Num) \, = \, \BCeFOadom(<)$\, on \,$\Cl$\, over \,$\UU$. 
    \mbox{ } \fertig
\end{enumerate} 
\end{theorem_ohne}%
In Section~\ref{section:translation_BCEFO} below we will show that
the duplicator can indeed translate strategies for the {$\BCeFOKleiner$}-game into 
strategies for the {$\BCeFO(<,\allowbreak\Arb)$}-game on $\ClNemb$ over $\UU$, 
for every linearly ordered infinite universe $\UU$. 
However, we first show a lemma that
will help us avoid some technical difficulties in the translation proof.
%
%
\subsection{A Technical Lemma Similar to Lemma~\ref{lemma:nuetzlich}}\label{section:lemma_BCEFO}
The following lemma is an analogue of Lemma~\ref{lemma:nuetzlich}.
Note, however, that the mappings $\alpha$ and $\beta$ now depend on the
number $r$ of moves in the game.
%
\begin{lemma_mit}\label{lemma:nuetzlich_BCEFO}
Let \,$P \deff \set{p_1<p_2<p_3<\cdots}$\, be a countable, infinitely increasing
sequence, and let
$\SuccP$ \index{succP@$\SuccP$} be the binary \emph{successor relation} 
\index{successor relation} 
on $P$, i.e., \,$\SuccP\deff\setc{(p_j,p_{j+1})}{j\in\NNpos}$.
Let $\tau$ be a signature, and let $\A$ and $\B$ be two $\NN$-embeddable 
$\tau$-structures over linearly ordered universes.
\\
For every $r\in\NN$ there exist ${<}$-preserving mappings
\,$\alpha:\adom(\A)\rightarrow P$\, and 
\,$\beta:\adom(\B)\rightarrow P$\, such that the following is true:
If
\,$\struc{\adom(\A),<,\tau^{\A}} 
   \dwinsBCEFO_r   
   \struc{\adom(\B),<,\allowbreak \tau^{\B}}$,
then also
\,$\Abig \deff \struc{P,<,p_1,\SuccP,\alpha(\tau^{\A})} 
   \dwinsBCEFO_r 
   \struc{P,<,p_1,\SuccP,\beta(\tau^{\B})} \ffed \Bbig$.
\end{lemma_mit}%
\begin{proof_mit}
The main idea is to define the mappings $\alpha$ and $\beta$ in such a way that there is
a large gap between any two active domain elements.
Precisely, given \,$P = \set{p_1<p_2<p_3<\cdots}$\, it suffices to move the
active domain elements of $\A$ and $\B$ onto the positions
\,$p_{2r}<p_{4r}<p_{6r}<\cdots$.
I.e.: $\alpha:\adom(\A)\rightarrow P$\, and \,$\beta:\adom(\B)\rightarrow P$\, 
map, for every $j$, the $j$-th smallest
element in $\adom(\A)$ and $\adom(\B)$, respectively, 
onto the position $p_{2rj}$. 
From the presumption we know that a ``virtual duplicator'' wins the single-round $r$-move game
on \,$\struc{\adom(\A),<,\tau^{\A}}$\, and \,$\struc{\adom(\B),<,\tau^{\B}}$,\, i.e.,
\begin{eqnarray*}
  (*):\qquad
  \struc{\adom(\A),<,\tau^{\A}} & \dwinsBCEFO_r &
  \struc{\adom(\B),<,\tau^{\B}}.
\end{eqnarray*}
Obviously, this remains valid if $r$ is replaced with a number $s\leq r$.
The game $(*)$ will henceforth be called \emph{the small game}.
\par
The aim is to find a winning strategy for the duplicator in the single-round $r$-move game
on 
\,$\Abig \deff \struc{P,<,p_1,\SuccP,\alpha(\tau^{\A})}$\, and 
\,$\Bbig \deff \struc{P,<,p_1,\SuccP,\beta(\tau^{\B})}$.
This game will henceforth be called \emph{the big game}.
\par
Assume that the spoiler chooses the elements \,$\abig_1,\twodots,\abig_r$\, in the 
universe of $\Abig$ (if he chooses the elements \,$\bbig_1,\twodots,\bbig_r$\, in the 
universe of $\Bbig$, we can proceed in the according way, interchanging the roles
of $\Abig$ and $\Bbig$).
Some --- possibly all, or none --- of the elements \,$\abig_1,\twodots,\abig_r$\, belong to
\,\allowbreak$\alpha\big(\adom(\A)\big)$.
Let $s$ be the number of these elements and let, without loss of generality, 
\,$\abig_1,\twodots,\abig_s\in\alpha\big(\adom(\A)\big)$\, and 
\,$\abig_{s+1},\twodots,\abig_r\not\in\alpha\big(\adom(\A)\big)$.
Furthermore, we may assume that \,$\abig_1 < \cdots < \abig_s$.
\\
Of course there exist positions \,$a_{1}<\cdots<a_{s}$\, in $\adom(\A)$ such that
\,$\abig_{1} = \alpha(a_{1})$, \ldots, $\abig_{s} = \alpha(a_{s})$.
These elements \,$a_{1},\twodots,a_{s}$\, are the moves for a 
``virtual spoiler'' in the small game.
A ``virtual duplicator'' who plays according to her winning strategy in the small game
will find answers \,$b_{1}<\cdots<b_{s}$\, in $\adom(\B)$.
We can translate these answers into moves \,$\bbig_{1}<\cdots<\bbig_{s}$\, in $\Bbig$
via \,$\bbig_{1} \deff \beta(b_{1})$, \ldots, $\bbig_{s} \deff \beta(b_{s})$.
The mapping \,$\abig_1,\twodots,\abig_s\mapsto\bbig_1,\twodots,\bbig_s$\,
obviously is a partial isomorphism between $\Abig$ and $\Bbig$.
\parno
The elements \,$\bbig_1,\twodots,\bbig_s$\, will belong to the duplicator's
answers in the big game. However, the duplicator also has to find elements 
\,{$\bbig_{s+1},\twodots,\bbig_{r} \not\in\beta\big(\adom(\B)\big)$}\, such that, for all
\,$\nu,\nu'\in\set{1,\twodots,r}$,\, we have 
{
\[
(**):\quad
\bbig_{\nu} = p_1 \mbox{ iff } \abig_{\nu} = p_1, \ \ 
\bbig_{\nu} < \bbig_{\nu'} \mbox{ iff } \abig_{\nu} < \abig_{\nu'}, \ \ 
\SuccP\big( \bbig_{\nu},\bbig_{\nu'}\big) \mbox{ iff } 
  \SuccP\big( \abig_{\nu},\abig_{\nu'}\big).
\]
}%
%
For every \,$i<s$,\, $\bbig_i$ is of the form $p_{2rj}$ and 
$\bbig_{i+1}$ is of the form $p_{2rj'}$, for suitable \,$j<j'\in\NNpos$.
In particular, there are at least \,$2r{-}1$\, different elements in $P$ between
$\bbig_i$ and $\bbig_{i+1}$. Therefore, it is straightforward to find elements 
\,$\bbig_{s+1},\twodots,\bbig_r$\, such that the condition $(**)$ is satisfied
by \,$\bbig_1,\twodots,\bbig_s,\bbig_{s+1},\twodots,\bbig_r$.
With these answers, the duplicator wins the big game, and hence the proof of
Lemma~\ref{lemma:nuetzlich_BCEFO} is complete.
\end{proof_mit}%
%

\subsection[How to Win the $\BCeFO(<,\Arb)$-Game]{How to Win the $\bs{\BCeFO(<,\Arb)}$-Game}
\label{section:translation_BCEFO}
%
%
\begin{theorem_mit}[$\bs{\BCeFO(<,\Arb)}$-Game over $\bs{\UU}$]
\label{theorem:BCeFO_Arb-game}
\mbox{ }\\
Let $\struc{\UU,<}$ be an infinite linearly ordered structure, and let
$\Arb$ be the collection of arbitrary, i.e., all, predicates on $\UU$.
\\
The duplicator can translate strategies for the $\BCeFOKleiner$-game into 
strategies for the $\BCeFO(<,\Arb)$-game on $\ClNemb$ over $\UU$.
\end{theorem_mit}%
The overall proof idea is an adaption and extension of a proof developed 
in the context of the \emph{Crane Beach conjecture} \cite{BILST} 
for the specific context of finite strings instead of arbitrary structures. 
\index{Crane Beach Conjecture!FO<Mon@$\FO(<,\Mon)$}%
We make use of the following variant
of \emph{Ramsey's Theorem}: \index{Ramsey Theorem}
\begin{theorem_mit}\label{theorem:Ramsey_extended}
Let $\struc{\UU,<}$ be an infinite linearly ordered structure.
Let $r\in\NNpos$, and let $C_{1},\twodots,C_{r}$ be finite sets. Each set
$C_{h}$ serves as a set of possible colors for $h$-element subsets of $\UU$.
I.e., for every \,$h\in\set{1,\twodots,r}$,\, let every $h$-element subset
\,$Y_h = \set{y_1<\cdots< y_h}\subseteq\UU$\, be colored with an element
\,$\Color_h(Y_h) \in C_h$.\\
If $\struc{\UU,<}$ contains an infinitely increasing sequence, then there exists
an infinitely increasing set \,$P = \set{p_1<p_2<\cdots}\subseteq\UU$\, that satisfies 
the following condition $(*)$:
\begin{quote}
  For every \,$h\in\set{1,\twodots,r}$\, there exists a color $c_h\in C_h$ such that
  \emph{every} $h$-element subset $Y_h\subseteq P$ has the color $\Color_h(Y_h) = c_h$.
\end{quote}
Otherwise, if $\struc{\UU,<}$ does not contain an infinitely increasing sequence, 
then there exists an infinitely decreasing set 
\,$P = \set{p_1>p_2>\cdots}\subseteq \UU$\, that satisfies the condition $(*)$.
\end{theorem_mit}%
\begin{proof_mit}
The idea is to apply the following ``classical'' Ramsey Theorem successively for 
\,$h=1,2,\twodots,r$.
\begin{theorem_mit}[{Ramsey, cf., \cite[Theorem\,9.1.2]{Diestel}}]\label{theorem:Ramsey_h}\index{Ramsey Theorem}
Let $X$ be an infinite set and let $h\in\NNpos$.
Let $C_h$ be a finite set such that every $h$-element set $Y_h\subseteq X$
is colored with an element  
\,$\Color_h(Y_h) \in C_h$.\\
There exists an infinite set $X'\subseteq X$ and a color $c_h\in C_h$ such that
\emph{every} $h$-element subset $Y_h\subseteq X'$ has the color $\Color_h(Y_h) = c_h$.
\end{theorem_mit}%
For the proof of Theorem~\ref{theorem:Ramsey_extended} let us first assume that
$\UU$ contains a countable, infinitely increasing subset $X_0$. 
For \,$X\deff X_0$\, and \,$h\deff 1$,\, the
above Ramsey Theorem~\ref{theorem:Ramsey_h} gives us an infinite set 
\,$X_1\deff X'\subseteq X_0$\, and a 
color \,$c_1\in C_1$\, such that all 1-element subsets of $X_1$ have the color $c_1$.
Another application of the Ramsey Theorem 
for \,$X\deff X_1$\, and \,$h\deff 2$\, yiels an
infinite set \,$X_2\subseteq X_1$\, and a color \,$c_2\in C_2$\, such that all
2-element subsets of $X_2$ have the color $c_2$. Iterating this process for
\,$h= 1,2,\twodots,r$\, leads to sets 
\,$X_1 \supseteq \cdots \supseteq X_{r}$\, and to colors
\,$c_1\in C_1$, \ldots, $c_{r}\in C_{r}$\, such that $X_{r}$ is an 
infinitely increasing set and, for every \,$h\in\set{1,\twodots,r}$,\,
every $h$-element subset of $X_{r}$ has the color $c_h$.
Consequently, the set \,$P\deff X_{r}$\, is the desired set of the form
\,$\set{p_1<p_2<\cdots}$\, that satisfies the condition $(*)$.
\par
It remains to consider the case where $\UU$ does \emph{not} contain a countable, 
infinitely increasing subset $X_0$. In this case, since $\UU$ is infinite and linearly ordered,
there must exist an infinitely \emph{de}creasing subset $X_0$ 
(see Fact~\ref{fact:Analysis}).
Starting with this particular set $X_0$, the same argumentation as above now leads to the
desired set \,$P\deff X_{r}$ of the form \,$\set{p_1>p_2>\cdots}$. 
Altogether, this completes the proof of Theorem~\ref{theorem:Ramsey_extended}.
\end{proof_mit}%
\\
\parno
%
\begin{proofc_mit}{of Theorem~\ref{theorem:BCeFO_Arb-game} ($\bs{\BCeFO(<,\Arb)}$-Game over $\bs{\UU}$)}\mbox{ }\\
We concentrate on the case where $\struc{\UU,<}$ contains an infinitely \emph{in}creasing 
sequence (at the end of the proof we will indicate how the arguments can be modified 
for the case that $\struc{\UU,<}$ contains no such sequence).
\par
Let $\Bip'$ be a \emph{finite} subset of $\Arb$. 
Let $\tau$ be a signature, and let $\kappa\in\NN$ be the number of constant symbols in $\tau$. 
For every number $k\in\NN$ of moves in the $\BCeFO(<,\Bip')$-game we choose 
\,$r \deff r(k) \deff  2k + \kappa$\, 
to be the according number of moves in the $\BCeFOKleiner$-game.
\\
Let \,$\A = \struc{\UU,\tau^{\A}}$\, and 
\,$\B = \struc{\UU,\tau^{\B}}$\, be two $\NN$-embeddable structures
on which the duplicator wins the single-round $r$-move $\BCeFOKleiner$-game, i.e., 
\begin{eqnarray*}
  (*):\qquad\bigstruc{\adom(\A),<,\tau^{\A}}
& \dwinsBCEFO_{r}
& \bigstruc{\adom(\B),<,\tau^{\B}}.
\end{eqnarray*}
We have to find ${<}$-preserving mappings \,$\alpha : \adom(\A)\rightarrow \UU$\, and
\,$\beta : \adom(\B)\rightarrow \UU$\, such that the duplicator wins the
single-round $k$-move $\BCeFO(<,\Bip')$-game on $\alpha(\A)$ and $\beta(\B)$, i.e.,
\,$\struc{\UU,<,\Bip',\alpha\big(\tau^{\A}\big)}
   \dwinsBCeFO_k
   \struc{\UU,<,\Bip',\beta\big(\tau^{\B}\big)}$. 
\\
\parno
\textbf{\underline{Step 1:}} \ 
We first choose a suitable infinite set \,$P = \set{p_1<p_2<\cdots}$\, onto which the 
active domain elements of $\A$ and $\B$ will be moved via ${<}$-preserving mappings 
$\alpha$ and $\beta$.
To find this set $P$ we use the above Ramsey Theorem~\ref{theorem:Ramsey_extended}.
The precise choice of the sets of colors $C_1,\twodots,C_{r}$ is quite elaborate.
For better accessibility of the proof it might be helpful to skip this at first reading,
to continue with Step 2, and to return to the precise choice of the coloring afterwards,
i.e., after having seen the duplicator's strategy for the 
single-round $k$-move $\BCeFO(<,\Bip')$-game on $\alpha(\A)$ and $\beta(\B)$.
\par
Let \,$h\in\set{1,\twodots,r}$\, and let 
\,$Y_h = \set{\abig'_1<\cdots<\abig'_h} \subseteq \UU$\,
be an $h$-element subset of $\UU$.
For every \,$(\abig_1,\twodots,\abig_k)\in\UU^k$\, we define
\,\(
   \type{=,<,\Bip'}\big(\abig_1,\twodots,\abig_k,\abig'_1,\twodots,\abig'_h\big)
\)\,
to be the \emph{complete atomic type} of \index{complete atomic type}
\,$(\abig_1,\twodots,\abig_k,\abig'_1,\twodots,\abig'_h)$\, with respect to the relations
\,$\set{=,<}\cup\Bip'$.
Precisely, this means the following: We use first-order variables \,$x_1,\twodots,x_k$\,
and \,$y_1,\twodots,y_h$,\, and we consider all 
atomic $(\set{=,<}\cup\Bip')$-formulas over these variables.
$\type{=,<,\Bip'}\big(\abig_1,\twodots,\abig_k,\abig'_1,\twodots,\abig'_h\big)$\, is defined to
be the set of exactly those atomic formulas $\varphi$ that
are satisfied when interpreting the variables \,$x_1,\twodots,x_k$\, and \,$y_1,\twodots,y_h$\,
with the elements \,$\abig_1,\twodots,\abig_k$\, and \,$\abig'_1,\twodots,\abig'_h$,\,
respectively.
It should be clear that 
\,$\type{=,<,\Bip'}(\abig_1,\twodots,\abig_k,\abig'_1,\twodots,\abig'_h) =
   \type{=,<,\Bip'}(\bbig_1,\twodots,\bbig_k,\bbig'_1,\twodots,\bbig'_h)$\,
if and only if the mapping 
\,$\big(${$\abig_1,\twodots,\abig_k,\abig'_1,\twodots,\abig'_h$} $\mapsto$
          {$\bbig_1,\twodots,\bbig_k,\bbig'_1,\twodots,\bbig'_h$}$\big)$\, 
is a partial automorphism of the structure \,$\struc{\UU,<,\Bip'}$.
\par
To apply the Ramsey Theorem~\ref{theorem:Ramsey_extended}, we color every
$h$-element set \,$Y_h = \set{\abig'_1<\cdots<\abig'_h}\subseteq \UU$\, with the collection of
all complete atomic types that are realizable with \,$\abig'_1,\twodots,\abig'_h$.
Precisely, this means that
\begin{eqnarray*}
  \Color_h(Y_h) & \deff &
  \bigsetc{
    \,\type{=,<,\Bip'}\big(\abig_1,\twodots,\abig_k,\abig'_1,\twodots,\abig'_h\big)\,}{
    \,(\abig_1,\twodots,\abig_k)\in\UU^k\,}.
\end{eqnarray*}
Since $\Bip'$ is \emph{finite}, the number of complete atomic types over the variables
\,$x_1,\twodots,x_k,\allowbreak y_1,\twodots,y_h$\, is finite.
Consequently, also the set of colors used for $h$-element subsets of $\UU$, i.e., the set
\,$C_h  \deff 
   \bigsetc{\,\Color_h(Y_h)\,}{\,\mbox{$Y_h$ is an $h$-element subset of $\UU$}\,}$,
must be finite.
\\ 
We use these colorings $C_h$, for all \,$h\in\set{1,\twodots,r}$,\, and apply the Ramsey
Theorem~\ref{theorem:Ramsey_extended}. Hence we obtain 
an infinitely increasing set \,$P = \set{p_1<p_2<\cdots}\subseteq\UU$\, that satisfies the
following condition:
  For every \,$h\in\set{1,\twodots,r}$\, there exists a color $c_h\in C_h$ such that
  \emph{every} $h$-element subset $Y_h\subseteq P$ has the color $\Color_h(Y_h) = c_h$.
%
\\
In the following we will use the elements of $P$ as ``special positions'' onto which 
the active domain of the given structures $\A$ and $\B$ will be moved.
\\
\parno
\textbf{\underline{Step 2:}} \ 
From the presumption $(*)$ and from Lemma~\ref{lemma:nuetzlich_BCEFO} we obtain 
${<}$-preserving mappings \,$\alpha:\adom(\A)\rightarrow P$\, and 
\,$\beta:\adom(\B)\rightarrow P$\, such
that a ``virtual duplicator'' has a winning strategy for the single-round $r$-move game on
\,$\struc{\,P,\,<,\,p_1,\,\SuccP,\,\alpha(\tau^{\A})\,}$\, and  
\,$\struc{\,P,\,<,\,p_1,\,\SuccP,\,\beta(\tau^{\B})\,}$, i.e., 
{\begin{eqnarray*}
  (**):\quad\Abig'\ \deff\ \struc{\,P,\,<,\allowbreak \,p_1,\,\SuccP,\,\alpha(\tau^{\A})\,}
& \dwinsBCEFO_{r}
& \struc{\,P,\,<,\,p_1,\,\SuccP,\,\beta(\tau^{\B})\,} \ffed\ \Bbig'\,.
\end{eqnarray*}}%
Obviously, $(**)$ remains valid if $r$ is replaced by a number $h\leq r$.
\\
We now describe a winning strategy for the duplicator, showing that 
$\Abig \deff \struc{\UU,<,\Bip',\alpha\big(\tau^{\A}\big)} \allowbreak
   \dwinsBCEFO_k\allowbreak
   \struc{\UU,<,\Bip',\beta\big(\tau^{\B}\big)} \ffed \Bbig$.
Assume that the spoiler chooses the elements 
\,$\vek{\abig} = \abig_1,\twodots,\abig_k$\, in the universe of $\Abig$
(if he chooses the elements \,$\vek{\bbig} = \bbig_1,\twodots,\bbig_k$\, in the universe of 
$\Bbig$, we can proceed in the according way, interchanging the roles of $\Abig$ and 
$\Bbig$).
To find appropriate answers \,$\vek{\bbig} = \bbig_1,\twodots,\bbig_k$\, for the duplicator,
we proceed as follows: 
We determine, for every $i\in\set{1,\twodots,k}$, the unique elements in
$P$ that are the closest elements to $\abig_i$. 
Precisely, if \,$p_j\leq \abig_i<p_{j+1}$\, then $p_j$ and $p_{j+1}$ are these closest
elements, and we fix the 2-element set \,$I_i\deff \set{p_j,p_{j+1}}$.
Accordingly, if \,$\abig_i<p_1$\, then $p_1$ is the closest element, and we fix the
singleton set \,$I_i\deff \set{p_1}$.
Of course, the set \,$I\deff I_1\cup\cdots\cup I_k$\, has cardinality $\leq 2k$.
Consequently, the union of $I$ with the set of all constants of $\Abig'$ is a set of
the form \,$\set{\abig'_1<\cdots<\abig'_h}\subseteq P$,\, for a suitable $h\leq 2k+\kappa = r$.
The elements \,$\abig'_1,\twodots,\abig'_h$\, are the moves for a 
``virtual spoiler'' in the game on $\Abig'$ and $\Bbig'$. 
A ``virtual duplicator'' who plays according to her winning strategy in the game 
$(**)$ will find answers \,$\bbig'_1<\cdots<\bbig'_h$\, in $\Bbig'$.
\\
Since \,$\set{\abig'_1<\cdots<\abig'_h}$\, and \,$\set{\bbig'_1<\cdots<\bbig'_h}$\, are
$h$-element subsets of $P$, and since $P$ was chosen according to 
Step~1, they must have the same color $c_h\in C_h$.
Due to the particular definition of the colors, as fixed in Step~1,
there hence must be elements 
\,$\vek{\bbig} = \bbig_1,\twodots,\bbig_k$\, in $\UU$ such that
\begin{eqnarray*}
 (*{**}):\qquad \type{=,<,\Bip'}(\vek{\bbig},\bbig'_1,\twodots,\bbig'_h)
& =
& \type{=,<,\Bip'}(\vek{\abig},\abig'_1,\twodots,\abig'_h)\,.
\end{eqnarray*}
We choose exactly these elements \,$\bbig_1,\twodots,\bbig_k$\, to be the duplicator's
answers in $\Bbig$.
\\
\parno
\textbf{\underline{Step 3:}} \ 
It remains to verify that the duplicator has indeed won the game on $\Abig$ and $\Bbig$.
I.e., we have to show that the mapping $\pi$ defined via
\[
  \pi\ :\ \left\{
    \begin{array}{rcll}
       \alpha(c^{\A}) & \mapsto & \beta(c^{\B}) & \mbox{for all constant symbols $c\in\tau$} \\
       \abig_i & \mapsto & \bbig_i & \mbox{for all $i\in\set{1,\twodots,k}$}
    \end{array}
  \right\}
\]
is a partial isomorphism between the structures
\,$\Abig = \struc{\UU,\,<,\,\Bip',\,\alpha(\tau^{\A})}$\, and
\,$\Bbig = \struc{\UU,\,<,\allowbreak\,\Bip',\,\beta(\tau^{\B})}$.
\par
\emph{Claim 1: $\pi$ is a partial automorphism of
$\struc{\UU,<,\Bip'}$.}
\\  
By definition, all the constants of $\Abig$
belong to the sequence \,$\abig'_1,\twodots,\abig'_h$.
Since the ``virtual duplicator'' wins the game $(**)$, all the constants of $\Bbig$ 
must occur in the sequence \,$\bbig'_1,\twodots,\bbig'_h$.
Consequently, the above property $({**}*)$
tells us that $\pi$ is a partial automorphism of
$\struc{\UU,<,\Bip'}$.
\par
\emph{Claim 2: \,$\abig_i\in P$\, iff \,$\bbig_i\in P$\, 
(for all $i\in\set{1,\twodots,k}$).}\\
To show this, we will essentially use that the strategy of the ``virtual duplicator''
in the game $(**)$ preserves the successor relation $\SuccP$ on $P$.
\\
For the ``only if'' direction let $\abig_i\in P$, and show that $\bbig_i\in P$:
Since $\abig_i\in P= \set{p_1<p_2<\cdots}$, there is an index $j$ such that 
$\abig_i=p_j$. By the definition of the set 
\,$\set{\abig'_1<\cdots<\abig'_h}$\, we have \,$\abig_i = p_j = \abig'_{\nu}$\,
for some $\nu\in\set{1,\twodots,h}$.
From $(*{**})$ we obtain that $\bbig_i = \bbig'_{\nu}\in P$.
\\
For the ``if'' direction let $\abig_i\not\in P$, and show that $\bbig_i\not\in P$:
If \,$\abig_i<p_1$\, then, by the definition of the set 
\,$\set{\abig'_1<\cdots<\abig'_h}$,\, we have \,$\abig_i < p_1 = \abig'_{1}$.
Since the ``virtual duplicator'' wins the game $(**)$, we know that 
$\bbig'_1 = p_1$. 
Furthermore, from $(*{**})$ 
we obtain that
\,$\bbig_i <\bbig'_1 = p_1$,\, and consequently, $\bbig_i\not\in P$.
\\
If 
there is a $j$ such that \,$p_j < \abig_i < p_{j+1}$, then,
by the definition of the set 
\,$\set{\abig'_1<\cdots<\abig'_h}$, we know that there is an index
\,$\nu<h$\, such that \,$\abig'_{\nu} = p_j$\, and \,$\abig'_{\nu+1} = p_{j+1}$.\,
In particular, $\abig'_{\nu}$ and $\abig'_{\nu+1}$ are successors in $P$, i.e.,
$\SuccP(\abig'_{\nu},\abig'_{\nu+1})$.
Since the ``virtual duplicator'' wins the game $(**)$, we know that also 
$\SuccP(\bbig'_{\nu},\bbig'_{\nu+1})$.
Furthermore, from $(*{**})$ 
we obtain that \,$\bbig'_{\nu}<\bbig_i<\bbig'_{\nu+1}$.\, In particular, this implies that
$\bbig_i\not\in P$.
Altogether, the proof of Claim~2
is complete. 
\par
All that remains to do is to consider the relations in $\tau$.
Let $R$ be a relation symbol in $\tau$ of arity, say, $m$
and let $\vek{x}\deff
(x_1,\twodots,x_m)$ be in the domain of $\pi$. We need to show that 
\,$\vek{x}\in \alpha(R^{\A})$\, iff \,$\pi(\vek{x})\in \beta(R^{\B})$.
If at least one of the elements in $\vek{x}$, say $x_j$, does not belong to $P$, then we know that 
\,$\vek{x}\not\in \alpha(R^{\A})\subseteq P^m$. 
From Claim 2 we furthermore know that also $\pi(x_j)$ does not belong to $P$. Consequently,
also \,$\pi(\vek{x})\not\in \beta(R^{\B})\subseteq P^m$.
If all the elements in $\vek{x}$ belong to $P$, then the following is true: 
By the definition of the set \,$\set{\abig'_1<\cdots<\abig'_h}$\, of moves for the
``virtual spoiler'' we know that all the elements in $\vek{x}$ belong to 
\,$\set{\abig'_1<\cdots<\abig'_h}$.\, 
I.e., there are indices \,$i_1,\twodots,i_m$\, such that 
\,$(x_1,\twodots,x_m) = (\abig'_{i_1},\twodots,\abig'_{i_m})$.
Since the ``virtual duplicator'' wins the game $(**)$, we know that
\,$(x_1,\twodots,x_m) = (\abig'_{i_1},\twodots,\abig'_{i_m})\in \alpha\big( R^{\A} \big)$\,
iff
\,$(\bbig'_{i_1},\twodots,\bbig'_{i_m}) \in \beta\big( R^{\B} \big)$.
Furthermore, from $(*{**})$ 
we obtain that 
\,$\big(\pi(x_1),\twodots,\pi(x_m)\big) = (\bbig'_{i_1},\twodots,\bbig'_{i_m})$.
Consequently, we have shown that 
\,$\vek{x}\in \alpha(R^{\A})$\, iff \,$\pi(\vek{x})\in \beta(R^{\B})$.
\par
Together with Claim 1 we obtain that $\pi$ is a partial isomorphism between the structures
$\Abig$ and $\Bbig$, and thus the duplicator has won the single-round $k$-move game
on $\Abig$ and $\Bbig$. 
Altogether, this completes the proof of Theorem~\ref{theorem:BCeFO_Arb-game} for the 
case where the structure $\struc{\UU,<}$ contains an infinitely \emph{in}creasing
sequence.
\par
For the remaining case where $\struc{\UU,<}$ does not contain an infinitely \emph{in}creasing
sequence, we know from Fact~\ref{fact:Analysis}
that $\UU$ must contain an infinitely \emph{de}creasing sequence.
With the same coloring as in Step~1 above, the Ramsey Theorem~\ref{theorem:Ramsey_extended}
gives us an infinitely decreasing set \,$P=\set{p_1>p_2>\cdots}$.
Concerning the given $\struc{\UU,\tau}$-structures $\A$ and $\B$, we know that 
$\A$ and $\B$ are $\NN$-embeddable. In particular, $\adom(\A)$ and $\adom(\B)$ must be
\emph{finite}, since otherwise they would constitute an infinitely \emph{in}creasing 
sequence in $\UU$. Consequently, it is possible to embed $\A$ and $\B$ in $P$ in such a
way that Lemma~\ref{lemma:nuetzlich_BCEFO} is valid if replacing $\SuccP$ with the
predecessor relation $\PredP$. \index{predecessor relation}\index{predP@$\PredP$} 
The rest can be taken almost verbatim from Step 2 and Step 3
above.
\\
Altogether, the proof of Theorem~\ref{theorem:BCeFO_Arb-game} is complete.
\end{proofc_mit}%
%



\section[How to Lift Collapse Results]{How to Lift Collapse Results}\label{section:Lift}
\begin{summary}
In this section we develop the notion of \emph{$\NN$-representable} structures,
which is a natural generalization of the notion of finitely
representable (i.e., order constraint) databases.\index{order constraint databases}
Following the spirit of \cite{BST99}'s lifting from finite to finitely representable 
databases, we
show that any collapse result for first-order logic on $\NN$-embeddable structures can be 
lifted to the analogous collapse result on $\NN$-representable structures.
\end{summary}
%
%
\subsection{The Lifting Method}\label{section:Outline}
It is by now quite a common method in database theory to
\emph{lift} results from one class of
databases to another. 
This \emph{lifting method} can be described as follows:\index{lifting method}%
\begin{center}
\small
\fbox{
\parbox[t]{11cm}{\itshape
\begin{description}
 \item[\ \ Known:\ ] 
    A result for a class of ``easy'' databases.
 \item[\ \ Wanted:]
    The analogous result for a class of ``complicated''
    databases.
 \item[\ \  Method:] \mbox{  }
    \begin{enumerate}[(1.)\ ] 
    \item 
      Show that all the relevant information about a
         ``complicated'' database can be represented by an
         ``easy'' database. 
    \item 
      Show that the translation from the
         ``complicated'' to the ``easy'' database (and vice versa) can be
         performed in an appropriate way (e.g., via an efficient
      algorithm or via $\FO$-formulas).
    \item 
      Use this to translate the known result for the
         ``easy'' databases into the desired result for the ``complicated''
         databases.
    \end{enumerate}
\end{description}
}
\quad}
\medskip
\end{center}
In the literature the ``easy'' database which represents a
``complicated'' database is often called
\emph{the invariant} of the ``complicated'' database.
Table~\ref{table:Papers} gives a listing of recent papers in which
the lifting method has been used:\\
%
\begin{table}[h!tbp]
{\footnotesize
\begin{center}
\begin{tabular}{|c||c|c|c|c|}
\hline
& \bf ``compl.'' dbs
& \bf ``easy'' dbs
& \bf \parbox{2.6cm}{\rule{0mm}{3mm}result for \\ ``easy'' dbs \rule[-1.5mm]{0mm}{1.5mm}
}
& \bf \parbox{2.1cm}{result for \\``compl.'' dbs}
\\ \hline\hline
  \parbox[t]{1cm}{\centerline{ }
  \centerline{{\cite{SV98}}}\centerline{ }} 
& \parbox[t]{1.5cm}{\vspace{0.1mm}\centerline{planar spatial}\centerline{dbs}}
& \parbox[t]{1cm}{\centerline{ }
  \centerline{finite dbs}\centerline{ }} 
& \parbox[t]{2.6cm}{\rule{0mm}{3mm}evaluation of \\ fixpoint+counting \\ queries\rule[-1.5mm]{0mm}{1.5mm}}
& \parbox[t]{2.1cm}{evaluation of \\ top.\;$\FO(<)$-\\ queries over $\RR$}
\\ \hline 
  \parbox[t]{1cm}{\centerline{ }
  \centerline{{\cite{KV99}}}\centerline{ }} 
& \parbox[t]{1.5cm}{\centerline{ }
  \centerline{{region dbs}}\centerline{ }} 
& \parbox[t]{1cm}{\centerline{ }
  \centerline{finite dbs}\centerline{ }} 
& \parbox[t]{2.6cm}{\rule{0mm}{3mm}collapse from \\{${<}$-gen.$\FO(\mbox{\small $<,+,\times$})$}\\ to $\FOadom(<)$ \\ over $\RR$\rule[-1.5mm]{0mm}{1.5mm}} 
& \parbox[t]{2.1cm}{collapse from \\ top.\;$\FO(<,+,\times)$ \\ to  
  top.\;$\FO(<)$ \\ over $\RR$}  
\\ \hline 
  \parbox[t]{1cm}{\centerline{ }
  \centerline{\cite{GraedelKreutzer99}}\centerline{ }} 
& \parbox[t]{1.5cm}{\centerline{finitely}
  \centerline{representable}\centerline{dbs}} 
& \parbox[t]{1cm}{\centerline{ }
  \centerline{finite dbs}\centerline{ }} 
& \parbox[t]{2.6cm}{\rule{0mm}{3mm}logical character-\\ ization of complexity classes\rule[-1.5mm]{0mm}{1.5mm}}
& \parbox[t]{2.1cm}{complexity of \\ query evaluation}
\\ \hline
  \parbox[t]{1cm}{\centerline{ }
  \centerline{\cite{BST99}}\centerline{ }}  
& \parbox[t]{1.5cm}{\centerline{finitely}
  \centerline{representable}\centerline{dbs}}
& \parbox[t]{1cm}{\centerline{ }
  \centerline{finite dbs}\centerline{ }} 
& \parbox[t]{2.6cm}{\rule{0mm}{3mm}natural generic\\ collapse
    over\\ $\struc{\UU,<,\Rel}$\rule[-1.5mm]{0mm}{1.5mm}}
& \parbox[t]{2.1cm}{natural generic\\ collapse
    over\\ $\struc{\UU,<,\Rel}$}
\\ \hline
\end{tabular}
\caption{\small Some papers that use the lifting method.}\label{table:Papers}
\end{center}
}
\end{table}
\\
Segoufin \index{Segoufin, Luc} and Vianu \index{Vianu, Victor} \cite{SV98} represent a spatial
database (of a certain kind) by a finite database called \emph{the topological
  invariant} of the spatial database. They concentrate on the
evaluation of topological $\FO(<)$-queries against spatial \index{topological query}
databases over $\RR$. One of their results is that a topological query against
the spatial database can be efficiently translated into a 
fixpoint+counting query against the
topological invariant. This shows that efficient query evaluation for
the topological invariants leads to efficient query evaluation for
spatial databases.
\par
Kuijpers \index{Kuijpers, Bart} and Van den Bussche \index{Van den Bussche, Jan} \cite{KV99} 
show that all topological 
$\FO(<,\allowbreak +,\times)$-queries over so-called \emph{(fully 2D) region databases} over $\RR$ 
can already be expressed in $\FO(<)$. A crucial step in their proof is to represent
region databases by \emph{finite} databases, to which the natural generic collapse of
\cite{BDLW} applies, i.e., the collapse from ${<}$-generic $\FO(<,+,\times)$ to
$\FOadom(<)$ on finite databases over $\RR$.
\par
Belegradek, \index{Belegradek, Oleg V.}
Stolboushkin, \index{Stolboushkin, Alexei P.}
and Taitslin \index{Taitslin, Michael A.}
\cite{BST99} and
Gr\"adel \index{Gr\"{a}del, Erich}
and Kreutzer \index{Kreutzer, Stephan} 
\cite{GraedelKreutzer99} consider \emph{finitely representable} databases
(also known as \emph{order constraint} databases), which are defined as follows:
\begin{definition_mit}[finitely representable]\label{definition:finrep}\index{finitely representable}\mbox{ }\\
Let $\struc{\UU,<}$ be a dense linear ordering without endpoints.\footnote{I.e., $<$ is a
linear ordering that has no maximal and no minimal element in $\UU$, and for any two elements
$u,v\in\UU$ with $u<v$ there is an element $w\in\UU$ with $u<w<v$. }\index{dense linear ordering without endpoints}
\begin{enumerate}[(a)\ ]
\item
A relation \,$R\subseteq\UU^m$\, is called \emph{finitely representable} iff it can be
explicitly defined by a $\FO$-formula that makes use of the linear ordering and of 
finitely many constants in $\UU$. Precisely this means that there are a 
number $k\in\NN$, elements
\,$s_1,\twodots,s_k\in\UU$, and a \,$\FO(<,s_1,\twodots,s_k)$-formula $\varphi(x_1,\twodots,x_m)$
such that \,$R=\setc{\vek{a}\in\UU^m}{\struc{\UU,<,s_1,\twodots,s_k}\models\varphi(\vek{a})}$.
Due to quantifier elimination $\varphi$ can, without loss of generality, be chosen 
{quantifier free}.
\item
For a signature $\tau$, a $\struc{\UU,\tau}$-structure $\A$ is called 
\emph{finitely representable} iff each of $\A$'s relations is.
\end{enumerate}
We use $\Clfinrep$ \index{C1finrep@$\Clfinrep$}
to denote the class of all finitely representable structures.
\end{definition_mit}%
In \cite{BST99} and \cite{GraedelKreutzer99} it was shown 
that all the relevant information about a {finitely representable} database
can be represented by a \emph{finite} database, and that
the translation from finitely representable to finite (and vice versa, in \cite{BST99})
can be done by a first-order interpretation.
Gr\"adel and Kreutzer use this translation to carry over logical
characterizations of complexity classes to results on the data
complexity of query evaluation. They lift, e.g., the
well-known logical characterization 
``PTIME = FO+LFP on ordered finite structures'' to the result stating
that the polynomial time computable queries 
against finitely representable databases are exactly the
FO+LFP-definable queries. 
Belegradek, Stolboushkin, and Taitslin use their $\FO$-translations from 
finitely representable databases to finite databases, and vice versa,
to lift collapse results for finite databases to collapse
results for finitely representable databases.
Precisely, they obtain the following \emph{lifting theorem} \cite[Theorem\;4.10]{BST99}:
\begin{theorem_mit}[BST's lifting from finite to finitely representable]\label{theorem:BSTLift}\index{lifting theorem}\mbox{ }\\
Let $\struc{\UU,<,\Rel}$ be a context structure.
If
$\OrderGen\FO(<,\Rel) = \!\OrderGen\FO(<)$\allowbreak on $\Clfinite$ over $\UU$,
then
$\OrderGen\FO(<,\Rel) = \OrderGen\FO(<)\mbox{ on }\Clfinrep \mbox{ over }\UU$.
\end{theorem_mit}%
Note that the collapse to $\FOadom(<)$ is not possible over $\Clfinrep$, since the  
${<}$-generic query \ 
{\em ``Does the active domain have an upper bound in $\UU$?''} \ 
is definable in $\FO(<)$, but not in $\FOadom(<)$.
\par
In the previous sections of this paper we obtained collapse results not only for the 
class $\Clfinite$, but even for the
larger class $\ClNemb$ of structures whose active domain is \emph{$\NN$-embeddable}.
In the present section we will lift these collapse results to a larger class of structures
that we call \emph{$\NN$-representable}. 
\index{w-representable$\omega$-representable|see{N-representable@$\NN$-representable}}
\index{omega-representable$\omega$-representable|see{N-representable@$\NN$-representable}}
The resulting lifting theorem was presented in the conference contribution 
\cite{Schweikardt_CSL01}. There, the according structures were called 
\emph{$\omega$-representable}. The author now thinks that the name \emph{$\NN$-representable}
is more appropriate.
%
%
%
\subsection[A Generalization of Finitely Representable: ${\NN}$-Representable]{A Generalization
   of Finitely Representable: $\bs{\NN}$-Representable}\label{section:NN-representable}
%
\index{N-representable@$\NN$-representable}
%
%
%
%
\subsubsection{An Informal Approach}\label{subsection:InformalApproach}
\begin{figure}[!htbp]
\bigskip
\begin{center}
\fbox{
\scalebox{0.35}{
\begin{pspicture}(-2,-2)(14.8,9.8)
%
%
%
\pspolygon*[fillcolor=gray,linecolor=gray,fillstyle=solid](-1,4)(1,4)(1,1)(0,1)(0,3)(-1,3)%
\pspolygon*[fillcolor=gray,linecolor=gray,fillstyle=solid](1,0)(3,0)(3,3)(1,1)%
\pspolygon*[fillcolor=gray,linecolor=gray,fillstyle=solid](3,1)(4,1)(4,3)(3,3)%
\pspolygon*[fillcolor=gray,linecolor=gray,fillstyle=solid](4,3)(4,1)(14,1)(14,3)%
\pspolygon*[fillcolor=gray,linecolor=gray,fillstyle=solid](4,0)(4,-1)(14,-1)(14,0)%
\psline[linecolor=gray,linewidth=3pt](1,4)(1,9)%
\psline[linecolor=gray,linewidth=3pt](4,4)(9,9)%
\psdot[fillcolor=gray,linecolor=gray,dotsize=6pt](3,4)%
\pspolygon*[fillcolor=gray,linecolor=gray,fillstyle=solid](4,4)(9,9)(14,9)(14,4)%
\psset{linestyle=dotted}%
%
\pnode(-1,-1){DiagUnten}%
\pnode(9,9){DiagOben}%
\ncline{DiagUnten}{DiagOben}%
%
\rput(0,-1.3){\rnode[t]{s1_unten}{\LARGE \rule{0cm}{2ex}$s_1$}}%
\pnode(0,9){s1_oben}%
\ncline[arrowsize=8pt,linestyle=solid]{->}{s1_unten}{s1_oben}%
\rput(1,-1.3){\rnode[t]{s2_unten}{\LARGE \rule{0cm}{2ex}$s_2$}}%
\pnode(1,9){s2_oben}%
\ncline{s2_unten}{s2_oben}%
\rput(3,-1.3){\rnode[t]{s3_unten}{\LARGE \rule{0cm}{2ex}$s_3$}}%
\pnode(3,9){s3_oben}%
\ncline{s3_unten}{s3_oben}%
\rput(4,-1.3){\rnode[t]{s4_unten}{\LARGE \rule{0cm}{2ex}$s_4$}}%
\pnode(4,9){s4_oben}%
\ncline{s4_unten}{s4_oben}%
%
%
\rput(-1.3,0){\rnode[r]{s1_links}{\LARGE $s_1$\rule{1ex}{0cm}}}%
\pnode(14,0){s1_rechts}%
\ncline[arrowsize=8pt,linestyle=solid]{->}{s1_links}{s1_rechts}%
\rput(-1.3,1){\rnode[r]{s2_links}{\LARGE $s_2$\rule{1ex}{0cm}}}%
\pnode(14,1){s2_rechts}%
\ncline{s2_links}{s2_rechts}%
\rput(-1.3,3){\rnode[r]{s3_links}{\LARGE $s_3$\rule{1ex}{0cm}}}%
\pnode(14,3){s3_rechts}%
\ncline{s3_links}{s3_rechts}%
\rput(-1.3,4){\rnode[r]{s4_links}{\LARGE $s_4$\rule{1ex}{0cm}}}%
\pnode(14,4){s4_rechts}%
\ncline{s4_links}{s4_rechts}%
\end{pspicture}%
}
}
\caption{\small A finitely representable binary relation ${R}$. The grey regions
  are those that belong to ${R}$. Essentially, $R$ consists of a finite number of  
  ``rectangular'' regions.}\label{figure:FinRepRelation}%
\end{center}%
\end{figure}
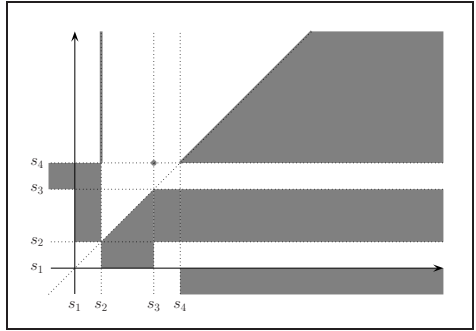

To find an adequate generalization, let us first point out what \emph{finitely representable}
structures look like.
Let $\tau$ consist, for the moment, of a single {binary} relation symbol, and
let \,$\A = \struc{\UU,{R}}$\, be a {finitely representable} 
$\struc{\UU,\tau}$-structure.
This means that the relation \,${R}\subseteq \UU^2$\, is definable by a
$\FO(<,s_1,\twodots,s_k)$-formula $\varphi_{{R}}(x_1,x_2)$.
Due to quantifier elimination $\varphi_{{R}}$ is, without loss of generality, a
Boolean combination of atomic formulas over the relations ${<},{=}$, the variables
$x_1,x_2$, and the constants $s_1,\twodots,s_k$.
In other words: The constants \,$s_1,\twodots,s_k$, 
together with the diagonal ``$x_1{=}x_2$'', impose a finite grid on the
plane $\UU^2$; and the formula $\varphi_{{R}}$ expresses, for each region $M$ in the
grid, whether \,$M\subseteq {R}$\, or \,$M\cap {R} = \emptyset$.
Such a relation ${R}$ is illustrated in Figure~\ref{figure:FinRepRelation}.
\\
In general, a binary relation ${R}$ is definable in $\FO(<,s_1,\twodots,s_k)$ if and only
if ${R}$ is \emph{constant}, in the sense of the following 
Definition~\ref{definition:R_constant_on_M}, on all the regions of the grid that is
defined by \,$s_1,\twodots,s_k$\, and the diagonal ``$x_1{=}x_2$''. 
%
%
%

%
\begin{definition_mit}[$\bs{R}$ constant on $\bs{M}$]\label{definition:R_constant_on_M}
Let $m\in\NNpos$.\\
We say that a relation $R\subseteq \UU^m$ is \emph{constant} on a set
$M\subseteq \UU^m$ if either all elements of $M$ belong to $R$ or no
element of $M$ belongs to $R$.
\end{definition_mit}%
In the proof of their Lifting Theorem~\ref{theorem:BSTLift}, Belegradek et al.\ represent
a $\FO(<,\allowbreak s_1,\twodots,s_k)$-definable $\struc{\UU,\tau}$-structure $\A$ by a
structure $\rep(\A)$ with {active domain $\set{s_1,\twodots,s_k}$}, and
they show that the translations from $\A$ to $\rep(\A)$, and vice versa, can be done
via first-order interpretations. 
%
%
In their lifting theorem they have available the collapse over $\Clfinite$, i.e., the
collapse over the representations $\rep(\A)$, for $\A\in\Clfinrep$.
\par
In the \emph{present} situation we have available the collapse over $\ClNemb$.
Thus, as representatives $\rep(\A)$, we may use structures whose active domain is
$\NN$-embeddable, i.e., of the form \,$\set{s_1<s_2<s_3<\cdots}$\, and unbounded in $\UU$.
Of course, the constants \,$s_1,s_2,s_3,\ldots$\, and the diagonal ``$x_1{=}x_2$'' impose an
\emph{infinite} grid on the plane $\UU^2$.
Consequently,
it seems reasonable to say that a relation \,$R\subseteq \UU^2$\, is 
\emph{$\NN$-representable} via \,$\set{s_1<s_2<\cdots}$\, if and only if
$R$ is constant on all the regions of the infinite grid that is defined by
$s_1,s_2,\ldots$ and the diagonal ``$x_1{=}x_2$''.
%
These relations are exactly the relations definable by
\emph{infinitary} Boolean combinations of atomic formulas over the 
relations ${<},{=}$, the variables $x_1,x_2$, and the constants 
$s_1,s_2,\twodots$. 
We will see that we can even allow infinitary formulas with quantifiers, i.e.,
$\Linf(<,\allowbreak s_1,s_2,\ldots)$-formulas to define such relations.
%
%
\subsubsection[Formalization: $\Linf$ and ${\NN}$-Representable Structures]{Formalization: $\bs{\Linf}$ and $\bs{\NN}$-Representable Structures}\label{subsection:PreciseNrep}
\index{infinitary logic $\Linf$}\index{Linf@$\Linf$} 
%
%
\emph{Infinitary logic} $\Linf$ is defined in the same way as first-order
logic, except that \emph{arbitrary} (i.e., possibly infinite) disjunctions
and conjunctions are allowed. \\
What we need in the present section is the following: 
Let $S$ be a possibly infinite set of constant symbols.
The logic $\Linf(<,S)$ is given by the following
clauses:
   It contains all atomic formulas $x{=}y$ and $x{<}y$, where $x$ and
   $y$ are variable symbols or elements in $S$.
   If it contains $\varphi$, then it contains also $\nicht\varphi$.
   If it contains $\varphi$ and if $x$ is a variable symbol,
   then it contains also $\exists x \varphi$ and $\forall x \varphi$.
   If $\Phi$ is a (possibly infinite) set of $\Linf(<,S)$-formulas,
   then $\Oder\Phi$ and $\Und \Phi$ are formulas in
   $\Linf(<,S)$. 
The semantics is a direct extension of the semantics of first-order
logic, where $\Oder\Phi$ is true if there is some $\varphi\in\Phi$
which is true; and $\Und \Phi$ is true if every $\varphi\in\Phi$ is true. 
We will always identify the set $S$ of constant symbols with a set $S\subseteq \UU$,
where $\UU$ is the universe of the underlying context structure $\struc{\UU,<,\Rel}$.
\begin{definition_mit}[$\bs{\NN}$-representable]\label{definition:Nrep} 
Let $\struc{\UU,<}$ be a dense linear ordering without endpoints.
\begin{enumerate}[(a)\ ]
\item
A relation \,$R\subseteq\UU^m$\, is called \emph{$\NN$-representable} iff it can be
explicitly defined by a $\Linf$-formula that makes use of the linear ordering and of an
$\NN$-embeddable set of constants in $\UU$. 
Precisely this means that there are a $\NN$-embeddable set 
\,$S = \set{s_1<s_2<\cdots} \subseteq\UU$\, 
and a \,$\Linf(<,S)$-formula \,$\varphi(x_1,\twodots,x_m)$\,
such that \,$R=\setc{\vek{a}\in\UU^m}{\struc{\UU,<,s_1,s_2,\ldots}\,\models\,\varphi(\vek{a})}$.
\\
Below we will see that $\varphi$ can, without loss of generality, be chosen 
\emph{quantifier free} and in the normal form obtained in Proposition~\ref{prop:Qeli}.
\item
For a signature $\tau$, a $\struc{\UU,\tau}$-structure $\A$ is called 
\emph{$\NN$-representable} iff each of $\A$'s relations is.
\end{enumerate}
We use $\ClNrep$ \index{C1Nrep@$\ClNrep$} 
to denote the class of all $\NN$-representable structures.
\end{definition_mit}%
%
%
%
\subsubsection[A Normal Form for ${\Linf(<,S)}$-Formulas]{A Normal Form for $\bs{\Linf(<,S)}$-Formulas}\label{section:InfinitaryLogic} 
From now on let $\struc{\UU,<}$ always be a dense linear ordering without endpoints.
\\
%
%
It is well-known that $\FO(<,S)$ allows
quantifier elimination over $\UU$, for every set of constants $S\subseteq \UU$. 
In this section we show that also $\Linf(<,S)$
allows quantifier elimination over $\UU$, provided that $S$ is \emph{$\NN$-embeddable}.
However, our aim is not only to show that $\Linf(<,S)$ allows
quantifier elimination, but to give an {explicit characterization} of
the quantifier free formulas.
\par
Before giving the formalization of the quantifier elimination let us
fix some notations:
For the rest of this section let 
$S\subseteq \UU$ always be $\NN$-embeddable.
We write $S(i)$ to denote the $i$-th smallest element in $S$. 
For infinite $S$ we define \,$S(0)\deff -\infty$\, and \,$N(S)\deff
\NN$.
For finite $S$ we define \,$S(0)\deff -\infty$, \,$N(S)\deff \set{0,\twodots,|S|}$, \,and
\,$S(|S|{+}1)\deff +\infty$.
For \,$m\in\NNpos$\, and \,$\vek{\imath}= (i_1,\twodots,i_m) \in N(S)^m$\, we define
\,$S(\vek{\imath}) \deff (S(i_1),\twodots,S(i_m))$, and
\index{CubeSi@$\Cube{S;\vek{\imath}}$}%
\,$\Cube{S;\vek{\imath}}
   \deff
   \intro{S(i_1),S(i_1{+}1)} \,\times\, \cdots \,\times\, \intro{S(i_m),S(i_m{+}1)}$\,
(where $\intro{{-}\infty,r}\deff\setc{r'\in\UU}{r'<r}$). 
We say that $S(\vek{\imath})$ are the \emph{coordinates} of the cube 
$\Cube{S;\vek{\imath}}$.
Obviously, $\UU^m$ is the disjoint union of the sets $\Cube{S;\vek{\imath}}$ for all
$\vek{\imath}\in N(S)^m$.
\\
Let \,$\vek{a} = (a_1,\twodots,a_m) \in\UU^m$.
The \,$\type{\vek{a};S;\vek{\imath}}$\, \index{typeaSi@$\type{\vek{a};S;\vek{\imath}}$}
of $\vek{a}$ with respect to \,$\Cube{S;\vek{\imath}}$\, is the conjunction
of all atoms in
\,\(
  \bigsetc{y_i{=}x_i,\ y_i{<}x_i,\ x_i{=}x_j,\
  x_i{<}x_j }{ i,j\in{\set{1,\twodots,m},\; i{\neq} j}}
\)\,
which are satisfied if one interprets the variables \,$x_1,\twodots,x_m,y_1,\twodots,y_m$\, by
the elements \,$a_1,\twodots,a_m,S(i_1),\twodots,S(i_m)$. 
I.e., $\type{\vek{a};S;\vek{\imath}}$\, describes the relative position of $\vek{a}$ with
respect to $\Cube{S;\vek{\imath}}$.
We define \,$\types_m$\, \index{typeaSim@$\types_m$}
to be the set of all \emph{complete conjunctions}
of atoms in
\,\(
   \setc{y_i{=}x_i,\ y_i{<}x_i,\ \allowbreak x_i{=}x_j,\
   x_i{<}x_j }{ i,j\in\set{1,\twodots,m},\; i\neq j} 
\),
i.e., the set of all
conjuctions $t$ where, for all \,$i,j\in\set{1,\twodots,m}$\, with
\,$i{\neq}j$, either \,$y_i{=}x_i$\, or \,$y_i{<}x_i$\, occurs in $t$, and
either \,$x_i{=}x_j$\, or \,$x_i{<}x_j$\, or \,$x_j{<}x_i$\, occurs in $t$.
Of course, \,$\types_m$\, is finite, and \,$\type{\vek{a};S;\vek{\imath}}\in\types_m$.
Analogously, we define \,$\Types_m$\, \index{TypeRSim@$\Types_m$}
to be the set of all subsets of 
\,$\types_m$, i.e., \,$\Types_m = \setc{T}{T\subseteq \types_m}$.
For a relation \,$R\subseteq \UU^m$\, we define \index{TypeRSi@$\Type{R;S;\vek{\imath}}$} 
\,$\Type{R;S;\vek{\imath}} 
   \deff 
   \setc{\type{\vek{a};S;\vek{\imath}} }{\vek{a}\in R\cap \Cube{S;\vek{\imath}}}$\,
to be the set of all types occurring in the restriction of $R$
to $\Cube{S;\vek{\imath}}$. 
We say that \,$\Type{R;S;\vek{\imath}}$\, is the \emph{type of
  $\Cube{S;\vek{\imath}}$ in $R$}. 
Of course, $\Type{R;S;\vek{\imath}} \in\Types_m$.
%
\par
In the formalization of the quantifier elimination we further use the following notation:
If $\varphi$ is a $\Linf(<,S)$-formula with free variables \,$\vek{x}\deff
x_1,\twodots,x_k$\, and \,$\vek{y} \deff y_1,\twodots,y_m$, then
we write \,$\varphi(\vek{y}/ S(\vek{\imath}))$\, to denote the
formula one obtains by replacing the variables \,$y_1,\twodots,y_m$\, by
the elements \,$S(i_1),\twodots,S(i_m)$. 
%
\begin{proposition_mit}[Quantifier Elimination for $\bs{\Linf(<,S)}$]\label{prop:Qeli}
\index{quantifier elimination}\index{Linf<S@$\Linf(<,S)$} 
Let $\struc{\UU,<}$ be a dense linear ordering without endpoints, 
let $S\subseteq \UU$ be $\NN$-embeddable, and let $m\in\NNpos$.
Every formula \,$\varphi(x_1,\twodots,x_m)$\, in $\Linf(<,S)$
is equivalent over $\UU$ to the formula 
\[ 
   \tilde{\varphi}(\vek{x}) \quad \deff\quad 
   \Oder_{\vek{\imath}\,\in\, N(S)^m}\ 
   \Oder_{t\in\Type{R;S;\vek{\imath}}}
   \Big( 
      t(\vek{y}/ S(\vek{\imath}))\ \und\ 
      \Und_{j=1}^{m} S(i_j)\leq x_j < S(i_j{+}1) 
   \Big) 
\]
where $R\subseteq \UU^m$ is the relation defined by
$\varphi(\vek{x})$, i.e., 
{
\begin{eqnarray*}
   R 
 &  \ =\ 
 & \setc{\,\vek{a}\in\UU^m}{\,\big\langle{\,\UU,\,<,\,S(1),\,S(2),\ldots}\big\rangle\;\models\;
   \varphi(\vek{a})\,}
\\
 & \ =\  
 & \setc{\,\vek{a}\in\UU^m}{\,\big\langle{\,\UU,\,<,\,S(1),\,S(2),\ldots}\big\rangle\;\models\;
   \tilde{\varphi}(\vek{a})\,}.
\end{eqnarray*}
}\vspace{-5ex}\\ \mbox{ }
\end{proposition_mit}%
\begin{proof_mit}
The proof is similar to the quantifier elimination
for $\FO(<,S)$ over $\UU$.
For simplicity, we write $N$ instead of $N(S)$.
\medskip\\ \indent
(1): We first show that the proposition is valid in the special case where
$\varphi$ is quantifier free.
Let $\tilde{R}$ be the relation defined by
$\tilde{\varphi}$. We need to show that $R=\tilde{R}$.
Let $\vek{a}\in \UU^m$, let $\vek{\imath}\in N^m$ such that
$\vek{a}\in\Cube{S;\vek{\imath}}$, and let $t\deff 
\type{\vek{a};S;\vek{\imath}}$. By definition we know that 
$t(\vek{y}/S(\vek{\imath}))$ is satisfied if one interprets $\vek{x}$
by $\vek{a}$.
\par
For showing that $R\subseteq \tilde{R}$, assume that $\vek{a}\in
R$. From the definition of $\Type{R;S;\vek{\imath}}$
we know that $t \in \Type{R;S;\vek{\imath}}$. Hence, $\tilde{\varphi}$
is satisfied if one interprets $\vek{x}$ by $\vek{a}$, i.e.,
$\vek{a}\in\tilde{R}$.
\par
For showing that $R\supseteq \tilde{R}$, assume that $\vek{a}\in
\tilde{R}$, i.e., $\tilde{\varphi}$ is satisfied when interpreting
$\vek{x}$ by $\vek{a}$. Of course, $\vek{\imath}$ is the only element in $N^m$
with $\vek{a}\in \Cube{S;\vek{\imath}}$, and $t$ is the only
element in $\types_m$ that is satisfied when interpreting $\vek{x}$ by
$\vek{a}$ and $\vek{y}$ by $S(\vek{\imath})$. We conclude that
$t$ must be an element of $\Type{R;S;\vek{\imath}}$. 
Thus there must be some
$\vek{b}\in R\cap \Cube{S;\vek{\imath}}$ such that 
$\type{\vek{b};S;\vek{\imath}} = t$.
One can easily see that every atomic formula in
\[
\setc{s{=}x_i\,,\ s{<}x_i\,,\ x_i{=}x_j\,,\ x_i{<}x_j\ }{\ s\in S,\
  i,j\in\set{1,\twodots,m},\ i\neq j}
\]
is satisfied if one 
interprets $\vek{x}$ by $\vek{a}$ if and only if it is satisfied if
one interprets $\vek{x}$ by $\vek{b}$. Since $\varphi$ is a (possibly
infinitary) Boolean combination of such atomic formulas, we conclude
that $\varphi$ is satisfied if one interprets $\vek{x}$ by $\vek{a}$
if and only if it is satisfied if one interprets $\vek{x}$ by $\vek{b}$. 
Since $\vek{b}\in R$ we hence obtain that also $\vek{a} \in R$. 
\\
Altogether, we have shown that $R=\tilde{R}$, which completes our
proof of (1).
\medskip\\ \indent
(2): We now show that the proposition is valid in the special case
where $\varphi$ is of the form 
\[
 (*):\quad\ \ 
 \displaystyle
 \exists x_{m+1}\  \big( \Und_{i=1}^{p} x_{m+1} = u_i     \big)
                   \ \und\ 
                   \big( \Und_{j=1}^{q}     v_j < x_{m+1} \big)
                   \ \und\ 
                   \big( \Und_{k=1}^{r} x_{m+1} < w_k     \big)\,,
\]
where $p,q,r\in\NN$ and $\set{u_1,\twodots,u_q,v_1,\twodots,v_q,w_1,\twodots,w_r}
\subseteq \set{x_1,\twodots,x_m}\cup S$.
\par
In case that $p\neq 0$, we can replace $x_{m+1}$ by $u_1$ and obtain that
$\varphi$ is equivalent (over $\UU$) to 
$\big( \Und_{i=1}^{p} u_1 = u_i     \big)
 \ \und\  
 \big( \Und_{j=1}^{q} v_j < u_1 \big)
 \ \und\ 
 \big( \Und_{k=1}^{r} u_1 < w_k     \big)$. 
\par
In case that $p=0$ and $q$ and $r$ are both different from $0$,
$\varphi$ says that there exists an element which is larger than each
$v_j$ and smaller that each $w_k$. Since ${<}$ is \emph{dense}, $\varphi$ is equivalent
(over $\UU$) to $\Und_{j=1}^{q}\Und_{k=1}^{r} v_j < w_k$.
\par
In case that $p=0$ and $r=0$, $\varphi$ says that there
exists an element which is larger than each $v_j$ --- which is
true since ${<}$ has no endpoints.  
Analogously, in case that $p=0$ and $q=0$,  $\varphi$ says that there
exists an element which is smaller than each $w_k$ --- which, again, is
true in since ${<}$ has no endpoints. Hence, in both cases 
$\varphi$ is equivalent to a formula which is
always true (e.g., the formula $x_1=x_1$).
\\
Altogether, we have seen that a formula $\varphi$ of the form $(*)$ is
equivalent to a quantifier free formula.
Thus we can use (1) to conclude that $\varphi$ is
equivalent to $\tilde{\varphi}$.
\medskip\\ \indent
(3): We are now ready to show, by induction on the construction of
$\varphi$, that the proposition is valid for all $\varphi$ in
$\Linf(<,S)$.
\par
If $\varphi$ is quantifier free, the claim follows from (1). 
If $\varphi$ is of the form $\nicht \psi$ or $\Oder \Phi$, the
induction step is obvious.
If $\varphi$ is of the form $\exists x_{m+1}\,
\psi(x_1,\twodots,x_{m+1})$ then we show 
\begin{quote}
 $(**)$:\quad
 \parbox[t]{65ex}{$\varphi$ is equivalent to a formula
 $\Oder \Phi$, where $\Phi$ is a set of formulas of the form $\xi\und
 \eta$, such that $\xi(x_1,\twodots,x_m)$ is of the form
 $(*)$ and $\eta(x_1,\twodots,x_m)$ is quantifier free.}
\end{quote}
Making use of $(**)$ and (2), we obtain that $\varphi$ is equivalent
to the quantifier free formula $\Oder\setc{\tilde{\xi}\und\eta}{\xi\und\eta \in\Phi}$.
According to (1) we thus conclude that $\varphi$ is equivalent to
$\tilde{\varphi}$.
\par
It remains to show $(**)$.
By the induction hypothesis,
$\psi$ is equivalent to $\tilde{\psi}$, which is, by definition, 
the disjunction of the formulas 
\[
 \chi_{\vek{\imath};\vek{a}}\ \ \deff\ \  
 \type{\vek{a};S;\vek{\imath}}(\vek{y}/ S(\vek{\imath}))\ \und\ 
 \big( \Und_{j=1}^{m+1} S(i_j)\leq x_j < S(i_j{+}1) \big)\,,
\] 
for all
$\vek{\imath} \in N^{m+1}$ and all
$\vek{a} \in R\cap \Cube{S;\vek{\imath}}$.
Since $\varphi$ is equivalent to $\exists x_{m+1}\, \tilde{\psi}$, it 
also is equivalent to
the disjunction of the formulas
\,$\exists x_{m+1}\,\chi_{\vek{\imath};\vek{a}}$.
\par
We transform each $\chi_{\vek{\imath};\vek{a}}$ into a finite disjunction
of finite conjunctions $\lambda_{\vek{\imath};\vek{a};j}$ of unnegated
atoms of the form $u{=}v$ and
$u{<}v$, where $u$ and $v$ are distinct elements in
$\set{x_1,\twodots,x_{m+1}}\cup S$, as follows:
For not necessarily distinct $u$ and $v$, 
we replace each negated atom of the form $(\nicht u{=}v)$ by
$(u{<}v\,\oder\, v{<}u)$, we replace each negated atom of the form $(\nicht u{<}v)$ by
$(v{<}u\,\oder\, v{=}u)$, and we replace each atom of the form
$(u{\leq}v)$ by $(u{<}v\,\oder\, u{=}v)$. Afterwards we repeatedly use
the distributive law ``$\alpha \und (\beta\oder\gamma)$ is equivalent
to $(\alpha\und\beta) \oder (\alpha\und\gamma)$'', to transform
$\chi_{\vek{\imath};\vek{a}}$ into a disjunction of conjunctions of
unnegated atoms of the form $u{=}v$ and $u{<}v$. Finally, we remove
each conjunction where there occurs an atom of the form $u{<}u$; and
in the remaining conjunctions we remove each atom of the form $u{=}u$.
This gives us that each $\chi_{\vek{\imath};\vek{a}}$ is equivalent to
a finite disjunction of finite conjunctions
$\lambda_{\vek{\imath};\vek{a};j}$ of unnegated atoms of the form $u{=}v$ and
$u{<}v$, where $u$ and $v$ are distinct elements in
$\set{x_1,\twodots,x_{m+1}}\cup S$.
\par
Since $\varphi$ is equivalent to the disjunction of the formulas 
$\exists x_{m+1}\,\chi_{\vek{\imath};\vek{a}}$, it is also equivalent to the
disjunction of the formulas $\exists x_{m+1}\,\lambda_{\vek{\imath};\vek{a};j}$.
Let $\zeta_{\vek{\imath};\vek{a};j}$ be the conjunction of all atoms
in $\lambda_{\vek{\imath};\vek{a};j}$ which do involve the variable
$x_{m+1}$, and let $\eta_{\vek{\imath};\vek{a};j}$ be the conjunction
of all other atoms in $\lambda_{\vek{\imath};\vek{a};j}$. Clearly, 
$\lambda_{\vek{\imath};\vek{a};j}$ is equivalent to
$\zeta_{\vek{\imath};\vek{a};j} \und \eta_{\vek{\imath};\vek{a};j}$.
Hence $\varphi$ is equivalent to the disjunction of the formulas
$\exists x_{m+1} \big( \zeta_{\vek{\imath};\vek{a};j} \und
 \eta_{\vek{\imath};\vek{a};j}  \big)$
which, in turn, is equivalent to the disjunction of the formulas
$\big(\exists x_{m+1} \zeta_{\vek{\imath};\vek{a};j}\big) \und
 \eta_{\vek{\imath};\vek{a};j}$.
This means that $\varphi$ is equivalent to the disjunction of the
formulas $\xi_{\vek{\imath};\vek{a};j}\und
\eta_{\vek{\imath};\vek{a};j}$, where
$\xi_{\vek{\imath};\vek{a};j}\deff \exists x_{m+1}
\zeta_{\vek{\imath};\vek{a};j}$ is of the form $(*)$ and where
$\eta_{\vek{\imath};\vek{a};j}$ is quantifier free.
\\
This completes the proof of $(**)$ and thus also the proof of
Proposition~\ref{prop:Qeli}. 
\end{proof_mit}
%
%
%
%
%
%
\subsection{The Lifting Theorem and its Proof}\label{section:LiftingTheorem}
\begin{figure}[!htbp]
\bigskip
\begin{center}
\fbox{
\scalebox{0.55}{
\begin{pspicture}(0,0)(12,8.7)
%
\rput[l](0.35,8){\Large \bf $\bs{\NN}$-representable}%
\rput[l](7.8,8){\Large \bf $\bs{\NN}$-embeddable}%
\psline(0.35,7.35)(12,7.35)%
\psline(0.35,7.25)(12,7.25)%
\rput[r](4.2,5.7){\LARGE $\A$}\rput[l](0.37,6.6){\large definable in $\Linf(<,S)$}%
\rput[l](7.8,5.7){\LARGE $\rep(\A)$}\rput[l](7.8,6.6){\large active domain $S$}%
\psline[arrowsize=6pt]{<-}(4.5,5.9)(7.5,5.9)\rput[l](6,6.2){\Large $\Phi$}%
\psline[arrowsize=6pt]{->}(4.5,5.5)(7.5,5.5)\rput[l](6,5.2){\Large $\Phi'$}%
\psline(0.35,4.5)(12,4.5)%
\rput(2,3.5){\large $\FO(<,\Rel)$:}%
\rput(4,3.5){\Large $\varphi$}%
\psline[arrowsize=6pt]{->}(4.5,3.5)(7.5,3.5)\rput[l](6,3.8){\Large $\Phi$}%
\rput(8,3.5){\Large $\varphi'$}%
\psline[arrowsize=6pt]{->}(8,3)(8,1.5)%
\rput[l](8.3,2.25){\large collapse for $\ClNemb$}
\rput(2,1){\large $\FO(<)$:}%
\rput(4,1){\Large $\psi$}%
\psline[arrowsize=6pt]{<-}(4.5,1)(7.5,1)\rput[l](6,0.7){\Large $\Phi'$}%
\rput(8,1){\Large $\psi'$}%
\end{pspicture}
}
}
\caption{\small The overall proof idea for the Lifting Theorem~\ref{theorem:Lift}.
}\label{figure:LiftProof}
\end{center}
\end{figure}
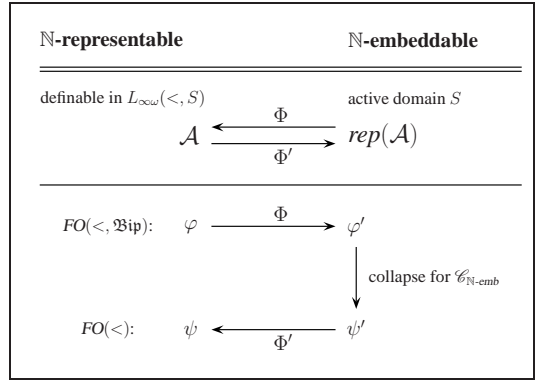

\begin{theorem_mit}[Lifting from $\bs{\NN}$-embeddable to $\bs{\NN}$-representable]\label{theorem:Lift}\index{lifting theorem}\mbox{ }\\
Let $\struc{\UU,<,\Rel}$ be a context structure where $<$ is a dense linear ordering 
without endpoints. If 
\,$\OrderGen\FO(<,\Rel) = \OrderGen\FO(<)$ on $\ClNemb$ over $\UU$, then 
\,$\OrderGen\FO(<,\Rel) = \OrderGen\FO(<)$ on $\ClNrep$ over $\UU$. 
\end{theorem_mit}%
The proof will be given throughout the following subsections:
In Section~\ref{section:OmegaRepresentableRelations}
we show how all the relevant information about an
$\NN$-repre\-sentable structure $\A$ can be represented by an $\NN$-embeddable structure 
$\rep(\A)$.  
In Section~\ref{section:FOInterpretations} we show that the translation from $\A$ to $\rep(\A)$, and vice versa, can be done via 
first-order interpretations $\Phi$ and $\Phi'$. 
As shown in Section~\ref{section:Proofs}, this will enable us
to prove Theorem~\ref{theorem:Lift}.
The overall proof idea is visualized in Figure~\ref{figure:LiftProof}.
%
%
%
%
%
\\
Let us mention that the proof presented here does \emph{not} work when replacing 
$\FO$ with the sublogic $\BCeFO$. The main objection is that the $\FO$-interpretations
contain several alternations of quantifiers.
It therefore remains open whether the Lifting Theorem can be proved for logics weaker than
$\FO$ and, in particular, for $\BCeFO$.
%
%
%
%
\subsubsection[$\NN$-Representations of Relations and Structures]{$\bs{\NN}$-Representations of 
  Relations and Structures}\label{section:OmegaRepresentableRelations}
\begin{definition_mit}[$\bs{S}$ sufficient for defining $\bs{R}$]\label{def:SetSufficientForR}
\mbox{ }\\
Let $R\subseteq \UU^m$.
A set $S\subseteq \UU$ is called \emph{sufficient for defining $R$} \,iff\,
$S$ is $\NN$-embeddable and $R$ is definable in $\Linf(<,S)$ over $\UU$. 
\end{definition_mit}
\begin{remark_mit}[$\bs{S}$ sufficient for defining $\bs{R}$]\label{remark:ConstantOnTypes}
From Proposition~\ref{prop:Qeli} we obtain that a $\NN$-embeddable set $S\subseteq \UU$ 
is sufficient for defining $R$ if and only if $R$ is
constant, in the sense of Definition~\ref{definition:R_constant_on_M}, on the sets 
\,$\Cube{S;\vek{\imath};t}\deff
   \setc{\vek{b} \in \Cube{S;\vek{\imath}}}{\type{\vek{b};S;\vek{\imath}} = t}$,\,
for all \,$\vek{\imath}\in N(S)^m$\, and all \,$t\in\types_m$.
\end{remark_mit}%
Let \,$R\subseteq \UU^m$\, be $\NN$-representable and let
\,$S\subseteq\UU$\, be sufficient for defining $R$.
From Remark~\ref{remark:ConstantOnTypes} we know, 
for all \,$\vek{\imath}\in N(S)^m$\, and all \,$t\in\types_m$, that
either \,$R\cap\Cube{S;\vek{\imath};t} = \emptyset$\, or
\,$R\supseteq\Cube{S;\vek{\imath};t}$.
This means that if we know, for each \,$\vek{\imath}\in N(S)^m$\, and each
\,$t\in\types_m$, whether
or not $R$ contains an element of \,$\Cube{S;\vek{\imath};t}$,
then we can reconstruct the entire relation $R$.
\\
For \,$i_j\neq 0$\, we represent the interval \,$\intro{S(i_j), S(i_j{+}1)} \subseteq \UU$\, by
the element $S(i_j)$. Consequently, for \,$\vek{\imath}\in (N(S)\setminus
\set{0})^m$, we can represent
\,$\Cube{S;\vek{\imath};t} \subseteq \UU^m$\, by the
tuple \,$S(\vek{\imath}) \in S^m$. 
The information whether or not $R$ contains an element of
\,$\Cube{S;\vek{\imath};t}$\, can be represented by the relation 
\begin{eqnarray*}
  R_{S;t} 
& \deff 
& \setc{\,S(\vek{\imath})\ }{\ \vek{\imath}\in (N(S)\setminus\set{0})^m \mbox{ \ and \ }
  R\cap\Cube{S;\vek{\imath};t} \neq \emptyset\,}.
\end{eqnarray*}
In general, we would like to represent every \,$\Cube{S;\vek{\imath};t}$,
for \emph{every} $\vek{\imath}\in N(S)^m$, by a tuple in $S^m$.
Unfortunately, the case where \,$i_j=0$\, must be treated separately, because 
\,$S(0)= -\infty \not\in S$. 
There are various possibilities for solving this technical
problem. Here we propose the following solution:
Use $S(1)$ to represent the interval \,$\intro{S(0),S(1)}$.
With every tuple \,$\vek{\imath} \in N(S)^m$\, we associate 
a \emph{characteristic tuple} \,$\Char(\vek{\imath}) \deff
(c_1,\twodots,c_m) \in\set{0,1}^m$\, and a tuple \,$\vek{\imath'}\in
(N(S)\setminus\set{0})^m$\, via 
\,$c_j \deff 0$\, and \,$i'_j \deff 1$\, if \,$i_j = 0$, \,and \,$c_j \deff 1$\, and
\,$i'_j \deff i_j$\, if \,$i_j \neq 0$.
Now \,$\Cube{S;\vek{\imath};t}$\, can be represented by the tuple
\,$S(\vek{\imath'})\in S^m$. 
The information whether or not $R$ contains an element of
\,$\Cube{S;\vek{\imath};t}$\, can be represented by the relations
\begin{eqnarray*}
  R_{S;t;\vek{u}} 
& \deff 
& \setc{\,S(\vek{\imath'})\ }{\ \vek{\imath}\in
  N(S)^m,\ \ \Char(\vek{\imath}) = \vek{u}, \ \mbox{ and } \ 
  R\cap\Cube{S;\vek{\imath};t} \neq \emptyset\,},
\end{eqnarray*}
for all \,$\vek{u}\in\set{0,1}^m$.  
This leads to the following definition:
\begin{definition_ohne}[$\boldsymbol{\NN}$-Representation of a
  Relation]\label{definition:Relation-Representation}
\mbox{ }\\
Let $R\subseteq \UU^m$ be $\NN$-representable, and
let $S\subseteq \UU$ be sufficient for defining $R$.
\begin{enumerate}[(a)]
\item 
We represent the $m$-ary relation $R$ over $\UU$ by a finite number of
$m$-ary relations over $S$ as follows:
The $\NN$-representation of $R$ with respect to $S$
is the collection \index{repSR@$\rep_S(R)$}
\begin{eqnarray*}
  \rep_S(R)
& \ \deff\ 
& \big\langle\,  R_{S;t;\vek{u}} \,\big\rangle_{t\in\types_m,\,\vek{u}\in\set{0,1}^m}\,,
\end{eqnarray*}
where 
\,$R_{S;t;\vek{u}} \deff 
 \setc{S(\vek{\imath'})}{\vek{\imath}\in N(S)^m,\ 
                         \Char(\vek{\imath}) = \vek{u}, \mbox{ and }
                         R\cap\Cube{S;\vek{\imath};t} \neq
                         \emptyset}$.
\\
Here, for \,{
$\vek{\imath}\in N(S)^m$}\, we define \,{
$\vek{\imath'}$}\, and
\,{
$\Char(\vek{\imath})$}\, via \,{
$i'_j \deff 1$}\, and
\,{
$\big(\Char(\vek{\imath})\big)_j \deff 0$}\, if \,$i_j = 0$,  
\,and \,{
$i'_j \deff i_j$}\,
and \,{
$\big(\Char(\vek{\imath})\big)_j \deff 1$}\, if \,$i_j \neq 0$.
\item 
For $\vek{x}\in \Cube{S;\vek{\imath};t}$ we say that
 \begin{itemize}
   \item
     $\vek{u}\deff\Char(\vek{\imath})$\, is the \,\emph{characteristic tuple}\, of \,$\vek{x}$\,
     w.r.t.\ $S$, \index{characteristic tuple of $\vek{x}$ w.r.t.\ $S$}
   \item
     $\vek{y}\deff S(\vek{\imath'})$\, is the \,\emph{representative}\, of \,$\vek{x}$\, w.r.t.\
     $S$, \ and \index{representative of $\vek{x}$ w.r.t.\ $S$}
   \item
     $t$\, is the \,\emph{type}\, of \,$\vek{x}$\, w.r.t.\ $S$.
     \index{type of $\vek{x}$ w.r.t.\ $S$}
 \end{itemize}
From Remark~\ref{remark:ConstantOnTypes} we obtain that
\,$\vek{x}\in R$ \ iff \ $\vek{y} \in R_{S;t;\vek{u}}$.
\mbox{ } \fertig
\end{enumerate}
\end{definition_ohne}%
%
%
%
We now tranfer the notion of \emph{$\NN$-representation} from
relations to $\tau$-structures.\\ 
Recall from Definition~\ref{definition:Nrep} that a
$\struc{\UU,\tau}$-structure $\A$ is called $\NN$-representable iff
each of $\A$'s relations is.
\begin{definition_ohne}[$\bs{S}$ sufficient for defining $\bs{\A}$]\label{def:SetSufficientForA}
\mbox{ }\\
Let $\A$ be a $\struc{\UU,\tau}$-structure.
A set $S\subseteq \UU$ is called \emph{sufficient for defining $\A$}
\,iff
\begin{itemize}
 \item 
   $S$ is $\NN$-embeddable, 
 \item 
   $c^{\A} \in S$, for every constant symbol $c\in\tau$, \ and
 \item 
   $S$ is sufficient for defining $R^{\A}$, for every relation symbol
   $R\in\tau$.
   \mbox{ } \fertig
\end{itemize}
\end{definition_ohne}%
Let $\A$ be a $\struc{\UU,\tau}$-structure and let $S$ be a set
sufficient for defining $\A$. According to
Definition~\ref{definition:Relation-Representation}, each of $\A$'s
relations $R^{\A}$ of arity, say, $m$ can be represented by
a finite collection 
$\rep_{S}(R^{\A}) = 
 \big\langle R^{\A}_{S;t;\vek{u}}\big\rangle_{t\in\types_{m},\,\vek{u}\in\set{0,1}^m}$ 
of relations over $S$.
I.e., $\A$ can be represented by a structure $\rep_S(\A)$ with active
domain $S$ as follows:
\begin{definition_ohne}[$\boldsymbol{\NN}$-Representation of $\bs{\A}$]\label{definition:Structure-Representation} 
Let $\tau$ be a signature.
\begin{enumerate}[(a)]
 \item 
   The \emph{type extension} $\tau'$ of $\tau$ \index{type extension $\tau'$ of $\tau$} 
   is the signature which
   consists of
   \begin{itemize}
     \item 
       the same constant symbols as $\tau$, 
     \item 
       a unary relation symbol $S$, \ and 
     \item 
       a relation symbol $R_{t;\vek{u}}$ of arity \,$m\deff\ar(R)$, 
       for every relation symbol \,$R\in\tau$, 
       every \,$t\in\types_{m}$, \,and every 
       \,$\vek{u}\in\set{0,1}^{m}$.
   \end{itemize} 
 \item 
   Let $\A$ be an $\NN$-representable $\struc{\UU,\tau}$-structure
   and let $S$ be a set sufficient for defining $\A$.
   We represent 
   $\A$ by the $\struc{\UU,\tau'}$-structure \,$\rep_S(\A)$\, which satisfies
   \index{repSA@$\rep_S(\A)$}
   \begin{itemize}
    \item 
      $c^{\rep_S(\A)} = c^{\A}$\,,\quad for each $c\in\tau'$,
    \item 
      $S^{\rep_S(\A)} = S$\,,\quad for the unary relation symbol
      $S\in\tau'$,\quad and
    \item 
      $R_{t;\vek{u}}^{\rep_S(\A)} = R^{\A}_{S;t;\vek{u}}$\,,\quad 
      for each \,$R\in\tau$\, of arity \,$m\deff\ar(R)$,\, \\ 
      each \,$t\in\types_{m}$,\, and each
      \,$\vek{u}\in\set{0,1}^{m}$. 
      \mbox{ }\fertig
   \end{itemize}
\end{enumerate}
\end{definition_ohne}%
%
%
%
%
%
%
\subsubsection[$\FO$-Interpretations]{$\bs{\FO}$-Interpretations}\label{section:FOInterpretations}\label{section:FO-Interpretations}
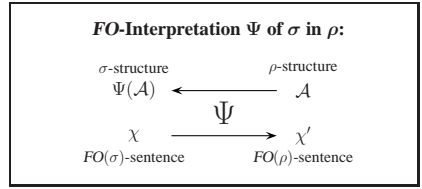
\begin{figure}[!htbp]
\bigskip
\begin{center}
\fbox{
\scalebox{0.5}{
\begin{pspicture}(-0.9,-2.7)(9.25,+1.7)
%
\rput(4.25,1.2){\Large \bf {$\bs{\FO}$-Interpretation $\bs{\Psi}$ of $\bs{\sigma}$ in $\bs{\rho}$:}}%
%
\psline[arrowsize=6pt]{<-}(3,-0.4)(5.8,-0.4)%
\psline[arrowsize=6pt]{->}(3,-1.6)(5.8,-1.6)%
\rput(4.4,-1){\Huge $\Psi$}
%
\rput(2,0.2){\large $\sigma$-structure}%
\rput(2,-0.4){\Large $\Psi(\A)$}%
\rput(2,-1.6){\Large $\chi$}%
\rput(2,-2.2){\large $\FO(\sigma)$-sentence}%
\rput(6.5,0.2){\large $\rho$-structure}%
\rput(6.5,-0.4){\Large $\A$}%
\rput(6.5,-1.6){\Large $\chi'$}%
\rput(6.5,-2.2){\large $\FO(\rho)$-sentence}%
\end{pspicture}
}
}
\caption{\small The effect of a $\FO$-interpretation $\Phi$ of $\sigma$ in $\rho$.
For every $\rho$-structure $\A$, $\Psi$ defines a $\sigma$-structure $\Psi(\A)$.
For every $\FO(\sigma)$-sentence $\chi$, $\Psi$ defines a $\FO(\rho)$-sentence
$\chi'$ such that \ $\A\models\chi'$ \ iff \ $\Psi(\A)\models \chi$.}\label{figure:FOInterpret}
\end{center}
\end{figure}

The concept of \emph{first-order interpretations} (or, reductions) is
well-known in mathematical logic (cf., e.g.,
\cite{EbbinghausFlum}). In the present section we consider the following
easy version:
\begin{definition_ohne}[$\bs{FO}$-Interpretation of 
  $\boldsymbol{\sigma}$ in $\boldsymbol{\rho}$]\label{definition:FO-interpretation}\index{FO-interpretation@$\bs{\FO}$-interpretation}\mbox{ }\\
Let $\sigma$ and $\rho$ be signatures.
A \,\emph{$\FO$-interpretation of $\sigma$ in $\rho$}\, is a collection
\begin{eqnarray*}
 \Psi
& =
& \Big\langle\,
    \big( \varphi_c(x) \big)_{c\in\sigma},\ 
    \big( \varphi_{R}(x_1,\twodots,x_{\ar(R)}) \big)_{R\in\sigma}
  \;\Big\rangle
\end{eqnarray*}
of $\FO(\rho)$-formulas.
For every $\struc{\UU,\rho}$-structure $\A$, the
$\struc{\UU,\sigma}$-structure $\Psi(\A)$ is given via
\begin{itemize}
 \item 
    $\set{c^{\Psi(\A)}} =
    \setc{a\in\UU}{ \A \models\varphi_c(a)}$,\quad
    for each constant symbol $c\in\sigma$, 
 \item 
    $R^{\Psi(\A)} = 
    \setc{\vek{a}\in\UU^{\ar(R)}}{ \A \models \varphi_{R}(\vek{a})}$,\quad
   for each relation symbol $R\in\sigma$. 
   \mbox{ }\fertig
\end{itemize} 
\end{definition_ohne}%
The effect of a $\FO$-interpretation is visualized in Figure~\ref{figure:FOInterpret}.
\\
Making use of a $\FO$-interpretation of $\sigma$ in $\rho$, one can
translate $\FO(\sigma)$-formulas into $\FO(\rho)$-formulas (cf.,
\cite[Exercise 11.2.4]{EbbinghausFlum}):
\begin{lemma_mit}\label{lemma:FO-interpretation}
Let $\sigma$ and $\rho$ be signatures and let $\Psi$ be a
$\FO$-interpretation of $\sigma$ in $\rho$.
\\
For every $\FO(\sigma)$-sentence $\chi$ there is a
$\FO(\rho)$-sentence $\chi'$ 
such that \ 
``$ \A \,\models\, \chi'
   \mbox{ \ iff \ }
   \Psi(\A) \,\models\, \chi$'' \ 
is true for every
$\struc{\UU,\rho}$-structure $\A$.
\mbox{ }
\end{lemma_mit}%
\begin{proof_mit}
$\chi'$ is obtained from $\chi$ by
replacing every atomic formula $R(\vek{x})$ (respectively, $x{=}c$) by
the formula $\varphi_{R}(\vek{x})$ (respectively, by the formula $\varphi_c(x)$).%
\end{proof_mit}%
\\
%
%
%
%

%
\parno
The following lemma shows that $\A$ is
first-order definable in $\rep_S(\A)$.
In other words: All relevant information about $\A$ can
be reconstructed from the structure  
$\rep_S(\A)$ (if $\A$ is $\NN$-representable and $S$ is
sufficient for defining $\A$).
%
\begin{lemma_mit}[{$\bs{\A \stackrel{\Phi}{\longleftarrow} \rep_S(\A)}$}]\label{lemma:step2}
There is a $\FO$-interpretation $\Phi$ of $\tau$ in $\tau'\cup\set{<}$
such that $\Phi(\struc{\rep_S(\A),<}) = \A$, for every
$\NN$-representable $\struc{\UU,\tau}$-structure $\A$ and every set
$S$ which is sufficient for defining $\A$.
\end{lemma_mit}%
\begin{proof_mit}
For every constant symbol $c\in\tau$ we define $\varphi_c(x)\deff
x{=}c$.\\
For every relation symbol
$R\in\tau$ of arity, say, $m$ we construct a formula
$\varphi_R(\vek{x})$ which expresses that $\vek{x}\in R$:
From Definition~\ref{definition:Relation-Representation}\;(b) we know that \,$\vek{x} \in R$\,
iff \,$\vek{y}\in R_{S;t;\vek{u}}$,\, where \,$\vek{y}$, $t$, and $\vek{u}$\,
are the representative, the type, and the characteristic tuple,
respectively, of \,$\vek{x}$\, w.r.t.\ $S$.
It is straightforward to construct, for fixed \,$t\in\types_m$\,
and \,$\vek{u}\in\set{0,1}^m$,\, a $\FO(\tau',<)$-formula
\,$\psi_{t,\vek{u}}(\vek{x})$\, which expresses that
\begin{itemize}
 \item 
   $\vek{x}$\, has type $t$\, w.r.t.\ $S$,
 \item 
   $\vek{u}$\, is the characteristic tuple of \,$\vek{x}$\, w.r.t.\ $S$, and
 \item 
   for the representative \,$\vek{y}$\, of \,$\vek{x}$\, w.r.t.\ $S$ it holds that
   $R_{t;\vek{u}}(\vek{y})$.
\end{itemize}
The disjunction of the formulas
\,$\psi_{t;\vek{u}}(\vek{x})$,\, for all \,$t\in\types_m$\,
and all \,$\vek{u}\in\set{0,1}^m$,\, gives us the desired formula
\,$\varphi_R(\vek{x})$\, which expresses that \,$\vek{x} \in R$.
\end{proof_mit}%
\\
\parno
We now want to show the converse of Lemma~\ref{lemma:step2}, i.e., we
want to show that the $\NN$-representation of $\A$ is 
first-order definable in $\A$.
Up to now the $\NN$-representation \,$\rep_S(\A)$\, was parameterized
by a set $S$ which is sufficient for defining $\A$.
For the current step we need the existence of a \emph{canonical}, first-order definable
set $S$. For this canonization we can use the following result of
Gr\"adel and Kreutzer \cite[Definition~6 and Lemmas 7 and 8]{GraedelKreutzer99}:
\begin{lemma_mit}[Canonical set $\bs{S_R}$ sufficient for defining
  $\boldsymbol{R}$; formula $\bs{\zeta_R(x)}$]\label{lemma:DefbarkeitVonS}\mbox{ }\\
Let $\struc{\UU,<}$ be a dense linear ordering without endpoints.
Let $R\subseteq \UU^m$ be $\NN$-representable and let $S_R$ be the
set of all elements $s\in\UU$ which satisfy the following condition
($*$):
\begin{quote}
 There are \,$a_1,\twodots,a_m,s_{\links},s_{\rechts}\in\UU$\, with \,$s_{\links}<s<s_{\rechts}$,
 such that one of the following holds:
 \begin{enumerate}[$\bullet$]
   \item 
     For all $s'\in\intoo{s_{\links}, s}$ and for
     no $s'\in\intoo{s , s_{\rechts}}$ we have
     $R\big(\vek{a}[s/s']\big)$.
     Here $\vek{a}[s/s']$ means that all components $a_j{=}s$
     are replaced by $s'$.
   \item 
     For no $s'\in\intoo{s_{\links}, s}$ and for
     all $s'\in\intoo{s , s_{\rechts}}$ we have
     $R\big(\vek{a}[s/s']\big)$.
   \item 
     $R\big(\vek{a}[s/s']\big)$ holds for all
     $s'\in\intoo{s_{\links} , s_{\rechts}}
     \setminus\set{s}$, but not for $s'= s$.
   \item 
     $R\big(\vek{a}[s/s']\big)$ holds for $s'=s$, but not for any  
     $s'\in\intoo{s_{\links}, s_{\rechts}}\setminus\set{s}$.
 \end{enumerate}
\end{quote}
The following holds true:
\begin{enumerate}[(1.)]
 \item 
   $S_R$ is included in every set $S\subseteq\UU$ which is
   sufficient for defining $R$.
 \item 
   $S_R$ is sufficient for defining $R$.
\end{enumerate}
The set $S_{R}$ is called \emph{the canonical set sufficient for
defining $R$}.
\\
It is straightforward to formulate a $\FO(R,<)$-formula \,$\zeta_R(x)$\,
which expresses condition ($*$). Consequently we have, 
for every $\NN$-representable $m$-ary relation $R$, that
\ $S_R\, = \,\setc{\,s\in\UU\,}{\,\struc{\UU,R,<}\,\models\,\zeta_R(s)\,}$.
\end{lemma_mit}
\begin{definition_mit}[Canonical Representation of $\bs{\A}$]\label{definition:CanonoicalRep} 
Let $\tau$ be a signature and let $\A$ be a $\NN$-repre\-sentable
$\struc{\UU,\tau}$-structure. 
The set 
\,$ S_{\A}\deff\setc{c^{\A}}{c\in\tau} \cup \bigcup_{R\in\tau} S_{R^{\A}}$\,
is called the \emph{canonical set sufficient for defining $\A$}.
The representation \,$\repcan(\A) \deff \rep_{S_{\A}}(\A)$\, is
called \emph{the canonical representation of $\A$}.
\index{repA canonical@$\repcan(\A)$ (canonical representation of $\A$)} 
\end{definition_mit}
\begin{remark_mit}\label{remark:rep_alpha}
It is straightforward to see that
\,``$\alpha\,\big(\repcan\,(\A)\big) = \repcan\,\big( \alpha\,(\A) \big)$''\,
is true for every $\NN$-representable $\struc{\UU,\tau}$-structure $\A$ and for
every ${<}$-preserving mapping \,$\alpha:\adom(\A)\rightarrow \UU$.
\end{remark_mit}%
We are now ready to prove the converse of Lemma~\ref{lemma:step2}.
\begin{lemma_mit}[{$\bs{\A \stackrel{\Phi'}{\longrightarrow} \repcan(\A)}$}]\label{lemma:step4}
There is a $\FO$-interpretation $\Phi'$ of $\tau'$ in $\tau\cup\set{<}$ 
such that $\Phi'(\struc{\A,<}) = \repcan(\A)$, for every
$\NN$-representable $\struc{\UU,\tau}$-structure $\A$.
\end{lemma_mit}%
\begin{proof_mit}
For every constant symbol \,$c\in\tau'$\, we define \,$\varphi_c(x)\deff
x{=}c$.\\
For every relation symbol \,$R\in\tau$\, let \,$\zeta_R(x)$\, be the formula
from Lemma~\ref{lemma:DefbarkeitVonS} describing the canonical
set sufficient for defining $R^{\A}$.
Obviously, the formula
\ $\varphi_S(x) \deff
  \Oder_{c\in\tau} x{=}c\ \oder\ \Oder_{R\in\tau} \zeta_R(x)$ \ 
describes the canonical set sufficient for defining $\A$. 
\\
For every relation symbol \,$R_{t;\vek{u}}\in\tau'$\, of arity, say, $m$ we
construct a formula \,$\varphi_{R_{t;\vek{u}}}(\vek{y})$\, which expresses
that \,$\vek{y} \in R_{t;\vek{u}}$.
We make use of Definition~\ref{definition:Relation-Representation}\;(b). 
I.e., \,$\varphi_{R_{t;\vek{u}}}$\, states that \,$y_1,\twodots,y_m$\,
satisfy $\varphi_S$ and that there is some 
\,$\vek{x}$\, such that
\begin{itemize}
 \item 
   $\vek{y}$\, is the representative of \,$\vek{x}$\, w.r.t.\ $S_{\A}$,
 \item 
   $R(\vek{x})$,
 \item 
   $\vek{x}$\, has type $t$\, w.r.t.\ $S_{\A}$,\, and
 \item 
   $\vek{u}$\, is the characteristic tuple of \,$\vek{x}$\, w.r.t.\ $S_{\A}$.
\end{itemize}
It is straightforward to formalize this in first-order logic.
\end{proof_mit}
%
%
%
%
%
%
%
\subsubsection{The Proof of the Lifting Theorem}\label{section:Proofs}
We are now ready to prove the lifting theorem, which allows to lift collapse results 
for $\NN$-embeddable structures to collapse results for 
$\NN$-representable structures.
An illustration of the overall proof idea is given in Figure~\ref{figure:LiftProof}.
\\
\parno
\begin{proofc_mit}{of Theorem~\ref{theorem:Lift} {
(Lifting from $\bs{\NN}$-embeddable to $\bs{\NN}$-representable)}}\mbox{ }\\
Let $\struc{\UU,<,\Rel}$ be a context structure where $<$ is a dense linear ordering without 
endpoints, and let 
\,$\OrderGen\FO(<,\Rel) = \OrderGen\FO(<)$ on $\ClNemb$ over $\UU$. 
Our aim is to show that
\,$\OrderGen\FO(<,\Rel) = \OrderGen\FO(<)$ on $\ClNrep$ over $\UU$.
%
\par
Let $\tau$ be a signature, let $\varphi$ be a
$\FO(\tau,<,\Rel)$-sentence, and let $\K$ be the class of all $\NN$-representable
$\struc{\UU,\tau}$-structures on which $\varphi$ is ${<}$-generic.
We need to find a $\FO(\tau,<)$-sentence $\psi$ such that, for all $\A\in\K$,
{
\begin{eqnarray*} 
  \struc{\,\A,\,<,\,\Rel\,}\ \models\ \varphi
& \mbox{\ iff \ }
&  \struc{\,\A,\,<\,}\ \models\  \psi.
\end{eqnarray*}}%
Let $\tau'$ be the type extension of $\tau$.
We first use Lemma~\ref{lemma:step2}, which gives us a
$\FO$-interpretation $\Phi$ of $\tau$ in $\tau'\cup\set{<}$ 
such that
\,$\Phi(\struc{\repcan(\A),<}) = \A$, for all $\A\in\K$.
From Lemma~\ref{lemma:FO-interpretation} we obtain a
$\FO(\tau',<,\Rel)$-sentence $\varphi'$ such that, for all $\A\in\K$,
{
\[
\begin{array}{rcl}
  \struc{\,\repcan(\A),\,<,\,\Rel\,}\ \models\ \varphi'
& \mbox{\ iff \ }
& \big\langle\, \Phi\big(\struc{\repcan(\A),<}\big),\,<,\,\Rel\,\big\rangle\ \models\ \varphi
\\
& \mbox{\ iff \ } 
&  \struc{\,\A,\,<,\,\Rel\,}\ \models\ \varphi.
\end{array}
\]
}%
From our presumption we know that the natural
generic collapse over $\struc{\UU,<,\allowbreak \Rel}$ is true for the class of
$\NN$-embeddable structures. 
Of course $\repcan(\A)$ is $\NN$-embeddable.
Furthermore, with Remark~\ref{remark:rep_alpha} we obtain that
$\varphi'$ is ${<}$-generic on $\repcan(\A)$, \,for all $\A\in\K$.
Hence there must be a $\FO(\tau',<)$-sentence $\psi'$ such that, for all $\A\in\K$, 
{
\begin{eqnarray*}
  \struc{\,\repcan(\A),\,<,\,\Rel\,}\ \models\ \varphi'
& \mbox{\ iff \ } 
& \struc{\,\repcan(\A),\,<\,}\ \models\ \psi'.
\end{eqnarray*}}%
We now use Lemma~\ref{lemma:step4}, which gives us a
$\FO$-interpretation $\Phi'$ of $\tau'$ in $\tau\cup\set{<}$
such that  
\,$\Phi'(\struc{\A,<}) = \repcan(\A)$, for all $\A\in\K$.
According to Lemma~\ref{lemma:FO-interpretation}, we can transform $\psi'$ into
a $\FO(\tau,<)$-sentence $\psi$ such that, for all $\A\in\K$, 
\[
\struc{\,\A,\,<\,}\ \models\ \psi
\quad\mbox{iff}\quad
 \big\langle\,\Phi'\big(\struc{\A,<}\big),\,<\,\big\rangle\ \models\ \psi'
\quad\mbox{iff}\quad
\struc{\,\repcan(\A),\,<\,}\ \models\ \psi'.
\]
Obviously, $\psi$ is the desired sentence, and hence the proof of 
Theorem~\ref{theorem:Lift} is complete.
\end{proofc_mit}
%
%
%
%
%
%
\subsection[${\ZZ}$-Representable instead of ${\NN}$-Representable]{$\bs{\ZZ}$-Representable instead of $\bs{\NN}$-Representable}\index{Z-representable@$\ZZ$-representable}\label{section:ConsequencesLift}
\index{Z-embeddable@$\ZZ$-embeddable}\index{Z-representable@$\ZZ$-representable}
It is straightforward to modify the proof of Theorem~\ref{theorem:Lift} in such a way
that collapse results for the class of \emph{$\ZZ$-embeddable} structures
can be lifted to the class $\ClZrep$ of structures which are \emph{$\ZZ$-representable}
in the following sense: A structure is called $\ZZ$-representable if all its relations
are $\ZZ$-representable, i.e., definable in \,$\Linf(<,S)$\, for a 
\emph{$\ZZ$-embeddable} set $S$.
\begin{corollary_mit}[Lifting from $\bs{\ZZ}$-embeddable to $\bs{\ZZ}$-representable]\label{corollary:Lift}\index{lifting theorem}\mbox{ }\\
Let $\struc{\UU,<,\Rel}$ be a context structure where $<$ is a dense linear ordering 
without endpoints. If 
\,$\OrderGen\FO(<,\Rel) = \OrderGen\FO(<)$ on $\ClZemb$ over $\UU$\,
then 
\,$\OrderGen\FO(<,\Rel) = \OrderGen\FO(<)$ on $\ClZrep$ over $\UU$. 
\end{corollary_mit}%
%
%
%

%


\section{Conclusion and Open Questions}\label{section:Conclusion}
Aiming at natural generic collapse results for potentially infinite databases we 
developed the notion of \emph{$<$-genericity} which coincides both, with the classical notion
of \emph{order-genericity} on the densely ordered context universes $\QQ$ and $\RR$ and with the
notion of \emph{local genericity} on the discretely ordered context universes $\NN$ and $\ZZ$ 
(Definition~\ref{definition:kleiner-generic}).
We presented the \emph{translation of winning strategies for the duplicator in the
Ehrenfeucht-Fra\"\i{}ss\'{e} game} as a new method for proving natural generic collapse results
and showed that, at least in principle, all collapse results can be proved by this method
(Theorem~\ref{theorem:EF-Collapse}).
In the Theorems~\ref{theorem:collapse_Monadic_UU},
\ref{theorem:Variation_Plus_MonQ}, and \ref{theorem:BCeFO_Arb-game}
we explicitly showed how the duplicator can translate
winning strategies for the Ehrenfeucht-Fra\"\i{}ss\'{e} game in the presence of particular 
built-in predicates. Via Theorem~\ref{theorem:EF-Collapse} this directly gives us the following
natural generic collapse results:
%
%
%
\begin{corollary_ohne}\label{corollary:collapse}
Let $\Mon$ be the class of all built-in \emph{monadic} predicates on the 
respective context universe.
Let $Q\subseteq \NN$ satisfy the conditions $W(\omega)$ 
(cf., Definition~\ref{definition:conditions_W_omega}), 
and let $\Mon_Q$ be the class of all
subsets of $Q$. Let $\Groups$ be the class of all subsets of $\RR$ 
that contain the number $1$ and that are groups with respect to $+$.
\begin{enumerate}[(a)]
\item
   $\OrderGen\FO(<,\Mon)$ = $\FOadom(<)$ on $\ClZemb$ over $\UU$, 
   for any linearly ordered infinite context universe $\UU$.
   {\rm In particular, for $\UU=\ZZ$ this implies the natural generic 
   collapse on {arbitrary} databases over $\struc{\ZZ,<,\Mon}$.}
\item
   $\OrderGen\FO(<,+,Q,\Mon_Q)\;{=}\;  \FOadom(<)$ 
   on $\Clarb$ over $\NN$ and on $\ClNemb$ over $\ZZ$.
\item
   $\OrderGen\FO(<,+,Q,\Mon_Q,\Groups) =  \FOadom(<)$  on $\ClNemb$ over $\RR$.
   {\rm
   In particular, this implies the natural generic collapse on $\NN$-embeddable databases
   over the context structures $\struc{\RR,<,+,\ZZ,\QQ}$ and $\struc{\QQ,<,+,\ZZ}$.}
\item
   $\OrderGen\BCeFO(<,\Bip) = \BCeFOadom(<)$ on $\ClNemb$ over $\UU$, 
   for any linearly ordered infinite context structure $\struc{\UU,<,\Bip}$.
   {\rm In particular, for $\UU=\NN$ this implies the natural generic collapse for the
   logic $\BCeFO$ on {arbitrary} databases.}
\mbox{ }\fertig
\end{enumerate}
\end{corollary_ohne}%
Theorem~\ref{theorem:Lift} (and Corollary~\ref{corollary:Lift}) allows us to lift 
collapse results 
from the class of $\NN$-embeddable (respectively, $\ZZ$-embeddable) 
databases to the larger class of
$\NN$-representable (respectively, $\ZZ$-representable) databases, 
provided that the context universe is equipped with a
dense linear orderings without endpoints.
\begin{corollary_ohne}\label{corollary:Lift_collapse}
\hspace{4cm}
\begin{enumerate}[(a)]
\item
  $\OrderGen\FO(<,\Mon) =  \OrderGen\FO(<)$ \,on $\ClZrep$ over $\UU$, 
  if $\struc{\UU,<}$ is a dense linear ordering without endpoints. 
\item
  $\OrderGen\FO(<,+,Q,\Mon_Q,\Groups) =  \OrderGen\FO(<)$ \,on $\ClNrep$ over $\RR$ and
  over $\QQ$.
  {\rm
  I.e., the natural generic collapse is true for the class of all 
  {$\NN$-representable} databases over the context structures
   \,$\struc{\QQ,<,+,\ZZ}$, \,$\struc{\RR,<,+,\ZZ,\QQ}$, \,and
   \,$\struc{\RR,<,+,Q,\Mon_Q,\Groups}$.}
  \mbox{ }\fertig
\end{enumerate}
\end{corollary_ohne}%
%
%
%
In the present paper we investigated collapse results from a \emph{logical} point of view.
From the point of view of \emph{computer science}, especially \emph{constructive} collapse
proofs are interesting, i.e., proofs which lead to a ``collapse algorithm'' that transforms
a ${<}$-generic input formula $\varphi\in\FO(<,\Rel)$ into an equivalent output formula
$\varphi'\in\FO(<)$.
Benedikt and Libkin 
\index{Benedikt, Michael}\index{Libkin, Leonid} \cite{BL00b} presented such an algorithm for the collapse
from \,$\OrderGen\allowbreak\FO(<,\Rel)$ to $\FOadom(<)$ on the class $\Clfinite$ over 
\emph{o-minimal} context structures.
Other deep {natural generic collapse} proofs for the class $\Clfinite$, such as the 
collapse results for context structures that have the Isolation Property 
\cite{BST99} or finite VC-dimension \cite{BB98}, are non-constructive.
Also, our Ehrenfeucht-Fra\"\i{}ss\'{e} game approach does not necessarily lead to a
collapse algorithm.
However, the lifting theorem~\ref{theorem:Lift} does preserve constructiveness.
Precisely, this means the following: 
Assume that we are given an algorithm that produces, for every input sentence
\,$\varphi'\in\FO(<,\Rel)$, an output sentence \,$\psi'\in\FO(<)$\, such that
{
\begin{eqnarray*}
  \struc{\,\UU,\,<,\,\Rel,\,\tau^{\A}\,}\ \models\ \varphi'
& \mbox{ iff }
& \struc{\,\UU,\,<,\,\tau^{\A}\,}\ \models\ \psi'
\end{eqnarray*}}%
is true for all \emph{$\NN$-embeddable} structures $\struc{\UU,\tau^{\A}}$
on which $\varphi'$ is ${<}$-generic.
Making use of this algorithm and of the $\FO$-interpretations $\Phi$ and $\Phi'$ from the
Lemmas~\ref{lemma:step2} and \ref{lemma:step4}, one directly obtains an
algorithm that produces, for every input sentence
\,$\varphi\in\FO(<,\Rel)$, an output sentence \,$\psi\in\FO(<)$\, such that
{
\begin{eqnarray*}
  \struc{\,\UU,\,<,\,\Rel,\,\tau^{\A}\,}\ \models\ \varphi
& \mbox{ iff }
& \struc{\,\UU,\,<,\,\tau^{\A}\,}\ \models\ \psi
\end{eqnarray*}}%
is true for all \emph{$\NN$-representable} structures 
$\struc{\UU,\tau^{\A}}$ on which $\varphi$ is ${<}$-generic.
\subsubsection*{Open questions:}
It remains open whether the natural generic collapse for 
$\NN$-embeddable databases
is valid over context structures other than
$\struc{\UU,<,\allowbreak \Mon}$, $\struc{\ZZ,<,+,Q,\Mon_Q}$, 
$\struc{\RR,<,\allowbreak +,Q,\Mon_Q,\allowbreak \Groups}$.
For example: Is it valid over $\struc{\RR,<,+,\times}$, over all
\emph{o-minimal} context structures, or even over all context structures that have
\emph{finite VC-dimension}? In other words: Can Theorem~\ref{theorem:collapse_VCdim} be 
generalized from $\Clfinite$ to $\ClNemb$ (or even to $\ClZemb$)?
Recall, however, from Section~\ref{section:Databases} that it cannot be generalized to 
$\Clarb$ since the natural generic collapse is not valid for arbitrary databases over the
context structure $\struc{\QQ,<,+}$.
\par
We also may ask whether the collapse results proved in this paper remain valid for even larger
classes of databases, e.g.: Is the collapse still valid for \emph{arbitrary} databases over
every context structure $\struc{\UU,<,\Mon}$ where $\Mon$ is the class of \emph{monadic}
predicates over $\UU$? Is the collapse still valid for \emph{arbitrary} databases over
$\struc{\ZZ,<,+}$ or for \emph{$\ZZ$-embeddable} databases over 
$\struc{\RR,<,+,Q,\Mon_Q,\allowbreak \Groups}$? 
\par
Another approach is to restrict the complexity of the formulas that may be used to formulate
queries. We know that the collapse over the context structure $\struc{\NN,<,+,\times}$ is 
not valid for full first-order logic, but that it is valid for Boolean combinations of purely 
existential first-order formulas. It remains open how many quantifier alternations are
necessary to defeat the collapse. A task to start with would be, e.g., to try to lift 
Theorem~\ref{theorem:BCeFO_Arb-game} from $\BCeFO$ to $\Sigma_2^0\cap\Pi_2^0$.
\par
Le us also mention a potential application concerning
\emph{topological queries}: \index{topological query}
\mbox{Kuijpers} and Van den Bussche \cite{KV99} 
\index{Kuijpers, Bart}\index{Van den Bussche, Jan} 
used the natural generic collapse on
$\Clfinite$ over $\struc{\RR,<,+,\times}$
to obtain a collapse result for
\emph{topological} first-order definable queries.
One step of their proof is to encode spatial databases \index{spatial database} 
(of a certain kind) by \emph{finite} databases, to which the natural generic collapse over 
$\struc{\RR,<,\allowbreak +,\times}$
can be applied. Here the question arises whether there is an interesting class of spatial
databases that can be encoded by $\NN$-embeddable 
structures in such a way that our collapse results for $\ClNemb$ 
help to obtain collapse results for topological queries. 
%
%

%
{\small

}

\end{document}